\documentclass[a4paper,11pt]{article}

\usepackage{jheppub}
\usepackage[T1]{fontenc}
\usepackage{lmodern}

\usepackage{bm}
\usepackage{mathrsfs}
\usepackage{booktabs}
\usepackage{placeins}
\usepackage{tikz}
\usetikzlibrary{decorations.pathmorphing}
\usepackage{etoolbox}

\usepackage{mathtools}
\usepackage{amsthm}
\newtheorem{theorem}{Theorem}[section]
\newtheorem{proposition}[theorem]{Proposition}
\newtheorem{lemma}[theorem]{Lemma}
\newtheorem{corollary}[theorem]{Corollary}
\usepackage{physics}
\usepackage{tensor}
\usepackage{subcaption}
\usepackage{enumitem}
\usepackage{placeins}

\usepackage{pgffor}
\foreach \x in {A,B,...,Z,a,b,...,z}{
    \expandafter\xdef\csname cal\x\endcsname{\noexpand\mathcal{\x}}
    \expandafter\xdef\csname scr\x\endcsname{\noexpand\mathscr{\x}}
    \expandafter\xdef\csname bb\x\endcsname{\noexpand\mathbb{\x}}
    \expandafter\xdef\csname rm\x\endcsname{\noexpand\mathrm{\x}}
    \expandafter\xdef\csname it\x\endcsname{\noexpand\mathit{\x}}
    \expandafter\xdef\csname bf\x\endcsname{\noexpand\mathbf{\x}}
    \expandafter\xdef\csname sf\x\endcsname{\noexpand\mathsf{\x}}
    \expandafter\xdef\csname tt\x\endcsname{\noexpand\mathtt{\x}}
    \expandafter\xdef\csname fr\x\endcsname{\noexpand\mathfrak{\x}}
    \expandafter\xdef\csname bm\x\endcsname{\noexpand\bm{\x}}
}

\newcommand{\bmell}{\bm{\ell}}
\newcommand{\id}{\mathrm{id}}

\newcommand{\GM}{\mathrm{GM}}
\renewcommand{\Im}{\operatorname{Im}}
\renewcommand{\Re}{\operatorname{Re}}
\renewcommand{\det}{\operatorname{det}}
\renewcommand{\Tr}{\operatorname{Tr}}
\renewcommand{\div}{\operatorname{div}}
\DeclareMathOperator{\diag}{diag}
\DeclareMathOperator{\ord}{ord}
\DeclareMathOperator{\ind}{ind}

\DeclareMathOperator{\Span}{span}
\DeclareMathOperator{\Area}{Area}
\DeclareMathOperator{\Spin}{Spin}

\graphicspath{{fig/}}

\subheader{\today}
\title{\boldmath R\'enyi multi-entropy from Riemann surfaces}

\author{Kenya Tasuki}

\affiliation{
    Center for Gravitational Physics and Quantum Information,\\
    Yukawa Institute for Theoretical Physics, Kyoto University,\\
	Kitashirakawa Oiwakecho, Sakyo-ku, Kyoto 606-8502, Japan
}

\emailAdd{kenya.tasuki@yukawa.kyoto-u.ac.jp}

\abstract{We compute vacuum tripartite R\'enyi multi-entropy in a two-dimensional compact-boson conformal field theory (CFT).
For every integer R\'enyi index \texorpdfstring{$n\geq2$}{n >= 2}, we evaluate the partition function on the replica Riemann surface for disjoint intervals, obtaining the twist four-point function in terms of a ratio of theta functions.
An independent calculation of the partition function on a Fermat curve yields the adjacent three-point coefficient.
In the fusion limit, the Riemann surface for disjoint intervals degenerates into two Fermat curves, and the residue of the four-point function reproduces the squared three-point coefficient, as required by the operator product expansion.
The genuine multi-entropy therefore approaches twice its strictly adjacent value.
The CFT predictions agree with lattice calculations for the critical XXZ and Ising spin chains and a harmonic chain.
We also exhibit a unitary CFT in which the canonical replica construction yields complex R\'enyi multi-entropy.}

\begin{document}

\begin{flushright}
    YITP-26-169
\end{flushright}

% --------------------------------------------------
\maketitle
\flushbottom

\section{Introduction and summary}\label{sec:introduction}
Quantum entanglement probes properties of quantum critical systems.
For integer $n\geq2$, the R\'enyi entanglement entropy of a subsystem $A$ in a pure state can be computed from a contraction of $n$ copies of the full density matrix, with a cyclic permutation acting on $A$ and the identity acting on its complement.
In a two-dimensional conformal field theory (CFT), the leading logarithmic scaling of the vacuum single-interval entropy is fixed by the central charge $c$~\cite{Holzhey:1994we,Calabrese:2004eu}.
Entanglement entropies of disjoint intervals also probe the operator spectrum~\cite{Furukawa:2008uk,Calabrese:2009ez,Calabrese:2010he}.
These quantities describe entanglement across a bipartition, but a division into three or more regions raises a different question: how is entanglement shared among the multiple parties?

We study vacuum tripartite entanglement among two intervals $A,B$, and their complement $O$ in a two-dimensional CFT.
The tripartite R\'enyi multi-entropy $S_n^{(3)}(A:B:O)$ generalizes the replica construction to $n^2$ copies of the density matrix, with a prescribed set of distinct permutations acting on $A$, $B$, and $O$~\cite{Gadde:2022cqi,Penington:2022dhr,Gadde:2023zzj,Gadde:2024taa}.~\footnote{Multi-entropy and related multi-invariants have been studied in holography~\cite{Iizuka:2024pzm,Yuan:2024yfg,Gadde:2024taa,Iizuka:2025ioc,Iizuka:2025bcc,Iizuka:2025caq,Iizuka:2025elr,Anegawa:2025prn,Iizuka:2026ahd,Fujiki:2026qdt,Ju:2026lst,Anegawa:2026yee,Iizuka:2026kbo,Anegawa:2026npk,Anegawa:2026hjf}, CFTs~\cite{Harper:2024ker,Harper:2025uui,Ma:2026frw,Ma:2026wrv}, gapped theories~\cite{DelZotto:2026fpw,Iizuka:2026qqg,Akella:2026rbe,Chang:2026uds} and Lifshitz theories~\cite{Berthiere:2025toi}.
For finite-dimensional systems, see Refs.~\cite{Iizuka:2025pqq,Akella:2026xza,Akella:2026bci,Czech:2026dgm,Czech:2026jil,Ma:2026wpv}.}

The multi-entropy can be nonzero even for a tensor product of bipartite pure states shared between pairs of parties.
To probe entanglement beyond these independent bipartite layers, we use the genuine multi-entropy~\cite{Liu:2024ulq,Harper:2024ker,Iizuka:2025ioc,Iizuka:2025caq,Gadde:2026msg},
\begin{equation}
	\GM_n^{(3)}(A:B:O) \coloneqq S_n^{(3)}(A:B:O) - \frac12 \left[S_n(A:A^c)+S_n(B:B^c)+S_n(O:O^c)\right].
	\label{eq:intro-genuine-subtraction}
\end{equation}
This combination vanishes for all such tensor products.
A nonzero value therefore rules out this pairwise structure.

In two-dimensional CFTs, the replica moments are partition functions on Riemann surfaces whose genus depends on the interval configuration and the R\'enyi index.
For adjacent intervals, the surface has genus $g_F=(n-1)(n-2)/2$.
It is a sphere at $n=2$ and a triangular torus at $n=3$.
The sphere fixes the length-independent strictly adjacent genuine entropy to $\kappa_2^{(3)}=(c/4)\log 2$ for a compact unitary CFT with $c_L=c_R=c$ and a unique vacuum~\cite{Harper:2024ker,Harper:2025uui}.
The torus result already depends on the seed CFT partition function~\cite{Penington:2022dhr,Harper:2024ker}, while the higher-genus covers at $n\geq4$ can probe data beyond the torus partition function.
For disjoint intervals, the surface has genus $g_d=(n-1)^2$ and depends on the cross ratio of the four endpoints.
We ask how these amplitudes depend on the seed CFT at general integer $n$, and how the adjacent coefficient is related to the fusion limit of the disjoint correlator.

We compute the disjoint twist four-point function and the adjacent three-point coefficient for a real compact boson of radius parameter $\eta=R^2$ at integer $n\geq2$.
Varying the radius changes the operator spectrum at fixed $c=1$.
Our three main results are as follows.

\paragraph{The four-point function for disjoint intervals.}
Theorem~\ref{prop:disjoint-four-point} gives the tripartite twist four-point function throughout $0<x<1$, where $x=z_{12}z_{34}/(z_{13}z_{24})$ for ordered endpoints $z_1<z_2<z_3<z_4$ and $z_{ij}\coloneqq z_i-z_j$.
After division by the two-point functions of the same endpoint twists on the individual intervals, the correlation ratio of Eq.~\eqref{eq:tripartite-reduced-function} is
\begin{equation}
	\calR_n^{(3)}(x;\eta) = \frac{\Theta(\bm0\mid T_\eta(\Omega))}{|\theta[\delta_*](\bm0\mid\Omega)|^2}, \qquad \Omega=\Omega_n^{(3)}(x).
	\label{eq:intro-disjoint-theta-ratio}
\end{equation}
Here $\Omega$ is the period matrix of the genus-$(n-1)^2$ replica surface.
The numerator is the momentum and winding sum, whose radius dependence enters through $T_\eta$ defined in Eq.~\eqref{eq:classical-theta-matrix}.
The denominator is a theta constant with the characteristic $\delta_*$ of Eq.~\eqref{eq:assembly-reference-characteristic} in the same integral symplectic basis.

We construct that basis and evaluate its periods.
The Ward--Takahashi calculation~\cite{Dixon:1986qv,Dijkgraaf:1987vp,Zamolodchikov:1987ae} expresses the known two-angle quantum factor~\cite{Anastasopoulos:2011gn} in these replica conventions.
Identity-channel normalization fixes its overall constant, and the abelian-cover Thomae formula~\cite{KopeliovichZemel:2019thomae} gives the denominator above.
This extends the known $n=2$ realignment-torus result~\cite{Yin:2022toc,Harper:2024ker,Harper:2025uui} to every integer $n\geq2$.
Together with the ordinary two-interval ratio $\calR_n^{(2)}$~\cite{Furukawa:2008uk,Calabrese:2009ez}, the ratio $\calR_n^{(3)}$ determines the genuine multi-entropy.

\paragraph{The three-point coefficient for adjacent intervals.}
Theorem~\ref{thm:adjacent-three-point} evaluates the adjacent coefficient $C_n^{(3)}(\eta)$ independently from the partition function on the Fermat curve $u^n+v^n=1$.
Writing $g_F=(n-1)(n-2)/2$ for its genus~\cite{Penington:2022dhr,Iizuka:2026mlw} and $\Omega_{F,n}$ for its period matrix, we obtain
\begin{equation}
	C_n^{(3)}(\eta) = \frac{(4\pi)^{g_F/2}}{n^{(3n-4)/4}}\, \left[\det\Im\Omega_{F,n} \prod_{k=1}^{n-1}\Gamma(k/n)^{3(2k-n)}\right]^{1/2} \Theta\!\left(\bm0\mid T_\eta(\Omega_{F,n})\right).
	\label{eq:intro-adjacent-gamma-coefficient}
\end{equation}
The Fermat periods determine the winding sum, while the scalar Laplacian determinant, constant-mode measure and external twist normalization determine the Gamma prefactor.
The coefficient is positive and gives the adjacent genuine multi-entropy $\kappa_n^{(3)}(\eta)=\log C_n^{(3)}(\eta)/[n(1-n)]$.
It reproduces the known sphere and triangular-torus coefficients~\cite{Harper:2024ker} and evaluates the higher-genus amplitudes for general $n$.

The anomaly factorization of these Fermat covers was obtained in Ref.~\cite{Ma:2026frw}, with the seed CFT partition function left as dynamical input.
Here we evaluate the complete coefficient for the compact boson.

\paragraph{The fusion limit.}
Theorem~\ref{thm:absolute-fusion} explicitly shows that the four-point fusion residue equals the squared magnitude of the three-point coefficient, as required by the twist operator product expansion (OPE).
Take the continuum limit before bringing the two inner endpoints together, so that $x\to1^-$.
The leading singularity has exponent $\Delta_n^{(3)}=(n^2-1)/12$.
Removing this singular factor defines the finite function $\calF_n^{(3)}\coloneqq(1-x)^{\Delta_n^{(3)}}\calR_n^{(3)}$, whose limit is
\begin{equation}
	\lim_{x\to1^-}\calF_n^{(3)}(x;\eta) = |C_n^{(3)}(\eta)|^2.
	\label{eq:intro-four-point-residue}
\end{equation}
The replica surface degenerates into two Fermat components carrying conjugate three-point amplitudes.
For the compact boson, each contributes one strictly adjacent value to the genuine entropy, so the fusion limit of the disjoint genuine multi-entropy is twice the strictly adjacent value,
\begin{equation}
	\lim_{x\to1^-}\mathrm{GM}_{n,\mathrm{disj}}^{(3)}(x;\eta) = 2\kappa_n^{(3)}(\eta).
	\label{eq:intro-fusion-genuine-limit}
\end{equation}
This explanation resolves the apparent mismatch between the disjoint fusion limit and the strictly adjacent result encountered in Ref.~\cite{Harper:2024ker}.

The compact-boson theory is relevant to the low-energy physics of a variety of critical lattice systems.
Finite-size lattice calculations support the continuum amplitudes derived here.
The critical XXZ chain tests their dependence on the compactification radius while the XX and Ising comparisons test the higher-index amplitudes.
The harmonic chain approaches the noncompact finite part when its Gaussian zero mode is matched to the continuum prescription.
The determinant and Pfaffian methods for the spin moments, together with the distinct observables defined by fermion-mode partitions, will be developed separately~\cite{Tasuki:2026fss}.

Beyond the compact boson, we examine which positivity properties follow from replica sewing.
For disjoint intervals, reflection positivity establishes a continuum-vacuum case of the coarse-graining conjecture~\cite{Gadde:2023zni}.
It bounds the genuine entropy below by minus half the ordinary R\'enyi mutual information.
The canonical adjacent coefficient need not be positive even in a unitary CFT.
For a particular heterotic Narain theory, the triangular-torus trace gives $C_3^{(3)}<0$, and thus the canonical R\'enyi multi-entropy at strict adjacency is complex.
% An independent finite-dimensional counterexample is provided by the antisymmetric three-qutrit state in Section~\ref{sec:finite-replica-definition}.

\paragraph{Organization of the paper.}
Sections~\ref{sec:replica-construction} and~\ref{sec:cft-framework} define the replica contractions and their twist correlators.
Sections~\ref{sec:replica-geometry} and~\ref{sec:common-boson} provide the common geometric data, compact-boson harmonic factor and external twist normalization.
The independent disjoint and adjacent calculations occupy Sections~\ref{sec:disjoint-boson} and~\ref{sec:adjacent-boson}.
Their fusion relation is established in Section~\ref{sec:analytic-comparisons}.
Section~\ref{sec:low-index-amplitudes} compares the general formulas with known sphere and torus amplitudes, Section~\ref{sec:model-comparisons} presents the lattice tests, and Section~\ref{sec:general-seed-amplitudes} studies reflection bounds and signs of replica amplitudes.
Section~\ref{sec:discussion} discusses the implications and extensions.

Appendix~\ref{sec:math-appendix} collects mathematical conventions and supporting cycle and period calculations.
Appendix~\ref{sec:twist-normalization} contains the determinant inputs, alternative normalization calculations and auxiliary Dirac sector used for the Ising amplitudes.
Appendix~\ref{sec:lattice-numerics} gives the TTN contractions, signed Gaussian kernels and harmonic-chain determinants used in the lattice comparisons.

\section{Replica moments and R\'enyi multi-entropy}\label{sec:replica-construction}
A multi-invariant is a local-unitary invariant obtained by contracting replicated quantum states according to a permutation of the copies assigned to each party.
Choosing cyclic shifts along independent replica directions gives the R\'enyi multi-entropy assignment~\cite{Szalay:2011vxa,Penington:2022dhr,Gadde:2022cqi,Gadde:2023zzj,Gadde:2024taa}.

\subsection{Replica construction of R\'enyi multi-entropy}\label{sec:finite-replica-definition}
For a normalized pure state on $q\geq2$ parties,
\begin{equation}
    \ket\psi\in\bigotimes_{i=1}^q\calH_{A_i},\qquad\braket\psi=1,
    \label{eq:finite-state}
\end{equation}
choose an integer $N\geq2$ and let $\frS_N$ be the symmetric group of its replica labels, with identity $\id$.
A replica moment assigns one permutation $\pi_i\in\frS_N$ to each party.
The tuple $\bm\pi\coloneqq(\pi_1,\ldots,\pi_q)$ specifies the contraction of $N$ copies.
We use the source-to-target action
\begin{equation}
    P_X(\pi)\bigotimes_{r=1}^N\ket{i_r}_X \coloneqq \bigotimes_{r=1}^N\ket{i_{\pi^{-1}(r)}}_X,
    \label{eq:permutation-action}
\end{equation}
so input slot $r$ moves to output slot $\pi(r)$.
With $(\alpha\beta)(r)=\alpha(\beta(r))$, this gives
\begin{equation}
    P_X(\alpha)P_X(\beta) = P_X(\alpha\beta).
    \label{eq:permutation-representation-law}
\end{equation}

Operators on different parties commute, and the multi-invariant is
\begin{equation}
    \widehat Z_N(\bm\pi;\psi) \coloneqq \Tr\!\left[\ketbra\psi^{\otimes N}\prod_{i=1}^qP_{A_i}(\pi_i)\right].
    \label{eq:general-multi-invariant}
\end{equation}
The same unitary applied to every copy of a party commutes with its permutation, proving local-unitary invariance~\cite{Szalay:2011vxa,Gadde:2025csh}.

\paragraph{Relative sewings and replica symmetry.}
A common permutation of all parties only reorders identical whole-state copies:
\begin{equation}
    \left[\prod_iP_{A_i}(\alpha)\right]\ket\psi^{\otimes N} = \ket\psi^{\otimes N}.
    \label{eq:whole-replica-invariance}
\end{equation}
The bra and ket can therefore be relabeled independently,
\begin{equation}
    \widehat Z_N((\pi_i);\psi) = \widehat Z_N((\alpha\pi_i\beta^{-1});\psi), \qquad \alpha,\beta\in\frS_N.
    \label{eq:pure-state-relabeling}
\end{equation}

Write $\bm\pi\sim\bm\pi'$ for this equivalence.
Choosing $\alpha=\id$ and $\beta=\pi_q$ makes the last party identity-sewn:
\begin{equation}
	\bm\pi\sim(\widehat\pi_1,\ldots,\widehat\pi_{q-1},\id), \qquad \widehat\pi_i\coloneqq\pi_i\pi_q^{-1}, \qquad \widehat\pi_q=\id.
	\label{eq:relative-party-permutations}
\end{equation}
This independent left and right relabeling uses purity.
A general mixed state guarantees only simultaneous conjugation.

Further relabeling preserves the identity sewing only if $\alpha=\beta$.
The relative sewings generate the subgroup
\begin{equation}
	K \coloneqq \langle\widehat\pi_1,\ldots,\widehat\pi_{q-1}\rangle \leq \frS_N.
	\label{eq:relative-permutation-group}
\end{equation}
The tuple still specifies which generator belongs to which party.~\footnote{Without choosing the identity gauge, the subgroup generated by the tuple need not be invariant under replica relabeling.
For $N=3$ and $q=2$, $(\id,(123))\sim((12),(23))$, but the first tuple generates a cyclic group of order three $\bbZ_3$, whereas the second generates $\frS_3$.}

A replica symmetry preserves every party sewing~\cite{Gadde:2024taa}.
In the identity gauge it must therefore commute with every generator of $K$.
Thus the symmetry group is the centralizer,
\begin{equation}
	C_{\frS_N}(K) \coloneqq \{\alpha\in\frS_N:\alpha\widehat\pi_i=\widehat\pi_i\alpha\quad\forall i<q\}.
	\label{eq:replica-symmetry-centralizer}
\end{equation}

\paragraph{The regular cyclic assignment.}
The R\'enyi multi-entropy construction uses one independent cyclic replica direction for each party except the identity-sewn party~\cite{Gadde:2022cqi,Penington:2022dhr}.~\footnote{Some works use the term multi-entropy for general permutation expectation values; see Ref.~\cite{Sohal:2026tpv}.}
Fix an integer $n\geq2$ and label the $N=n^{q-1}$ copies by
\begin{equation}
	\bmr \coloneqq (r_1,\ldots,r_{q-1})\in(\bbZ_n)^{q-1}, \qquad \bbZ_n \coloneqq \bbZ/n\bbZ.
	\label{eq:replica-label}
\end{equation}
Write $\sfg$ for the unit cyclic shift and $\sfe$ for the identity in $\bbZ_n$.
The generator $\sfg_i$ shifts only coordinate $i$,
\begin{equation}
	\sfg_i\cdot\bmr = (r_1,\ldots,r_i+1,\ldots,r_{q-1}), \qquad i<q,
	\label{eq:generator-action}
\end{equation}
and the ordered assignment is
\begin{equation}
	\bm\pi_n^{(q)} \coloneqq (\sfg_1,\ldots,\sfg_{q-1},\sfe^{[q-1]}), \qquad \sfe^{[q-1]} = (\sfe,\ldots,\sfe).
	\label{eq:multi-entropy-assignment}
\end{equation}

Its relative sewings generate the regular translation group $K\cong(\bbZ_n)^{q-1}$.
Every permutation commuting with these translations is fixed by its value at the zero label:
\begin{equation}
	\alpha\cdot\bmr = \bmr+\alpha\cdot\bm0, \qquad C_{\frS_N}(K) = K.
	\label{eq:translation-centralizer-proof}
\end{equation}
This proves that the replica symmetries are precisely the same translations.

The full permutation and its moment are
\begin{align}
	P_n^{(q)}(A_1:\cdots:A_q) &\coloneqq \prod_{i=1}^{q-1}P_{A_i}(\sfg_i),
	\label{eq:full-permutation}\\
	\widehat Z_n^{(q)}(A_1:\cdots:A_q) &\coloneqq \Tr\!\left[\ketbra\psi^{\otimes n^{q-1}}P_n^{(q)}(A_1:\cdots:A_q)\right].
	\label{eq:replica-moment-definition}
\end{align}

This replica moment is real for every state.~\footnote{The same conjugation proof applies when the pure-state projector $\ketbra{\psi}$ is replaced by a density operator $\rho$.}
The replica-inversion argument of Ref.~\cite{Penington:2022dhr} for $q=3$ extends to arbitrary $q$.
Indeed, the common replica inversion $\iota:\bmr \mapsto - \bmr$ satisfies  $\iota \sfg_i \iota^{-1} = \sfg_i^{-1}$.
Conjugating all party permutations by the inversion leaves the replicated state unchanged and replaces $P_n^{(q)}$ by its adjoint.
Hence
\begin{equation}
	\left(\widehat{Z}_n^{(q)}\right)^* = \Tr\!\left[\ketbra{\psi}^{\otimes n^{q-1}} P_n^{(q)\dagger}\right] = \Tr\!\left[\ketbra{\psi}^{\otimes n^{q-1}} P_n^{(q)}\right] = \widehat{Z}_n^{(q)} \in \bbR.
	\label{eq:regular-replica-reality}
\end{equation}

The R\'enyi multi-entropy is defined by the logarithm of the replica moment~\cite{Gadde:2022cqi,Penington:2022dhr},
\begin{equation}
	S_n^{(q)}(A_1:\cdots:A_q) \coloneqq \frac{n}{1-n} \frac{1}{n^{q-1}} \log{\widehat{Z}_n^{(q)}(A_1:\cdots:A_q)}.
	\label{eq:general-multi-entropy}
\end{equation}
The entropy is real for positive moments, requires a complex logarithm branch for negative moments, and has no finite value at a zero moment.

\paragraph{The bipartite R\'enyi entanglement entropy.}
For $q=2$, the single cyclic shift reproduces the bipartite R\'enyi entropy.
Writing $\rho_X \coloneqq \Tr_{X^c} \ketbra{\psi}$, the cyclic contraction gives
\begin{equation}
	\widehat{Z}_n^{(2)}(X:X^c) = \Tr\!\left[\ketbra{\psi}^{\otimes n} P_X(\sfg)\right] = \Tr\!\left[\rho_X^{\otimes n} P_X(\sfg)\right] = \Tr\rho_X^n,
	\label{eq:ordinary-renyi-moment}
\end{equation}
and hence
\begin{equation}
	S_n(X) \coloneqq S_n^{(2)}(X:X^c) = \frac{1}{1-n} \log\Tr\!{\rho_X^n}.
	\label{eq:ordinary-renyi-entropy}
\end{equation}

\paragraph{The tripartite contraction.}
For $(A,B,O)$ with $O=(A\cup B)^c$, the two generators are $\sfg_1=(\sfg,\sfe)$ and $\sfg_2=(\sfe,\sfg)$.
Tracing the identity-sewn party gives
\begin{align}
	\widehat Z_n^{(3)}(A:B:O) &= \Tr\!\left[\rho_{AB}^{\otimes n^2}P_A(\sfg_1)P_B(\sfg_2)\right], \qquad \rho_{AB}\coloneqq\Tr_O\ketbra\psi,
	\label{eq:tripartite-replica-moment}\\
	S_n^{(3)}(A:B:O) &= \frac{\log\widehat Z_n^{(3)}(A:B:O)}{n(1-n)}.
	\label{eq:multi-entropy}
\end{align}

The pure-state invariant is symmetric under permutations of $A,B,O$.
Exchanging $A$ and $B$ exchanges the replica coordinates.
Exchanging $A$ and $O$ and restoring identity sewing on the last party gives the tuple $(\sfg_1^{-1},\sfg_2\sfg_1^{-1},\sfe^{[2]})$.
The replica relabeling $(r_1,r_2)\mapsto(-r_1-r_2,r_2)$ conjugates the original generators into this pair, establishing the symmetry.

\paragraph{A negative replica moment.}
Positivity of the replica moment for general states was left open in Refs.~\cite{Penington:2022dhr,Gadde:2023zni}.
One of the counterexamples is the normalized antisymmetric state of three qutrits.~\footnote{The same counterexample is recorded in Ref.~\cite{Omega:2026mpf}.}
With the basis ordered as $A,B,O$, it is
\begin{equation}
	\ket{\psi_\wedge}\coloneqq\frac1{\sqrt6} \bigl(\ket{012}+\ket{120}+\ket{201}-\ket{021}-\ket{102}-\ket{210}\bigr).
	\label{eq:antisymmetric-qutrit-state}
\end{equation}
An explicit calculation gives $\widehat Z_3^{(3)}(A:B:O)=0$, while at $n=4$ the replica moment is
\begin{equation}
	\widehat Z_4^{(3)}(A:B:O) = -\frac{5\times 7}{2^{12}\times3^{15}}<0.
	\label{eq:antisymmetric-qutrit-moment}
\end{equation}
Section~\ref{sec:cft-amplitude-signs} discusses the sign of the tripartite replica moment in a continuum CFT.

\subsection{Genuine multi-entropy and layerwise separability}\label{sec:genuine-subtraction}

For $\ket\psi=\ket\phi_{AB}\otimes\ket\chi_O$ with normalized factors, common relabeling of the $A,B$ copies cancels the $B$ sewing and replaces the $A$ sewing by $\sfg_1\sfg_2^{-1}$.
This translates $(r_1,r_2)$ by $(1,-1)$ and decomposes the $n^2$ copies into $n$ cycles of length $n$.
Each cycle contributes $\Tr\rho_A^n$, so
\begin{equation}
	\widehat Z_n^{(3)}(A:B:O)=(\Tr\rho_A^n)^n, \qquad S_n^{(3)}(A:B:O) = S_n(A) = S_n(B).
	\label{eq:spectator-check}
\end{equation}

For an entangled pair with an unentangled third party, $S_n^{(3)}$ counts the pair entropy once and the sum of the three bipartite entropies counts it twice.
Within the symmetric linear combinations $S_n^{(3)}-a[S_n(A)+S_n(B)+S_n(O)]$, cancellation of this contribution fixes $a=1/2$, giving the genuine multi-entropy~\cite{Liu:2024ulq,Harper:2024ker,Iizuka:2025ioc},
\begin{equation}
	\GM_n^{(3)}(A:B:O) \coloneqq S_n^{(3)}(A:B:O) - \frac12\sum_{X=A,B,O}S_n^{(2)}(X:X^c).
	\label{eq:tripartite-genuine-multi-entropy}
\end{equation}

\paragraph{Layer decompositions and entanglement signals.}
Suppose the Hilbert space of each party admits a tensor factorization
\begin{equation}
	\calH_X = \bigotimes_{\ell=1}^{m}\calH_{X,\ell}, \qquad X=A,B,O,
	\label{eq:layer-hilbert-factorization}
\end{equation}
with $m\geq1$ factors.
A layer decomposition of the state is a factorization
\begin{equation}
	\ket{\psi} = \bigotimes_{\ell=1}^{m}\ket{\psi_\ell}, \qquad \ket{\psi_\ell}\in\calH_{A,\ell}\otimes\calH_{B,\ell}\otimes\calH_{O,\ell}.
	\label{eq:layer-state-factorization}
\end{equation}
Each normalized factor $\ket{\psi_\ell}$ is called a layer.

A pure state is layerwise-separable if it admits a layer decomposition in which every layer is separable across a nontrivial partition of the parties~\cite{Gadde:2026msg}.
For three parties, each layer therefore has an unentangled party, which may differ between layers.
The full state need not factorize across any bipartition of $A,B,O$.
For example, one layer can entangle $A$ with $B$ and another $B$ with $O$.

Replica moments multiply over independent layers, so the entropies are additive when the layer moments are positive.
For a layerwise-separable state, Eq.~\eqref{eq:spectator-check} gives a positive moment and zero $\GM_n^{(3)}$ for every layer, and hence for their product.
Together with local-unitary invariance, this makes $\GM_n^{(3)}$ a multipartite entanglement signal~\cite{Gadde:2026msg}.
A nonzero value rules out layerwise separability.

\subsection{The coarse-graining conjecture}\label{sec:second-index-reflected-entropy}
At $n=2$, a reflected purity determines the tripartite moment and gives the coarse-graining bound for every finite-dimensional pure state~\cite{Penington:2022dhr,Liu:2024ulq}.
Define
\begin{equation}
    I_n(A:B)\coloneqq S_n(A)+S_n(B)-S_n(A\cup B).
    \label{eq:state-renyi-mutual-information}
\end{equation}
Let $A^*,B^*$ denote conjugate copies of the Hilbert spaces of $A,B$.
Let $\ket{\rho_{AB}}$ be the vectorization of $\rho_{AB}$ in $AA^*BB^*$, and set
\begin{equation}
    \ket{\widehat\rho_{AB}}\coloneqq\frac{\ket{\rho_{AB}}}{\sqrt{\Tr\rho_{AB}^2}},\qquad
    \sigma_{AA^*}\coloneqq\Tr_{BB^*}\ketbra{\widehat\rho_{AB}}.
    \label{eq:discussion-reflected-purification}
\end{equation}
Regrouping the four-copy contraction gives
\begin{equation}
    \widehat Z_2^{(3)}(A:B:O)=(\Tr\rho_{AB}^2)^2\Tr\sigma_{AA^*}^2>0.
    \label{eq:state-realignment-four-copy}
\end{equation}
The normalized vector $\ket{\widehat\rho_{AB}}$ purifies $\rho_{AB}^2/\Tr\rho_{AB}^2$, so the reflected entropy has indices $(2,2)$~\cite{Dutta:2019gen,Liu:2024ulq,Berthiere:2023gkx}.
Consequently
\begin{equation}
    S^R_{2,2}(A:B)=-\log\Tr\sigma_{AA^*}^2 = 2\bigl[S_2^{(3)}(A:B:O)-S_2(A\cup B)\bigr]\geq0,
    \label{eq:discussion-reflected-purity}
\end{equation}
and hence
\begin{equation}
    \GM_2^{(3)}(A:B:O) = \frac12\bigl[S^R_{2,2}(A:B)-I_2(A:B)\bigr] \geq -\frac12 I_2(A:B).
    \label{eq:discussion-reflected-genuine}
\end{equation}
For a pure state, the corresponding gap at any index is
\begin{equation}
    \GM_n^{(3)}(A:B:O)+\frac12 I_n(A:B) = S_n^{(3)}(A:B:O)-S_n(A\cup B).
    \label{eq:discussion-coarse-graining-gap}
\end{equation}
Its nonnegativity is the tripartite case of the coarse-graining conjecture~\cite{Gadde:2023zni},
\begin{equation}
    S_n^{(q)}(A_1:\cdots:A_q) \geq S_n^{(q-1)}(A_1:\cdots:A_{q-2}:A_{q-1}\cup A_q),\qquad q\geq3,
    \label{eq:discussion-coarse-graining-conjecture}
\end{equation}
on the positive-moment domain.
Section~\ref{sec:discussion-crossing} proves the tripartite bound at general integer $n$ for the continuum vacuum sewing considered here.

\section{Multi-entropy in two-dimensional CFT}\label{sec:cft-framework}
Replica twists reduce the interval problem to two normalized quantities: the three-point coefficient $C_n^{(3)}$ for strict adjacency and the four-point correlation ratio $\calR_n^{(3)}(x)$ for disjoint intervals~\cite{Calabrese:2004eu,Lunin:2000yv,Harper:2024ker}.
We derive the corresponding genuine multi-entropies and then use their OPE channels to relate the fusion limit to the adjacent coefficient.

\subsection{Normalized replica twists}\label{sec:interval-geometry}
We consider a unitary two-dimensional CFT $\calC$ in its pure vacuum $\ket{\psi}$ on the infinite spatial line, with equal left and right central charges $c$.
The interval endpoints lie on the real axis of the Euclidean $z$ plane.

Replica sewing expresses the normalized multi-invariant as a ratio of partition functions on replica Riemann surfaces.
The path integral over the lower half-plane prepares an unnormalized vacuum ket $\ket{\Psi}$, and the upper half-plane prepares its conjugate bra.
Gluing them along the real axis gives the norm as a path integral over the full complex plane.
With common ultraviolet and infrared regulators, write
\begin{equation}
	\begin{aligned}
		Z &\coloneqq \braket{\Psi}, \qquad \ket{\psi} \coloneqq Z^{-1/2}\ket{\Psi},\\
		Z_N(\bm{\pi}) &\coloneqq \Tr\!\left[\ketbra{\Psi}^{\otimes N} \prod_{i=1}^{q}P_{A_i}(\pi_i)\right].
	\end{aligned}
	\label{eq:unnormalized-replica-partition}
\end{equation}

Let $\Sigma_N(\bm\pi)$ denote the replica Riemann surface obtained by sewing these copies of the complex plane along the chosen party regions.
Its path integral uses the metric, local regulator and state preparation inherited from the copies.
\begin{subequations}
	\label{eq:replica-surface-normalization}
	\begin{align}
		Z_N(\bm\pi) &= Z[\Sigma_N(\bm\pi)],
		\label{eq:replica-surface-normalization-partition}\\
		\widehat{Z}_N(\bm{\pi};\psi) &= \frac{Z_N(\bm\pi)}{Z^N}.
		\label{eq:replica-surface-normalization-moment}
	\end{align}
\end{subequations}

The relative sewing on the two sides of an endpoint fixes its monodromy.
By the source-to-target convention of Eq.~\eqref{eq:permutation-action}, the lower bank of copy $r$ is sewn to the upper bank of copy $\pi_i(r)$ on $A_i$.
Let $\pi_L$ and $\pi_R$ be the sewing permutations immediately to the left and right of an endpoint.
Start a counterclockwise circuit just below the real axis on its right.
The circuit crosses the right seam from lower to upper bank and the left seam from upper to lower bank, giving
\begin{equation}
	r \longmapsto \pi_R(r) \longmapsto (\pi_L^{-1}\pi_R)(r).
	\label{eq:endpoint-monodromy}
\end{equation}
Thus the endpoint monodromy is $\pi_L^{-1}\pi_R$.
With party $A_i$ on the left and $A_j$ on the right, the identity gauge of Eq.~\eqref{eq:relative-party-permutations} gives the conjugate monodromy $\widehat\pi_i^{-1}\widehat\pi_j=\pi_q(\pi_i^{-1}\pi_j)\pi_q^{-1}\in K$.
Below we use these gauge-fixed sewings and suppress their hats.

The endpoint monodromies define twisted sectors of the permutation orbifold $\calC^{\otimes N}/K$, where $K$ is generated by the relative sewings in Eq.~\eqref{eq:relative-permutation-group}~\cite{Bantay:1999us,Lunin:2000yv,Gadde:2024taa}.
We keep the permutation labels fixed in the twist correlators, with no conjugacy-class sum.
The local sewing and the cover vacuum define the regular replica twist $\calT_\pi(z,\bar z)$~\cite{Lunin:2000yv,Caraglio:2008pk}.
In radial quantization, the monodromy $\pi$ specifies the boundary conditions on a small circle around the insertion.
The path integral inside its branched cover, with the identity at each ramification point, prepares a state in this twisted sector.
The state--operator correspondence defines $\calT_\pi$ from this state, up to normalization.
Thus the endpoint $z_a$ between the two sewings carries
\begin{equation}
	\calT_{\pi_L^{-1}\pi_R}(z_a,\bar z_a), \qquad \pi_L,\pi_R\in K.
	\label{eq:interface-twist}
\end{equation}

The product of endpoint twists represents the replica moment in the continuum limit~\cite{Gadde:2024taa,Calabrese:2004eu}.
Write $\pi_{L,a}$ and $\pi_{R,a}$ for the sewings on either side of $z_a$.
The leading continuum relation is
\begin{equation}
	\widehat{Z}_N(\bm{\pi};\psi) = \frac{Z_N(\bm{\pi})}{Z^N} \propto \left\langle \prod_a \calT_{\pi_{L,a}^{-1}\pi_{R,a}}(z_a,\bar z_a) \right\rangle.
	\label{eq:twist-correlator-representation}
\end{equation}
The expectation value is in the normalized vacuum of the replicated theory.
The proportionality factor contains the cutoff powers and local endpoint normalizations that survive division by the state norm.

\paragraph{Twist primarity and conformal dimensions.}
The regular replica twists are primary, with conformal weights determined by the cycle lengths of their permutations.
The total holomorphic stress tensor of the product CFT is
\begin{equation}
	T_{\mathrm{tot}}(z) \coloneqq \sum_{r=1}^{N}T_r(z),
	\label{eq:replica-stress-tensor}
\end{equation}
with central charge $Nc$.

The singular terms in the stress-tensor OPE can be computed separately for each cycle of $\pi$.
Consider one cycle of length $m$.
Near its ramification point, the $m$ sheets are uniformized by the local coordinates
\begin{equation}
	w_r \coloneqq e^{2\pi i r/m}z^{1/m}, \qquad r=0,\ldots,m-1.
	\label{eq:local-cycle-uniformization}
\end{equation}
Here $r$ labels the sheets locally within this cycle.
A counterclockwise circuit sends $w_r$ to $w_{r+1}$, with $r$ understood modulo $m$.
The Schwarzian transformation gives~\cite{Calabrese:2004eu}
\begin{equation}
	T_r(z) = (w_r'(z))^2 T(w_r) + \frac{c}{12}\{w_r,z\} = (w_r'(z))^2 T(w_r) + \frac{c(1-m^{-2})}{24z^2}.
	\label{eq:stress-schwarzian-transformation}
\end{equation}

The identity insertion prepares the cover vacuum in radial quantization, so $T(w_r)$ acting on this state is regular at $w_r=0$.
Its transformed contribution is at worst of order $z^{-2+2/m}$.
The sum of stress tensors over sheets is single-valued, so its Laurent powers are integers and this contribution has at most a simple pole.
Summing the Schwarzian terms over the $m$ sheets gives the double pole $c(m-m^{-1})/(24z^2)$.

The singular OPE for a general permutation is the sum of its cycle contributions.
Let $m_1,\ldots,m_s$ be the cycle lengths, with fixed points counted as length-one cycles.
The double-pole coefficients add, and the translation Ward identity fixes the simple pole to the derivative of the insertion:
\begin{equation}
	T_{\mathrm{tot}}(z) \calT_\pi(0) = \frac{c}{24z^2} \sum_{a=1}^{s} \left(m_a-\frac1{m_a}\right) \calT_\pi(0) + \frac{1}{z} \partial\calT_\pi(0) + O(1).
	\label{eq:twist-stress-ope}
\end{equation}

The absence of higher poles makes the twist state primary, and the double-pole coefficient gives its holomorphic weight.
The antiholomorphic calculation gives the other weight.
Writing $\Delta_\pi\coloneqq h_\pi+\bar h_\pi$, the cycle contributions give
\begin{equation}
	\sum_{a=1}^{s} m_a = N, \qquad h_\pi = \bar h_\pi = \frac{c}{24} \sum_{a=1}^{s} \left(m_a - \frac1{m_a}\right), \qquad \Delta_\pi = \frac{c}{12} \sum_{a=1}^{s} \left(m_a - \frac1{m_a}\right).
	\label{eq:general-permutation-dimension}
\end{equation}

\paragraph{Unit two-point normalization.}
\label{sec:twist-correlators-normalization}
For any $\pi\in\frS_N$, the regular twists are normalized by
\begin{subequations}
	\label{eq:unit-twist-two-point}
	\begin{align}
		\calT_{\pi}^\dagger &= \calT_{\pi^{-1}},
		\label{eq:unit-twist-two-point-adjoint}\\
		\left\langle \calT_\pi(z,\bar z) \calT_{\pi^{-1}}(0,0) \right\rangle &= \abs{z}^{-2\Delta_\pi}.
		\label{eq:unit-twist-two-point-norm}
	\end{align}
\end{subequations}
Their phases are fixed by the replica sewing and we use trivial discrete torsion.
Since the twists have zero spin, define their insertion at infinity by the conformal transformation $z\mapsto1/z$,
\begin{equation}
	\calT_\pi(\infty) \coloneqq \lim_{\abs{z}\to\infty} \abs{z}^{2\Delta_\pi} \calT_\pi(z,\bar z).
	\label{eq:twist-infinity-insertion}
\end{equation}

\subsection{Endpoint normalization and adjacent intervals}\label{sec:twist-correlators}

\begin{figure}[t]
	\centering
	\begin{tikzpicture}[x=1cm,y=1cm,>=latex,font=\small]
		\node[anchor=west] at (0,1.15) {(a) Disjoint intervals};
		\draw[->,line width=0.45pt] (0,0) -- (11.7,0) node[right] {$\Re z$};
		\draw[line width=1.5pt] (1.6,0) -- (4.1,0);
		\draw[line width=1.5pt] (7.0,0) -- (10.1,0);
		\foreach \x/\i in {1.6/1,4.1/2,7.0/3,10.1/4}{
			\fill (\x,0) circle[radius=2pt];
			\node[below=5pt] at (\x,0) {$z_{\i}$};
			\node at (\x,0.65) {$\sigma_{\i}$};
		}
		\node at (0.65,0.26) {$O$};
		\node at (2.85,0.26) {$A$};
		\node at (5.55,0.26) {$O$};
		\node at (8.55,0.26) {$B$};
		\node at (10.85,0.26) {$O$};

		\begin{scope}[yshift=-2.1cm]
			\node[anchor=west] at (0,1.15) {(b) Adjacent intervals};
			\draw[->,line width=0.45pt] (0,0) -- (11.7,0) node[right] {$\Re z$};
			\draw[line width=1.5pt] (1.6,0) -- (10.1,0);
			\foreach \x/\i in {1.6/1,5.85/2,10.1/3}{
				\fill (\x,0) circle[radius=2pt];
				\node[below=5pt] at (\x,0) {$z_{\i}$};
			}
			\node at (1.6,0.65) {$\varsigma_1$};
			\node at (5.85,0.65) {$\varsigma_2=\sigma_2\sigma_3$};
			\node at (10.1,0.65) {$\varsigma_3$};
			\node at (0.65,0.26) {$O$};
			\node at (3.725,0.26) {$A$};
			\node at (7.975,0.26) {$B$};
			\node at (10.85,0.26) {$O$};
		\end{scope}
	\end{tikzpicture}
	\caption{Counterclockwise replica monodromies for (a) disjoint and (b) adjacent intervals on the infinite line.
	The complement $O$ occupies the remaining portions of the real axis.
	The adjacent junction carries the product $\varsigma_2=\sigma_2\sigma_3=\sfg_1^{-1}\sfg_2$ of the two inner endpoint monodromies.}
	\label{fig:interval-geometries}
\end{figure}
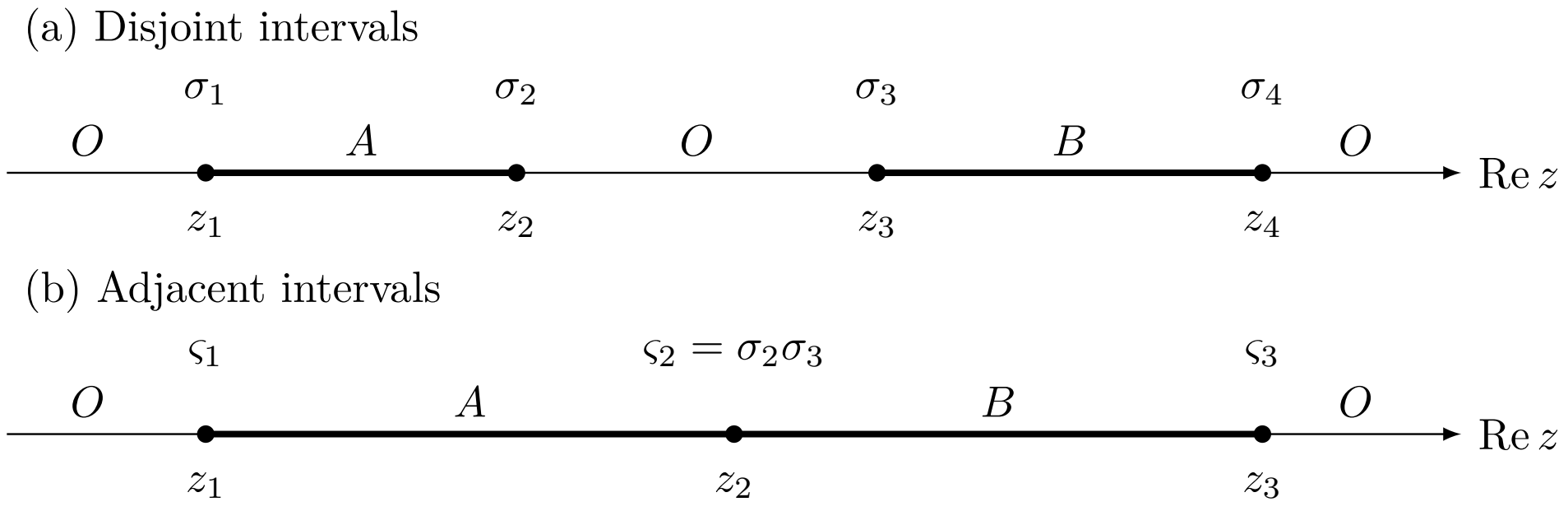

The cycle structure fixes the endpoint dimensions and regulator factors for both interval geometries shown in Fig.~\ref{fig:interval-geometries}.
For $q=2,3$ and integer $n\geq2$, the translations of Eq.~\eqref{eq:multi-entropy-assignment} specify
\begin{equation}
	N = n^{q-1}, \qquad \pi_i = \sfg_i \quad (1\leq i<q), \qquad \pi_q = \sfg_q \coloneqq \sfe^{[q-1]}.
	\label{eq:multi-entropy-cft-specialization}
\end{equation}
Here $\sfg_q$ denotes the identity sewing.
For this assignment, abbreviate
\begin{equation}
	Z_n^{(q)} \coloneqq Z_N(\bm\pi_n^{(q)}), \qquad \Sigma_n^{(q)} \coloneqq \Sigma_N(\bm\pi_n^{(q)}), \qquad \widehat{Z}_n^{(q)} \coloneqq \widehat{Z}_N(\bm\pi_n^{(q)};\psi).
	\label{eq:multi-entropy-surface-specialization}
\end{equation}

All interfaces between distinct parties have twists of the same scaling dimension.
The interface permutation $\sfg_i^{-1}\sfg_j$ shifts at least one replica coordinate by $+1$ or $-1$.
A label therefore returns after $\ell$ applications exactly when $\ell=0$ modulo $n$.
Each cycle has length $n$, so the $n^{q-1}$ labels split into $n^{q-2}$ cycles.
Substitution into Eq.~\eqref{eq:general-permutation-dimension} gives
\begin{equation}
	\Delta_n^{(q)} \coloneqq \frac{c\,n^{q-2}}{12} \left(n-\frac1n\right), \qquad h_{\sfg_i^{-1}\sfg_j} = \bar h_{\sfg_i^{-1}\sfg_j} = \frac12 \Delta_n^{(q)}.
	\label{eq:replica-twist-dimension}
\end{equation}
In particular,
\begin{subequations}
	\label{eq:two-three-party-twist-dimensions}
	\begin{align}
		\Delta_n^{(2)} &= \frac{c}{12}\left(n-\frac1n\right),
		\label{eq:two-three-party-twist-dimensions-bipartite}\\
		\Delta_n^{(3)} &= \frac{c}{12}(n^2-1) = n \Delta_n^{(2)}.
		\label{eq:two-three-party-twist-dimensions-tripartite}
	\end{align}
\end{subequations}

These cycle multiplicities also determine the endpoint factors relating regulated moments to normalized twist correlators.
Let $\epsilon$ be the common ultraviolet length and $c_n$ the positive nonuniversal coefficient of a state-normalized single-interval moment, with $c_1=1$~\cite{Calabrese:2004eu,Calabrese:2009ez}.
We assume a permutation-symmetric regulator for which the endpoint factors multiply over disjoint cycles, each length-$n$ cycle contributing $c_n^{1/2}$.
Each $q$-party interface has $n^{q-2}$ such cycles, so $p$ distinct endpoints give
\begin{equation}
	\widehat{Z}_n^{(q)} \approx \left(c_n^{n^{q-2}/2}\epsilon^{\Delta_n^{(q)}}\right)^p \left\langle \prod_{a=1}^{p} \calT_{\pi_{L,a}^{-1}\pi_{R,a}}(z_a,\bar z_a) \right\rangle.
	\label{eq:normalized-moment-endpoint-factors}
\end{equation}
For a nonzero leading correlator, $\approx$ denotes a ratio tending to one as $\epsilon\to0$ at fixed distinct endpoints and fixed $n$.

\paragraph{Single-interval normalization.}
The single-interval moment fixes the normalization used for all endpoint twists.
For $A=[z_1,z_2]$, write $z_{ij} \coloneqq z_i-z_j$.
Conformal symmetry and Eq.~\eqref{eq:unit-twist-two-point-norm} give
\begin{subequations}
	\label{eq:single-interval-correlator}
	\begin{align}
		\left\langle \calT_{\sfg}(z_1) \calT_{\sfg^{-1}}(z_2) \right\rangle &= \frac{1}{\abs{z_{12}}^{2\Delta_n^{(2)}}},
		\label{eq:single-interval-correlator-twist}\\
		\widehat{Z}_n^{(2)}(A:A^c) &\approx \frac{\left(c_n^{1/2}\epsilon^{\Delta_n^{(2)}}\right)^2}{\abs{z_{12}}^{2\Delta_n^{(2)}}}.
		\label{eq:single-interval-correlator-moment}
	\end{align}
\end{subequations}
Taking the R\'enyi logarithm reproduces the universal one-interval result of Ref.~\cite{Calabrese:2004eu}, whose $n\to1$ limit agrees with the geometric-entropy calculation of Ref.~\cite{Holzhey:1994we}:
\begin{equation}
	S_n^{(2)}(A:A^c) = \frac{c}{6}\left(1+\frac1n\right) \log\frac{\abs{z_{12}}}{\epsilon} + \frac{\log c_n}{1-n} + o(1).
	\label{eq:single-interval-entropy-normalized}
\end{equation}

\paragraph{Adjacent three-point function.}
Two adjacent intervals and their complement produce three twist insertions, including one at the shared endpoint.
Set $(A_1,A_2,A_3)=(A,B,O)$, with $O=\bbR\setminus(A\cup B)$, and choose
\begin{equation}
	z_1 < z_2 < z_3, \qquad A \coloneqq [z_1,z_2], \qquad B \coloneqq [z_2,z_3].
	\label{eq:adjacent-geometry}
\end{equation}
The spatial sequence is $O,A,B,O$, and the endpoint permutations are
\begin{equation}
	\bm{\varsigma} \coloneqq (\varsigma_1,\varsigma_2,\varsigma_3) \coloneqq (\sfg_1,\sfg_1^{-1}\sfg_2,\sfg_2^{-1}).
	\label{eq:three-permutation-tuple}
\end{equation}
The three insertions are $\calT_{\varsigma_i}(z_i,\bar z_i)$, each with dimension $\Delta_n^{(3)}$.
In particular, the junction between $A$ and $B$ carries the relative permutation
\begin{equation}
	\varsigma_2\cdot(r_1,r_2) = (r_1-1,r_2+1).
	\label{eq:adjacent-junction-action}
\end{equation}

Global conformal symmetry fixes the position dependence of three scalar primaries up to one coefficient~\cite{DiFrancesco:1997nk}.
With the unit two-point normalization of Eq.~\eqref{eq:unit-twist-two-point-norm}, we define
\begin{equation}
	C_n^{(3)} \coloneqq \left\langle \calT_{\varsigma_1}(0) \calT_{\varsigma_2}(1) \calT_{\varsigma_3}(\infty) \right\rangle.
	\label{eq:adjacent-three-point-coefficient}
\end{equation}
The three-point function at arbitrary positions is then
\begin{equation}
	\left\langle \calT_{\varsigma_1}(z_1) \calT_{\varsigma_2}(z_2) \calT_{\varsigma_3}(z_3) \right\rangle = \frac{C_n^{(3)}}{\abs{z_{12}z_{23}z_{31}}^{\Delta_n^{(3)}}}.
	\label{eq:adjacent-three-point-correlator}
\end{equation}
Restoring the endpoint factors gives the leading adjacent moment,
\begin{equation}
	\widehat{Z}_{n,\mathrm{adj}}^{(3)}(A:B:O) \approx \frac{\left(c_n^{n/2}\epsilon^{\Delta_n^{(3)}}\right)^3 C_n^{(3)}}{\abs{z_{12}z_{23}z_{31}}^{\Delta_n^{(3)}}},
	\label{eq:adjacent-normalized-moment}
\end{equation}
and the corresponding adjacent R\'enyi multi-entropy is
\begin{equation}
	S_{n,\mathrm{adj}}^{(3)}(A:B:O) = \frac{c}{12}\left(1+\frac1n\right) \log\frac{\abs{z_{12}z_{23}z_{31}}}{\epsilon^3} + \frac{\log C_n^{(3)}}{n(1-n)} + \frac{3\log{c_n}}{2(1-n)} + o(1).
	\label{eq:adjacent-renyi-multi-entropy}
\end{equation}
We distinguish the two geometries by the subscripts $\mathrm{adj}$ and $\mathrm{disj}$ on their moments and entropies.

\paragraph{The adjacent genuine multi-entropy.}
The same regulator appears in the three-point moment and in the three ordinary moments.
Purity identifies the moment on $O$ with that on $A\cup B=[z_1,z_3]$.
Using $\Delta_n^{(3)}=n\Delta_n^{(2)}$, their ratio gives
\begin{equation}
	\lim_{\epsilon\to0} \frac{\widehat Z_{n,\mathrm{adj}}^{(3)}(A:B:O)}{[\widehat Z_n^{(2)}(A:A^c)\widehat Z_n^{(2)}(B:B^c) \widehat Z_n^{(2)}(O:O^c)]^{n/2}} = C_n^{(3)}.
	\label{eq:adjacent-genuine-moment-ratio}
\end{equation}
The powers of all three distances, $\epsilon$ and $c_n$ cancel separately.
For $C_n^{(3)}>0$, the genuine subtraction therefore leaves an ultraviolet-finite constant,
\begin{equation}
	\GM_{n,\mathrm{adj}}^{(3)}(A:B:O) = \kappa_n^{(3)}+o(1), \qquad \kappa_n^{(3)} \coloneqq \frac{\log C_n^{(3)}}{n(1-n)}.
	\label{eq:adjacent-genuine-multi-entropy}
\end{equation}
Three points have no cross ratio, so this value is independent of the interval lengths.

\subsection{Correlation ratios for disjoint intervals}\label{sec:disjoint-correlators}

For disjoint intervals, conformal symmetry leaves a four-point function with nontrivial cross-ratio dependence.
Normalizing each four-point moment by its disconnected single-interval product cancels the local endpoint factors.

\paragraph{Four endpoints and the cross ratio.}
Figure~\ref{fig:interval-geometries}(a) shows the disjoint tripartition, with spatial sequence $O,A,O,B,O$.
Take $O=(A\cup B)^c$ and
\begin{equation}
	z_1<z_2<z_3<z_4, \qquad A \coloneqq [z_1,z_2], \qquad B \coloneqq [z_3,z_4].
	\label{eq:disjoint-geometry}
\end{equation}
The cross ratio is
\begin{equation}
	x \coloneqq \frac{z_{12}z_{34}}{z_{13}z_{24}}\in(0,1), \qquad 1-x = \frac{z_{14}z_{23}}{z_{13}z_{24}},
	\label{eq:interval-cross-ratio}
\end{equation}
with $\bar x = x$ for equal-time intervals.
The M\"obius map
\begin{equation}
	f(z) \coloneqq \frac{(z-z_1)z_{34}}{(z-z_4)z_{31}}
	\label{eq:four-point-mobius-map}
\end{equation}
sends $(z_1,z_2,z_3,z_4)$ to $(0,x,1,\infty)$.

\paragraph{Replica sewings.}
For the bipartition $(A\cup B):O$, the same cyclic permutation $\sfg$ sews the $n$ replicas on both intervals.
The endpoint permutations are $(\sfg,\sfg^{-1},\sfg,\sfg^{-1})$, and the four twists have dimension $\Delta_n^{(2)}$.
Define
\begin{subequations}
	\label{eq:bipartite-reduced-function}
	\begin{align}
		\calG_n^{(2)}(x,\bar x) &\coloneqq \left\langle \calT_{\sfg}(0) \calT_{\sfg^{-1}}(x,\bar x) \calT_{\sfg}(1) \calT_{\sfg^{-1}}(\infty) \right\rangle,
		\label{eq:bipartite-reduced-function-correlator}\\
		\calR_n^{(2)}(x,\bar x) &\coloneqq |x|^{2\Delta_n^{(2)}} \calG_n^{(2)}(x,\bar x).
		\label{eq:bipartite-reduced-function-ratio}
	\end{align}
\end{subequations}

For the tripartition $A:B:O$, independent translations on $A$ and $B$ sew $n^2$ replicas.
Equation~\eqref{eq:endpoint-monodromy} gives the endpoint tuple
\begin{equation}
	\bm{\sigma} \coloneqq (\sigma_1,\sigma_2,\sigma_3,\sigma_4) \coloneqq (\sfg_1,\sfg_1^{-1},\sfg_2,\sfg_2^{-1}),
	\label{eq:four-permutation-tuple}
\end{equation}
whose twists have dimension $\Delta_n^{(3)}$.
The corresponding correlator and ratio are
\begin{subequations}
	\label{eq:tripartite-reduced-function}
	\begin{align}
		\calG_n^{(3)}(x,\bar x) &\coloneqq \left\langle \calT_{\sigma_1}(0) \calT_{\sigma_2}(x,\bar x) \calT_{\sigma_3}(1) \calT_{\sigma_4}(\infty) \right\rangle,
		\label{eq:tripartite-reduced-function-correlator}\\
		\calR_n^{(3)}(x,\bar x) &\coloneqq |x|^{2\Delta_n^{(3)}} \calG_n^{(3)}(x,\bar x).
		\label{eq:tripartite-reduced-function-ratio}
	\end{align}
\end{subequations}

\paragraph{Regulated four-point moments.}
Applying the primary-field Jacobian of Eq.~\eqref{eq:four-point-mobius-map} and $z_{12}z_{34}=x z_{13}z_{24}$ to the two sewings gives~\cite{Calabrese:2009ez,Harper:2024ker}
\begin{subequations}
\label{eq:four-point-factorizations}
	\begin{align}
		\left\langle \calT_{\sfg}(z_1) \calT_{\sfg^{-1}}(z_2) \calT_{\sfg}(z_3) \calT_{\sfg^{-1}}(z_4) \right\rangle
		&= \frac{\calG_n^{(2)}(x,\bar x)}{\abs{z_{13}z_{24}}^{2\Delta_n^{(2)}}}
		= \frac{\calR_n^{(2)}(x,\bar x)}{\abs{z_{12}z_{34}}^{2\Delta_n^{(2)}}},
		\label{eq:bipartite-four-point-factorizations}\\
		\left\langle \calT_{\sigma_1}(z_1) \calT_{\sigma_2}(z_2) \calT_{\sigma_3}(z_3) \calT_{\sigma_4}(z_4) \right\rangle
		&= \frac{\calG_n^{(3)}(x,\bar x)}{\abs{z_{13}z_{24}}^{2\Delta_n^{(3)}}}
		= \frac{\calR_n^{(3)}(x,\bar x)}{\abs{z_{12}z_{34}}^{2\Delta_n^{(3)}}}.
		\label{eq:tripartite-four-point-factorizations}
	\end{align}
\end{subequations}

Each bipartite endpoint has one length-$n$ cycle, whereas each tripartite endpoint has $n$.
Restoring their local factors therefore gives the state-normalized moments
\begin{subequations}
\label{eq:disjoint-replica-moments}
	\begin{align}
		\widehat{Z}_{n,\mathrm{disj}}^{(2)}(A\cup B:O)
		&\approx \left(c_n^{1/2}\epsilon^{\Delta_n^{(2)}}\right)^4 \frac{\calR_n^{(2)}(x,\bar x)}{\abs{z_{12}z_{34}}^{2\Delta_n^{(2)}}},
		\label{eq:bipartite-disjoint-moment}\\
		\widehat{Z}_{n,\mathrm{disj}}^{(3)}(A:B:O)
		&\approx \left(c_n^{n/2}\epsilon^{\Delta_n^{(3)}}\right)^4 \frac{\calR_n^{(3)}(x,\bar x)}{\abs{z_{12}z_{34}}^{2\Delta_n^{(3)}}}.
		\label{eq:tripartite-disjoint-moment}
	\end{align}
\end{subequations}
The resulting entropies are
\begin{subequations}
	\label{eq:disjoint-replica-entropies}
	\begin{align}
		S_{n,\mathrm{disj}}^{(2)}(A\cup B:O) &= \frac{c}{6}\left(1+\frac1n\right) \log\frac{\abs{z_{12}z_{34}}}{\epsilon^2} + \frac{\log\calR_n^{(2)}(x,\bar x)}{1-n} + \frac{2\log c_n}{1-n} + o(1),
		\label{eq:bipartite-disjoint-entropy}\\
		S_{n,\mathrm{disj}}^{(3)}(A:B:O) &= \frac{c}{6}\left(1+\frac1n\right) \log\frac{\abs{z_{12}z_{34}}}{\epsilon^2} + \frac{\log\calR_n^{(3)}(x,\bar x)}{n(1-n)} + \frac{2\log c_n}{1-n} + o(1).
		\label{eq:tripartite-disjoint-entropy}
	\end{align}
\end{subequations}

\paragraph{Correlation ratios from replica moments.}
The endpoint factors cancel in the following ratios of the replica moments in Eqs.~\eqref{eq:bipartite-disjoint-moment} and~\eqref{eq:tripartite-disjoint-moment}:
\begin{subequations}
	\label{eq:disjoint-moment-ratios}
	\begin{align}
		\calR_n^{(2)}(x,\bar x) &= \lim_{\epsilon\to0} \frac{\widehat{Z}_{n,\mathrm{disj}}^{(2)}(A\cup B:O)}{\widehat{Z}_n^{(2)}(A:A^c)\widehat{Z}_n^{(2)}(B:B^c)},
		\label{eq:bipartite-disjoint-moment-ratio}\\
		\calR_n^{(3)}(x,\bar x) &= \lim_{\epsilon\to0} \frac{\widehat{Z}_{n,\mathrm{disj}}^{(3)}(A:B:O)}{[\widehat{Z}_n^{(2)}(A:A^c)\widehat{Z}_n^{(2)}(B:B^c)]^n}.
		\label{eq:tripartite-disjoint-moment-ratio}
	\end{align}
\end{subequations}

\paragraph{Disjoint genuine multi-entropy.}
Purity identifies the complementary moments,
\begin{equation}
	\widehat{Z}_{n,\mathrm{disj}}^{(2)}(O:O^c) = \widehat{Z}_{n,\mathrm{disj}}^{(2)}(A\cup B:O).
\end{equation}
Hence at finite cutoff the subtraction is
\begin{equation}
	\GM_{n,\mathrm{disj}}^{(3)}(A:B:O) = \frac{1}{n(1-n)} \log\frac{\widehat{Z}_{n,\mathrm{disj}}^{(3)}(A:B:O)}{[\widehat{Z}_n^{(2)}(A:A^c)\widehat{Z}_n^{(2)}(B:B^c)\widehat{Z}_{n,\mathrm{disj}}^{(2)}(A\cup B:O)]^{n/2}}.
	\label{eq:disjoint-genuine-moment-ratio}
\end{equation}
In the continuum limit, Eq.~\eqref{eq:disjoint-moment-ratios} gives
\begin{equation}
	\GM_{n,\mathrm{disj}}^{(3)}(x) \coloneqq \frac{1}{n(1-n)} \log{\frac{\calR_n^{(3)}(x,x)}{[\calR_n^{(2)}(x,x)]^{n/2}}} = \lim_{\epsilon\to0} \GM_{n,\mathrm{disj}}^{(3)}(A:B:O).
	\label{eq:disjoint-genuine-reduced-function}
\end{equation}

\subsection{OPE channels of the twist correlator}\label{sec:cft-ope-limits}

The limits $x\to0$ and $x\to1$ probe different OPE channels of the four-point function.
In the $s$ channel, the endpoint pairs $(12)(34)$ become $(0,x)(1,\infty)$, and $x\to0$ fuses the twists at $0$ and $x$.
In the $t$ channel, the pairing $(14)(23)$ becomes $(0,\infty)(x,1)$, and $x\to1$ fuses the two inner endpoints.~\footnote{The remaining pairing $(13)(24)$ is the $u$ channel, with limit $x\to\infty$.}
The lowest-dimension exchanged operators determine the leading endpoint powers.

\paragraph{The identity-channel limit.}
A unique normalizable vacuum dominating the identity channel makes the correlation ratios tend to one as the two intervals separate.
The unit-normalized inverse twists obey
\begin{equation}
	\calT_\pi(x,\bar x)\calT_{\pi^{-1}}(0) \sim \abs{x}^{-2\Delta_\pi}\mathbf{1}, \qquad \lim_{x,\bar x\to0} \calR_n^{(q)}(x,\bar x) = 1.
	\label{eq:R-identity-normalization}
\end{equation}
Together with Eq.~\eqref{eq:bipartite-disjoint-moment-ratio}, this gives
\begin{equation}
	\lim_{\epsilon\to0} \left[S_n(A) + S_n(B) - S_{n,\mathrm{disj}}^{(2)}(A\cup B:O)\right] = \frac{1}{n-1} \log{\calR_n^{(2)}(x,x)} \xrightarrow{x\to0} 0.
	\label{eq:disjoint-mutual-information-limit}
\end{equation}
At fixed interval lengths, $x\to0$ means increasing the separation, and factorization into the two single-interval moments makes the mutual information vanish.

Since both $\calR_n^{(2)}$ and $\calR_n^{(3)}$ tend to one in the $s$ channel, Eq.~\eqref{eq:disjoint-genuine-reduced-function} gives
\begin{equation}
	\lim_{x\to0} \GM_{n,\mathrm{disj}}^{(3)}(x) = 0.
	\label{eq:disjoint-genuine-separated-limit}
\end{equation}
The two disjoint intervals retain their individual entanglement with $O$, but those bipartite contributions cancel in the genuine subtraction.

\paragraph{The adjacent fusion channel.}
The adjacent limit combines the two inner endpoint permutations into the junction permutation of the adjacent geometry.
For the tripartite assignment, the disjoint and adjacent tuples have trivial total monodromy,
\begin{equation}
	\sigma_1 \sigma_2 \sigma_3 \sigma_4 = \varsigma_1 \varsigma_2 \varsigma_3 = \sfe^{[2]}.
	\label{eq:endpoint-product-identity}
\end{equation}
The individual endpoints are related by
\begin{equation}
	\varsigma_1 = \sigma_1, \qquad \varsigma_2 = \sigma_2\sigma_3, \qquad \varsigma_3 = \sigma_4.
	\label{eq:endpoint-fusion}
\end{equation}
For the bipartition the inner permutations are inverses of each other, so their product is the identity, while for the tripartition it is the nontrivial $\varsigma_2$.

For the bipartite correlator, the adjacent limit is governed by identity exchange.
\begin{equation}
	\calT_{\sfg^{-1}}(x,\bar x) \calT_{\sfg}(1) \sim \abs{1-x}^{-2\Delta_n^{(2)}} \mathbf{1}.
	\label{eq:bipartite-crossed-twist-ope}
\end{equation}
The remaining two-point function is unit normalized, giving
\begin{equation}
	\lim_{x\to1^-} (1-x)^{2\Delta_n^{(2)}} \calG_n^{(2)}(x,x) = 1.
	\label{eq:bipartite-crossed-residue}
\end{equation}

In the convergent OPE, assume that the junction twist has $C_n^{(3)}\neq0$ and is the unique lowest-dimension state in its sector, separated from the other states by a positive gap.
The unit-normalized OPE coefficient is obtained by inserting the inverse junction twist:
\begin{equation}
	\left\langle\calT_{\varsigma_1^{-1}}(0)\calT_{\varsigma_3^{-1}}(1)\calT_{\varsigma_2^{-1}}(\infty)\right\rangle = (C_n^{(3)})^*.
	\label{eq:crossed-inverse-junction-coefficient}
\end{equation}
Reordering the zero-spin insertions gives the conjugate of Eq.~\eqref{eq:adjacent-three-point-coefficient}, with the sewing phases fixed in Section~\ref{sec:twist-correlators-normalization}.
\begin{equation}
	\calT_{\sigma_2}(x,\bar x) \calT_{\sigma_3}(1) \sim (C_n^{(3)})^* \abs{1-x}^{-\Delta_n^{(3)}} \calT_{\varsigma_2}(1).
	\label{eq:crossed-twist-ope-normalization}
\end{equation}
The remaining three-point function contributes $C_n^{(3)}$, so
\begin{equation}
	\lim_{x\to1^-} (1-x)^{\Delta_n^{(3)}} \calG_n^{(3)}(x,x) = |C_n^{(3)}|^2.
	\label{eq:crossed-residue}
\end{equation}

\paragraph{Finite scaling functions.}
Removing the $t$-channel power defines~\footnote{These definitions agree with the standard two-interval conventions at $q=2$.
The function $\calF_n^{(2)}$ is the scaling function of Ref.~\cite{Calabrese:2009ez}, and $\calR_n^{(2)}$ is the ratio $R_{2,n}$ of Ref.~\cite{Coser:2013qda}.}
\begin{subequations}
	\label{eq:finite-correlation-functions}
	\begin{align}
		\calF_n^{(2)}(x,\bar x) &\coloneqq \abs{1-x}^{2\Delta_n^{(2)}} \calR_n^{(2)}(x,\bar x) = \abs{x(1-x)}^{2\Delta_n^{(2)}} \calG_n^{(2)}(x,\bar x),
		\label{eq:finite-bipartite-function}\\
		\calF_n^{(3)}(x,\bar x) &\coloneqq \abs{1-x}^{\Delta_n^{(3)}} \calR_n^{(3)}(x,\bar x) = \abs{x}^{2\Delta_n^{(3)}} \abs{1-x}^{\Delta_n^{(3)}} \calG_n^{(3)}(x,\bar x).
		\label{eq:finite-tripartite-function}
	\end{align}
\end{subequations}
The preceding OPE limits give
\begin{subequations}
	\label{eq:finite-function-endpoint-limits}
	\begin{align}
		\lim_{x\to0} \calF_n^{(q)}(x,x) &= 1 \quad (q=2,3),
		\label{eq:finite-function-endpoint-limits-identity}\\
		\lim_{x\to1^-} \calF_n^{(2)}(x,x) &= 1,
		\label{eq:finite-function-endpoint-limits-bipartite-crossed}\\
		\lim_{x\to1^-} \calF_n^{(3)}(x,x) &= |C_n^{(3)}|^2.
		\label{eq:finite-function-endpoint-limits-tripartite-crossed}
	\end{align}
\end{subequations}

On the ordered real line we abbreviate
\begin{equation}
	\begin{gathered}
		\calG_n^{(q)}(x) \coloneqq \calG_n^{(q)}(x,x),\qquad
		\calR_n^{(q)}(x) \coloneqq \calR_n^{(q)}(x,x),\qquad
		\calF_n^{(q)}(x) \coloneqq \calF_n^{(q)}(x,x),\\
		0<x<1,\qquad q\in\{2,3\}.
	\end{gathered}
	\label{eq:real-cross-ratio-functions}
\end{equation}
The definitions above give the real-line conversions
\begin{subequations}
	\label{eq:real-cross-ratio-relations}
	\begin{align}
		\calG_n^{(2)}(x) &= \frac{\calR_n^{(2)}(x)}{x^{2\Delta_n^{(2)}}} = \frac{\calF_n^{(2)}(x)}{[x(1-x)]^{2\Delta_n^{(2)}}},
		\label{eq:real-cross-ratio-relations-2}\\
		\calG_n^{(3)}(x) &= \frac{\calR_n^{(3)}(x)}{x^{2\Delta_n^{(3)}}} = \frac{\calF_n^{(3)}(x)}{x^{2\Delta_n^{(3)}}(1-x)^{\Delta_n^{(3)}}}.
		\label{eq:real-cross-ratio-relations-3}
	\end{align}
\end{subequations}

\paragraph{The factor of two in the adjacent limit.}
The adjacent limit of the disjoint genuine multi-entropy is twice the strictly adjacent value when $C_n^{(3)}>0$.
To see this, first use $\Delta_n^{(3)}=n\Delta_n^{(2)}$.
The powers removed from $\calR$ cancel in the genuine ratio, so at every $0<x<1$,
\begin{equation}
	\GM_{n,\mathrm{disj}}^{(3)}(x) = \frac{1}{n(1-n)} \log{\frac{\calF_n^{(3)}(x)}{[\calF_n^{(2)}(x)]^{n/2}}}.
	\label{eq:disjoint-genuine-finite-function}
\end{equation}
With the OPE assumptions above, the finite endpoint limits give
\begin{equation}
	\lim_{x\to1^-} \frac{\calR_n^{(3)}(x)}{[\calR_n^{(2)}(x)]^{n/2}} = \lim_{x\to1^-} \frac{\calF_n^{(3)}(x)}{[\calF_n^{(2)}(x)]^{n/2}} = |C_n^{(3)}|^2.
	\label{eq:disjoint-genuine-adjacent-residue}
\end{equation}
The residue therefore gives
\begin{equation}
	\lim_{\epsilon\to0} \GM_{n,\mathrm{adj}}^{(3)}(A:B:O) = \kappa_n^{(3)} = \frac12 \lim_{x\to1^-} \GM_{n,\mathrm{disj}}^{(3)}(x).
	\label{eq:adjacent-half-disjoint-genuine-limit}
\end{equation}
The limit in Eq.~\eqref{eq:adjacent-half-disjoint-genuine-limit} takes the continuum limit with distinct endpoints before fusion.\footnote{At $n=2$, Refs.~\cite{Harper:2024ker,Harper:2025uui} recover the strictly adjacent value using a vertex-removal prescription. In Ref.~\cite{Harper:2024ker}, the interval being removed is assigned length $\epsilon/4$, rather than $\epsilon$ as in the ordinary two-interval problem.}

\paragraph{Crossing of the labelled correlators.}
The permutation data explain why the endpoint values need not agree.
Under $z\mapsto1-z$, the tuple attached to the ordered points $(0,x,1,\infty)$ becomes
\begin{equation}
	(\sfg_1,\sfg_1^{-1},\sfg_2,\sfg_2^{-1}) \longmapsto (\sfg_2,\sfg_1^{-1},\sfg_1,\sfg_2^{-1})
	\label{eq:discussion-crossed-tuple}
\end{equation}
at $(0,1-x,1,\infty)$.
The first pair in the original tuple has trivial monodromy, whereas the first pair in the crossed tuple does not.
A simultaneous replica relabelling cannot change that property.
Crossing therefore relates different labelled correlators and does not require $\calF_n^{(3)}(x)=\calF_n^{(3)}(1-x)$.

\section{Replica geometry}\label{sec:replica-geometry}
The compact-boson winding sum requires the period matrix in a complete integral symplectic basis of $H_1(\Sigma,\bbZ)$, also called a Torelli marking.
We construct these data for the disjoint four-point ratio $\calR_n^{(3)}(x;\eta)$ and the adjacent three-point coefficient $C_n^{(3)}(\eta)$.
The two replica covers share root coordinates, character forms and the square-cycle construction of Section~\ref{sec:replica-common-cycles}, but have different genera and require different complete markings.

\subsection{Replica covers}\label{sec:replica-common-geometry}
Fix an integer $n\geq2$, and put $y=1-x$ with $0<x<1$ for the disjoint configuration.
The twist positions and physical cross ratio are specified in $z$; $w$ and its roots $u,v$ provide coordinates for the forms and cycles.
We choose the positive root rotations for geometric coordinates
\begin{equation}
    \zeta_n\coloneqq e^{2\pi i/n}, \qquad \sfh_1(u,v)\coloneqq (\zeta_nu,v), \qquad \sfh_2(u,v)\coloneqq (u,\zeta_nv).
    \label{eq:common-root-rotations}
\end{equation}
These generate the common deck group.
A root character $\bmell$ has eigenvalues $\zeta_n^{\ell_j}$ under $\sfh_j^*$, while a physical character $\bmk$ has eigenvalues $\zeta_n^{k_j}$ under $\sfg_j^*$.
Both are coordinates on the same character group $(\bbZ_n)^2$.
Each pair of root equations below has $n^2$ sheets.

\paragraph{The cover for disjoint intervals.}
The physical coordinate $z$ places the four insertions at $0,x,1,\infty$.
Their monodromies are $(\sfg_1,\sfg_1^{-1},\sfg_2,\sfg_2^{-1})$.
The M\"obius transformation to geometric coordinates
\begin{subequations}
    \label{eq:common-base-coordinate}
    \begin{gather}
    w\coloneqq\frac{z-x}{yz}
    \label{eq:common-base-coordinate-forward}\\
    z=\frac{x}{1-yw}
    \label{eq:common-base-coordinate-inverse}
    \end{gather}
\end{subequations}
sends these points to $\infty,0,1,y^{-1}$.
The two root equations are
\begin{equation}
    u^n=w, \qquad v^n=\frac{1-w}{1-yw}, \qquad u^n+v^n-yu^nv^n=1.
    \label{eq:disjoint-cover-equations}
\end{equation}
They define the degree-$n$ maps
\begin{equation}
	f_u(u)\coloneqq u^n, \qquad f_v(v)\coloneqq \frac{1-v^n}{1-yv^n}, \qquad f_u,f_v:\bbP^1\longrightarrow\bbP^1_w,
	\label{eq:disjoint-cover-maps}
\end{equation}
with disjoint branch sets $\{0,\infty\}$ and $\{1,y^{-1}\}$.
Their fiber product
\begin{equation}
	\Sigma_n^{(3)}(x) \coloneqq \bbP^1_u\times_{\bbP^1_w}\bbP^1_v
	\label{eq:disjoint-cover-definition}
\end{equation}
is compact and smooth because at least one factor map is unramified at every point of the fiber product.
The defining equation therefore has a nonzero derivative in one variable.
The same compact surface admits the cyclic projection $v^n=(1-u^n)/(1-yu^n)$ to the $u$ sphere, with $n$ distinct zeros and $n$ distinct poles.
After moving one pole to infinity, it belongs to the cyclic family of
Refs.~\cite{Enolski:2003hia, Enolski:2006sdf}, with their parameters $N=n$ and $m=n-1$.
The physical projection to the $z$ sphere still has degree $n^2$.

The local parameters over $w=0,1,y^{-1},\infty$ are respectively $u,v,v^{-1},u^{-1}$.
Each fiber consists of $n$ points of ramification index $n$, and there is no ramification elsewhere.

In the common root frame the physical replica generators act by
\begin{equation}
	\sfg_1:(u,v)\longmapsto(\zeta_n^{-1}u,v), \qquad \sfg_2:(u,v)\longmapsto(u,\zeta_nv), \qquad \sfh_1=\sfg_1^{-1},\quad \sfh_2=\sfg_2.
	\label{eq:disjoint-deck-action}
\end{equation}
Thus root sheet $(i,j)$ has replica label $\bmr=(-i,j)$.
Reading the zeros and poles of Eq.~\eqref{eq:disjoint-cover-equations} in the physical insertion order gives the endpoint monodromies~\cite{Harper:2024ker}.

\paragraph{The cover for adjacent intervals.}
The adjacent three-point configuration has no cross ratio.
After sending its physical endpoints to $0,1,\infty$, we use the same coordinate for the geometric presentation and set $w=z$.
The endpoint monodromies are
\begin{equation}
    (\varsigma_1,\varsigma_2,\varsigma_3) = (\sfg_1,\sfg_1^{-1}\sfg_2,\sfg_2^{-1}), \qquad \varsigma_1\varsigma_2\varsigma_3=1.
    \label{eq:fermat-monodromy}
\end{equation}
Their compact cover is the smooth projective Fermat curve
\begin{equation}
	F_n \coloneqq \bigl\{[U:V:W]\in\bbP^2\mid U^n+V^n=W^n\bigr\}, \qquad u\coloneqq U/W,\quad v\coloneqq V/W,
	\label{eq:fermat-projective-curve}
\end{equation}
with affine presentation
\begin{equation}
	u^n=w, \qquad v^n=1-w, \qquad u^n+v^n=1.
	\label{eq:fermat-affine-cover}
\end{equation}
The defining polynomial has no simultaneous zero of its three partial derivatives in projective space.
The projective curve normalizes the root fiber product.
Its shared branch value at infinity resolves into the $n$ points $[1:\rho\zeta_n^j:0]$, $0\leq j<n$, where $\rho\coloneqq e^{\pi i/n}$~\cite{TorresNova:2018fermat}.
Near the point $[1:\rho\zeta_n^j:0]$, introduce
\begin{equation}
    p\coloneqq\frac1u,\qquad h\coloneqq\frac vu,
    \qquad 1+h^n=p^n,\qquad w^{-1}=p^n.
    \label{eq:fermat-infinity-chart}
\end{equation}
The implicit-function theorem gives an analytic branch $h=h_j(p)$ with $h_j(0)=\rho\zeta_n^j$.
Here $p=0$ is the chosen point at infinity, and $p$ is a local coordinate because the derivative with respect to $h$ is nonzero there.
The relation $w^{-1}=p^n$ gives ramification index $n$.
The same index occurs above $0$ and $1$, with local coordinates $u$ and $v$, and there is no other ramification.

The physical sheet assignment is
\begin{equation}
	(u_{\bmr},v_{\bmr}) = (\zeta_n^{r_1+r_2}u,\zeta_n^{r_2}v), \qquad
	(r_1,r_2)=(i-j,j).
	\label{eq:fermat-sheet-labels}
\end{equation}
Consequently
\begin{equation}
	\sfg_1:(u,v)\longmapsto(\zeta_nu,v), \qquad \sfg_2:(u,v)\longmapsto(\zeta_nu,\zeta_nv), \qquad \sfh_1=\sfg_1,\quad \sfh_2=\sfg_1^{-1}\sfg_2.
	\label{eq:fermat-deck-action}
\end{equation}
The positive monodromies in $w$ are $\sfh_1,\sfh_2,(\sfh_1\sfh_2)^{-1}$, as required by Eq.~\eqref{eq:fermat-monodromy}.

\paragraph{Genera and character dictionaries}
In both covers the positive loops about $w=0$ and $w=1$ generate $\sfh_1$ and $\sfh_2$.
Their monodromy action is transitive on the $n^2$ sheets, proving connectedness.
The $n^2$ root rotations then exhaust the deck group of the connected cover.
Each branch fiber contains $n$ points of index $n$.
The Riemann--Hurwitz formula for the two degree-$n^2$ covers therefore gives
\begin{subequations}
    \label{eq:common-cover-genera}
    \begin{align}
        2g_d-2&=-2n^2+4n(n-1),&g_d&=(n-1)^2,
        \label{eq:disjoint-cover-genus}\\
        2g_F-2&=-2n^2+3n(n-1),&g_F&=\frac{(n-1)(n-2)}2.
        \label{eq:fermat-genus}
    \end{align}
\end{subequations}

The sheet assignments above give the dual character dictionaries.
For disjoint intervals,
\begin{equation}
    \bmell=(-k_1,k_2), \qquad \bmk=(-\ell_1,\ell_2).
    \label{eq:disjoint-character-dictionary}
\end{equation}
For adjacent intervals,
\begin{equation}
    \bmell=(k_1,k_2-k_1), \qquad \bmk=(\ell_1,\ell_1+\ell_2).
    \label{eq:fermat-character-dictionary}
\end{equation}
Both maps are invertible modulo every $n$ and preserve the character pairing $\bmk\cdot\bmr=\ell_1 i+\ell_2 j$ modulo $n$.
Table~\ref{tab:replica-cover-comparison} collects these covering and character data.

\begin{table}[t]
    \centering
    \begin{tabular}{@{}p{.24\linewidth}p{.36\linewidth}p{.32\linewidth}@{}}
        \toprule
        & Disjoint intervals & Adjacent intervals\\
        \midrule
        Branch values in $w$ & $0,1,y^{-1},\infty$ & $0,1,\infty$\\
        Covering equations & $u^n=w$, $v^n=(1-w)/(1-yw)$ & $u^n=w$, $v^n=1-w$\\
        Monodromies & $\sfh_1,\sfh_2,\sfh_2^{-1},\sfh_1^{-1}$ & $\sfh_1,\sfh_2,(\sfh_1\sfh_2)^{-1}$\\
        Replica label of $(i,j)$ & $(-i,j)$ & $(i-j,j)$\\
        Genus & $g_d=(n-1)^2$ & $g_F=(n-1)(n-2)/2$\\
        Characters & $1\leq \ell_1,\ell_2\leq n-1$
        & $\ell_1,\ell_2\geq1$, $\ell_1+\ell_2<n$\\
        Period integrals & Hypergeometric functions & Beta functions\\
        Modulus & Physical cross ratio $x$ & None\\
        \bottomrule
    \end{tabular}
    \caption{Replica covers in common root coordinates.
    The physical disjoint insertion order is $0,x,1,\infty$ in $z$, whose images in $w$ are $\infty,0,1,y^{-1}$.
    Both covers have degree $n^2$ and deck group $(\bbZ_n)^2$.
    The character ranges label bases of holomorphic one-forms.}
    \label{tab:replica-cover-comparison}
\end{table}

\paragraph{A common character frame.}
Define
\begin{equation}
	\omega_{\bmell}(y) \coloneqq \frac{u^{\ell_1-1}v^{\ell_2-n}}{1-yu^n}\dd{u}, \qquad
	\sfh_1^*\omega_{\bmell} = \zeta_n^{\ell_1}\omega_{\bmell}, \qquad
	\sfh_2^*\omega_{\bmell} = \zeta_n^{\ell_2}\omega_{\bmell}.
	\label{eq:common-character-forms}
\end{equation}
Here $\bmell=(\ell_1,\ell_2)$ uses integer representatives of a root character, and $y=0$ in this formula denotes the Fermat curve.
Regularity at the branch fibers and at Fermat infinity selects different character sets on the two covers.

\begin{samepage}
	\begin{lemma}[Holomorphic character bases]\label{lem:common-holomorphic-bases}
		The character forms give the complete decompositions
		\begin{subequations}\label{eq:common-holomorphic-bases}
			\begin{align}
				H^{1,0}(\Sigma_n^{(3)}(x)) &= \bigoplus_{\ell_1,\ell_2=1}^{n-1} \bbC\,\omega_{\bmell}(y),
				\label{eq:disjoint-holomorphic-basis}\\
				H^{1,0}(F_n) &= \bigoplus_{\substack{\ell_1,\ell_2\geq1\\\ell_1+\ell_2<n}} \bbC\,\omega_{\bmell}(0).
				\label{eq:fermat-complete-basis}
			\end{align}
		\end{subequations}
	\end{lemma}
\end{samepage}

Differentiating the covering equation gives
\begin{equation}
	\omega_{\bmell}(y) = -\frac{u^{\ell_1-n}v^{\ell_2-1}}{1-yv^n}\dd{v}.
	\label{eq:fermat-form-at-one}
\end{equation}
For the disjoint cover, Table~\ref{tab:disjoint-differential-orders} lists the local coordinates and the orders of the character forms at the four branch fibers.
\begin{table}[t]
    \centering
    \begin{tabular}{@{}lcccc@{}}
        \toprule
        Branch value $w(P)$ & $0$ & $1$ & $y^{-1}$ & $\infty$\\
        Local coordinate & $u$ & $v$ & $v^{-1}$ & $u^{-1}$\\
        $\ord_P\omega_{\bmell}(y)$
        & $\ell_1-1$ & $\ell_2-1$
        & $n-\ell_2-1$ & $n-\ell_1-1$\\
        \bottomrule
    \end{tabular}
    \caption{Local orders of the character differentials on the disjoint cover.
    Each branch fiber contains $n$ points with the displayed local behavior.}
    \label{tab:disjoint-differential-orders}
\end{table}
For example, at $w=y^{-1}$ the parameter $v^{-1}$ has a simple zero, $1-yu^n$ has order $n$, and $\dd{u}$ has order $n-1$.
All four orders are nonnegative exactly when $1\leq \ell_1,\ell_2\leq n-1$.

On Fermat, the finite branch fibers give the same first two orders.
The forms are
\begin{equation}
	\omega_{\bmell}(0) = u^{\ell_1-1}v^{\ell_2-n}\dd{u} = \frac1n w^{\ell_1/n-1}(1-w)^{\ell_2/n-1}\dd{w}, \qquad \ell_1,\ell_2\geq1,\quad \ell_1+\ell_2<n.
	\label{eq:fermat-holomorphic-forms}
\end{equation}
At the $j$th point at infinity, Eq.~\eqref{eq:fermat-infinity-chart} gives
\begin{equation}
    \omega_{\bmell}(0)
    =-h_j(p)^{\ell_2-n}p^{n-\ell_1-\ell_2-1}\dd{p}.
    \label{eq:fermat-form-at-infinity}
\end{equation}
Since $h_j(0)\ne0$, regularity at all $n$ points at infinity is equivalent to $\ell_1+\ell_2<n$.
On Fermat the physical replica characters are
\begin{equation}
	\sfg_1^*\omega_{\bmell}(0)=\zeta_n^{\ell_1}\omega_{\bmell}(0),\qquad
	\sfg_2^*\omega_{\bmell}(0)=\zeta_n^{\ell_1+\ell_2}\omega_{\bmell}(0).
	\label{eq:fermat-form-characters}
\end{equation}
Thus the chosen adjacent physical representatives are $\bmk=(\ell_1,\ell_1+\ell_2)$ with $1\leq k_1<k_2\leq n-1$.

Averaging a linear relation against each root character isolates its coefficient, proving independence of the allowed forms.
Their counts equal the respective genera, proving completeness in Lemma~\ref{lem:common-holomorphic-bases}.
The Fermat basis is empty at $n=2$ and consists of $\dd{u}/v^2$ at $n=3$.

\subsection{Common square cycles}\label{sec:replica-common-cycles}
The lifted interval $0\leq w\leq1$ gives a common family of closed squares.
Their completion to a full integral homology basis depends on the remaining branch values.

\paragraph{Closing the lifted arc.}
Choose positive roots over $0<w<1$, and orient the lifted arc $\lambda$ from $(0,1)$ to $(1,0)$.
Its translates satisfy
\begin{equation}
	\lambda_{i,j} \coloneqq \sfh_1^i\sfh_2^j\lambda: b_j=(0,\zeta_n^j)\longrightarrow w_i=(\zeta_n^i,0), \qquad (i,j)\in(\bbZ_n)^2.
	\label{eq:fermat-pham-arc}
\end{equation}
There is one edge for every pair of endpoints, so their union is an embedded $K_{n,n}$.
Taking differences in both root directions cancels the endpoints and gives the closed Pham square
\begin{equation}
    e_{i,j}\coloneqq(1-\sfh_1)(1-\sfh_2)\lambda_{i,j}
    =\lambda_{i,j}-\lambda_{i+1,j}-\lambda_{i,j+1}+\lambda_{i+1,j+1}.
    \label{eq:fermat-pham-square}
\end{equation}
On the Fermat curve this is the negative of the Pochhammer generator used in Ref.~\cite{Ejder:2018fermat}.
The corresponding Pochhammer contour is a commutator of loops about the two branch values.
Its lift closes because the two monodromies commute.
To specify its orientation, fix a base point below the interval, access paths in the lower half-plane, and the reference sheet by continuation from $\lambda$.
Let $\eta_0,\eta_1$ be positive based loops about $w=0,1$, respectively.
With rightmost-first composition, the word
\begin{equation}
	\eta_0^{-1}\eta_1^{-1}\eta_0\eta_1
	\label{eq:common-square-word}
\end{equation}
first circles the right endpoint positively, then the left endpoint positively, then the right endpoint negatively, and finally the left endpoint negatively.
Its lifted traversals visit the root sheets
\begin{equation}
	(i,j) \longrightarrow (i,j+1) \longrightarrow (i+1,j+1) \longrightarrow (i+1,j) \longrightarrow (i,j).
	\label{eq:common-square-sheet-sequence}
\end{equation}
Shrinking the endpoint circles leaves the four oriented interval lifts in
Eq.~\eqref{eq:fermat-pham-square}.
The access-path coefficients are $(1-\sfh_1)(1-\sfh_2)$ at the right endpoint and its negative at the left, as derived in Appendix~\ref{sec:replica-intersections}.

\paragraph{Periods of the common square.}
The period of the square reduces to the integral on its reference arc.
In the character frame of Eq.~\eqref{eq:common-character-forms}, this is
\begin{align}
	L_{\bmell}(y) \coloneqq \int_\lambda\omega_{\bmell}(y) &= \frac1n\int_0^1 w^{\ell_1/n-1}(1-w)^{\ell_2/n-1}(1-yw)^{-\ell_2/n}\dd{w} \notag\\
	&= \frac1n B(\ell_1/n,\ell_2/n) {}_2F_1\!\left(\ell_1/n,\ell_2/n;(\ell_1+\ell_2)/n;y\right).
	\label{eq:common-euler-period}
\end{align}
The Pochhammer traversal therefore multiplies this open-arc period by $(1-\zeta_n^{\ell_1})(1-\zeta_n^{\ell_2})$, and translation by $(i,j)$ multiplies it by $\zeta_n^{i\ell_1+j\ell_2}$.
In physical character coordinates, the square multiplier is $(1-\zeta_n^{-k_1})(1-\zeta_n^{k_2})$ for disjoint intervals and $(1-\zeta_n^{k_1})(1-\zeta_n^{k_2-k_1})$ for adjacent intervals.
On the triangular root-character set, $y=0$ gives the Fermat beta period.

\begin{figure}[t]
	\centering
	\begin{tikzpicture}[x=1.05cm,y=0.9cm,>=stealth,font=\footnotesize]
	\node at (1.75,3.15) {(a) Closed square};
	\coordinate (bj) at (0.55,2.25);
	\coordinate (wi) at (0.55,0.35);
	\coordinate (bjp) at (3.05,0.35);
	\coordinate (wip) at (3.05,2.25);
	\node[above=3pt] at (bj) {$b_j$};
	\node[below=3pt] at (bjp) {$b_{j+1}$};
	\node[below=3pt] at (wi) {$w_i$};
	\node[above=3pt] at (wip) {$w_{i+1}$};

	\draw[->,shorten <=2.4pt,shorten >=2.4pt]
	(bj) -- node[left] {$\lambda_{i,j}$} (wi);
	\draw[<-,shorten <=2.4pt,shorten >=2.4pt]
	(bjp) -- node[below] {$\lambda_{i,j+1}$} (wi);
	\draw[<-,shorten <=2.4pt,shorten >=2.4pt]
	(bj) -- node[above] {$\lambda_{i+1,j}$} (wip);
	\draw[->,shorten <=2.4pt,shorten >=2.4pt]
	(bjp) -- node[right] {$\lambda_{i+1,j+1}$} (wip);
	\fill (bj) circle (2.4pt) (bjp) circle (2.4pt);
	\draw[fill=white] (wi) circle (2.4pt) (wip) circle (2.4pt);

	\node at (6.75,3.15) {(b) $u=0$};
	\coordinate (ub) at (6.75,1.35);
	\foreach \ang/\lab in {20/{\lambda_{i,j}},110/{\lambda_{i+1,j}},200/{\lambda_{i+2,j}},290/{\cdots}}{
		\draw (ub)--++(\ang:0.85);
		\path (ub) ++(\ang:1.0) node {$\lab$};
	}
	\fill (ub) circle (2.6pt);
	\draw[->] (ub) ++(20:0.58) arc[start angle=20,end angle=110,radius=0.58];
	\node[below=5pt] at (ub) {$b_j$};
	\node at (10.75,3.15) {(c) $v=0$};
	\coordinate (uw) at (10.75,1.35);
	\foreach \ang/\lab in {20/{\lambda_{i,j}},110/{\lambda_{i,j+1}},200/{\lambda_{i,j+2}},290/{\cdots}}{
		\draw (uw)--++(\ang:0.85);
		\path (uw) ++(\ang:1.0) node {$\lab$};
	}
	\draw[fill=white] (uw) circle (2.6pt);
	\draw[->] (uw) ++(20:0.58) arc[start angle=20,end angle=110,radius=0.58];
	\node[below=5pt] at (uw) {$w_i$};
\end{tikzpicture}
	\caption{The common closed square and its ribbon order.
	The arrows in panel (a) follow the four signed edges of Eq.~\eqref{eq:fermat-pham-square}.
	Panels (b,c) use the local $u$ and $v$ coordinates, where increasing the incident root label turns counterclockwise.
	The same rule applies on both replica covers.}
	\label{fig:fermat-pham-cycles}
\end{figure}
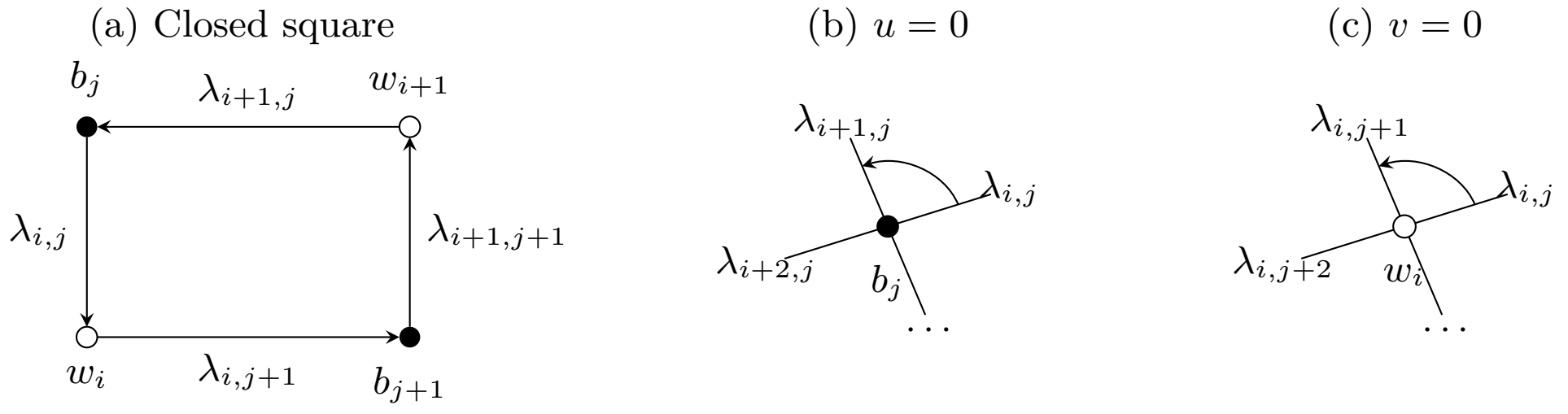

\paragraph{Intersections from the ribbon order.}
The local parameters at the two types of vertex are $u$ and $v$ in both covers.
Increasing the incident root label therefore turns counterclockwise at each vertex (Fig.~\ref{fig:fermat-pham-cycles}), fixing a common ribbon order and intersection kernel.

The local half-edge rule for these two vertex families is Eq.~\eqref{eq:fermat-ribbon-pairing} in Appendix~\ref{sec:replica-intersections}.
Substituting the four coefficients of $e_{0,0}$ and $e_{p,q}$ gives the kernel $I(p,q)\coloneqq e_{0,0}\circ e_{p,q}$.
\begin{align}
	I(p,q)
	&= \frac12(\delta_{p,1}-\delta_{p,-1}) (2\delta_{q,0}-\delta_{q,1}-\delta_{q,-1})
	+ \frac12(2\delta_{p,0}-\delta_{p,1}-\delta_{p,-1}) (\delta_{q,1}-\delta_{q,-1})
	\nonumber\\
	&= \delta_{p,0}(\delta_{q,1}-\delta_{q,-1})
	+ \delta_{q,0}(\delta_{p,1}-\delta_{p,-1})
	- \delta_{p,1}\delta_{q,1}
	+ \delta_{p,-1}\delta_{q,-1}.
	\label{eq:fermat-intersection-kernel}
\end{align}
All deltas are modulo $n$.
The two terms in the first line are respectively the contributions from the black and white vertices.
Thus the positive nearest-neighbour pairings have displacements $(1,0)$ and $(0,1)$, whereas the diagonal displacement $(1,1)$ has negative sign.
Deck invariance supplies the remaining pairings.
\begin{equation}
	e_{i,j} \circ e_{k,l}=I(k-i,l-j).
	\label{eq:fermat-orbit-intersection}
\end{equation}
Adding coincident displacement classes gives the vanishing kernel at $n=2$.

\paragraph{The common neighborhood and its two completions.}
A small base disc enclosing $[0,1]$ lifts to a connected regular neighborhood of the graph.
Its Euler characteristic is $2n-n^2$.
The boundary monodromy $\sfh_1\sfh_2$ has $n$ cycles of length $n$, so there are $n$ boundary components and the neighborhood has genus $g_F$.
On Fermat the complementary disc contains only the branch value at infinity, and each of its $n$ lifts is a disc.
On the cover for disjoint intervals the complementary base disc contains $y^{-1}$ and $\infty$, whose two independent monodromies give another connected surface.
Its Euler characteristic is again $2n-n^2$, and its boundary count is again $n$, so its genus is $g_F$.
Gluing the two neighborhoods along all $n$ boundary components gives
\begin{equation}
	g_d=2g_F+n-1.
	\label{eq:common-genus-decomposition}
\end{equation}

At finite $x$, this decomposes the surface for disjoint intervals into two bordered pieces of genus $g_F$.
The graph has cycle rank $(n-1)^2$, and its $n$ oriented boundary cycles have one relation.
Capping these boundaries on Fermat leaves rank $2g_F$.
The disjoint cover requires cycles through the complementary piece to complete its rank-$2g_d$ homology.
The compact Fermat components themselves arise only in the degeneration of Section~\ref{sec:adjacent-degeneration}.

\subsection{Integral cycles on the cover for disjoint intervals}\label{sec:disjoint-integral-homology}

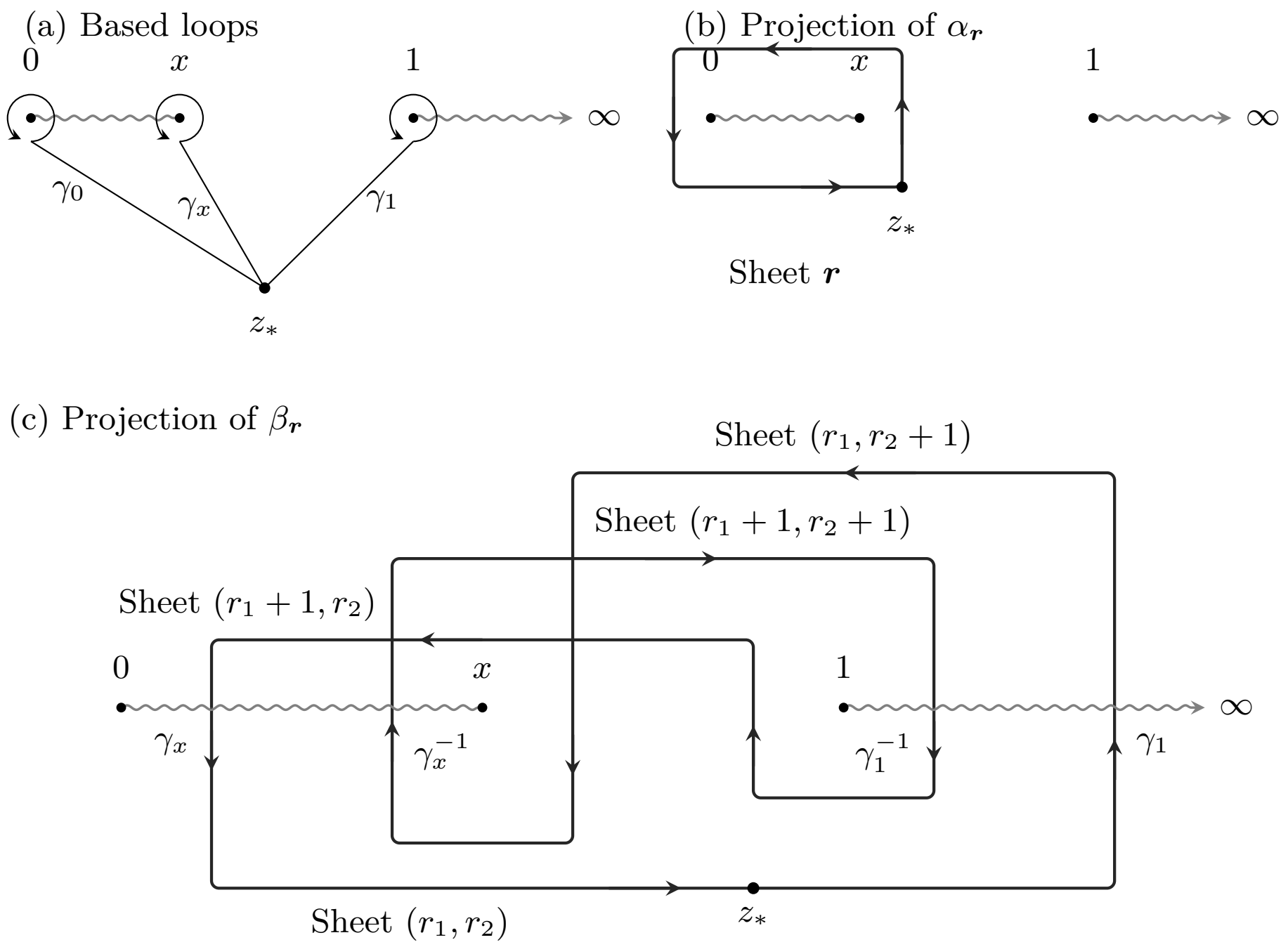
\begin{figure}[t]
	\centering
	\begin{tikzpicture}[x=1cm,y=1cm,>=stealth,font=\footnotesize,
	cut/.style={draw=black!50,line width=.65pt,decorate,
	decoration={snake,amplitude=.6pt,segment length=5pt}},
	contour/.style={draw=black!85,line width=.8pt,rounded corners=2pt}]
	\begin{scope}[xshift=.15cm]
		\node[anchor=west] at (-.2,.85) {(a) Based loops};
		\draw[cut] (0,0)--(1.4,0);
		\draw[cut,->,decoration={snake,amplitude=.6pt,segment length=5pt,post length=3pt}]
		(3.6,0)--(5.1,0);
		\foreach \a/\lab in {0/0,1.4/x,3.6/1}{
			\fill (\a,0) circle (1.3pt);
			\node[above=9pt] at (\a,0) {$\lab$};
			\draw[->] (\a,-.22) arc[start angle=-90,end angle=245,radius=.22];
		}
		\node[right] at (5.1,0) {$\infty$};
		\coordinate (base) at (2.2,-1.6);
		\draw (base)--(0,-.22);
		\draw (base)--(1.4,-.22);
		\draw (base)--(3.6,-.22);
		\fill (base) circle (1.5pt);
		\node[below=4pt] at (base) {$z_*$};
		\node[left] at (.65,-.7) {$\gamma_0$};
		\node[left] at (1.85,-.85) {$\gamma_x$};
		\node[right] at (3.02,-.78) {$\gamma_1$};
	\end{scope}

	% The deformed alpha representative stays on its starting sheet.
	\begin{scope}[xshift=6.55cm]
		\node[anchor=west] at (-.4,.85) {(b) Projection of $\alpha_{\bmr}$};
		\draw[cut] (0,0)--(1.4,0);
		\draw[cut,->,decoration={snake,amplitude=.6pt,segment length=5pt,post length=3pt}]
		(3.6,0)--(4.9,0);
		\foreach \a/\lab in {0/0,1.4/x,3.6/1}{
			\fill (\a,0) circle (1.3pt);
			\node[above=9pt] at (\a,0) {$\lab$};
		}
		\node[right] at (4.9,0) {$\infty$};
		\draw[contour] (-.35,-.65)--(1.8,-.65)--(1.8,.65)
		--(-.35,.65)--cycle;
		\foreach \xa/\ya/\xb/\yb in {
			.85/-.65/1.25/-.65,1.8/-.25/1.8/.2,
			.95/.65/.5/.65,-.35/.2/-.35/-.25
		}{
			\draw[black!85,line width=.8pt,->] (\xa,\ya)--(\xb,\yb);
		}
		\fill (1.8,-.65) circle (1.5pt);
		\node[below=4pt] at (1.8,-.65) {$z_*$};
		\node[below] at (.7,-1.2) {Sheet $\bmr$};
	\end{scope}

	% Four portions, with sheet changes only at the wavy cuts.
	\node[anchor=west] at (-.2,-2.85) {(c) Projection of $\beta_{\bmr}$};
	\begin{scope}[shift={(1,-5.55)},x=1.7cm,y=.85cm]
		\draw[contour] (.5,0)--(.5,-2)--(5.5,-2)--(5.5,0);
		\draw[contour] (5.5,0)--(5.5,2.6)--(2.5,2.6)
		--(2.5,-1.5)--(1.5,-1.5)--(1.5,0);
		\draw[contour] (1.5,0)--(1.5,1.65)--(4.5,1.65)--(4.5,0);
		\draw[contour] (4.5,0)--(4.5,-1)--(3.5,-1)
		--(3.5,.75)--(.5,.75)--(.5,0);
		\foreach \xa/\ya/\xb/\yb in {
			2.7/-2/3.1/-2,5.5/-.8/5.5/-.35,4.4/2.6/4/2.6,
			2.5/-.3/2.5/-.75,1.5/-.6/1.5/-.15,2.9/1.65/3.3/1.65,
			4.5/-.15/4.5/-.6,3.5/-.65/3.5/-.2,2/.75/1.65/.75,
			.5/-.25/.5/-.7
		}{
			\draw[black!85,line width=.8pt,->] (\xa,\ya)--(\xb,\yb);
		}
		\draw[cut] (0,0)--(2,0);
		\draw[cut,->,decoration={snake,amplitude=.6pt,segment length=5pt,post length=3pt}]
		(4,0)--(6,0);
		\foreach \xx/\lab in {0/0,2/x,4/1}{
			\fill (\xx,0) circle (1.3pt);
			\node[above=4pt] at (\xx,0) {$\lab$};
		}
		\node[right] at (6,0) {$\infty$};
		\node[above] at (4,2.65) {Sheet $(r_1,r_2+1)$};
		\node[above] at (3.5,1.7) {Sheet $(r_1+1,r_2+1)$};
		\node[above] at (.7,.8) {Sheet $(r_1+1,r_2)$};
		\node[below] at (1.6,-2.05) {Sheet $(r_1,r_2)$};
		\fill (3.5,-2) circle (1.6pt);
		\node[below] at (3.5,-2.05) {$z_*$};
		\node[below left=2pt] at (.5,-.05) {$\gamma_x$};
		\node[below right=2pt] at (1.5,-.05) {$\gamma_x^{-1}$};
		\node[below left=2pt] at (4.5,-.05) {$\gamma_1^{-1}$};
		\node[below right=2pt] at (5.5,-.05) {$\gamma_1$};
	\end{scope}
\end{tikzpicture}
	\caption{The physical $z$-plane representatives used to transport the
	disjoint marking.
	(a) The based positive loops have access paths below
	the cuts.
	(b) Alpha surrounds $[0,x]$ counterclockwise on sheet $\bmr$.
	(c) Beta follows the four traversals in Eq.~\eqref{eq:homology-lifted-words-beta}, with physical replica labels on each portion.
	In the common root labels it is the negative translated square of Eq.~\eqref{eq:homology-root-transport-square}.}
	\label{fig:disjoint-cycle-lifts}
\end{figure}

\paragraph{Closed lifts and their root labels.}
Choose $z_*$ in the lower half-plane and access paths to positive small circuits $\gamma_0,\gamma_x,\gamma_1$ around $0,x,1$ as in
Fig.~\ref{fig:disjoint-cycle-lifts}.
Products follow the rightmost path first, consistently with Eq.~\eqref{eq:permutation-representation-law}.
The increments of the physical replica label are
\begin{equation}
	\gamma_0:(1,0),\qquad \gamma_x:(-1,0),\qquad \gamma_1:(0,1).
	\label{eq:homology-loop-increments}
\end{equation}
Define the closed lifts starting at sheet $\bmr$ by
\begin{subequations}
	\label{eq:homology-lifted-words}
	\begin{gather}
	\alpha_{\bmr} \coloneqq \left(\widetilde{\gamma_0\gamma_x}\right)_{\bmr}
	\label{eq:homology-lifted-words-alpha}\\
	\beta_{\bmr} \coloneqq \left(\widetilde{\gamma_x\gamma_1^{-1}\gamma_x^{-1}\gamma_1}\right)_{\bmr}.
	\label{eq:homology-lifted-words-beta}
	\end{gather}
\end{subequations}
Figure~\ref{fig:disjoint-cycle-lifts}(a) fixes the access paths.
Panel (b) shows alpha as a counterclockwise contour around $[0,x]$ on its starting sheet.
Panel (c) shows the four beta traversals $\gamma_1,\gamma_x^{-1},\gamma_1^{-1},\gamma_x$ on the successive sheets $\bmr$, $\bmr+(0,1)$, $\bmr+(1,1)$ and $\bmr+(1,0)$.
Both words return to their starting sheet.
For the reference sheet obtained by continuing the positive common arc through the lower half-plane, contraction of the beta contour gives
\begin{equation}
    e_{i,j}=-\beta_{-i-1,j}.
    \label{eq:homology-root-transport-square}
\end{equation}
Appendix~\ref{sec:replica-intersections} derives this shift and sign from the access chains.

\paragraph{Intersections of the lifted cycles.}
Write $\delta_n(a,b)$ for the Kronecker delta modulo $n$.
The lifted contours have the intersection numbers
\begin{subequations}
    \label{eq:homology-elementary-intersections}
    \begin{align}
        \alpha_{\bmr}\circ\alpha_{\bmr'}&=0,
        \label{eq:homology-alpha-alpha}\\
        \alpha_{\bmr}\circ\beta_{\bmr'}
        &=-[\delta_n(r_1,r_1'+1)-\delta_n(r_1,r_1')]
        [\delta_n(r_2,r_2'+1)-\delta_n(r_2,r_2')],
        \label{eq:homology-alpha-beta}\\
        \beta_{\bmr}\circ\beta_{\bmr'}
        &=\delta_n(r_1,r_1'+1)\delta_n(r_2,r_2')
        -\delta_n(r_1,r_1'+1)\delta_n(r_2,r_2'-1)\notag\\
        &\quad+\delta_n(r_1,r_1')\delta_n(r_2,r_2'-1)
        -\delta_n(r_1,r_1'-1)\delta_n(r_2,r_2')\notag\\
        &\quad+\delta_n(r_1,r_1'-1)\delta_n(r_2,r_2'+1)
        -\delta_n(r_1,r_1')\delta_n(r_2,r_2'+1).
        \label{eq:homology-beta-beta}
    \end{align}
\end{subequations}
For example, the lower-left crossing of $\alpha_{0,0}$ with $\beta_{0,0}$ has a rightward first tangent and a downward second tangent, giving $-1$ in the complex orientation.
The beta pairing follows from the common square kernel and Eq.~\eqref{eq:homology-root-transport-square}.
Appendix~\ref{sec:appendix-disjoint-intersections} derives all the pairings both from the cut-plane crossings and from the ribbon incidences.

\paragraph{Integral completeness of the selected cycles.}
The $2n^2$ closed lifts are redundant, since the homology rank is only $2g_d=2(n-1)^2$.
Consider the $2g_d$ curves
\begin{equation}
	\alpha_{r_1,r_2}, \beta_{r_1,r_2},\qquad 0 \leq r_1,r_2 \leq n-2.
	\label{eq:homology-selected-curves}
\end{equation}
Order the pairs with the first index varying fastest, place all alpha curves before the beta curves, and call the list $C_1,\ldots,C_{2g_d}$.
Their integer span is a subgroup of $H_1(\Sigma_n^{(3)}(x),\bbZ)$, and their intersection matrix is
\begin{equation}
	J_{\mu\nu} \coloneqq C_\mu \circ C_\nu, \qquad 1 \leq \mu,\nu \leq 2g_d.
	\label{eq:homology-intersection-matrix}
\end{equation}
On the selected range, the cyclic deltas in Eq.~\eqref{eq:homology-elementary-intersections} reduce to ordinary deltas.
Define the $(n-1)\times(n-1)$ matrix
\begin{equation}
	D_{ab} \coloneqq \delta_{a,b+1} - \delta_{a,b}, \qquad 0\leq a,b\leq n-2.
	\label{eq:homology-difference-matrix}
\end{equation}
It is lower bidiagonal with diagonal entries $-1$, so $\det D=(-1)^{n-1}$.

Let $M$ and $L$ be the matrices with entries $M_{\bmr,\bmr'}=\alpha_{\bmr}\circ\beta_{\bmr'}$ and $L_{\bmr,\bmr'}=\beta_{\bmr}\circ\beta_{\bmr'}$ on this range.
The vanishing alpha--alpha intersections and antisymmetry then give
\begin{equation}
	J = \mqty( 0 & M \\ -M^{\top} & L ), \qquad M = -D \otimes D, \qquad \det J = (\det M)^2 = 1.
	\label{eq:homology-selected-determinant}
\end{equation}

Any relation among the selected curves would give a null vector of $J$, so its nonzero determinant proves their independence.
For $2g_d$ independent integral curves, unimodularity of the ambient homology pairing makes $\sqrt{\det J}$ the index of their integer span, as shown in Appendix~\ref{sec:math-riemann}.
Equation~\eqref{eq:homology-selected-determinant} gives index one, so this span is all of $H_1(\Sigma_n^{(3)}(x),\bbZ)$.

To recover the relations among all lifts, sum either index in the cyclic-difference formulas.
Each resulting sum has zero intersection with the selected integral basis and hence vanishes in homology.
In particular,
\begin{equation}
	\sum_{r_1=0}^{n-1}\alpha_{r_1,r_2} = \sum_{r_2=0}^{n-1}\alpha_{r_1,r_2} = \sum_{r_1=0}^{n-1}\beta_{r_1,r_2} = \sum_{r_2=0}^{n-1}\beta_{r_1,r_2} = 0,
	\label{eq:homology-orbit-relations}
\end{equation}
with the unsummed component held fixed.
These relations eliminate the last row and column of either family with integer coefficients, leaving the basis in Eq.~\eqref{eq:homology-selected-curves}.
They therefore generate all relations among the raw curves.
For each family, an independent choice consists of the $n$ row relations and the first $n-1$ column relations, with the last column recovered as the row sum minus the other columns.
The resulting quotient is free abelian of rank $2g_d$.

\paragraph{Canonical cycles from cumulative sums.}
The mixed intersection in Eq.~\eqref{eq:homology-alpha-beta} is a product of two finite differences.
Rectangular cumulative sums of alpha curves invert these differences and become dual to the individual beta curves.
The beta curves still intersect one another but adding alpha curves removes those intersections without changing the mixed pairing.
For $1\leq I,J\leq n-1$, define
\begin{subequations}
	\label{eq:homology-prefix-cycles}
	\begin{align}
		A_{I,J} &\coloneqq -\sum_{r_1=0}^{I-1}\sum_{r_2=0}^{J-1}\alpha_{r_1,r_2},
		\label{eq:homology-prefix-A}\\
		B_{I,J} &\coloneqq -e_{-I,J-1} + A_{I,J-1} - A_{I-1,J} + A_{I-1,J+1}.
		\label{eq:homology-prefix-B}
	\end{align}
\end{subequations}
Here $-e_{-I,J-1}=\beta_{I-1,J-1}$, so the B-cycle formula uses the physical marking in Eq.~\eqref{eq:homology-selected-curves}.
Extend the cumulative sums to $0\leq I,J\leq n$.
The empty sums give $A_{0,J}=A_{I,0}=0$, while Eq.~\eqref{eq:homology-orbit-relations} gives $A_{n,J}=A_{I,n}=0$ in homology.
These identities therefore hold for all periods.

\begin{lemma}[Complete disjoint integral marking]\label{lem:disjoint-complete-integral-marking}
	The cycles $A_{I,J},B_{I,J}$ in Eq.~\eqref{eq:homology-prefix-cycles}, with the boundary conventions above and the first index varying fastest, form a complete integral symplectic basis of $H_1(\Sigma_n^{(3)}(x),\bbZ)$.
\end{lemma}

Summing Eq.~\eqref{eq:homology-alpha-beta} over the rectangle in Eq.~\eqref{eq:homology-prefix-A} leaves only its selected corner, with the overall minus sign giving
\begin{equation}
	A_{I,J} \circ \beta_{I'-1,J'-1} = \delta_{II'}\delta_{JJ'}, \qquad A_{I,J} \circ A_{I',J'} = 0.
	\label{eq:homology-prefix-pairing}
\end{equation}
The uncorrected beta curves are therefore already dual to the A-cycles.
To cancel their mutual intersection matrix $L$, write the correction to the $(I,J)$th beta curve as
\begin{equation}
	\sum_{I',J'}F_{(I,J),(I',J')}A_{I',J'}.
\end{equation}
The corrections to the first and second curves contribute $F$ and $-F^\top$, respectively, so the B--B intersection matrix is $L+F-F^\top$.
We therefore need $F-F^{\top}=-L$ to have vanishing B--B intersections.

The six terms in Eq.~\eqref{eq:homology-beta-beta} come in three pairs of opposite sheet displacements.
For each pair, placing the negative of one coefficient in $F$ lets $-F^\top$ supply the term at the opposite displacement.
One such choice is
\begin{equation}
	F_{(I,J),(I',J')} \coloneqq \delta_{I',I}\delta_{J',J-1} - \delta_{I',I-1}\delta_{J',J} + \delta_{I',I-1}\delta_{J',J+1},
	\label{eq:homology-correction-matrix}
\end{equation}
where an index outside $1,\ldots,n-1$ gives zero.
In the shifted indexing $(r_1,r_2)=(I-1,J-1)$, this choice satisfies $L=F^\top-F$ and gives precisely the three A-cycle terms in Eq.~\eqref{eq:homology-prefix-B}.
Hence
\begin{subequations}\label{eq:homology-canonical-pairing}
	\begin{align}
		B_{I,J}\circ B_{I',J'} &= L_{(I,J),(I',J')} + F_{(I,J),(I',J')} - F_{(I',J'),(I,J)}=0,
		\label{eq:homology-canonical-pairing-B}\\
		A_{I,J}\circ B_{I',J'} &= \delta_{II'}\delta_{JJ'}.
		\label{eq:homology-canonical-pairing-mixed}
	\end{align}
\end{subequations}

The backward differences recover the original integral basis as
\begin{equation}
	\alpha_{I-1,J-1} = -A_{I,J}+A_{I-1,J}+A_{I,J-1}-A_{I-1,J-1}.
	\label{eq:homology-prefix-inverse}
\end{equation}
Equation~\eqref{eq:homology-prefix-B} then recovers each selected beta by subtracting its A-cycle correction.
Since the selected raw cycles already span the full integer homology, this inverse integral change completes the proof of Lemma~\ref{lem:disjoint-complete-integral-marking}.

\subsection{Periods of the cover for disjoint intervals}\label{sec:disjoint-period-matrix}

Canonical cycles, holomorphic differentials and period matrices for this cyclic family are given in Ref.~\cite{Enolski:2003hia}.
We evaluate the complete integral marking of Section~\ref{sec:disjoint-integral-homology} in the physical coordinate $z$, using the positive replica characters defined in Eq.~\eqref{eq:disjoint-character-dictionary}.

\subsubsection{Character periods}
In the physical frame, the two real cut integrals take Euler form. Define
\begin{equation}
	t\coloneqq y^{1/n}>0, \qquad u_0\coloneqq\frac1{tu}, \qquad v_0\coloneqq\rho^{-1}tv, \qquad u_0^n=\frac{z}{z-x}, \qquad v_0^n=z-1.
	\label{eq:disjoint-physical-frame}
\end{equation}
The subscript zero distinguishes these computational roots from the common geometric roots.
For $\bmk=(k_1,k_2)$ with $1\leq k_1,k_2\leq n-1$, put
\begin{equation}
	\varpi_{\bmk} \coloneqq \frac{\dd{z}}{(z-x)u_0^{n-k_1}v_0^{n-k_2}} = z^{k_1/n-1}(z-x)^{-k_1/n}(z-1)^{k_2/n-1}\dd{z}.
	\label{eq:disjoint-holomorphic-forms}
\end{equation}
They are related to the common forms by
\begin{equation}
	\omega_{n-k_1,k_2}(y) = Q_{\bmk}\varpi_{\bmk}, \qquad Q_{\bmk} \coloneqq \frac{\rho^{k_2-n}}{n}t^{k_1-k_2}, \qquad \bmell=(n-k_1,k_2).
	\label{eq:common-differential-transport}
\end{equation}
It follows by substituting Eq.~\eqref{eq:disjoint-physical-frame} and $\dd{z}/(z-x)=n\dd{u}/[u(1-yu^n)]$.
Thus the regularity and completeness of the common character basis also hold in this frame.
Its physical replica characters are
\begin{equation}
	\sfg_j^*\varpi_{\bmk}=\zeta_n^{k_j}\varpi_{\bmk}, \qquad j=1,2.
	\label{eq:disjoint-form-characters}
\end{equation}

Over the full square of character labels, the exponents of $y$ in the frame transformation sum to zero.
If its matrix, including the permutation of the first character, is $\sfQ$, then
\begin{subequations}\label{eq:common-period-frame-transport}
	\begin{gather}
		|\det\sfQ|=n^{-g_d}
		\label{eq:common-period-frame-transport-determinant}\\
		\sfA^{(\omega)}=\sfQ\sfA
		\label{eq:common-period-frame-transport-A}\\
		\sfB^{(\omega)}=\sfQ\sfB.
		\label{eq:common-period-frame-transport-B}
	\end{gather}
\end{subequations}
Here unadorned blocks will refer to $\varpi$ and the superscript to $\omega$.

\paragraph{Branch phases and Euler integrals.}
Set
\begin{equation}
	a \coloneqq \frac{k_1}{n}, \qquad b \coloneqq \frac{k_2}{n}, \qquad p_1 \coloneqq \zeta_n^{k_1}, \qquad p_2 \coloneqq \zeta_n^{k_2}.
	\label{eq:period-character-parameters}
\end{equation}
Thus $p_1,p_2$ are the pullback eigenvalues in Eq.~\eqref{eq:disjoint-form-characters}.
Fix the reference sheet over the interval $z \in (x, 1)$ by
\begin{subequations}\label{eq:period-reference-branch}
	\begin{gather}
	u_0(z) = \left(\frac{z}{z-x}\right)^{1/n} > 0
	\label{eq:period-reference-branch-u}\\
	v_0(z) = e^{-i\pi/n}(1-z)^{1/n}.
	\label{eq:period-reference-branch-v}
	\end{gather}
\end{subequations}
Continue these roots through the lower half-plane to the base point $z_*$ and along the connecting paths of Fig.~\ref{fig:disjoint-cycle-lifts}(a).
On the lower bank of the first cut, $0<z<x$, $u_0(z)=e^{i\pi/n}[z/(x-z)]^{1/n}$.
Equation~\eqref{eq:disjoint-holomorphic-forms} fixes the two bank values as
\begin{equation}
	\varpi_{\bmk} = \begin{cases}
		e^{i\pi(1+a-b)}z^{a-1}(x-z)^{-a}(1-z)^{b-1}\dd{z}, &0<z<x,\\
		e^{i\pi(1-b)}z^{a-1}(z-x)^{-a}(1-z)^{b-1}\dd{z}, &x<z<1.
	\end{cases}
	\label{eq:period-bank-phases}
\end{equation}

The real integrals that remain after these phases are separated are
\begin{subequations}\label{eq:period-real-integrals}
	\begin{align}
		I_{\bmk}(x) &\coloneqq \int_0^x z^{a-1}(x-z)^{-a}(1-z)^{b-1}\dd{z},
		\label{eq:period-real-I}\\
		J_{\bmk}(x) &\coloneqq \int_x^1 z^{a-1}(z-x)^{-a}(1-z)^{b-1}\dd{z}.
		\label{eq:period-real-J}
	\end{align}
\end{subequations}
Both are positive and convergent because $0<a,b<1$.
The substitution $z=xs$ and the Euler formulas in Appendix~\ref{sec:math-special} give
\begin{align}
	I_{\bmk}(x) &= \int_0^1 s^{a-1} (1-s)^{-a} (1-xs)^{b-1}\dd{s}\notag\\
	&= B(a,1-a)\,{}_2F_1(a,1-b;1;x) = \frac{\pi}{\sin\pi a}\,{}_2F_1(a,1-b;1;x).
	\label{eq:period-I-euler}
\end{align}
For the second integral use $z=1-(1-x)s$, obtaining
\begin{align}
	J_{\bmk}(x) &= (1-x)^{b-a}\int_0^1 s^{b-1} (1-s)^{-a} [1-(1-x)s]^{a-1}\dd{s}\notag\\
	&=(1-x)^{b-a}B(1-a,b)\,{}_2F_1(1-a,b;1-a+b;1-x).
	\label{eq:period-J-euler}
\end{align}
All hypergeometric functions here are evaluated on their real Euler branch.

\paragraph{Periods of the reference lifts.}
All lifted-cycle periods follow from two reference periods by deck translation.
Define
\begin{subequations}\label{eq:period-raw-definition}
	\begin{gather}
		\calA_{\bmk} \coloneqq\oint_{\alpha_{0,0}}\varpi_{\bmk}
		\label{eq:period-raw-definition-A}\\
		\calB_{\bmk} \coloneqq\oint_{\beta_{0,0}}\varpi_{\bmk}.
		\label{eq:period-raw-definition-B}
	\end{gather}
\end{subequations}
Deck translation gives
\begin{subequations}\label{eq:period-deck-translates}
	\begin{gather}
		\oint_{\alpha_{\bmr}}\varpi_{\bmk} = p_1^{r_1}p_2^{r_2}\calA_{\bmk}
		\label{eq:period-deck-translates-alpha}\\
		\oint_{\beta_{\bmr}}\varpi_{\bmk} = p_1^{r_1}p_2^{r_2}\calB_{\bmk}.
		\label{eq:period-deck-translates-beta}
	\end{gather}
\end{subequations}

To evaluate the reference periods, contract each small branch-point loop to its access path.
If its monodromy multiplies the differential by $\chi$, the outgoing and returning paths contribute $1-\chi$ times the outgoing integral.
For the inverse loop the factor is $1-\chi^{-1}$.
The small circular arcs give zero in the limit because the differential is holomorphic on the cover.

The monodromies of $\varpi_{\bmk}$ around $0,x,1$ are $p_1,p_1^{-1},p_2$.
The word $\gamma_0\gamma_x$ defining $\alpha_{\bmr}$ in Eq.~\eqref{eq:homology-lifted-words-alpha} traverses $\gamma_x$ and then $\gamma_0$.
Its two access contributions combine as
\begin{equation}
	\begin{aligned}
		\calA_{\bmk}
		&= (1-p_1^{-1})\int_{z_*}^{x}\varpi_{\bmk} + p_1^{-1}(1-p_1)\int_{z_*}^{0}\varpi_{\bmk}\\
		&= (1-p_1^{-1})e^{i\pi(1+a-b)}I_{\bmk}(x).
	\end{aligned}
	\label{eq:period-alpha-discontinuity}
\end{equation}

For $\beta_{0,0}=-e_{-1,0}$, the translation and orientation reversal in Eq.~\eqref{eq:homology-root-transport-square} multiply the root-square factor by $-p_1$, giving
\begin{equation}
	-p_1(1-p_1^{-1})(1-p_2)=(1-p_1)(1-p_2).
	\label{eq:homology-beta-character-multiplier}
\end{equation}
The reference bank phase in Eq.~\eqref{eq:period-bank-phases} then gives
\begin{equation}
	\calB_{\bmk}=(1-p_1)(1-p_2)e^{i\pi(1-b)}J_{\bmk}(x).
	\label{eq:period-beta-discontinuity}
\end{equation}
Using $1-e^{2\pi i a}=-2ie^{i\pi a}\sin\pi a$ and $1-e^{-2\pi i a}=2ie^{-i\pi a}\sin\pi a$, the explicit periods are
\begin{subequations}\label{eq:period-raw-euler-values}
	\begin{align}
		\calA_{\bmk} &= -2\pi i e^{-i\pi b}\,{}_2F_1(a,1-b;1;x),
		\label{eq:period-raw-euler-values-A}\\
		\calB_{\bmk} &= 4e^{i\pi a}\sin\pi a\sin\pi b\,J_{\bmk}(x).
		\label{eq:period-raw-euler-values-B}
	\end{align}
\end{subequations}

The reference ratio must use $-\alpha_{0,0}$ to match the orientation $A_{1,1}=-\alpha_{0,0}$.
Define
\begin{align}
	\tau_{\bmk}(x) \coloneqq -\frac{\calB_{\bmk}}{\calA_{\bmk}}
	&=-\frac{2i}{\pi}e^{i\pi(a+b)}\sin\pi a\sin\pi b\,B(1-a,b)(1-x)^{b-a}\nonumber\\
	&\qquad \times \frac{{}_2F_1(1-a,b;1-a+b;1-x)}{{}_2F_1(a,1-b;1;x)}.
	\label{eq:period-character-ratio}
\end{align}
The scalar ratio $\tau_{\bmk}$ uses the raw alpha and beta curves.

\subsubsection{Periods in the canonical basis}
\paragraph{Canonical A- and B-periods.}
The canonical periods follow by summing deck phases over the cumulative A-cycles and including the correction to the B-cycles.
Write $\bmI=(I_1,I_2)\in\{1,\ldots,n-1\}^2$ and $A_{\bmI}=A_{I_1,I_2}$, with the same convention for $B_{\bmI}$.
The rectangular A-cycle sums factor into one cumulative character sum for each replica direction.
Write
\begin{equation}
	S_k(I) \coloneqq \sum_{r=0}^{I-1}\zeta_n^{kr} = \frac{1-\zeta_n^{kI}}{1-\zeta_n^k},
	\qquad 0\leq I\leq n,
	\label{eq:period-prefix-sum}
\end{equation}
where $S_k(0)=S_k(n)=0$.
For $1\leq k\leq n-1$, the denominator is nonzero.
The vanishing sum at $I=n$ agrees with the orbit relations in Eq.~\eqref{eq:homology-orbit-relations}.
With differential labels on rows and cycle labels on columns, define
\begin{subequations}\label{eq:period-matrix-blocks}
	\begin{gather}
		\sfA_{\bmk,\bmI} \coloneqq \oint_{A_{\bmI}}\varpi_{\bmk}
		\label{eq:period-matrix-blocks-A}\\
		\sfB_{\bmk,\bmI} \coloneqq \oint_{B_{\bmI}}\varpi_{\bmk}.
		\label{eq:period-matrix-blocks-B}
	\end{gather}
\end{subequations}
Substituting the canonical cycles gives
\begin{subequations}\label{eq:period-canonical-blocks}
	\begin{align}
		\sfA_{\bmk,\bmI}
		&=-\calA_{\bmk}S_{k_1}(I_1)S_{k_2}(I_2),
		\label{eq:period-A-block}\\
		\sfB_{\bmk,\bmI}
		&=\calA_{\bmk}\left[
			-p_1^{I_1-1}p_2^{I_2-1}\tau_{\bmk}
			+\sfC_{\bmk,\bmI}\right],
		\label{eq:period-B-block}
	\end{align}
\end{subequations}
where the three A-cycle corrections in $B_{\bmI}$ combine to
\begin{align}
	\sfC_{\bmk,\bmI}
	&\coloneqq-S_{k_1}(I_1)S_{k_2}(I_2-1)
		+S_{k_1}(I_1-1)S_{k_2}(I_2)
		-S_{k_1}(I_1-1)S_{k_2}(I_2+1)\nonumber\\
	&=-S_{k_1}(I_1)S_{k_2}(I_2-1)
		-S_{k_1}(I_1-1)p_2^{I_2}.
	\label{eq:period-B-correction}
\end{align}
The second line uses $S_k(I+1)-S_k(I)=\zeta_n^{kI}$, including at the upper edge $I=n-1$.

\paragraph{Normalization of the A-periods.}
The row convention in Eq.~\eqref{eq:period-matrix-blocks} makes the normalization matrix $\sfA^{-1}$.
It can be written explicitly since finite differences invert the cumulative character sums.
Set
\begin{equation}
	D_k(I)\coloneqq\zeta_n^{-k(I-1)}-\zeta_n^{-kI}, \qquad 1\leq I\leq n-1.
	\label{eq:period-dual-difference}
\end{equation}
Expanding $S_k(J)$ and using $\sum_{k=1}^{n-1}\zeta_n^{km}=n\delta_n(m,0)-1$ yields
\begin{equation}
	\sum_{k=1}^{n-1}D_k(I)S_k(J) = n\sum_{r=0}^{J-1} [\delta_n(I-1,r)-\delta_n(I,r)] = n\delta_{IJ}.
	\label{eq:period-finite-inverse}
\end{equation}
This inverse acts independently in the two replica directions.
Since every $\calA_{\bmk}$ is nonzero by Eq.~\eqref{eq:period-raw-euler-values-A},
\begin{equation}
	(\sfA^{-1})_{\bmI,\bmk} = -\frac{D_{k_1}(I_1)D_{k_2}(I_2)}{n^2\calA_{\bmk}}.
	\label{eq:period-A-inverse}
\end{equation}

The inverse matrix now converts the character forms into normalized differentials,
\begin{subequations}\label{eq:period-normalized-forms}
	\begin{gather}
		\nu_{\bmI} \coloneqq \sum_{k_1,k_2=1}^{n-1}(\sfA^{-1})_{\bmI,\bmk}\varpi_{\bmk}
		\label{eq:period-normalized-forms-definition}\\
		\oint_{A_{\bmJ}}\nu_{\bmI} = \delta_{\bmI\bmJ}.
		\label{eq:period-normalized-forms-A}
	\end{gather}
\end{subequations}
Their B-periods define the marked period matrix by
\begin{equation}
	(\Omega_n^{(3)}(x))_{\bmI\bmJ} \coloneqq \oint_{B_{\bmJ}}\nu_{\bmI}, \qquad \Omega_n^{(3)}(x) = \sfA^{-1}\sfB.
	\label{eq:period-canonical-definition}
\end{equation}
An invertible change of the holomorphic differential basis left-multiplies $\sfA$ and $\sfB$ by the same matrix and cancels in $\sfA^{-1}\sfB$.
Thus the character basis is a computational choice, and the canonical cycles fix the marking of $\Omega$.
Equations~\eqref{eq:period-B-block} and~\eqref{eq:period-A-inverse} give the explicit matrix below.
\begin{proposition}[Marked disjoint period matrix]
\label{prop:disjoint-period-matrix}
	In the canonical integral marking, the period matrix has entries
	\begin{equation}
		\begin{aligned}
		(\Omega_n^{(3)}(x))_{\bmI\bmJ}
		&=\frac1{n^2}\sum_{k_1,k_2=1}^{n-1}
			D_{k_1}(I_1)D_{k_2}(I_2)\\
		&\quad\times\left[
			\zeta_n^{k_1(J_1-1)+k_2(J_2-1)}\tau_{\bmk}(x)
			-\sfC_{\bmk,\bmJ}\right].
		\end{aligned}
		\label{eq:period-canonical-explicit}
	\end{equation}
	% The free indices label canonical cycles: $D_k(I)$ inverts the cumulative sums, and $\sfC$ supplies the B-cycle correction.
	It is symmetric,
	\begin{equation}
		(\Omega_n^{(3)}(x))^{\top}=\Omega_n^{(3)}(x),
		\label{eq:period-symmetry}
	\end{equation}
    and its imaginary part is positive definite.
    Equivalently, $\Omega_n^{(3)}(x)$ belongs to the Siegel upper half-space of degree $g_d$.
\end{proposition}

\subsubsection{The Hodge pairing}
\paragraph{Symmetry of the period matrix.}
The Riemann bilinear relations in the complete integral marking of Lemma~\ref{lem:disjoint-complete-integral-marking} give Eq.~\eqref{eq:period-symmetry}.

\paragraph{Positive imaginary part.}
Deck invariance of the surface integral and the distinct characters of the differentials give
\begin{equation}
	\frac{i}{2}\int_{\Sigma_n^{(3)}(x)} \varpi_{\bmk}\wedge\overline{\varpi_{\bmm}} = \delta_{\bmk\bmm}h_{\bmk}, \qquad
	h_{\bmk} \coloneqq \frac{i}{2}\int_{\Sigma_n^{(3)}(x)} \varpi_{\bmk}\wedge\overline{\varpi_{\bmk}}.
	\label{eq:period-character-gram}
\end{equation}
Indeed, a deck translation multiplies an off-diagonal entry by $\zeta_n^{k_i-m_i}$, so that entry vanishes whenever $k_i\ne m_i$.
Using the explicit differential and summing the $n^2$ sheets gives
\begin{equation}
	h_{\bmk} = n^2\int_{\bbC} |z|^{2a-2}|z-x|^{-2a}|z-1|^{2b-2}\dd[2]{z}>0, \qquad \dd[2]{z} \coloneqq \frac{i}{2}\dd{z}\wedge\dd{\bar{z}}.
	\label{eq:period-character-positive-integral}
\end{equation}
The powers at $0,x,1$ are integrable because $0<a,b<1$.
At infinity the density decays as $|z|^{2b-4}$ and is integrable because $b<1$.

The Riemann bilinear relation~\eqref{eq:math-bilinear} converts these positive Hodge norms into the imaginary part of the period matrix.
Applied to the computed period blocks, it gives
\begin{equation}
	\diag(h_{\bmk})
	= \frac{i}{2}(\sfA\sfB^\dagger-\sfB\sfA^\dagger)
	= \sfA\,\Im\Omega_n^{(3)}(x)\,\sfA^\dagger,
	\label{eq:period-positive-form}
\end{equation}
where $\dagger$ denotes conjugate transpose and the second equality uses
Eq.~\eqref{eq:period-symmetry}.
Consequently the imaginary part of the explicit matrix is the Gram matrix
\begin{subequations}\label{eq:period-riemann-properties}
	\begin{align}
		\Im\Omega_n^{(3)}(x)
		&=\sfA^{-1}\diag(h_{\bmk}) (\sfA^{-1})^\dagger>0,
		\label{eq:period-riemann-properties-matrix}\\
		(\Im\Omega_n^{(3)}(x))_{\bmI\bmJ}
		&=\frac1{n^4}\sum_{k_1,k_2=1}^{n-1}
			\frac{h_{\bmk}}{|\calA_{\bmk}|^2}
			D_{k_1}(I_1)D_{k_2}(I_2)
			\overline{D_{k_1}(J_1)D_{k_2}(J_2)}.
		\label{eq:period-riemann-properties-entries}
	\end{align}
\end{subequations}
Every weight is positive and the character transform is invertible, establishing strict positive definiteness.

For $n=2$, the marked period reduces to $\Omega_2^{(3)}(x)=2iK(1-x)/K(x)$.
Section~\ref{sec:torus-comparison} derives this specialization and compares the corresponding torus amplitudes.

\subsection{Integral cycles on the Fermat curve}\label{sec:fermat-integral-homology}
Integral homology bases and intersection matrices for Fermat curves are known~\cite{Kamata:2002periods,Otsubo:2016fermat}.

Interval sums of the Pham cycles give one isotropic family, and their $\sfh_1$ translates give a second.
Cumulative sums invert the mixed pairing over the integers.
Label the $g_F$ pairs by $0\leq a\leq b\leq n-3$, in lexicographic order.
Their scalar index is
\begin{equation}
	\mu(a,b) \coloneqq 1+b+a(n-3)-\frac{a(a-1)}2, \qquad 1\leq\mu(a,b)\leq g_F.
	\label{eq:fermat-cycle-index}
\end{equation}
Define
\begin{equation}
	A_{a,b}\coloneqq\sum_{j=a}^{b}e_{a+b,j},\qquad
	\widetilde B_{a,b}\coloneqq\sfh_1A_{a,b},\qquad
	B_{a,b}\coloneqq\sum_{c=a}^{b}\sum_{d=b}^{n-3}\widetilde B_{c,d}.
	\label{eq:fermat-explicit-cycles}
\end{equation}
The indices of $e_{i,j}$ are read modulo $n$, and the geometric kernel in Eq.~\eqref{eq:fermat-intersection-kernel} applies to every translate.

\begin{lemma}[Complete Fermat integral marking]\label{lem:fermat-complete-integral-marking}
	The ordered cycles in Eq.~\eqref{eq:fermat-explicit-cycles} form a complete integral symplectic basis of $H_1(F_n,\bbZ)$.
	At $n=2$, this is the empty basis of the genus-zero curve.
\end{lemma}

The symmetry of the interval sums makes each family isotropic.
The interval in $A_{a,b}$ is symmetric under $j\mapsto a+b-j$.
The kernel obeys $I(p,p-q)=-I(p,q)$, so replacing every $e_{i,j}$ by $e_{i,i-j}$ reverses the pairing of two such cycles and leaves both cycles fixed.
Thus the $A$-cycles are mutually isotropic.
Deck invariance gives the same property for $\widetilde B=\sfh_1 A$.
Only the pairing between the two families needs to be corrected.
Substituting Eq.~\eqref{eq:fermat-intersection-kernel} gives
\begin{align}
	M_{(a,b),(c,d)}
	&\coloneqq A_{a,b}\circ\widetilde B_{c,d}
	= \sum_{j=a}^{b}\sum_{l=c}^{d}I(c+d+1-a-b,l-j)
	\nonumber\\
	&= \delta_{a,c}\delta_{b,d}-\delta_{a,c}\delta_{b-1,d}
	-\delta_{a-1,c}\delta_{b,d}+\delta_{a-1,c}\delta_{b-1,d}.
	\label{eq:fermat-explicit-cross-pairing}
\end{align}
The deltas in the final line are ordinary integer deltas, with pairs outside the triangle omitted.
The interval-cycle calculation in Appendix~\ref{sec:appendix-fermat-bases} derives this expression from the modular kernel, including the wraparound terms.

It remains to invert the mixed backward-difference matrix $M$ on the triangular index set.
Its inverse and the resulting canonical pairings are
\begin{subequations}\label{eq:fermat-explicit-duality}
	\begin{align}
		(M^{-1})_{(a,b),(c,d)} &= \mathbf{1}_{\{c\leq a\leq d\leq b\}},
		\label{eq:fermat-explicit-duality-inverse}\\
		A_{a,b}\circ B_{c,d} &= \delta_{a,c}\delta_{b,d},\qquad
		A_{a,b}\circ A_{c,d} = B_{a,b}\circ B_{c,d}=0.
		\label{eq:fermat-explicit-duality-canonical-pairings}
	\end{align}
\end{subequations}
Taking the two backward differences of the indicator gives $\delta_{a,c}\delta_{b,d}$, and multiplication by this inverse gives the last sum in Eq.~\eqref{eq:fermat-explicit-cycles}.

The corrected cycles have a canonical intersection matrix of determinant one, so the index formula of Appendix~\ref{sec:math-riemann} proves that their integer span is all of $H_1(F_n,\bbZ)$.

Appendix~\ref{sec:replica-intersections} gives an independent index-one argument for the rectangular Pham basis and its explicit integral change to these interval cycles.
The same interval sums on the cover for disjoint intervals span a primitive symplectic sublattice of rank $2g_F$, because their pairings are identical and unimodular.
Section~\ref{sec:replica-neck-periods} transports this sublattice to a component block of the degenerating disjoint marking and supplies the remaining cycles.

\subsection{Periods of the Fermat curve}\label{sec:fermat-periods}
The common Euler period specializes at $y=0$ to
\begin{equation}
	L_{\bmell}(0) = \int_\lambda\omega_{\bmell}(0)
	=\frac1n B(\ell_1/n,\ell_2/n),
	\qquad \ell_1,\ell_2\geq1,\quad \ell_1+\ell_2<n.
	\label{eq:fermat-beta-period}
\end{equation}
Write $\omega_{\bmell}=\omega_{\bmell}(0)$ within this subsection.
For the reference period, also write $\bmell=(k_1,k_2-k_1)$ with $1\leq k_1<k_2\leq n-1$ in the adjacent physical dictionary.
Applying the square and its root translates gives
\begin{subequations}\label{eq:fermat-orbit-periods}
	\begin{align}
		b_{\bmell}\coloneqq\oint_{e_{0,0}}\omega_{\bmell} &= \frac{(1-\zeta_n^{\ell_1})(1-\zeta_n^{\ell_2})}{n}
		B(\ell_1/n,\ell_2/n)
		\notag\\
		&= \frac{(1-\zeta_n^{k_1})(1-\zeta_n^{k_2-k_1})}{n} B(k_1/n,(k_2-k_1)/n)
		\label{eq:fermat-orbit-periods-reference}\\
		\oint_{e_{i,j}}\omega_{\bmell} &= b_{\bmell}\zeta_n^{i\ell_1+j\ell_2}.
		\label{eq:fermat-orbit-periods-translate}
	\end{align}
\end{subequations}
We use root indices in the calculations below.
These are the character periods in the present orientation~\cite{TorresNova:2018fermat}.
The integral cycles of Section~\ref{sec:fermat-integral-homology} now turn them into the marked Fermat period matrix.

\paragraph{Canonical period blocks.}
The integral-cycle algorithm of Ref.~\cite{Kamata:2002periods} constructs normalized periods for the same Fermat curves.
We evaluate them in the explicit interval marking of Section~\ref{sec:fermat-integral-homology}.
The orbit periods in Eq.~\eqref{eq:fermat-orbit-periods-reference} give $b_{\bmell}\ne0$ for every holomorphic character.
Set $\phi_{\bmell}\coloneqq\omega_{\bmell}/b_{\bmell}$ and use the canonical cycles of Eq.~\eqref{eq:fermat-explicit-cycles} to define
\begin{subequations}\label{eq:fermat-period-blocks}
	\begin{gather}
		\sfA_{\bmell,\mu}\coloneqq\oint_{A_\mu}\phi_{\bmell}
		\label{eq:fermat-period-blocks-A}\\
		\sfB_{\bmell,\mu}\coloneqq\oint_{B_\mu}\phi_{\bmell}.
		\label{eq:fermat-period-blocks-B}
	\end{gather}
\end{subequations}
Rows label forms and columns label cycles, as in Section~\ref{sec:disjoint-period-matrix}.

Write $(r,s)=(\ell_1,\ell_1+\ell_2)$ with $1\leq r<s\leq n-1$, ordered lexicographically.
The indices $r,s$ label the Vandermonde nodes, while $\bmr$ denotes a physical replica sheet.
This order is also the lexicographic order in $(\ell_1,\ell_2)$.
Put $\xi_j\coloneqq\zeta_n^j$.
Summing the orbit phases over the interval cycles in Eq.~\eqref{eq:fermat-explicit-cycles} makes each A-period a symmetric polynomial in $\xi_r,\xi_s$ and gives the concrete blocks
\begin{subequations}\label{eq:fermat-schur-periods}
	\begin{align}
	\sfA_{(r,s),(a,b)}
		&=\sum_{j=a}^{b}\xi_r^{a+b-j}\xi_s^j
		=\frac{\xi_r^a\xi_s^{b+1}-\xi_r^{b+1}\xi_s^a}{\xi_s-\xi_r},
		\label{eq:fermat-schur-periods-A}\\
		\sfB_{(r,s),(c,d)}
		&=\frac{\xi_r}{\xi_s-\xi_r}
		\sum_{p=c}^{d}\sum_{q=d}^{n-3} \bigl(\xi_r^p\xi_s^{q+1}-\xi_r^{q+1}\xi_s^p\bigr).
		\label{eq:fermat-schur-periods-B}
	\end{align}
\end{subequations}
The factor $\xi_r$ in the second line is the $\sfh_1$ character.

\begin{proposition}[Marked Fermat period matrix]\label{prop:fermat-period-matrix}
	The A-period block is invertible.
	The normalized forms and period matrix are
	\begin{subequations}\label{eq:fermat-period-matrix}
		\begin{gather}
			\nu_\mu\coloneqq\sum_{\bmell}(\sfA^{-1})_{\mu,\bmell}\phi_{\bmell}
			\label{eq:fermat-period-matrix-forms}\\
			\Omega_{F,n}\coloneqq\sfA^{-1}\sfB
			\label{eq:fermat-period-matrix-matrix}\\
			\oint_{B_\lambda}\nu_\mu=(\Omega_{F,n})_{\mu\lambda}.
			\label{eq:fermat-period-matrix-B}
		\end{gather}
	\end{subequations}
	In particular, $\oint_{A_\lambda}\nu_\mu=\delta_{\mu\lambda}$.
	The explicit blocks in Eq.~\eqref{eq:fermat-schur-periods} and inverse in Eq.~\eqref{eq:fermat-period-block-inverse} give entries in $\bbQ(\zeta_n)$.
	For $n\geq3$, the matrix is symmetric and has positive-definite imaginary part.
\end{proposition}

\paragraph{Vandermonde inversion of the period block.}
The numerator in Eq.~\eqref{eq:fermat-schur-periods-A} is a $2\times2$ Vandermonde minor.
Inverting the full period block therefore reduces to inverting one $(n-1)\times(n-1)$ matrix.
Define
\begin{equation}
	V_{r,a} \coloneqq \xi_r^a, \qquad (V^{-1})_{a,r}=\frac{\xi_r^{-a}-\xi_r}{n}, \qquad 1\leq r\leq n-1,\quad 0\leq a\leq n-2.
	\label{eq:fermat-vandermonde-inverse}
\end{equation}
Finite root orthogonality gives
\begin{equation}
	\frac1n\sum_{r=1}^{n-1}(\xi_r^{-a}-\xi_r)\xi_r^b = \delta_{ab}.
	\label{eq:fermat-vandermonde-inverse-check}
\end{equation}
Appendix~\ref{sec:period-identities} gives the exponent ranges and the exterior-square calculation.
The column pairs $(a,b+1)$ exhaust all pairs of distinct exponents in $0,\ldots,n-2$.
Thus the matrix with entries $(\xi_s-\xi_r)\sfA_{(r,s),(a,b)}$ is the exterior square of $V$.
The inverse exterior square is the exterior square of $V^{-1}$, giving
\begin{equation}
	(\sfA^{-1})_{(a,b),(r,s)} = (\xi_s-\xi_r) \bigl[ (V^{-1})_{a,r}(V^{-1})_{b+1,s} - (V^{-1})_{a,s}(V^{-1})_{b+1,r} \bigr].
	\label{eq:fermat-period-block-inverse}
\end{equation}
Multiplying this inverse by the $B$-block in Eq.~\eqref{eq:fermat-schur-periods-B} gives $\Omega_{F,n}=\sfA^{-1}\sfB$ explicitly.

\paragraph{The Hodge pairing in the canonical basis.}
The canonical cycles are a full integral basis, so the bilinear relations of Appendix~\ref{sec:math-riemann} apply.
In the rescaled differential frame their Hermitian pairing is diagonal by deck invariance, with
\begin{equation}
	H_{\bmell\bmell} = \frac1{|b_{\bmell}|^2}\int_{\bbC} |w|^{2\ell_1/n-2}|1-w|^{2\ell_2/n-2}\dd[2]{w}>0.
	\label{eq:fermat-positive-hodge}
\end{equation}
The $n^2$ sheets cancel the squared factor $1/n^2$ in the differential, and convergence is precisely the triangular character restriction.
Hence
\begin{subequations}\label{eq:fermat-siegel-properties}
	\begin{gather}
	\Omega_{F,n}^{\top} = \Omega_{F,n}
	\label{eq:fermat-siegel-properties-symmetry}\\
	\Im\Omega_{F,n} = \sfA^{-1}H\,\overline{\sfA}^{\,-\top}>0.
	\label{eq:fermat-siegel-properties-positivity}
	\end{gather}
\end{subequations}

For $n=3$, the marked matrix is $\Omega_{F,3}=\zeta_3$.
Section~\ref{sec:low-genus-boson-checks} relates this marking to the triangular torus and compares the adjacent coefficients.

\section{The compact boson on a Riemann surface}\label{sec:common-boson}
The complete integral markings of Section~\ref{sec:replica-geometry} supply the period matrices and full winding lattices of both replica curves.
We first separate a real compact field into a harmonic configuration and a zero-period fluctuation, then evaluate the harmonic sum on a general marked surface.
The external two-point twist norms determine how this sum combines with the connected constant-mode measure.
Sections~\ref{sec:disjoint-boson} and~\ref{sec:adjacent-boson} use this common factor and fix the remaining quantum determinants and finite constants in the two geometries.

\subsection{Winding sectors}\label{sec:classical-winding}

\paragraph{Compactification convention.}
On a Riemann surface $\Sigma$, take a real scalar with
\begin{subequations}
	\label{eq:ward-real-action}
	\begin{align}
		S_{\bbR}[\phi]
		&=\frac1{4\pi}\int_\Sigma\dd{\phi}\wedge*\dd{\phi}
		=\frac1\pi\int_\Sigma\dd[2]{z}\,\partial\phi\,\bar\partial\phi,
		\label{eq:common-real-action}\\
		\phi&\sim\phi+2\pi R,\qquad\eta\coloneqq R^2>0.
		\label{eq:common-compactification}
	\end{align}
\end{subequations}
Here $\partial=\partial_z$, $\bar\partial=\partial_{\bar z}$ and $\dd[2]{z}=(i/2)\dd{z}\wedge\dd{\bar{z}}$ in a local complex coordinate.
In particular, momentum $m\in\bbZ$ and winding $w\in\bbZ$ have weights
\begin{subequations}
	\label{eq:ward-radius-spectrum}
	\begin{align}
		h_{m,w}&=\frac14\left(\frac{m}{R}+wR\right)^2,
		\label{eq:common-left-weight}\\
		\bar h_{m,w}&=\frac14\left(\frac{m}{R}-wR\right)^2,
		\label{eq:common-right-weight}\\
		h_{m,w}+\bar h_{m,w}&=\frac12\left(\frac{m^2}{\eta}+\eta w^2\right).
		\label{eq:common-scaling-dimension}
	\end{align}
\end{subequations}
Thus $\eta$ is the compactification parameter of the ordinary two-interval function, with T-duality $\eta\leftrightarrow\eta^{-1}$ and self-dual value $\eta=1$.~\footnote{The radius conventions of Refs.~\cite{Calabrese:2009ez,Furukawa:2008uk} are related by $\eta=R^2=gR_{\rm CCT}^2=2\pi R_{\rm FPS}^2$.}

\paragraph{Integer winding data.}
Let $\Sigma$ be a compact connected Riemann surface of genus $g$, with a full integral symplectic basis $A_i,B_i$ and holomorphic forms normalized as in Eq.~\eqref{eq:classical-normalized-periods}.
Write
\begin{equation}
	\Omega=\calR+i\calI,\qquad \calI>0,
	\qquad \bm\nu=(\nu_i)_{i=1}^g.
	\label{eq:classical-period-data}
\end{equation}
The real symmetric matrices $\calR$ and $\calI$ are the real and imaginary parts of the period matrix.
The disjoint and Fermat markings in Section~\ref{sec:replica-geometry} provide these data with $g=g_d$ and $g=g_F$, respectively.
In genus zero the lattice is empty, with unit sum and unit empty determinant.

A compact field is a map from $\Sigma$ to the circle $S^1$.
A local real lift may change by an integer multiple of $2\pi R$ around a closed curve, while its differential is globally defined.
Its winding sector is therefore specified by integer vectors $\bmw_A,\bmw_B\in\bbZ^g$ through the periods of the harmonic representative $\phi_{\rm cl}$ selected below:
\begin{subequations}
	\label{eq:classical-winding-periods}
	\begin{align}
		\frac1{2\pi R}\oint_{A_i}\dd{\phi_{\rm cl}} &= w_{A,i},
		\label{eq:common-A-windings}\\
		\frac1{2\pi R}\oint_{B_i}\dd{\phi_{\rm cl}} &= w_{B,i}.
		\label{eq:common-B-windings}
	\end{align}
\end{subequations}
The full integral basis makes these $2g$ windings independent: the sectors form $H^1(\Sigma,\bbZ)\simeq\bbZ^{2g}$, with each counted once.
They label integral cohomology, whereas the two deck-character coordinates label finite Fourier sectors.
Minimizing the action in a winding class gives a unique harmonic differential~\cite{Dijkgraaf:1987vp,Calabrese:2009ez}.

\paragraph{Harmonic configurations and fluctuations.}
For each fixed winding class, write $\phi=\phi_{\rm cl}+\phi_{\rm qu}$, with $\dd{\phi_{\rm cl}}$ its harmonic representative and $\phi_{\rm qu}$ single-valued.
The fluctuation consequently satisfies
\begin{equation}
    \oint_C\dd{\phi_{\rm qu}}=0
    \label{eq:ward-zero-winding}
\end{equation}
on every closed curve of the cover.
Since $\dd{*\dd{\phi_{\rm cl}}}=0$, integration by parts on the compact surface gives a vanishing cross term with the exact fluctuation.
These zero-period constraints will determine the quantum Green functions
in Section~\ref{sec:quantum-ward}.
After fixing the additive constant of the harmonic representative, we integrate the constant field mode separately and take the nonconstant fluctuation to have zero mean.
The connected Gaussian measure is fixed in Appendix~\ref{sec:math-conical-determinants}.
Section~\ref{sec:common-external-normalization} tracks its constant-mode contribution to the normalized quotients.

\subsection{The harmonic factor}\label{sec:common-harmonic-factor}

\paragraph{Action of the harmonic representative.}
A real harmonic one-form is a sum of holomorphic differentials and their complex conjugates.
Parameterize it as
\begin{equation}
	\frac{\dd{\phi_{\rm cl}}}{2\pi R}
	=\frac1{2i}\left(\bmc^{\top}\bm\nu - \overline{\bmc}^{\top}\overline{\bm\nu}\right),
	\qquad \bmc=\bmc_{\bmw_A,\bmw_B}\in\bbC^g.
	\label{eq:classical-harmonic-ansatz}
\end{equation}
Its A- and B-periods give $\Im\bmc=\bmw_A$ and $\Im(\Omega\bmc)=\bmw_B$, using the symmetry of $\Omega$.
Since $\calI>0$, these conditions determine
\begin{equation}
	\bmc = \calI^{-1}(\bmw_B-\calR\bmw_A)+i\bmw_A = \calI^{-1}(\bmw_B-\overline\Omega\bmw_A).
	\label{eq:classical-harmonic-coefficients}
\end{equation}

The pairing~\eqref{eq:classical-hodge-pairing} and Hodge-star convention~\eqref{eq:math-hodge-signs} give
\begin{equation}
	\int_\Sigma\dd{\phi_{\rm cl}} \wedge *\dd{\phi_{\rm cl}} = (2\pi R)^2\overline{\bmc}^{\top}\calI\bmc.
\end{equation}
The action in Eq.~\eqref{eq:common-real-action} is therefore
\begin{equation}
	S_{\rm cl}(\bmw_A,\bmw_B;\Omega,\eta)
	=\pi\eta\,\overline{\bmc}^{\top}\calI\bmc
	=\pi\eta\left[
		(\bmw_B-\calR\bmw_A)^{\top}\calI^{-1}
		(\bmw_B-\calR\bmw_A)+\bmw_A^{\top}\calI\bmw_A
	\right].
	\label{eq:classical-winding-action}
\end{equation}
Both quadratic terms are nonnegative, and the action vanishes only in the zero-winding sector.
Define the classical factor of the single real boson by
\begin{equation}
	Z_{\rm cl}^{\bbR}(\Omega;\eta)
	\coloneqq \sum_{\bmw_A,\bmw_B\in\bbZ^g} e^{-S_{\rm cl}(\bmw_A,\bmw_B;\Omega,\eta)}.
	\label{eq:classical-winding-sum}
\end{equation}

\paragraph{Theta representation of the harmonic factor.}
Introduce integers $\bmm\in\bbZ^g$ dual to the $B$-cycle windings $\bmw_B$; on a torus they are the momentum integers.
With the Riemann theta convention in Eq.~\eqref{eq:classical-theta-definition}, order the lattice vector as $\bmN=\binom{\bmw_A}{\bmm}$ and define
\begin{equation}
	T_\eta(\Omega)\coloneqq
	\begin{pmatrix}
		i\eta\calI&\calR\\
		\calR&i\eta^{-1}\calI
	\end{pmatrix}.
	\label{eq:classical-theta-matrix}
\end{equation}
The matrix is symmetric and
\begin{equation}
	\Im T_\eta= \begin{pmatrix}\eta\calI&0\\0&\eta^{-1}\calI\end{pmatrix}>0.
	\label{eq:classical-theta-positivity}
\end{equation}
The theta series therefore converges absolutely.

\begin{proposition}[The common harmonic factor]\label{prop:common-harmonic-factor}
	The winding sum has the theta representation
	\begin{equation}
		Z_{\rm cl}^{\bbR}(\Omega;\eta)
		= \eta^{-g/2}(\det\calI)^{1/2}
		\Theta(\bm0\mid T_\eta(\Omega)).
		\label{eq:classical-theta-result}
	\end{equation}
	The associated harmonic factor is
	\begin{equation}
		\calH_\Sigma(\eta)
		\coloneqq \eta^{g/2}Z_{\rm cl}^{\bbR}(\Omega;\eta)
		= \sqrt{\det\calI}\,\Theta(\bm0\mid T_\eta(\Omega)).
		\label{eq:common-harmonic-factor}
	\end{equation}
	It is invariant under integral symplectic changes of marking.
	Radius inversion gives
	\begin{subequations}
		\label{eq:classical-radius-duality}
		\begin{align}
			\Theta(\bm0\mid T_{1/\eta}(\Omega))
			&=\Theta(\bm0\mid T_\eta(\Omega)),
			\label{eq:common-theta-radius-duality}\\
			Z_{\rm cl}^{\bbR}(\Omega;1/\eta)
			&=\eta^g Z_{\rm cl}^{\bbR}(\Omega;\eta).
			\label{eq:common-classical-radius-duality}
		\end{align}
	\end{subequations}
	The harmonic factor therefore obeys $\calH_\Sigma(1/\eta)=\calH_\Sigma(\eta)$.
\end{proposition}

\paragraph{Poisson resummation.}
At fixed $\bmw_A$, resum the $\bmw_B$ sum using Eq.~\eqref{eq:math-shifted-poisson}, with $Q=\eta\calI^{-1}$ and $\bma=\calR\bmw_A$.
The determinant factor is $(\det Q)^{-1/2}=\eta^{-g/2}(\det\calI)^{1/2}$,
giving
\begin{equation}
	Z_{\rm cl}^{\bbR}(\Omega;\eta)
	=\eta^{-g/2}(\det\calI)^{1/2} \sum_{\bmw_A,\bmm\in\bbZ^g}
	\exp\left[ -\pi\eta\bmw_A^{\top}\calI\bmw_A-\pi\eta^{-1}\bmm^{\top}\calI\bmm + 2\pi i\bmm^{\top}\calR\bmw_A\right].
	\label{eq:classical-resummed-lattice}
\end{equation}
The mixed phase retains the real part of the period matrix, while its imaginary part controls the Gaussian suppression.
The off-diagonal blocks of $T_\eta$ reproduce this phase, proving Eq.~\eqref{eq:classical-theta-result}.

An integral symplectic change of cycles relabels $H^1(\Sigma,\bbZ)$ bijectively, leaving each harmonic field and its action unchanged.
This proves marking invariance of $\calH_\Sigma$.
Its determinant and theta factors need not be invariant separately.

Radius inversion follows from a different relabeling.
Exchanging $\bmw_A$ and $\bmm$ in Eq.~\eqref{eq:classical-resummed-lattice} proves Eq.~\eqref{eq:classical-radius-duality} and completes Proposition~\ref{prop:common-harmonic-factor}.

\subsection{Normalization by external twists}\label{sec:common-external-normalization}

The external two-point norms turn the cover path integrals into the disjoint ratio and adjacent coefficient.
Let $Z_d,Z_F$ be the connected disjoint and Fermat path integrals, $Z_0$ the base-sphere path integral, and $Z_u,Z_v,Z_p$ the connected cyclic reference path integrals in the respective round pullback metrics.
The disjoint references join the physical pairs $(0,x)$ and $(1,\infty)$ in $z$; in the geometric coordinate $w$ these are $(\infty,0)$ and $(1,y^{-1})$.
The adjacent reference is the antipodal two-point cover.
Since an external translation has $n$ cycles of length $n$, the quotients are
\begin{subequations}
	\label{eq:common-normalized-quotients}
	\begin{align}
		\calR_n^{(3)}(x;\eta) &= \frac{Z_dZ_0^{n^2}}{Z_u^nZ_v^n},
		\label{eq:common-disjoint-quotient}\\
		C_n^{(3)}(\eta) &= 2^{-\Delta_n^{(3)}} \frac{Z_FZ_0^{n^2/2}}{Z_p^{3n/2}}.
		\label{eq:common-adjacent-quotient}
	\end{align}
\end{subequations}
The external Weyl factors cancel pairwise in the disjoint ratio.
Equation~\eqref{eq:common-adjacent-quotient} converts the round three-point quotient to its plane coefficient.
Section~\ref{sec:direct-fermat-determinant} derives the factor $2^{-\Delta_n^{(3)}}$.

Each connected surface contributes one compact constant mode.
The net counts in these quotients are
\begin{subequations}
	\label{eq:common-component-counts}
	\begin{align}
		1+n^2-2n &= g_d,
		\label{eq:common-disjoint-component-count}\\
		1+\frac{n^2}{2}-\frac{3n}{2} &= g_F.
		\label{eq:common-adjacent-component-count}
	\end{align}
\end{subequations}
Thus the net radius powers in the normalized quotients are $R^{g_d}$ and $R^{g_F}$.
For either genus, $R^g=\eta^{g/2}$ combines with the winding sum into $\calH_\Sigma$.

\section{The disjoint four-point function}\label{sec:disjoint-boson}
For $0<x<1$, let $\Omega=\Omega_n^{(3)}(x)$ be the genus-$g$ period matrix in the canonical integral marking of Section~\ref{sec:disjoint-period-matrix}, with $g=(n-1)^2$.
In this marking, set $\delta_*=[\bm0;\tfrac12\bm1_g]$; Section~\ref{sec:theta-ratio} establishes the nonvanishing of its theta constant.
The matrix $T_\eta(\Omega)$ is defined in Eq.~\eqref{eq:classical-theta-matrix}.
\begin{theorem}[Normalized disjoint four-point ratio]\label{prop:disjoint-four-point}
	\begin{equation}
		\calR_n^{(3)}(x;\eta)
		=\frac{\Theta(\bm0\mid T_\eta(\Omega_n^{(3)}(x)))}
			{|\theta[\delta_*](\bm0\mid\Omega_n^{(3)}(x))|^2}.
		\label{eq:assembly-normalized-ratio}
	\end{equation}
\end{theorem}

The ratio tends to one at $x\to0$.
Equation~\eqref{eq:tripartite-four-point-factorizations} restores arbitrary ordered endpoints, and Eq.~\eqref{eq:tripartite-disjoint-moment} gives the leading regulated moment.

\subsection{Quantum fluctuations from the Ward--Takahashi identity}\label{sec:quantum-ward}

Moving $x$ changes the fluctuation determinant through a residue of the stress tensor.
Point splitting expresses that residue in terms of the Green function, whose local singularities are fixed by the twists and whose homogeneous terms are fixed by zero periods.
For the tripartite cover the resulting two-period system depends on the two independent twist angles.
This applies the Green-function method for orbifold twists~\cite{Dixon:1986qv} and ordinary replica twists~\cite{Calabrese:2009ez}.
The stress-tensor and global-monodromy method also appears in the intersecting-brane four-twist calculation with two independent angles~\cite{Cvetic:2003ch}.

\subsubsection{The two-angle twist correlator}
\label{sec:ward-two-angle}

\paragraph{An auxiliary complex field.}
For the fluctuation calculation, use a complex Gaussian field $X=\phi+i\chi$ with zero periods and the constant modes omitted.
Its symmetric action is
\begin{equation}
	\begin{aligned}
		S_{\bbR}[\phi]+S_{\bbR}[\chi]
		&=\frac1{4\pi}\int_\Sigma\dd{X}\wedge *\dd{\bar X}\\
		&=\frac1{2\pi}\int_\Sigma\dd[2]{z}\,\bigl(\partial X\,\bar\partial\bar X + \partial\bar X\,\bar\partial X\bigr).
	\end{aligned}
	\label{eq:ward-complex-action}
\end{equation}
The mixed terms cancel, leaving two independent real Gaussian fields, so the auxiliary complex field has central charge $c=2$.
The plane contractions fixing the local singular terms and the corresponding stress tensor are
\begin{subequations}\label{eq:ward-field-normalization}
	\begin{gather}
		\expval{\phi(z)\phi(z')} = -\log|z-z'|
		\label{eq:ward-field-normalization-real-propagator}\\
		\expval{X(z)\bar X(z')} = -\log|z-z'|^2
		\label{eq:ward-field-normalization-complex-propagator}\\
		T_{\bbC} = -:\partial X\,\partial\bar X:.
		\label{eq:ward-field-normalization-stress-tensor}
	\end{gather}
\end{subequations}
We attach $\bbC$ to the quantum factors of the complex field and $\bbR$ to those of the physical real theory.
Since the two real Gaussian integrals are independent, with the same treatment of the constant mode,
\begin{equation}
	Z_{\rm qu}^{\bbC}=(Z_{\rm qu}^{\bbR})^2.
	\label{eq:ward-real-complex-determinant}
\end{equation}
We take the positive square root after multiplying all character sectors, including self-conjugate characters, as in Ref.~\cite{Calabrese:2009ez}, and combine it with the single real-boson winding lattice of Section~\ref{sec:classical-winding}.

\paragraph{Fourier sectors and their twist fields.}
Let $X_{\bmr}$ denote the auxiliary field on sheet $\bmr=(r_1,r_2)\in(\bbZ_n)^2$.
The unitary deck Fourier transform is
\begin{equation}
	Y_{\bmk} \coloneqq \frac1n\sum_{\bmr\in(\bbZ_n)^2} \zeta_n^{-\bmk\cdot\bmr}X_{\bmr}, \qquad \bmk=(k_1,k_2)\in(\bbZ_n)^2.
	\label{eq:ward-fourier-fields}
\end{equation}
A continuation $X_{\bmr}\mapsto X_{\bmr+\bme_j}$ multiplies $Y_{\bmk}$ by the physical phase $p_j$ of Eq.~\eqref{eq:period-character-parameters}.
Here that convention extends to the axis characters: $k_j=0$ gives $p_j=1$.
For a mixed character it is exactly the eigenvalue of $\varpi_{\bmk}$ in Eq.~\eqref{eq:disjoint-form-characters}.
This diagonalization reduces the Ward--Takahashi problem to the same character sectors that supplied the period matrix.
The doubled modes $Y_{\bmk}$ are independent.

The label $\bmk$ refers to the physical replica generators.
Its relation to the root-character label of Section~\ref{sec:replica-geometry} is
\begin{equation}
	\bmell=(n-k_1,k_2)\pmod n.
	\label{eq:ward-root-character-dictionary}
\end{equation}
We use the physical-coordinate representatives $\varpi_{\bmk}$ of Eq.~\eqref{eq:disjoint-holomorphic-forms}; Eq.~\eqref{eq:common-differential-transport} relates them to the common root forms.
The insertions and the derivative with respect to $x$ remain in the coordinate $z$.
% For a single real field, the Fourier components instead obey $\widetilde\phi_{-\bmk}=\overline{\widetilde\phi_{\bmk}}$.

For a mixed character $1\leq k_1,k_2\leq n-1$, write
\begin{equation}
	a = \frac{k_1}{n}, \qquad b = \frac{k_2}{n}, \qquad \bmk^*=(n-k_1,n-k_2).
	\label{eq:ward-character-angles}
\end{equation}
The twists at $(0,x,1,\infty)$ impose the charges $\bm\beta=(1-a,a,1-b,b)$ on this complex mode.
With the convention that $\partial Y_{\bmk}$ has local behavior $(z-z_j)^{-\beta_j}$, their weights are
\begin{equation}
	h_j = \bar h_j = \frac12 \beta_j(1-\beta_j).
	\label{eq:ward-sector-weights}
\end{equation}
Let $Z_{{\rm qu},\bmk}^{\bbC}(x,\bar x)$ be their four-point quantum factor, with the insertion at infinity defined by its primary-field limit.
This is the two-angle orbifold Green-function problem of Refs.~\cite{Erler:1992gt,Stieberger:1992vb}.
Ref.~\cite{Erler:1992gt} uses zero global monodromy and two-point factorization to fix its quantum part.
We use our replica periods and phases, together with the compact winding lattice of Section~\ref{sec:classical-winding}.
We differentiate with $x$ and $\bar x$ independently and restrict to $\bar x=x\in(0,1)$ after solving both Ward--Takahashi equations.

The Euler periods of Section~\ref{sec:disjoint-period-matrix} enter through the bilinear
\begin{equation}
	\calI_{\bmk}(x,\bar x) \coloneqq \frac{\sin\pi a}{\pi} \left[I_{\bmk}(x)J_{\bmk^*}(\bar x) + J_{\bmk}(x)I_{\bmk^*}(\bar x)\right].
	\label{eq:ward-bilinear}
\end{equation}
This bilinear is positive on $0<x<1$.

The quantum factor has the corresponding two-angle form
\begin{equation}
	Z_{{\rm qu},\bmk}^{\bbC}(x,\bar x) = \calN_{\bmk} \frac{|x|^{-2a(1-a)} (1-x)^{-a(1-b)}(1-\bar x)^{-b(1-a)}}{\calI_{\bmk}(x,\bar x)}.
	\label{eq:ward-complex-sector-result}
\end{equation}
Here $\calN_{\bmk}$ is independent of $x,\bar x$.
Denoting the bilinear of Ref.~\cite[Eq.~(4.1)]{Anastasopoulos:2011gn} by $I_{\rm ABR}$, the Euler transformation gives
\begin{equation}
	2\pi I_{\rm ABR}(x,\bar x) = (1-x)^{a-b}(1-\bar x)^{b-a}\calI_{\bmk}(x,\bar x).
	\label{eq:ward-two-angle-dictionary}
\end{equation}
The powers on the right cancel the corresponding change in the kinematic prefactor, so the cited quantum factor is $2\pi$ times Eq.~\eqref{eq:ward-complex-sector-result} at $\calN_{\bmk}=1$.
Appendix~\ref{sec:period-identities} gives the Euler-period dictionary on the same continued branches.

\subsubsection{Solving the Ward--Takahashi identity}
\label{sec:ward-period-constraints}

\paragraph{Local Green functions.}
Define the normalized sector expectation by
\begin{equation}
	\langle\!\langle \calO \rangle\!\rangle_{\bmk} \coloneqq \frac{\langle\calO\prod_{j=1}^4\sigma_{\beta_j}(z_j) \rangle_{\rm qu}}{\langle\prod_{j=1}^4\sigma_{\beta_j}(z_j)\rangle_{\rm qu}}.
	\label{eq:ward-normalized-insertion}
\end{equation}
Define Green functions for the holomorphic and mixed insertions of the complex field,
\begin{subequations}\label{eq:ward-green-functions}
	\begin{align}
		g_{\bmk}(z,z') &\coloneqq - \langle\!\langle \partial Y_{\bmk}(z) \partial \bar Y_{\bmk}(z')\rangle\!\rangle_{\bmk},
		\label{eq:ward-green-functions-holomorphic}\\
		h_{\bmk}(\bar z,z') &\coloneqq - \langle\!\langle \bar\partial Y_{\bmk}(\bar z) \partial\bar Y_{\bmk}(z')\rangle\!\rangle_{\bmk}.
		\label{eq:ward-green-functions-mixed}
	\end{align}
\end{subequations}
In the normalization~\eqref{eq:ward-field-normalization-complex-propagator}, $g_{\bmk}(z,z')\sim(z-z')^{-2}$, whereas the mixed Green function has no such pole.~\footnote{These are the stress-tensor Green functions of Ref.~\cite{Calabrese:2009ez}, with its complex field rescaled by $1/\sqrt2$ to the normalization~\eqref{eq:ward-field-normalization-complex-propagator}.}

Write the physical-coordinate representatives as $\varpi_{\bmk}=w_{\bmk}(z)\dd{z}$.
The coefficient functions of complementary characters obey
\begin{equation}
	w_{\bmk}(z) = \frac1{(z-x)u_0(z)^{n-k_1}v_0(z)^{n-k_2}}, \qquad w_{\bmk}(z)w_{\bmk^*}(z) = \frac1{z(z-x)(z-1)}.
	\label{eq:ward-cut-differentials}
\end{equation}
We use the branches fixed in Eq.~\eqref{eq:period-reference-branch}.
Write $\bar w_{\bmk^*}(\bar z;\bar x)$ for the complex conjugate coefficient, including its branch, on the physical slice $\bar x=x^*$.
Away from this slice, it denotes the local continuation in the independent variable $\bar x$.

The kernel in the holomorphic ansatz is fixed by its singular part up to a homogeneous term.
With two accessory functions $A_{\bmk}(x,\bar x)$ and $B_{\bmk}(x,\bar x)$, the Green functions can be written as
\begin{subequations}\label{eq:ward-green-ansatz}
	\begin{align}
		g_{\bmk}(z,z') &= w_{\bmk}(z)w_{\bmk^*}(z') \left[\frac{P_{\bmk}(z,z')}{(z-z')^2}+A_{\bmk}\right],
		\label{eq:ward-green-ansatz-holomorphic}\\
		h_{\bmk}(\bar z,z') &= B_{\bmk}\, \bar w_{\bmk^*}(\bar z;\bar x)\,w_{\bmk^*}(z'),
		\label{eq:ward-green-ansatz-mixed}
	\end{align}
\end{subequations}
with
\begin{equation}
	P_{\bmk}(z,z') = (1-b)z(z-1)(z'-x) + a(z-x)z'(z'-1) + (b-a)z(z'-x)(z'-1).
	\label{eq:ward-polar-polynomial}
\end{equation}
The polynomial satisfies $P_{\bmk}(z,z)=z(z-x)(z-1)$ and
\begin{equation}
	\frac{\partial_{z'}P_{\bmk}(z,z')|_{z'=z}}{z(z-x)(z-1)}
	=\frac{a}{z}+\frac{1-a}{z-x}+\frac{b}{z-1}
	=-\partial_z\log w_{\bmk^*}(z).
	\label{eq:ward-simple-pole-cancellation}
\end{equation}
The second identity cancels the simple pole produced by expanding $w_{\bmk^*}(z')$, leaving precisely the required double pole.
The coefficient functions also give the required local powers at the four twist insertions.
After subtraction of the prescribed polar bidifferential, the lifted homogeneous bidifferential is holomorphic in each variable on the compact cover, with characters $\bmk$ and $\bmk^*$.
Their one-dimensional eigenspaces in Eq.~\eqref{eq:disjoint-holomorphic-basis} force its remaining coordinate dependence to be $w_{\bmk}(z)w_{\bmk^*}(z')$.
For the mixed kernel, the antiholomorphic first argument has character $\bmk$, whose eigenspace is spanned by $\bar\varpi_{\bmk^*}$; its coordinate dependence is therefore $\bar w_{\bmk^*}(\bar z)w_{\bmk^*}(z')$.
The two zero-period conditions below fix the coefficients $A_{\bmk}(x,\bar x)$ and $B_{\bmk}(x,\bar x)$ of these products.

\paragraph{Accessory coefficients from zero periods.}
Equation~\eqref{eq:ward-zero-winding}, applied to the normalized insertion
of $\partial\bar Y_{\bmk}(z')$, gives
\begin{equation}
	\oint_C g_{\bmk}(z,z')\dd{z} + \oint_C h_{\bmk}(\bar z,z')\dd{\bar z}=0.
	\label{eq:ward-global-monodromy}
\end{equation}
We use the two raw closed lifts $C_\alpha=\alpha_{0,0}$ and $C_\beta=\beta_{0,0}$ of Eq.~\eqref{eq:homology-lifted-words}; their base-plane words are twist-neutral.
Their deck translates multiply these conditions by character phases and contain the selected integral basis of Eq.~\eqref{eq:homology-selected-curves}, whose completeness is proved in Lemma~\ref{lem:disjoint-complete-integral-marking}.
Vanishing on the two raw lifts therefore enforces all periods in this fixed character sector.

Transport each closed lift with fixed homology class as $x$ varies.
Divide Eq.~\eqref{eq:ward-global-monodromy} by $w_{\bmk^*}(z')$ and take $z'\to\infty$.
Since $P_{\bmk}(z,z')/(z-z')^2\to bz-ax$, the result is
\begin{equation}
	A_{\bmk}\oint_C\varpi_{\bmk} + B_{\bmk}\oint_C\bar\varpi_{\bmk^*} = \oint_C[ax-bz]\,\varpi_{\bmk} = x(1-x) \partial_x\oint_C\varpi_{\bmk}.
	\label{eq:ward-inhomogeneous-periods}
\end{equation}
Here $\bar\varpi_{\bmk^*}=\bar w_{\bmk^*}\dd{\bar z}$.
Indeed,
\begin{equation}
	\partial_z[z(1-z)w_{\bmk}] = [ax-bz]w_{\bmk} -x(1-x)\partial_xw_{\bmk},
	\label{eq:ward-total-derivative}
\end{equation}
and the total derivative integrates to zero on a closed lift.

The raw periods $\calA_{\bmk},\calB_{\bmk}$ in Eq.~\eqref{eq:period-raw-euler-values} already contain the needed contour phases.
Their coefficients $\calA_{\bmk}/I_{\bmk}$ and $\calB_{\bmk}/J_{\bmk}$ are independent of $x$.
For real $x$, conjugating the complementary periods along the same oriented curves gives
\begin{subequations}\label{eq:ward-contour-period-table}\label{eq:ward-contour-phases}
	\begin{align}
		\overline{\calA_{\bmk^*}}&=\frac{\calA_{\bmk}}{I_{\bmk}}I_{\bmk^*},
		\label{eq:ward-contour-period-table-alpha}\\
		\overline{\calB_{\bmk^*}}&=-\frac{\calB_{\bmk}}{J_{\bmk}}J_{\bmk^*}.
		\label{eq:ward-contour-period-table-beta}
	\end{align}
\end{subequations}
Only the beta period acquires the relative minus sign.
Continuing the same branches extends these relations locally to independent $x,\bar x$.

Dividing the alpha equation by $\calA_{\bmk}/I_{\bmk}$ and the beta equation by $\calB_{\bmk}/J_{\bmk}$ yields
\begin{subequations}\label{eq:ward-two-period-system}
	\begin{align}
		A_{\bmk}I_{\bmk}(x)+B_{\bmk}I_{\bmk^*}(\bar x) &= x(1-x)I_{\bmk}'(x),
		\label{eq:ward-two-period-system-alpha}\\
		A_{\bmk}J_{\bmk}(x)-B_{\bmk}J_{\bmk^*}(\bar x) &= x(1-x)J_{\bmk}'(x).
		\label{eq:ward-two-period-system-beta}
	\end{align}
\end{subequations}
The determinant of this system is proportional to the bilinear in Eq.~\eqref{eq:ward-bilinear}.

Eliminating $B_{\bmk}$ from the two-period system fixes the accessory coefficient entering the stress tensor:
\begin{equation}
	A_{\bmk}=x(1-x)\partial_x\log\calI_{\bmk}.
	\label{eq:ward-accessory-solution}
\end{equation}
The same nonsingular system fixes $B_{\bmk}$ uniquely.
The factor $\sin(\pi a)/\pi$ retains the beta-function normalization of the Euler periods.

\paragraph{The Ward--Takahashi identity.}
Point splitting the holomorphic Green function gives the stress-tensor
insertion,
\begin{equation}
	\langle\!\langle T_{\bbC}(z)\rangle\!\rangle_{\bmk} = \lim_{z'\to z}\left[g_{\bmk}(z,z')-\frac1{(z-z')^2}\right].
	\label{eq:ward-point-splitting}
\end{equation}
The quadratic coefficient of $P_{\bmk}(z,z')$ at $z'=z$ is $bz-ax$.
Expanding the complementary coefficient function to the same order and using the linear cancellation in Eq.~\eqref{eq:ward-simple-pole-cancellation} gives the poles at the moving twist:
\begin{equation}
	\langle\!\langle T_{\bbC}(z)\rangle\!\rangle_{\bmk} = \frac{a(1-a)}{2(z-x)^2} + \frac1{z-x}\left[-\frac{a(1-a)}x + \frac{a(1-b)}{1-x}-\frac{A_{\bmk}}{x(1-x)}\right]+O(1).
	\label{eq:ward-stress-pole}
\end{equation}
The double pole reproduces the weight in Eq.~\eqref{eq:ward-sector-weights}.

The simple pole in the conformal Ward--Takahashi identity is $\partial_x\log Z_{{\rm qu},\bmk}^{\bbC}$, so using the accessory solution~\eqref{eq:ward-accessory-solution} gives
\begin{equation}
	\partial_x\log Z_{{\rm qu},\bmk}^{\bbC} = - \frac{a(1-a)}x + \frac{a(1-b)}{1-x} - \partial_x\log\calI_{\bmk}.
	\label{eq:ward-holomorphic-equation}
\end{equation}
The antiholomorphic calculation replaces $a,b$ by $1-a,1-b$ and $x$ by $\bar x$.
Using $\calI_{\bmk^*}(\bar x,x)=\calI_{\bmk}(x,\bar x)$ gives
\begin{equation}
	\partial_{\bar x}\log Z_{{\rm qu},\bmk}^{\bbC} = -\frac{a(1-a)}{\bar x}+\frac{b(1-a)}{1-\bar x} - \partial_{\bar x}\log\calI_{\bmk}.
	\label{eq:ward-antiholomorphic-equation}
\end{equation}

Integrating the holomorphic equation leaves an arbitrary function of $\bar x$; the antiholomorphic equation reduces it to the constant $\calN_{\bmk}$.
This recovers Eq.~\eqref{eq:ward-complex-sector-result}, including the relative powers of $1-x$ and $1-\bar x$ fixed by the contour phases.

\subsubsection{The real-boson determinant}
\label{sec:ward-real-determinant}
The $(n-1)^2=g_d$ mixed characters supply the bilinears.
Characters of type $(k_1,0)$ see only the twists at $0,x$, and those of type $(0,k_2)$ see only the twists at $1,\infty$.
The other sectors therefore give
\begin{subequations}\label{eq:ward-axis-sectors}
	\begin{gather}
		Z_{{\rm qu},(k_1,0)}^{\bbC} = \calN_{k_1,0}|x|^{-2a(1-a)}
		\label{eq:ward-axis-sectors-first-axis}\\
		Z_{{\rm qu},(0,k_2)}^{\bbC} = \calN_{0,k_2}
		\label{eq:ward-axis-sectors-second-axis}\\
		Z_{{\rm qu},(0,0)}^{\bbC} = \calN_{0,0}.
		\label{eq:ward-axis-sectors-trivial}
	\end{gather}
\end{subequations}

Independence of the auxiliary complex Fourier modes gives a product over all characters; write $\calN_n^{\bbC}$ for the product of their constants.
The powers follow from the finite sums
\begin{subequations}\label{eq:ward-character-power-sums}
	\begin{align}
		2n\sum_{k=1}^{n-1}\frac{k}{n}\left(1-\frac{k}{n}\right) &= \frac{n^2-1}{3}=2\Delta_{n,\bbC}^{(3)},
		\label{eq:ward-character-power-sums-x}\\
		\sum_{k_1,k_2=1}^{n-1}\frac{k_1}{n}\left(1-\frac{k_2}{n}\right) &= \frac{(n-1)^2}{4}=\frac{g_d}{4}.
		\label{eq:ward-character-power-sums-one-minus-x}
	\end{align}
\end{subequations}
Taking the positive Gaussian square root of the full character product, as in Eq.~\eqref{eq:ward-real-complex-determinant}, gives on the ordered real line
\begin{subequations}\label{eq:ward-real-quantum-factor}
	\begin{gather}
		Z_{\rm qu}^{\bbR}(x) = \calN_n^{\bbR} x^{-2\Delta_{n,\bbR}^{(3)}}(1-x)^{-g_d/4} \prod_{k_1,k_2=1}^{n-1}\calI_{\bmk}(x,x)^{-1/2},
		\label{eq:ward-real-quantum-factor-determinant}\\
		\Delta_{n,\bbR}^{(3)}=\frac{n^2-1}{12}, \qquad \Delta_{n,\bbC}^{(3)}=2\Delta_{n,\bbR}^{(3)}, \qquad \calN_n^{\bbR}=(\calN_n^{\bbC})^{1/2}.
		\label{eq:ward-real-quantum-factor-dimensions-and-normalization}
	\end{gather}
\end{subequations}
The explicit power of $x$ agrees with the $c=1$ twist dimension of Section~\ref{sec:interval-geometry}.
The adjacent exponent also receives a contribution from the bilinear product, whose asymptotics are evaluated in Eq.~\eqref{eq:endpoint-quantum-limit}.
% In real Fourier modes, $\bmk$ pairs with $-\bmk$, and each self-conjugate character $2\bmk=0$ contributes a single real mode, including at even $n$.

\paragraph{The equal-angle specialization.}
At $a=b$, set $F(x)={}_2F_1(a,1-a;1;x)$.
Both Euler periods have the factor $\pi/\sin\pi a$, so
\begin{equation}
	\calI_{\bmk}(x,\bar x) = \frac{\pi}{\sin\pi a} \left[F(x)F(1-\bar x)+F(\bar x)F(1-x)\right].
	\label{eq:ward-equal-angle-bilinear}
\end{equation}
The bracket is the ordinary replica-twist hypergeometric bilinear~\cite{Calabrese:2009ez}; the prefactor retains the beta-function normalization of our periods.

\subsection{Absolute four-point normalization}\label{sec:four-point-normalization}

Put $g=g_d=(n-1)^2$ and $\Omega=\Omega_n^{(3)}(x)$.
The Hodge determinant from the winding sum and the Ward bilinears from the fluctuation calculation encode the same pairing in different differential bases.

\paragraph{Winding and fluctuation factors.}
We write $Z_{\rm cl}^{\bbR}$ for the winding sum and $Z_{\rm qu}^{\bbR}$ for the nonconstant Gaussian contribution to the fixed-frame four-point function.
With the Ward--Takahashi integration constant $\calN_n^{\bbR}$ included in $Z_{\rm qu}^{\bbR}$, their product gives
\begin{equation}
	\calG_n^{(3)}(x,x;\eta) = \frac{\calN_{n,\eta}}{\calN_n^{\bbR}} Z_{\rm qu}^{\bbR}(x) Z_{\rm cl}^{\bbR}(\Omega_n^{(3)}(x);\eta).
	\label{eq:ward-factorized-correlator}
\end{equation}
Here $\calN_{n,\eta}$ collects the Ward integration constant, compact constant-mode measure, and state and twist normalizations.
These factors are independent of $x$.

Multiplying Eqs.~\eqref{eq:classical-theta-result} and~\eqref{eq:ward-real-quantum-factor-determinant} gives the fixed-frame four-point
function up to a constant:
\begin{equation}
	\calG_n^{(3)}(x,x;\eta)
	= \calN_{n,\eta}\,\eta^{-g/2} x^{-2\Delta_n^{(3)}}(1-x)^{-g/4} \frac{\sqrt{\det\Im\Omega} \Theta(\bm0\mid T_\eta(\Omega))}{\prod_{k_1,k_2=1}^{n-1}\calI_{\bmk}(x,x)^{1/2}}.
	\label{eq:assembly-unfixed-correlator}
\end{equation}

With unit two-point twist norms, the pair left at $1,\infty$ has expectation one, as in the ordinary two-interval calculation~\cite{Calabrese:2009ez}.
Equation~\eqref{eq:R-identity-normalization} therefore requires
\begin{equation}
	\lim_{x\to0^+}x^{2\Delta_n^{(3)}} \calG_n^{(3)}(x,x;\eta) = \lim_{x\to0^+}\calR_n^{(3)}(x;\eta) = 1.
	\label{eq:assembly-identity-condition}
\end{equation}

\paragraph{The A-period determinant.}
To isolate the $x$ dependence of the A-period determinant, define
\begin{equation}
	\scrA_n(x)\coloneqq \prod_{k_1,k_2=1}^{n-1} {}_2F_1\!\left(\frac{k_1}{n},1-\frac{k_2}{n};1;x\right).
	\label{eq:assembly-hypergeometric-product}
\end{equation}
Every factor is positive on $0<x<1$ and equals one at $x=0$.
In Eq.~\eqref{eq:period-A-block}, each row of the constant character matrix is multiplied by its reference A-period.
The beta factor and phase of each period cancel in the ratio to $x=0$.
Equation~\eqref{eq:period-I-euler} directly gives
\begin{equation}
	\frac{\det\sfA(x)}{\det\sfA(0)}=\scrA_n(x).
	\label{eq:assembly-A-determinant}
\end{equation}
Here $\sfA(0)$ denotes the finite limiting matrix, which is invertible by the finite character inverse \eqref{eq:period-finite-inverse}.

\paragraph{The Hodge determinant from character norms.}

\begin{lemma}[Hodge norms from the Ward bilinear]\label{lem:hodge-ward-identity}
	For a mixed physical character, put $a=k_1/n$, $b=k_2/n$ and
	$\bmk^*=(n-k_1,n-k_2)$ as in the Ward calculation.
	On $0<x<1$, the norm of $\varpi_{\bmk}$ defined in Eq.~\eqref{eq:period-character-gram} and the bilinear in Eq.~\eqref{eq:ward-bilinear} obey
	\begin{equation}
		\begin{aligned}
			h_{\bmk}
			&=n^2\sin(\pi a)(1-x)^{b-a}
				[I_{\bmk}J_{\bmk^*}+J_{\bmk}I_{\bmk^*}]\\
			&=\pi n^2(1-x)^{b-a}\calI_{\bmk}(x,x).
		\end{aligned}
		\label{eq:assembly-character-hodge-norm}
	\end{equation}
\end{lemma}

The Riemann bilinear relation~\eqref{eq:period-positive-form} evaluates the same Hodge pairing in the canonical frame.
Contracting the canonical blocks with the finite prefix and correction sums derived in Appendix~\ref{sec:period-identities} gives
\begin{equation}
	h_{\bmk} = n^2\sin(\pi a)I_{\bmk} \bigl[2J_{\bmk} + \sin(\pi a)[\cot(\pi a)-\cot(\pi b)]I_{\bmk}\bigr].
	\label{eq:assembly-hodge-period-contraction}
\end{equation}
The cotangent term comes from the A-cycle terms added to the raw B-cycles to enforce canonical intersections.
By Eq.~\eqref{eq:assembly-complementary-period-connection-J}, it changes one copy of $J_{\bmk}$ into $(1-x)^{b-a}J_{\bmk^*}$.
The companion identity $(1-x)^{b-a}I_{\bmk^*}=I_{\bmk}$ then gives the two terms of the Ward bilinear and proves Lemma~\ref{lem:hodge-ward-identity}.

Taking the determinant of the Gram identity gives
\begin{equation}
	|\det\sfA(x)|^2\det\Im\Omega(x)
	=\prod_{\bmk}h_{\bmk}
	=(\pi n^2)^g\prod_{\bmk}\calI_{\bmk}(x,x),
	\label{eq:assembly-hodge-determinant-step}
\end{equation}
because $\sum_{k_1,k_2}(b-a)=0$.
The blocks use the physical $\varpi$ frame.
The exact transport $\omega_{n-k_1,k_2}=Q_{\bmk}\varpi_{\bmk}$ changes each norm to $\pi y^{a-b}\calI_{\bmk}$ and multiplies both the Gram determinant and $|\det\sfA|^2$ by $|\det\sfQ|^2=n^{-2g}$.
Their ratio $\det\Im\Omega$ is therefore unchanged; the opposite character powers cancel in the full product in either frame.
Solving the determinant identity yields
\begin{equation}
	\frac{\sqrt{\det\Im\Omega(x)}}{\prod_{\bmk}\calI_{\bmk}(x,x)^{1/2}}
	=\frac{(\pi n^2)^{g/2}}{|\det\sfA(0)|\,\scrA_n(x)}.
	\label{eq:assembly-hodge-basis-ratio}
\end{equation}

\paragraph{The identity-channel limit.}
The Euler integrals in Eqs.~\eqref{eq:period-real-I}--\eqref{eq:period-real-J} obey
\begin{subequations}\label{eq:assembly-euler-cusp}
	\begin{gather}
		I_{\bmk}(x)=\frac{\pi}{\sin\pi a}+O(x)
		\label{eq:assembly-euler-cusp-I}\\
		J_{\bmk}(x)=\log(1/x)+O(1),\qquad a=\frac{k_1}{n}.
		\label{eq:assembly-euler-cusp-J}
	\end{gather}
\end{subequations}
The first relation follows from the regular hypergeometric series at $x=0$.
In the second integral, $t=xu$ makes the lower-endpoint region $t=O(x)$ finite, while the region $x\ll t\ll1$ has integrand $t^{-1}$ and supplies the logarithm.
The upper-endpoint region also contributes only a finite term.
The complementary character has the same leading terms, so Eq.~\eqref{eq:ward-bilinear} gives
\begin{equation}
	\calI_{\bmk}(x,x)=2\log(1/x)+O(1)
	\label{eq:assembly-bilinear-cusp}
\end{equation}
for every mixed character.

The degeneration of the period matrix follows from Eq.~\eqref{eq:period-character-ratio}.
Using the same physical phases,
\begin{equation}
	\tau_{\bmk}(x) = \frac{i\log(1/x)}{2\pi}(1-p_1)(1-p_2)+O(1).
	\label{eq:assembly-character-cusp}
\end{equation}
Define the $(n-1)\times(n-1)$ Cartan matrix of type $A_{n-1}$ by
\begin{equation}
	(C_{A_{n-1}})_{IJ} = 2\delta_{IJ}-\delta_{I,J+1}-\delta_{I,J-1}, \qquad 1\leq I,J\leq n-1.
	\label{eq:assembly-cartan-definition}
\end{equation}
The deltas here are ordinary, noncyclic Kronecker deltas.
Its determinant is $\det C_{A_{n-1}}=n$.
The finite transform that appears in Eq.~\eqref{eq:period-canonical-explicit} then reduces in each direction to
\begin{equation}
	\begin{aligned}
		\frac1n\sum_{k=1}^{n-1}D_k(I)\zeta_n^{k(J-1)}(1-\zeta_n^k)
		&= 2\delta_{IJ}-\delta_{I,J+1}-\delta_{I,J-1}\\
		&= (C_{A_{n-1}})_{IJ},\qquad 1\leq I,J\leq n-1,
	\end{aligned}
	\label{eq:assembly-cartan-transform}
\end{equation}
Indeed, the summand is $\zeta_n^{k(J-I)}(2-\zeta_n^k-\zeta_n^{-k})$, and Fourier orthogonality gives the three Kronecker deltas.
The A-cycle correction $\sfC_{\bmk,\bmJ}$ is independent of $x$ and contributes only to the finite term.
Hence the two replica directions give
\begin{equation}
	\Im\Omega_n^{(3)}(x) = \frac{\log(1/x)}{2\pi}(C_{A_{n-1}}\otimes C_{A_{n-1}}) + O(1).
	\label{eq:assembly-period-cusp}
\end{equation}

The determinant of the leading Cartan tensor product fixes the finite Hodge coefficient.
It follows from $\det C_{A_{n-1}} = n$ that
\begin{subequations}\label{eq:assembly-hodge-cusp}
	\begin{align}
		\det(C_{A_{n-1}}\otimes C_{A_{n-1}} )&= n^{2(n-1)},
		\label{eq:assembly-hodge-cusp-cartan-determinant}\\
		\sqrt{\det\Im\Omega_n^{(3)}(x)} &= n^{n-1}\left(\frac{\log(1/x)}{2\pi}\right)^{g/2} \bigl[1+O\bigl((\log(1/x))^{-1}\bigr)\bigr].
		\label{eq:assembly-hodge-cusp-hodge-determinant}
	\end{align}
\end{subequations}

The Cartan tensor product is positive definite.
Thus every eigenvalue of $\Im T_\eta$ grows linearly with $\log(1/x)$, suppressing every nonzero momentum--winding vector.
Bounding the absolute value of the nonzero terms by a positive Gaussian lattice sum therefore gives
\begin{equation}
	\Theta(\bm0\mid T_\eta(\Omega_n^{(3)}(x)))\longrightarrow1.
	\label{eq:assembly-theta-cusp}
\end{equation}
Combining Eqs.~\eqref{eq:assembly-bilinear-cusp} and~\eqref{eq:assembly-hodge-cusp-hodge-determinant}, the logarithmic powers cancel:
\begin{equation}
	\lim_{x\to0^+} \frac{\sqrt{\det\Im\Omega_n^{(3)}(x)}}{\displaystyle\prod_{k_1,k_2=1}^{n-1} \calI_{\bmk}(x,x)^{1/2}} = \frac{n^{n-1}}{(4\pi)^{g/2}}.
	\label{eq:assembly-identity-finite-part}
\end{equation}
The unit identity coefficient in Eq.~\eqref{eq:assembly-identity-condition} consequently fixes
\begin{equation}
	\calN_{n,\eta} = \eta^{g/2}\frac{(4\pi)^{g/2}}{n^{n-1}}.
	\label{eq:assembly-normalization-constant}
\end{equation}
The combined measure and normalization factors supply $\eta^{g/2}$, canceling the Poisson prefactor in Eq.~\eqref{eq:assembly-unfixed-correlator}.

\paragraph{The normalized ratio.}
The finite identity-channel ratio also fixes the constant in Eq.~\eqref{eq:assembly-hodge-basis-ratio}:
\begin{equation}
	\det\Im\Omega(x) = \frac{n^{2(n-1)}}{(4\pi)^g} \frac{\prod_{\bmk}\calI_{\bmk}(x,x)}{\scrA_n(x)^2}.
	\label{eq:assembly-hodge-identity}
\end{equation}
Substitution in the assembled path integral gives the two equivalent forms
\begin{align}
	\calR_n^{(3)}(x;\eta)
	&=\frac{(4\pi)^{g/2}}{n^{n-1}}(1-x)^{-g/4}
	\frac{\sqrt{\det\Im\Omega(x)}}{\prod_{\bmk}\calI_{\bmk}(x,x)^{1/2}}
	\Theta(\bm0\mid T_\eta(\Omega(x)))
	\label{eq:assembly-hodge-form}\\
	&=\frac{(1-x)^{-g/4}}{\scrA_n(x)}
	\Theta(\bm0\mid T_\eta(\Omega(x))).
	\label{eq:assembly-hypergeometric-ratio}
\end{align}
Appendix~\ref{sec:disjoint-quillen} derives the prefactor from character determinants.

\subsection{The theta-ratio representation}\label{sec:theta-ratio}

We express the A-period product in Eq.~\eqref{eq:assembly-hypergeometric-ratio} as a theta constant using the abelian-cover Thomae formula~\cite{KopeliovichZemel:2019thomae} in the monic algebraic differential frame.
Appendix~\ref{sec:math-thomae} identifies the non-special spin line and computes its characteristic in our marking.
Appendix~\ref{sec:fixed-spin-reference} gives a physical derivation of the same identity by bosonizing the auxiliary Dirac charge correlator.
The same spin line has zero characteristic in the cyclic marking of Ref.~\cite[Theorem~6.1]{Enolski:2006sdf}; see also the cyclic theta--period relation in Ref.~\cite[Theorem~5.6]{Enolski:2003hia} and the hyperelliptic two-interval reduction in Ref.~\cite{Zamolodchikov:1987ae}.

\paragraph{The reference spin line.}
Let $P_\xi$ denote the reduced sum of the $n$ points above the branch value $\xi\in\{0,x,1,\infty\}$.
These labels refer to the physical coordinate $z$; in the common geometric coordinate, $P_x$ lies above $w=0$ and $P_\infty$ above $w=y^{-1}$.
The meromorphic functions $u_0,v_0$ in Eq.~\eqref{eq:disjoint-physical-frame} express the same cover in this physical chart.
The meromorphic differential
\begin{equation}
	\varpi_*\coloneqq(u_0v_0)^{-(n-1)}\dd{z} = \rho^{n-1}(u/v)^{n-1}\dd{z}
	\label{eq:assembly-reference-differential}
\end{equation}
has even divisor
\begin{equation}
	\div\varpi_*=2D_*, \qquad D_*\coloneqq(n-1)P_x-P_\infty, \qquad \deg D_*=n(n-2)=g-1.
	\label{eq:assembly-reference-divisor}
\end{equation}
The invariant spin line $L_*\coloneqq\calO(D_*)$ is non-special: $H^0(\Sigma,L_*)=0$.
The valuation argument and quadratic-refinement calculation in Appendix~\ref{sec:math-thomae} give
\begin{equation}
	q_{\varpi_*}(A_{I,J})=0,\qquad
	q_{\varpi_*}(B_{I,J})=1,\qquad
	\delta_*=\begin{bmatrix}\bm0\\ \tfrac12\bm1_g\end{bmatrix}.
	\label{eq:assembly-reference-characteristic}
\end{equation}
This is the characteristic of $L_*$ in the transported canonical basis.
The root-coordinate change preserves that spin line and the integral cycles; its explicit divisor representative above uses the physical roots $u_0,v_0$.

\begin{lemma}[Thomae identity in the canonical marking]\label{lem:marked-thomae}
	On the marked real branch $0<x<1$, with $g=(n-1)^2$ and the A-period product of Eq.~\eqref{eq:assembly-hypergeometric-product},
	\begin{equation}
		\left|\theta[\delta_*](\bm0\mid\Omega(x))\right|^2
		=\theta[\delta_*](\bm0\mid\Omega(x))^2
		=(1-x)^{g/4}\scrA_n(x).
		\label{eq:assembly-thomae-identity}
	\end{equation}
\end{lemma}

\paragraph{The algebraic frame.}
The physical representatives $\varpi_{\bmk}$ of Eq.~\eqref{eq:disjoint-holomorphic-forms} have the monic normalization required by the abelian-cover Thomae formula:
\begin{equation}
	\begin{aligned}
		y_{\bmk} &\coloneqq (z-x)u_0^{n-k_1}v_0^{n-k_2},
		&\varpi_{\bmk} &= \frac{\dd{z}}{y_{\bmk}},\\
		y_{\bmk}^{\,n} &= z^{n-k_1}(z-x)^{k_1}(z-1)^{n-k_2}.
	\end{aligned}
	\label{eq:assembly-thomae-algebraic-basis}
\end{equation}
For each mixed character the sum of the four fractional branch orders is two, giving exactly one differential with no polynomial multiplier.
Nontrivial axis characters have sum one and contribute none.
Thus these are the same $g$ differentials used in the A-period determinant of Eq.~\eqref{eq:assembly-A-determinant}, with the physical root phases fixed as in Section~\ref{sec:disjoint-period-matrix}.

To keep the normalization at infinity explicit, temporarily place the fourth branch value at $\Lambda>1$ and use the monic family defined in Appendix~\ref{sec:math-thomae}.
Its normalized divisor satisfies the non-speciality criterion, and its differential frame tends to the physical one with
\begin{equation}
	\begin{aligned}
		\omega_{\bmk,\Lambda} &= (-\Lambda)^{-k_2/n}\bigl[\varpi_{\bmk}+O(\Lambda^{-1})\bigr],\\
		\det\sfA_\Lambda &= (-\Lambda)^{-g/2}\bigl[\det\sfA(x)+O(\Lambda^{-1})\bigr].
	\end{aligned}
	\label{eq:assembly-thomae-frame-limit}
\end{equation}
Here $x\in(0,1)$ and $n$ are fixed, and the same compact cycle representatives and roots are continued through the family.
The exponent follows from $\sum_{k_1,k_2=1}^{n-1}k_2/n=g/2$.
The source uses cycles as rows, so its period matrix $C$ is $\sfA_\Lambda^{\top}$ and has the same determinant.

\paragraph{The branch exponents.}
For the ordered monodromies $(\sfg_1,\sfg_1^{-1},\sfg_2,\sfg_2^{-1})$ at $(0,x,1,\Lambda)$, the abelian-cover theorem determines the exponents in $\theta^2/\det\sfA_\Lambda$~\cite[Corollary~6.4 and Theorem~6.6]{KopeliovichZemel:2019thomae}.
There are only two types of pair.
Independent directions have inertia subgroups intersecting only at the identity and exponent $g/4$; inverse directions have exponent zero, by the finite sums evaluated in Eq.~\eqref{eq:math-thomae-replica-exponents}.
Consequently the branch factor is
\begin{equation}
	(0-1)^{g/4}(0-\Lambda)^{g/4}(x-1)^{g/4}(x-\Lambda)^{g/4}.
	\label{eq:assembly-thomae-branch-product}
\end{equation}
The two factors involving $\Lambda$ grow as $(-\Lambda)^{g/2}$ and cancel the frame power in Eq.~\eqref{eq:assembly-thomae-frame-limit}.
Keeping the branch order and phases fixed, the limit $\Lambda\to\infty$ leaves $(1-x)^{g/4}$ times an $x$-independent phase~\cite[Proposition~7.5 and its proof]{KopeliovichZemel:2019thomae}.
Absorbing that phase and $\det\sfA(0)$ into one constant gives
\begin{equation}
	\theta[\delta_*](\bm0\mid\Omega(x))^2 = c_*\,(1-x)^{g/4}\scrA_n(x).
	\label{eq:assembly-thomae-structure}
\end{equation}

\paragraph{Cusp normalization.}
At $x\to0^+$, the positive logarithmic growth in Eq.~\eqref{eq:assembly-period-cusp} and the zero upper characteristic imply $\theta[\delta_*](\bm0\mid\Omega(x))\to1$: the zero lattice vector contributes one and the other terms are suppressed by a convergent Gaussian sum.
Since $\scrA_n(x)\to1$, Eq.~\eqref{eq:assembly-thomae-structure} fixes $c_*=1$.

On this marked real slice the positive right-hand side forces the nonzero theta constant to be real.
Continuation from its unit cusp limit fixes the positive sign.
This is the counterpart of the Thomae reduction of the hypergeometric product in the ordinary two-interval calculation~\cite{Calabrese:2009ez}.

\paragraph{The normalized theta ratio.}
Substituting Eq.~\eqref{eq:assembly-thomae-identity} into Eq.~\eqref{eq:assembly-hypergeometric-ratio} proves Theorem~\ref{prop:disjoint-four-point}.
All radius dependence lies in the numerator, so momentum--winding exchange gives T-duality.

Equation~\eqref{eq:real-cross-ratio-relations-3} then gives the fixed-frame correlator and the finite function.
In particular,
\begin{equation}
	\calF_n^{(3)}(x;\eta)
	=(1-x)^{\Delta_n^{(3)}}
	\frac{\Theta(\bm0\mid T_\eta(\Omega_n^{(3)}(x)))}
	{|\theta[\delta_*](\bm0\mid\Omega_n^{(3)}(x))|^2}.
	\label{eq:assembly-finite-function-power}
\end{equation}

\section{The adjacent three-point coefficient}\label{sec:adjacent-boson}
For adjacent intervals, the three labelled twists leave no cross ratio.
We calculate the compact-boson coefficient $C_n^{(3)}$ with the unit two-point norms of Eq.~\eqref{eq:adjacent-three-point-coefficient}.

The Fermat curve $F_n$ and its marked period matrix $\Omega_{F,n}$ were constructed in Section~\ref{sec:fermat-periods}.
Its genus is $g_F=(n-1)(n-2)/2$, and $\calH_{F_n}$ denotes the common harmonic factor of Eq.~\eqref{eq:common-harmonic-factor} evaluated on this curve.
Define the positive finite Gamma product
\begin{equation}
	\calP_n \coloneqq \prod_{k=1}^{n-1}\Gamma(k/n)^{3(2k-n)}.
	\label{eq:fermat-gamma-single-product}
\end{equation}
\begin{theorem}[Normalized adjacent three-point coefficient]\label{thm:adjacent-three-point}
	The plane-normalized adjacent coefficient is
	\begin{equation}
		C_{n}^{(3)}(\eta) = \frac{(4\pi)^{g_F/2}}{n^{(3n-4)/4}}\,\sqrt{\calP_n}\,\calH_{F_n}(\eta).
		\label{eq:fermat-direct-normalized-coefficient}
	\end{equation}
\end{theorem}
All radius dependence is contained in $\calH_{F_n}$.
The coefficient determines $\kappa_n^{(3)}$.

\subsection{The normalized Fermat path integral}\label{sec:direct-fermat-determinant}
We apply the common quotient prescription in Eq.~\eqref{eq:common-adjacent-quotient} to the connected Fermat cover.
The unnormalized three-point function is $Z_F/Z_0^{n^2}$.
Each external two-point norm contains $n$ connected cyclic covers and is $Z_p^n/Z_0^{n^2}$.
Dividing by the square roots of the three norms therefore gives $Z_FZ_0^{n^2/2}/Z_p^{3n/2}$ before conversion from the round sphere to the plane.
% The twists have unit two-point norms, regular caps and trivial discrete torsion.
The measure of Appendix~\ref{sec:math-conical-determinants} fixes the remaining finite factors.
A complementary covering-space formulation factors the adjacent coefficient into a geometric anomaly and the seed CFT partition function in a constant-curvature reference metric~\cite{Ma:2026frw}.

\paragraph{Cover metrics and reference determinants.}
On the adjacent cover, the geometric coordinate $w=u^n$ is also the physical three-point coordinate; denote it by $z$ in the determinant calculation.
Put the insertions at $0,1,\infty$, and equip this base sphere with the curvature-one round metric
\begin{subequations}\label{eq:direct-fermat-metrics}
	\begin{gather}
		m_0=\frac{4|\dd{z}|^2}{(1+|z|^2)^2},
		\label{eq:direct-fermat-metrics-base}\\
		f_n:F_n\longrightarrow\bbP^1,\quad f_n(u,v)=u^n,\qquad
		p_n:\bbP^1\longrightarrow\bbP^1,\quad p_n(t)=t^n.
		\label{eq:direct-fermat-metrics-maps}
	\end{gather}
\end{subequations}
The adjacent cover carries metric $m_{F,n}=f_n^*m_0$.
A two-point cover for one length-$n$ cycle is a sphere with two conical points, often called a football, with metric $m_{p,n}=p_n^*m_0$.
Since each external translation has $n$ disjoint length-$n$ cycles, its two-point cover has $n$ connected football components.
The areas are
\begin{equation}
	A_0=4\pi,\qquad A_{p,n}=4\pi n,\qquad A_{F,n}=4\pi n^2.
	\label{eq:direct-fermat-areas}
\end{equation}
The Fermat cover has $3n$ cones of angle $2\pi n$; each football has two.
In a centered chart near a cone, $z=\xi^n$ gives the leading metric $\dd{r}^2+n^2r^2\dd{(\arg\xi)}^2$, where $r$ is proper radius.

Let $D_0$, $D_{p,n}$ and $D_{F,n}$ denote the zeta determinants of the nonnegative scalar Laplacians for these three metrics.
On the conical surfaces we take the Friedrichs extension, obtained by closing the Dirichlet energy on functions supported away from the cones.
The prime removes the single constant eigenfunction on each connected surface.
\begin{subequations}\label{eq:direct-fermat-determinants}
	\begin{gather}
		D_0=\det{}'\Delta_{m_0}
		\label{eq:direct-fermat-determinants-sphere}\\
		D_{p,n}=\det{}'\Delta_{m_{p,n}}
		\label{eq:direct-fermat-determinants-football}\\
		D_{F,n}=\det{}'\Delta_{m_{F,n}}.
		\label{eq:direct-fermat-determinants-Fermat}
	\end{gather}
\end{subequations}
The spectral zeta function and the heat coefficient used below are recalled in Appendix~\ref{sec:math-conical-determinants}.

With $\Delta_n^{(3)}=(n^2-1)/12$, the normalized path integral reduces to
\begin{equation}
	C_{n}^{(3)}(\eta) = 2^{-\Delta_n^{(3)}}(4\pi)^{g_F/2}n^{-(3n-4)/4} \frac{D_{p,n}^{3n/4}}{D_{F,n}^{1/2}D_0^{n^2/4}} \calH_{F_n}(\eta).
	\label{eq:direct-fermat-spectral-coefficient}
\end{equation}

\paragraph{Scalar cap cancellation in the external quotient.}
Fix one smoothing profile in the proper-radius variable $r/\varepsilon$ and use it for every cone of angle $2\pi n$.
Scalar gluing separates local Dirichlet-cap and collar factors from the boundary operator carrying the exterior geometry~\cite{Kalvin:2019fcs}.
With these matched profiles and the same Friedrichs domain, the local factors cancel by the cone count
\begin{equation}
	3n-\frac{3n}{2}\,2=0.
	\label{eq:direct-fermat-cone-cancellation}
\end{equation}
This includes finite cap constants.
For scalar smoothing within the same conformal class, the zero-energy extension and fixed-cut boundary metric are unchanged, so the jump operator is unchanged~\cite[Lemma~2.10]{Kalvin:2019fcs}.
The character-weight smoothing used in Section~\ref{sec:fermat-absolute-determinant} changes the sector operator; that separate comparison uses the boundary estimate in Appendix~\ref{sec:character-cap-comparison}.

For the constant kernel, gluing compares $\det{}'\Delta/\Area$; its area factor combines with the Gaussian constant-mode measure.
The remaining bulk terms cancel because
\begin{subequations}\label{eq:direct-fermat-bulk-cancellation}
	\begin{gather}
		A_{F,n}+\frac{n^2}{2}A_0-\frac{3n}{2}A_{p,n}=0
		\label{eq:direct-fermat-bulk-cancellation-area}\\
		(2-2g_F)+n^2-3n=0.
		\label{eq:direct-fermat-bulk-cancellation-euler}
	\end{gather}
\end{subequations}
Thus the quotient can be evaluated with the closed conical determinants.

\paragraph{Normalization of the Fermat winding sum.}
Regular caps exclude peripheral windings, so the Fermat contribution is the full $H^1(F_n,\bbZ)$ sum of Eqs.~\eqref{eq:classical-winding-sum} and~\eqref{eq:classical-theta-result}, evaluated in the integral basis of Section~\ref{sec:fermat-integral-homology} with $\Omega=\Omega_{F,n}$.

\paragraph{The radius factor.}
The connected Gaussian measure of Eq.~\eqref{eq:direct-fermat-mode-measure} uses the same real-mode and boundary measures on the cover and reference spheres.
Each connected surface contributes one constant mode.
Their signed count in the external quotient is $1+n^2/2-3n/2=g_F$, as in Eq.~\eqref{eq:common-adjacent-component-count}, giving $(\sqrt{2\pi}R)^{g_F}$.
The remaining powers of $2\pi$ come from the nonconstant spectral measure.

The spectral part of Eq.~\eqref{eq:direct-fermat-mode-measure} contributes $(2\pi)^{\zeta_\Delta(0)/2}$ on each connected surface.
Inserting the areas and cone counts into the heat formula~\eqref{eq:math-conical-heat} gives
\begin{equation}
	\begin{aligned}
		\zeta_0(0)&=\frac13-1=-\frac23,\\
		\zeta_{p,n}(0)&=\frac n3+\frac2{12}\left(\frac1n-n\right)-1
		=\frac16\left(n+\frac1n\right)-1,\\
		\zeta_{F,n}(0)&=\frac{n^2}{3}+\frac{3n}{12}\left(\frac1n-n\right)-1
		=\frac{n^2-9}{12}.
	\end{aligned}
	\label{eq:direct-fermat-heat-values}
\end{equation}
The curvature terms are proportional to the area, the conical terms count the indicated branch points, and each final $-1$ removes one constant eigenfunction.

Combining the connected-component count with the weighted heat coefficient gives
\begin{subequations}\label{eq:direct-fermat-measure-cancellation}
	\begin{gather}
	\zeta_{F,n}(0)+\frac{n^2}{2}\zeta_0(0) - \frac{3n}{2}\zeta_{p,n}(0) = -g_F
	\label{eq:direct-fermat-measure-cancellation-heat}\\
	(\sqrt{2\pi}R)^{g_F}(2\pi)^{-g_F/2} = \eta^{g_F/2}.
	\label{eq:direct-fermat-measure-cancellation-radius}
	\end{gather}
\end{subequations}
Thus the connected and spectral measures supply precisely the factor $\eta^{g_F/2}$ in $\calH_{F_n}$.

\paragraph{The area factor.}
The constant eigenfunction also contributes its area normalization through $\det{}'\Delta/\Area$.
Using Eq.~\eqref{eq:direct-fermat-areas}, the surviving area ratio is
\begin{equation}
	\begin{aligned}
	\sqrt{A_{F,n}}A_0^{n^2/4}A_{p,n}^{-3n/4}
	&=(4\pi n^2)^{1/2}(4\pi)^{n^2/4}(4\pi n)^{-3n/4}\\
	&=(4\pi)^{g_F/2}n^{1-3n/4}.
	\end{aligned}
\end{equation}
The radius and area factors now give the measure prefactor in Eq.~\eqref{eq:direct-fermat-spectral-coefficient}.

\paragraph{Conversion to the plane coefficient.}
The reference norms in Eq.~\eqref{eq:common-adjacent-quotient} are round-sphere two-point values at antipodal points $0,\infty$.
Writing $m_0=e^{2\sigma}|\dd{z}|^2$, the length factors $e^\sigma$ in the local charts at $0,1,\infty$ are $2,1,2$, using $1/z$ as the chart at infinity.
For the scalar twist of dimension $\Delta_n^{(3)}$, the round three-point value is $2^{-2\Delta_n^{(3)}}C_n^{(3)}$, while its round antipodal two-point norm is $2^{-2\Delta_n^{(3)}}$.
Their quotient is therefore
\begin{equation}
	\frac{2^{-2\Delta_n^{(3)}}C_n^{(3)}}{\bigl(2^{-2\Delta_n^{(3)}}\bigr)^{3/2}} = 2^{\Delta_n^{(3)}}C_n^{(3)}.
    \label{eq:direct-fermat-round-quotient}
\end{equation}
Thus recovering the plane coefficient multiplies the round quotient by $2^{-\Delta_n^{(3)}}$, as in Eq.~\eqref{eq:common-adjacent-quotient}.

\subsection{The Fermat determinant quotient}\label{sec:fermat-absolute-determinant}

The radius-independent quotient entering Eq.~\eqref{eq:direct-fermat-spectral-coefficient} is fixed by the following scalar determinant identity.
\begin{proposition}[Fermat scalar determinant quotient]\label{prop:fermat-determinant-quotient}
	For every integer $n\geq2$, let $D_0,D_{p,n},D_{F,n}$ be the determinants in Eq.~\eqref{eq:direct-fermat-determinants} for the maps and unit round pullback metrics of Eq.~\eqref{eq:direct-fermat-metrics}.
	With the nonnegative Friedrichs scalar Laplacians and the single constant mode omitted on each connected surface,
	\begin{equation}
		\frac{D_{F,n}D_0^{n^2/2}}{D_{p,n}^{3n/2}} = 2^{-(n^2-1)/6}\calP_n^{-1}.
		\label{eq:fermat-absolute-gamma-identity}
	\end{equation}
	The defining Gamma product gives $\calP_2=1$.
\end{proposition}

\subsubsection{The character bundles}

The sphere and football determinants $D_0,D_{p,n}$ follow from the separated spectrum and Barnes zeta function in Eqs.~\eqref{eq:direct-fermat-football-spectrum} and~\eqref{eq:direct-fermat-football-determinant}.
For $D_{F,n}$ the Friedrichs Laplacian commutes with the root rotations $\sfh_1,\sfh_2$ of Eq.~\eqref{eq:common-root-rotations}.
A field of character $\bmell=(\ell_1,\ell_2)\in(\bbZ_n)^2$ descends to the punctured base with holonomies
\begin{equation}
	\left(e^{2\pi i\ell_1/n},e^{2\pi i\ell_2/n},e^{-2\pi i(\ell_1+\ell_2)/n}\right)
	\label{eq:direct-fermat-character-holonomies}
\end{equation}
around $0,1,\infty$.
Denote its round-sphere Laplacian by $\Delta_{\bmell}$, with the finite-energy behavior inherited from the cover.
The character projectors are orthogonal and exhaust the complexified scalar space, so additivity of the spectral zeta function gives
\begin{subequations}\label{eq:direct-fermat-character-product}
	\begin{gather}
		D_{F,n}=\prod_{\ell_1,\ell_2=0}^{n-1}\det{}'\Delta_{\bmell}
		\label{eq:direct-fermat-character-product-all}\\
		\det{}'\Delta_{(0,0)}=D_0.
		\label{eq:direct-fermat-character-product-trivial}
	\end{gather}
\end{subequations}
Only the trivial character has a zero mode.
The product includes both members of each conjugate pair and counts the eigenvalues of the real scalar operator once.
Its contribution to the real Gaussian path integral is $D_{F,n}^{-1/2}$.
For a character nontrivial at all three points, the required determinant is a three-holonomy determinant, not the two-cone scalar determinant in Eq.~\eqref{eq:direct-fermat-football-determinant-cover}.

\paragraph{Degrees of the character bundles.}
The bundle degree controls the cohomology norm entering the determinant comparison.
Choose representatives $0\leq\ell_1,\ell_2<n$ and write the root weights as
\begin{equation}
	a=\frac{\ell_1}{n},\qquad b=\frac{\ell_2}{n},\qquad
	m=\lceil a+b\rceil,\qquad c=m-a-b.
	\label{eq:fermat-character-weights}
\end{equation}
For the trivial character $m=0$.
The character generator $u^{\ell_1}v^{\ell_2}$ gives a local frame with squared norm $|z|^{2a}|1-z|^{2b}$.
To extend it across infinity, absorb its integer growth into the frame $z^{-m}u^{\ell_1}v^{\ell_2}$.
The remaining fractional weight is $c$, and the frame change identifies the extension as $\calO(-m)$, with metric and weights
\begin{equation}
	h_{a,b}=|z|^{2a}|1-z|^{2b},\qquad
	(\alpha_0,\alpha_1,\alpha_\infty)=(a,b,c),\qquad
	0\leq\alpha_p<1.
	\label{eq:fermat-character-metric}
\end{equation}
The constant factor $n^2$ in the fibre norm is removed by the unitary normalization of the character transform.
It does not change the sector operator.

There is one degree-zero sector, while $m=2$ precisely when $\ell_1+\ell_2>n$.
Counting these degrees gives the holomorphic pushforward decomposition
\begin{equation}
	f_{n*}\calO_{F_n} \simeq \calO \oplus \calO(-1)^{\oplus(n^2-g_F-1)} \oplus \calO(-2)^{\oplus g_F}.
	\label{eq:fermat-pushforward-splitting}
\end{equation}
Complementing both indices identifies the $g_F$ degree-minus-two characters with $\ell_1',\ell_2'>0$, $\ell_1'+\ell_2'<n$.
All nontrivial sectors have vanishing $H^0$.
The $m=1$ bundles also have vanishing $H^1$, whereas each $m=2$ bundle has a one-dimensional $H^1$, Serre-dual to the constant section of $K_{\bbP^1}\otimes\calO(2)$.
Thus precisely the characters complementary to the Fermat triangle carry the norm factors to be evaluated below.

The football reference has the corresponding decomposition $\calO\oplus\calO(-1)^{\oplus(n-1)}$, with weights $(k/n,1-k/n)$ at its two antipodal branch values.

Multiplication by the locally holomorphic factor $z^a(1-z)^b$ identifies the weighted operator $2\bar\partial_h^*\bar\partial$ with the flat-character scalar Laplacian away from the punctures.
This follows also directly by lifting the $L^2$ norm and Dirichlet energy sheet by sheet.
The closure of the energy selects the positive radial powers $r^{|j+\alpha|}$ in a local angular mode $j\in\bbZ$.
Thus the sector operator is the same Friedrichs operator as in Eq.~\eqref{eq:direct-fermat-character-product}.
Appendix~\ref{sec:character-cap-comparison} controls the passage from this singular weighted operator to the smooth comparison below.

\subsubsection{The smooth determinant comparison}

\paragraph{Cancellation of the character caps.}
Smooth each positive character weight $r^{2\alpha}$ with the same disk profile.
Zero weights are left unchanged.
Gluing separates a local Dirichlet cap factor from the boundary determinant coupling that cap to the exterior.
At fixed exterior geometry and cutting radius $\delta$, the character-cap comparison of Appendix~\ref{sec:character-cap-comparison} sends the relative boundary determinant to one for each nontrivial character as $\varepsilon\to0$.
Only the local factors $C_\alpha(\varepsilon,\delta)$ in Eq.~\eqref{eq:fermat-cap-determinant-comparison} remain.

For every $k=1,\ldots,n-1$, the weight $k/n$ occurs $n$ times at each Fermat branch value, hence $3n$ times in total.
In the denominator, each football contributes two occurrences and is raised to power $3n/2$.
The net multiplicity of its cap factor is consequently $3n-2(3n/2)=0$.
Thus every $C_{k/n}(\varepsilon,\delta)$ in Eq.~\eqref{eq:fermat-cap-determinant-comparison} cancels exactly, including its finite constant.
The trivial sector is unsmoothed and its constant mode is omitted identically.
We can therefore evaluate
\begin{equation}
	\log D_{F,n}+\frac{n^2}{2}\log D_0 -\frac{3n}{2}\log D_{p,n}
	\label{eq:fermat-relative-logdet}
\end{equation}
using smooth line-bundle determinants and then take $\varepsilon\to0$.

On $\calO(-m)$ use the smooth reference metric $h_m=(1+|z|^2)^m$ and set $\varphi=\log(h_{a,b}/h_m)$.
Appendix~\ref{sec:math-quillen} supplies the anomaly formula at the present scalar spectral scale and the smooth reference determinant values.
After separating the fixed-basis Gram terms, denote the anomaly on the right of Eq.~\eqref{eq:math-quillen-anomaly} by $\calA_m(\varphi)$.
These two inputs are
\begin{subequations}\label{eq:fermat-quillen-reference}
	\begin{gather}
		\calA_m(\varphi) = -\frac{1-m}{4\pi}\int\varphi\dd{A} - \frac1{8\pi}\int|\nabla\varphi|^2\dd{A}
		\label{eq:fermat-quillen-reference-anomaly}\\
		D_1^\circ=e^{-4\zeta_{\rm R}'(-1)},\quad D_2^\circ=D_0.
		\label{eq:fermat-quillen-reference-spectra}
	\end{gather}
\end{subequations}
These formulas are first applied to the smooth weights above.
The weighted sum of their singular limits is finite by the cap cancellation.

\paragraph{The Gamma product from cohomology norms.}
For an $m=2$ sector, fix the Serre-dual section $\dd{z}\otimes e_0^{-1}$ and the dual $H^1$ basis used in Appendix~\ref{sec:math-quillen}.
The squared norm of this section is proportional to
\begin{equation}
	J(a,b) = \int_{\bbC}\frac{\dd[2]{z}}{|z|^{2a}|1-z|^{2b}} = \pi\frac{\Gamma(1-a)\Gamma(1-b)\Gamma(a+b-1)}{\Gamma(a)\Gamma(b)\Gamma(2-a-b)}.
	\label{eq:fermat-quillen-beta-norm}
\end{equation}
The corresponding reference integral for $h_2$ is $\pi$, so their squared-norm ratio is $J(a,b)/\pi$.
Keeping the Serre pairing fixed makes the Gram norm of the dual $H^1$ basis vary inversely.
With $G_1$ as defined in Eq.~\eqref{eq:math-quillen-anomaly},
\begin{equation}
	\frac{\det G_1(h_{a,b})}{\det G_1(h_2)}=\frac{\pi}{J(a,b)}.
	\label{eq:fermat-serre-dual-gram-ratio}
\end{equation}
The resulting contribution to the scalar log determinant is $-\log[\pi/J(a,b)]=\log[J(a,b)/\pi]$.
For the smoothed weights, the Serre-dual squared norms converge to these beta integrals because a disk of radius $\varepsilon$ at weight $\alpha<1$ contributes only $O(\varepsilon^{2-2\alpha})$.

Complementing $(a,b)$ to $(1-a,1-b)$ puts the $g_F$ cohomology factors on the triangular index range.
Their product in the scalar determinant is $\calP_n^{-1}$, with the following triangular representation of the Gamma product:
\begin{equation}
	\calP_n =  \prod_{\substack{\ell_1,\ell_2\geq1\\\ell_1+\ell_2<n}} \frac{\Gamma(1-\ell_1/n)\Gamma(1-\ell_2/n)\Gamma((\ell_1+\ell_2)/n)}{\Gamma(\ell_1/n)\Gamma(\ell_2/n)\Gamma(1-(\ell_1+\ell_2)/n)}.
	\label{eq:fermat-gamma-product}
\end{equation}
In Eq.~\eqref{eq:fermat-gamma-product}, the total exponent of $\Gamma(k/n)$ is $3(2k-n)$.
Its numerator multiplicity is $3(k-1)$, and its denominator multiplicity is $3(n-k-1)$.
This identifies the triangular product with Eq.~\eqref{eq:fermat-gamma-single-product}.

\subsubsection{Finite normalization of the quotient}\label{sec:fermat-coefficient-assembly}

The cohomology norms supply $\calP_n^{-1}$ to the quotient of Eq.~\eqref{eq:fermat-relative-logdet}.
To determine its remaining geometric factor, we evaluate the weighted anomaly and combine it with the smooth reference spectra of Eq.~\eqref{eq:fermat-quillen-reference-spectra}.
Use the sphere Green logarithms $g_p$ of Appendix~\ref{sec:math-quillen}, with the pairings in Eq.~\eqref{eq:fermat-sphere-green-pairing}.
Here $g_0(\infty)=0$ and $g_0(1)=g_1(\infty)=-\log2$, and
\begin{equation}
	\varphi=a g_0+b g_1+c g_\infty+b\log2.
	\label{eq:fermat-quillen-potential}
\end{equation}
The self-energies are identical at all points under round rotations.
They cancel by the same weight multiplicities as the cap determinants.
The remaining cross terms contribute, to the weighted anomaly,
\begin{equation}
	\sum_{\ell_1,\ell_2}\left[ab+ac+bc-b(a+c)\log2\right] - \frac{3n}{2}\sum_{k=1}^{n-1}\frac{k}{n}\left(1-\frac{k}{n}\right).
	\label{eq:fermat-quillen-cross-sum}
\end{equation}
The sums run over $0\leq \ell_1,\ell_2\leq n-1$ and include a vanishing trivial term.

To evaluate the part independent of $\log2$, use $m^2=a^2+b^2+c^2+2(ab+ac+bc)$.
The required degree and weight sums are
\begin{subequations}\label{eq:fermat-quillen-degree-count}
	\begin{gather}
		\sum_{\ell_1,\ell_2}m^2 = n^2+3g_F-1
		\label{eq:fermat-quillen-degree-count-degree}\\
		\sum_{\ell_1,\ell_2}(a^2+b^2+c^2) = 3n\sum_{k=1}^{n-1}(k/n)^2.
		\label{eq:fermat-quillen-degree-count-weight}
	\end{gather}
\end{subequations}
Substitution into Eq.~\eqref{eq:fermat-quillen-cross-sum} gives $g_F$ for its constant part.
For the linear term of the anomaly, integrating Eq.~\eqref{eq:fermat-quillen-potential} with $\int g_p\dd{A}=-4\pi$ gives
\begin{equation}
	-\frac{1-m}{4\pi}\int\varphi\dd{A}=(1-m)(m-b\log2).
	\label{eq:fermat-quillen-linear-term}
\end{equation}
It vanishes for $m=1$ and equals $-2+b\log2$ for each of the $g_F$ sectors with $m=2$.
Its constant contribution is therefore $-2g_F$, leaving $-g_F$ after the cross terms are included.

For the coefficient of $\log2$, the linear terms contribute $\sum_{m=2}b$, while the cross terms contribute $-\sum b(a+c)=-\sum b(m-b)$.
Together they give
\begin{equation}
	\sum_{m=2}b-\sum_{\ell_1,\ell_2}b(m-b) = -\sum_{\ell_1,\ell_2}b(1-b)=-\frac{n^2-1}{6}.
	\label{eq:fermat-quillen-logtwo-count}
\end{equation}

The reference contribution follows from the bundle multiplicities in Eq.~\eqref{eq:fermat-pushforward-splitting}.
The Fermat sectors contribute $D_0(D_1^\circ)^{n^2-g_F-1}(D_2^\circ)^{g_F}$, while each football contributes $D_0(D_1^\circ)^{n-1}$.
Including the base spheres in the quotient and using $D_2^\circ=D_0$ leaves
\begin{equation}
	\frac{D_0(D_1^\circ)^{n^2-g_F-1}(D_2^\circ)^{g_F}D_0^{n^2/2}}
	{\bigl[D_0(D_1^\circ)^{n-1}\bigr]^{3n/2}}
	=\left(\frac{D_0}{D_1^\circ}\right)^{2g_F}.
\label{eq:fermat-reference-spectrum-quotient}
\end{equation}
Since $\log(D_0/D_1^\circ)=1/2$, the reference spectra contribute $+g_F$.
Combining the cohomology norms, the finite anomaly and these reference spectra gives
\begin{equation}
\begin{aligned}
	\log\frac{D_{F,n}D_0^{n^2/2}}{D_{p,n}^{3n/2}}
	&=-\log\calP_n-g_F-\frac{n^2-1}{6}\log2+g_F\\
	&=-\log\calP_n-\frac{n^2-1}{6}\log2.
\end{aligned}
\label{eq:fermat-quotient-contributions}
\end{equation}
This proves Proposition~\ref{prop:fermat-determinant-quotient}.

\paragraph{The adjacent coefficient.}
Substitution into Eq.~\eqref{eq:direct-fermat-spectral-coefficient} gives
\begin{equation}
\begin{aligned}
	C_n^{(3)}(\eta)
	&=2^{-\Delta_n^{(3)}}(4\pi)^{g_F/2}n^{1-3n/4}
	\left(2^{-(n^2-1)/6}\calP_n^{-1}\right)^{-1/2}
	\calH_{F_n}(\eta)\\
	&=(4\pi)^{g_F/2}n^{1-3n/4}\sqrt{\calP_n}\,\calH_{F_n}(\eta).
\end{aligned}
\label{eq:fermat-coefficient-substitution}
\end{equation}
The determinant factor cancels the round-to-plane factor because $\Delta_n^{(3)}=(n^2-1)/12$.
This proves Theorem~\ref{thm:adjacent-three-point}.
Its radius duality follows from Proposition~\ref{prop:common-harmonic-factor}.

The individual Gamma ratios in Eq.~\eqref{eq:fermat-gamma-product} are the local factors already present in earlier three-twist couplings.
They follow, after separating the angle-independent normalization, from the squared quantum prefactor in Ref.~\cite[Eq.~(5.20)]{Burwick:1990tu} by the Gamma reflection formula, and appear explicitly in the positive-triangle couplings of Refs.~\cite[Eq.~(4.7)]{Erler:1992gt} and~\cite[Eq.~(3.7)]{Stieberger:1992vb}.
The same ratios enter the two-angle OPE coefficients of Ref.~\cite[Eqs.~(4.24) and~(4.26)]{Anastasopoulos:2011gn}.
They also occur in quantum three-twist couplings extracted from intersecting-brane four-point amplitudes~\cite{Cvetic:2003ch,Lust:2004cx}.
The character multiplicities above assemble those local factors into the scalar determinant quotient.

\section{Fusion through degeneration}\label{sec:analytic-comparisons}
After the continuum limit at distinct endpoints, let $y=1-x\to0^+$, with $g_d=(n-1)^2$ and $g_F=(n-1)(n-2)/2$.
Here $\calH_{F_n}$ is the Fermat specialization of the common harmonic factor in Eq.~\eqref{eq:common-harmonic-factor} and $\calP_n$ is the Gamma product of Eq.~\eqref{eq:fermat-gamma-single-product}.
We will show the following:
\begin{theorem}[Fusion of the normalized replica amplitude]\label{thm:absolute-fusion}
	The continuum fusion coefficient is
	\begin{equation}
		\lim_{x\to1^-}\calF_{n}^{(3)}(x;\eta)
		=\frac{(4\pi)^{g_F}}{n^{(3n-4)/2}}\,\calP_n\,\calH_{F_n}(\eta)^2
		=\left|C_{n}^{(3)}(\eta)\right|^2.
		\label{eq:endpoint-matching-result}
	\end{equation}
\end{theorem}

\subsection{The two Fermat components}\label{sec:adjacent-degeneration}

\paragraph{Two charts of the same family.}
The common roots of Eq.~\eqref{eq:disjoint-cover-equations} make both Fermat components of the fusion limit explicit.
Put
\begin{equation}
	y=1-x=t^n,\qquad t>0,\qquad \rho=e^{\pi i/n}.
	\label{eq:degeneration-base-change}
\end{equation}
The near chart $(u,v)$ and the reciprocal far chart $(u_-,v_-)$ obey
\begin{subequations}\label{eq:degeneration-family}
	\begin{align}
		u^n+v^n-yu^nv^n&=1,
		\label{eq:degeneration-family-near}\\
		u_-=\frac1{tu},\qquad v_-=\frac1{tv},\qquad
		u_-^n+v_-^n-yu_-^nv_-^n&=1.
		\label{eq:degeneration-family-far}
	\end{align}
\end{subequations}
The transformation is an involution.
On the two base charts it is $w_-=u_-^n=1/(yw)$, so the same positive-root arc over $0\leq w_-\leq1$ describes the far region over $1/y\leq w\leq\infty$, with reversed base orientation.
Keeping either pair of root coordinates finite as $t\to0$ gives a Fermat curve:
\begin{subequations}\label{eq:degeneration-inner-component}
	\begin{gather}
		u^n+v^n=1
		\label{eq:degeneration-inner-component-plus}\\
		u_-^n+v_-^n=1.
		\label{eq:degeneration-inner-component-minus}
	\end{gather}
\end{subequations}
Denote their projective completions by $F_n^+$ and $F_n^-$ respectively.
Setting $y=0$ in only the near affine equation omits the second component at simultaneous $u,v$ infinity; it does not give the compact limiting surface.

\paragraph{Plumbing at Fermat infinity.}
Use the root convention of Eq.~\eqref{eq:common-root-rotations}.
Near a node define
\begin{equation}
	p_j=\frac1u,\qquad q_j=\frac1{u_-},\qquad r=\frac vu.
	\label{eq:degeneration-local-coordinates}
\end{equation}
The subscript $j$ selects one of the $n$ analytic branches of $r$.
The reciprocal chart relation fixes $p_jq_j=t$ exactly.
Expressing the family equation in these local parameters determines which infinity points are joined and fixes the phases of their node coordinates.

\begin{lemma}[Degeneration into two Fermat components]\label{lem:fermat-degeneration}
	After the base change $y=t^n$, the compact nodal limit consists of $F_n^+$ and $F_n^-$ joined at $n$ nodes.
	For $0\leq j<n$, the local smoothing is
	\begin{equation}
		p_jq_j=t, \qquad r^n=-\frac{1-p_j^n}{1-q_j^n}, \qquad r=\rho\zeta_n^j \left(\frac{1-p_j^n}{1-q_j^n}\right)^{1/n},
		\label{eq:degeneration-exact-plumbing}
	\end{equation}
	where the root in parentheses is analytic and tends to one at the origin.
	The paired node preimages are
	\begin{equation}
		P_j^+=[1:\rho\zeta_n^j:0], \qquad P_j^-=[1:\rho^{-1}\zeta_n^{-j}:0], \qquad 0\leq j<n.
		\label{eq:degeneration-node-pairs}
	\end{equation}
	For $n\geq3$ the central fiber is stable; at $n=2$ it is the semistable union of two spheres at two nodes, and there is no component homology.
\end{lemma}

Substituting the reciprocal charts of Eq.~\eqref{eq:degeneration-family} into the local coordinates gives Eq.~\eqref{eq:degeneration-exact-plumbing}.
At $q_j=0$, the projective near coordinates are $[1:r:p_j]$; at $p_j=0$, the far coordinates are $[1:r^{-1}:q_j]$.
These give the stated node preimages at Fermat infinity.
If both infinity fibers are indexed by the ratio $\rho\zeta_n^j$, their pairing is $j\mapsto-j-1$ modulo $n$.
The equation $p_jq_j=t$ identifies two punctured disks along a cylinder, with coefficient exactly $+1$ for every node.
Rescaling local parameters changes the finite regularized periods, so the displayed parameters and node phases are part of the period data.

The two genus-$g_F$ components contribute $2g_F$ handles, while the dual graph of $n$ connecting cylinders contributes $n-1$ independent cycles.
Hence the smoothed genus is $g_d=2g_F+n-1=(n-1)^2$, as required for the compact cover for disjoint intervals. The additional $n-1$ handles will carry the logarithmically growing neck periods.

\paragraph{Character limits.}
The limiting behavior of the common differentials separates the component sectors from the neck sector.
Substitution of the reciprocal roots into Eq.~\eqref{eq:common-character-forms} gives the exact identity
\begin{equation}
	\omega_{(\ell_1,\ell_2)}(y;u,v) = t^{n-\ell_1-\ell_2}\,\omega_{(n-\ell_1,n-\ell_2)}(y;u_-,v_-).
	\label{eq:degeneration-character-involution}
\end{equation}
The minus sign from $\dd{u}$ cancels against the transformed denominator.
The following limits on both components hold on compact subsets away from the nodes.
For $\ell_1+\ell_2<n$, the unscaled form limits to a holomorphic Fermat form on $F_n^+$ and vanishes on $F_n^-$.
For $\ell_1+\ell_2>n$, the form $t^{\ell_1+\ell_2-n}\omega_{(\ell_1,\ell_2)}(y)$ limits to the complementary holomorphic form on $F_n^-$ and vanishes on $F_n^+$.
The $n-1$ middle forms with $\ell_1+\ell_2=n$ have simple node poles.
The square index set $\{1,\ldots,n-1\}^2$ of mixed disjoint root characters therefore separates into the two Fermat triangles and the middle line, with counts $g_F,g_F,n-1$; the integral normalization of the neck forms requires the cycles of Section~\ref{sec:replica-neck-periods}.

\subsection{The period matrix near degeneration}\label{sec:replica-neck-periods}

The two Fermat components supply finite periods, while the $n-1$ additional handles supply logarithmic periods.
To use both in the winding sum, we construct a complete integral marking adapted to Lemma~\ref{lem:fermat-degeneration}.

\paragraph{Marked Fermat components.}
Fix the canonical Fermat cycles of Section~\ref{sec:fermat-integral-homology} on a reference curve $X^n+Y^n=Z^n$.
Identify this reference curve with $F_n^+$ by $[X:Y:Z]\mapsto[X:Y:Z]$, and with $F_n^-$ by the automorphism
\begin{equation}
	h:[X:Y:Z]\longmapsto[\rho X:Z:Y].
	\label{eq:degeneration-component-marking}
\end{equation}
Transport both cycles and normalized differentials by these maps, and choose cycle representatives away from the node disks.
Denote the resulting cycles by $A_\mu^\pm,B_\mu^\pm$, $1\leq\mu\leq g_F$.
This choice preserves the marked period matrix $\Omega_{F,n}$ on both components.
In the physical frame of Eq.~\eqref{eq:disjoint-physical-frame}, the near coordinates are $(u,v)=((tu_0)^{-1},\rho v_0/t)$, while
\begin{subequations}\label{eq:degeneration-physical-component-map}
	\begin{gather}
		(U,V)=(u_0v_0,\rho v_0)
		\label{eq:degeneration-physical-component-map-physical}\\
		(u_-,v_-)=\left(\frac{\rho U}{V},\frac1V\right).
		\label{eq:degeneration-physical-component-map-root}
	\end{gather}
\end{subequations}
The pair $(U,V)$ in Eq.~\eqref{eq:degeneration-physical-component-map-physical} is the physical outer Fermat chart at the limit.

\paragraph{Component characters in the transported marking.}
Write $\omega^F_{\bmell}=\omega_{\bmell}(0)$ for the unnormalized Fermat forms.
The disjoint physical character $\bmk$ has common root label $\bmell=(n-k_1,k_2)$.
Combining the character involution~\eqref{eq:degeneration-character-involution} with the component maps gives Table~\ref{tab:degeneration-component-characters}.
\begin{table}[tbp]
	\centering
	\begin{tabular}{llll}
		\toprule
		Physical character & Limiting chart & Root label & Reference physical label\\
		\midrule
		$1\leq k_2<k_1<n$ & Near & $(n-k_1,k_2)$ & $(n-k_1,n-k_1+k_2)$\\
		$1\leq k_1<k_2<n$ & Reciprocal far & $(k_1,n-k_2)$ & $(k_1,k_2)$\\
		$k_1=k_2$ & Node poles & $(n-k_1,k_1)$ & ---\\
		\bottomrule
	\end{tabular}
	\caption{Character labels in the two Fermat components and the neck sector.
	The root label is evaluated in the limiting chart before transport to the reference Fermat curve.
	The final column uses the physical generators in the transported reference marking.}
	\label{tab:degeneration-component-characters}
\end{table}
The two off-diagonal halves give the two holomorphic Fermat triangles; the $n-1$ diagonal forms give the neck differentials.
On the far component, the reference root label differs from the reciprocal-chart label because the marking uses $h$:
\begin{equation}
	h^*\omega^F_{(k_1,n-k_2)} = \rho^{k_1}\omega^F_{(k_1,k_2-k_1)}, \qquad 1\leq k_1<k_2<n.
	\label{eq:degeneration-far-form-phase}
\end{equation}
Indeed, in affine coordinates $h$ sends $(u,v)$ to $(\rho u/v,1/v)$; using $\dd{(u/v)}=v^{-n-1}\dd{u}$ gives the displayed scalar phase.

\paragraph{Integral neck cycles.}
Let $a_j$ be the counterclockwise waist in the $p_j$ coordinate.
The oriented boundary of the punctured plus component gives the single relation
\begin{equation}
	\sum_{j=0}^{n-1}a_j=0.
	\label{eq:degeneration-waist-relation}
\end{equation}
For $1\leq i<n$, choose a closed path $b_i$ crossing cylinder $i$ from $F_n^+$ to $F_n^-$ and returning through cylinder $0$.
Its intersections with the waists are
\begin{equation}
	a_j\circ b_i=\delta_{ji}-\delta_{j0},\qquad
	a_j\circ a_k=0.
	\label{eq:degeneration-neck-duality}
\end{equation}
The sign can be read locally: an inward radial path in the $p_i$ disk has positive intersection with its counterclockwise waist.
In particular, $a_1,\ldots,a_{n-1}$ pair integrally with $b_1,\ldots,b_{n-1}$.

The waist pairings are canonical, but arbitrary connecting paths can still intersect component cycles and one another.
To separate the winding integers into component and neck directions, we remove these intersections by integer corrections.
First, the projection in Eq.~\eqref{eq:math-integral-symplectic-projection} gives
\begin{equation}
	b_i^\perp = b_i-\sum_{\epsilon=\pm}\sum_{\mu=1}^{g_F} \left[(b_i\circ B_\mu^\epsilon)A_\mu^\epsilon - (b_i\circ A_\mu^\epsilon)B_\mu^\epsilon\right].
	\label{eq:degeneration-component-projection}
\end{equation}
This leaves the waist pairings unchanged and makes $b_i^\perp$ orthogonal to both component lattices.
Write $H_{ij}=b_i^\perp\circ b_j^\perp$ and define
\begin{subequations}\label{eq:degeneration-neck-correction}
	\begin{gather}
		\widehat A_i=a_i
		\label{eq:degeneration-neck-correction-A}\\
		\widehat B_i=b_i^\perp-\sum_{j<i}H_{ij}a_j,
		\qquad 1\leq i<n.
		\label{eq:degeneration-neck-correction-B}
	\end{gather}
\end{subequations}
The hats distinguish the additional handles from the component cycles $A_\mu^\pm,B_\mu^\pm$.
Indeed, for $i>j$ the added term contributes $-H_{ij}$ to $\widehat B_i\circ \widehat B_j$, canceling the original intersection; skew symmetry treats $i<j$.
The resulting cycles satisfy
\begin{equation}
	\widehat A_i\circ \widehat B_j = \delta_{ij},\qquad
	\widehat A_i\circ \widehat A_j = \widehat B_i\circ \widehat B_j = 0.
	\label{eq:degeneration-canonical-necks}
\end{equation}

Orient the dual-graph edge $e_j$ of cylinder $j$ from $F_n^+$ to $F_n^-$.
With the common plumbing parameter $p_jq_j=t$, factor out the common logarithmic length and take the oriented graph edges to be unit and orthogonal.
The paths $b_i$ project to $c_i=e_i-e_0$, so their Gram matrix in this Euclidean edge metric is
\begin{equation}
	(\widehat G_n)_{ik}=\delta_{ik}+1,\qquad
	\widehat G_n^{-1}=I_{n-1}-\frac1n\bm1\bm1^{\top},
	\qquad \det \widehat G_n=n.
	\label{eq:degeneration-graph-gram}
\end{equation}
The determinant follows from the eigenvalue $n$ on $\bm1$ and eigenvalue one on its orthogonal complement.
The shared return edge $e_0$ produces the off-diagonal ones in this positive Gram matrix.
Its determinant $n$ measures the logarithmic period growth; the alternating intersection matrix of the same neck handles has determinant one.

\paragraph{Asymptotics of the period matrix.}
Let $\widetilde\Omega_n^{(3)}(x)$ denote the period matrix with the plus and minus component cycles ordered first, followed by $\widehat A_i,\widehat B_i$.
Let $\nu_\mu^\pm$ be the normalized Fermat forms in the transported bases.
Define the component--neck Abel matrices by
\begin{subequations}\label{eq:degeneration-abel-blocks}
	\begin{gather}
		U^+_{\mu i}=\int_{P_0^+}^{P_i^+}\nu_\mu^+
		\label{eq:degeneration-abel-blocks-plus}\\
		U^-_{\mu i}=-\int_{P_0^-}^{P_i^-}\nu_\mu^-,
		\label{eq:degeneration-abel-blocks-minus}
	\end{gather}
\end{subequations}
where the relative paths include the component-cycle corrections in Eq.~\eqref{eq:degeneration-component-projection}.
The minus sign records the return path on the minus component.
The corrected paths fix the period-lattice representatives of the Abel integrals.

Let $C=C^{\top}$ denote the regularized neck integrals after subtraction of their local logarithms in the parameters $p_j=1/u$ and $q_j=1/u_-$.
It depends on these node coordinates and the corrected paths and need not vanish.
Put $Y_F=\Im\Omega_{F,n}$. For $n=2$ the component blocks and Abel matrices are empty, and $\det Y_F=1$.

\begin{proposition}[Period matrix in the degeneration]\label{prop:degeneration-marked-periods}
	For fixed integer $n\geq2$ and $y=1-x=t^n\to0^+$, the component cycles and corrected neck cycles above form a complete integral symplectic basis.
	In this marking,
	\begin{equation}
		\widetilde\Omega_n^{(3)}(x)=
		\begin{pmatrix}
			\Omega_{F,n}&0&U^+\\
			0&\Omega_{F,n}&U^-\\
			(U^+)^{\top}&(U^-)^{\top}&\displaystyle\frac{i}{2\pi n}\log\frac1{1-x}\,\widehat G_n+C
		\end{pmatrix}+o(1),
		\label{eq:degeneration-period-blocks}
	\end{equation}
	and
	\begin{equation}
		\det\Im\widetilde\Omega_n^{(3)}(x) = n\left[\frac1{2\pi n}\log\frac1{1-x}\right]^{n-1} (\det Y_F)^2\,[1+o(1)].
		\label{eq:degeneration-im-determinant}
	\end{equation}
\end{proposition}

The component and neck cycles have rank $4g_F+2(n-1)=2g_d$ and intersection determinant one.
By the index formula of Appendix~\ref{sec:math-riemann}, these cycles span the full integral homology and give the symplectic lattice splitting
\begin{equation}
	H_1(\Sigma_n^{(3)}(x),\bbZ) \simeq H_1(F_n^+,\bbZ) \oplus H_1(F_n^-,\bbZ) \oplus \bbZ^{2(n-1)},
	\label{eq:degeneration-integral-splitting}
\end{equation}
where each summand carries its canonical pairing.
The smooth surface remains connected, and finite off-diagonal periods can couple these summands.

Both this marking and the earlier disjoint marking are full bases of the same integer homology.
Their change of basis is therefore integral symplectic; its coefficients are the intersections with the earlier dual cycles.

In the convergence domain specified in Section~\ref{sec:adjacent-degeneration}, the normalized holomorphic forms associated with the two component blocks converge to the Fermat forms $\nu_\mu^\pm$ on their own component and to zero on the other.
Their integrals along the corrected connecting paths give the finite Abel blocks in Eq.~\eqref{eq:degeneration-abel-blocks}.

The $i$th normalized neck form converges to a meromorphic differential with simple poles at the node preimages.
Put $N_{ji}=\delta_{ji}-\delta_{j0}$, for $0\leq j<n$ and $1\leq i<n$.
Its residues on the plus and minus components are respectively $N_{ji}/(2\pi i)$ and $-N_{ji}/(2\pi i)$, and its component $A$-periods vanish.
These conditions fix the limiting differential.
On cylinder $j$ its singular part is $N_{ji}\dd{p_j}/(2\pi i p_j)$.
Integrating along $\widehat B_k$ therefore gives the logarithmic contribution
\begin{equation}
	\frac{\log t}{2\pi i}\sum_{j=0}^{n-1}N_{ji}N_{jk} = \frac{i}{2\pi n}\log\frac1{1-x}\,(\delta_{ik}+1).
	\label{eq:degeneration-log-coefficient}
\end{equation}
The factor $1/n$ follows from the base change, and the factor $1/(2\pi i)$ from unit waist periods.
For $n\geq3$, the normalized plumbing expansion of Ref.~\cite[Corollary~4.6]{HuNorton:2018periods} gives the finite Abel blocks and the regularized neck block $C$ after these logarithms have been subtracted.
Its differentials of residue one are divided by $2\pi i$ for our unit-waist convention, and its plumbing parameters are the exact $p_jq_j=t$ of Eq.~\eqref{eq:degeneration-exact-plumbing}.

For the semistable $n=2$ fiber, the same finite-remainder statement follows directly from the dual elliptic marking:
\begin{equation}
	-\frac1{\tau_2^{(3)}(1-y)} = \frac{iK(1-y)}{2K(y)} = \frac{i}{2\pi}\log\frac{16}{y}+o(1).
	\label{eq:degeneration-elliptic-neck}
\end{equation}
Its logarithmic term agrees with $\widehat G_2=(2)$ in Eq.~\eqref{eq:degeneration-period-blocks}; an integral shear of the connecting cycle only changes the finite real part.

The finite upper block of $\Im\widetilde\Omega_n^{(3)}$ tends to $Y_F\oplus Y_F>0$, while its Schur complement in the neck block is
\begin{equation}
	\frac1{2\pi n}\log\frac1{1-x}\,\widehat G_n+O(1).
	\label{eq:degeneration-schur-block}
\end{equation}
The finite Abel blocks affect only the bounded term and therefore do not alter the leading determinant.
Using $\det\widehat G_n=n$ gives Eq.~\eqref{eq:degeneration-im-determinant} and completes the proof.

\subsection{Factorization of the normalized amplitude}\label{sec:adjacent-factorization}

\paragraph{The harmonic factor in the adapted marking.}
By Proposition~\ref{prop:common-harmonic-factor}, evaluate the harmonic factor in the adapted integral marking of Section~\ref{sec:replica-neck-periods}, using the full lattice splitting~\eqref{eq:degeneration-integral-splitting}.

\begin{lemma}[Harmonic factor in the degeneration]\label{lem:harmonic-degeneration}
	As $y\to0^+$, uniformly for $\eta$ in compact subsets of $(0,\infty)$,
	\begin{equation}
		\begin{aligned}
			&\sqrt{\det\Im\Omega_n^{(3)}(1-y)}\,
			\Theta\!\left(\mathbf0\mid T_\eta(\Omega_n^{(3)}(1-y))\right)\\
			&\qquad=\sqrt n\left[\frac{\log(1/y)}{2\pi n}\right]^{(n-1)/2}
			\calH_{F_n}(\eta)^2[1+o(1)].
		\end{aligned}
		\label{eq:endpoint-harmonic-limit}
	\end{equation}
\end{lemma}

\paragraph{Neck suppression and the Fermat winding sums.}
Write $Y,V,\widehat Y$ for the component, mixed and neck blocks of $\Im\widetilde\Omega_n^{(3)}$ in Eq.~\eqref{eq:degeneration-period-blocks}.
The component block tends to $Y_F\oplus Y_F>0$, the mixed block is bounded, and the neck Schur complement is Eq.~\eqref{eq:degeneration-schur-block}.

Use the absolutely convergent momentum--winding sum in Eq.~\eqref{eq:classical-resummed-lattice}.
Its integers decompose as $\bmw_A=(\bmw_{A,+},\bmw_{A,-},\widehat{\bmw}_A)$ and $\bmm=(\bmm_+,\bmm_-,\widehat{\bmm})$.
The real part of the period matrix contributes only a phase.
To control the absolute value, complete the square for either of the two Gaussian quadratic forms:
\begin{equation}
	\begin{pmatrix}\bmv\\\widehat{\bmv}\end{pmatrix}^{\!\top}
	\begin{pmatrix}Y&V\\V^{\top}&\widehat Y\end{pmatrix}
	\begin{pmatrix}\bmv\\\widehat{\bmv}\end{pmatrix}
	=(\bmv+Y^{-1}V\widehat{\bmv})^{\top}Y(\bmv+Y^{-1}V\widehat{\bmv}) + \widehat{\bmv}^{\top}(\widehat Y-V^{\top}Y^{-1}V)\widehat{\bmv}.
	\label{eq:endpoint-gaussian-square}
\end{equation}

The shifted component Gaussian sum has a uniform bound.
Reduce the shift modulo the integer lattice to a compact fundamental cell and use the uniform positive lower bound on $Y$ for sufficiently small $y$.
Fix a compact set $K\subset(0,\infty)$ of radius parameters $\eta$.
Summing the absolute values over the component integers at fixed $(\widehat{\bmw}_A,\widehat{\bmm})$ is therefore bounded by
\begin{equation}
	C_K\exp\!\left[-c_K\log(1/y) \bigl(\|\widehat{\bmw}_A\|^2+\|\widehat{\bmm}\|^2\bigr)\right], \qquad c_K>0.
	\label{eq:endpoint-gaussian-bound}
\end{equation}

Summing the bound over the nonzero neck integers gives a vanishing contribution.
For zero neck integers the mixed Abel blocks and the finite neck block drop out.
The two component blocks converge to $\Omega_{F,n}$, so dominated convergence gives
\begin{equation}
	\Theta\!\left(\mathbf0\mid T_\eta(\widetilde\Omega_n^{(3)}(x))\right) = \Theta\!\left(\mathbf0\mid T_\eta(\Omega_{F,n})\right)^2+o(1).
	\label{eq:endpoint-theta-limit}
\end{equation}
The limiting sum and the tail estimate are uniform for $\eta\in K$.

Combining marking invariance, Eq.~\eqref{eq:endpoint-theta-limit}, and the determinant asymptotic~\eqref{eq:degeneration-im-determinant} proves Lemma~\ref{lem:harmonic-degeneration}.
The factor $\sqrt n$ comes from $\det\widehat G_n=n$; the factor $1/n$ inside the logarithmic coefficient comes from the exact local relation $p_jq_j=y^{1/n}$ in Eq.~\eqref{eq:degeneration-exact-plumbing}.

\paragraph{Connection coefficients of the quantum factor.}
For $a=k_1/n$ and $b=k_2/n$, write $I_{a,b}$ and $J_{a,b}$ for the Euler periods of Eqs.~\eqref{eq:period-I-euler} and~\eqref{eq:period-J-euler}.
In the real-slice Ward bilinear~\eqref{eq:ward-bilinear} the complementary character has angles $(1-a,1-b)$.

\begin{lemma}[Quantum factor in the degeneration]\label{lem:quantum-degeneration}
	As $y\to0^+$, the complete mixed-character product, with positive real square roots, satisfies
	\begin{equation}
		\prod_{k_1,k_2=1}^{n-1}\calI_{\bmk}(1-y,1-y)^{-1/2} = \calP_n\,y^{g_d/4-\Delta_n^{(3)}}[2\log(1/y)]^{-(n-1)/2}[1+o(1)].
		\label{eq:endpoint-quantum-limit}
	\end{equation}
\end{lemma}
Off-diagonal characters supply power singularities and finite Gamma factors, whereas diagonal characters supply logarithms.
The component-character dictionary in Section~\ref{sec:replica-neck-periods} identifies the two off-diagonal halves with the two Fermat components and the Ward diagonal with the neck forms.

For $a<b$, the product $I_{a,b}(1-y)J_{1-a,1-b}(1-y)$ in Eq.~\eqref{eq:ward-bilinear} gives the divergent contribution.
The connection formula in Appendix~\ref{sec:math-special} gives
\begin{subequations}\label{eq:endpoint-euler-asymptotics}
	\begin{align}
		I_{a,b}(1-y) &= \frac{\pi}{\sin(\pi a)}\frac{\Gamma(b-a)}{\Gamma(1-a)\Gamma(b)}+O(y^{b-a}),
		\label{eq:endpoint-I-asymptotic}\\
		J_{1-a,1-b}(1-y) &= B(a,1-b)y^{a-b}[1+O(y)].
		\label{eq:endpoint-J-asymptotic}
	\end{align}
\end{subequations}
Multiplying the two displayed leading terms by the Ward factor $\sin(\pi a)/\pi$ cancels their trigonometric prefactor.
Expanding $B(a,1-b)$ gives the Gamma ratio below with power $y^{a-b}$.
The other product in Eq.~\eqref{eq:ward-bilinear} is $O(1)$: $J_{a,b}=O(y^{b-a})$ and $I_{1-a,1-b}=O(y^{a-b})$.
For $a>b$ the roles of the two products are exchanged.
In both cases,
\begin{subequations}\label{eq:endpoint-offdiagonal-data}
	\begin{align}
		\calI_{a,b}(1-y,1-y)
		&= \Xi(a,b)y^{-|a-b|}[1+o(1)],\qquad a\ne b,
		\label{eq:endpoint-offdiagonal}\\
		\Xi(a,b)
		&\coloneqq \frac{\Gamma(|a-b|)\Gamma(\min(a,b))\Gamma(1-\max(a,b))}{\Gamma(\max(a,b))\Gamma(1-\min(a,b))\Gamma(1-|a-b|)}.
		\label{eq:endpoint-Xi}
	\end{align}
\end{subequations}
The same off-diagonal ratio appears in the three-twist OPE coefficients of Ref.~\cite{Anastasopoulos:2011gn}; the squared coefficient in that complex-boson normalization is $2\pi/\Xi(a,b)$.

For a diagonal character, the zero-balanced hypergeometric asymptotic~\eqref{eq:math-zero-balanced} instead gives $I_{a,a}(1-y)=\log(1/y)+O(1)$ and $J_{a,a}(1-y)=\pi/\sin(\pi a)+O(y)$.
Each of the two terms in the bilinear contributes one logarithm, and hence
\begin{equation}
	\calI_{a,a}(1-y,1-y) = 2\log(1/y)+O(1).
	\label{eq:endpoint-diagonal}
\end{equation}
There are exactly $n-1$ such characters, matching the number of independent neck directions.

The non-diagonal powers are fixed by counting ordered pairs:
\begin{equation}
	\sum_{k_1,k_2=1}^{n-1}\frac{|k_1-k_2|}{n} = \frac2n\sum_{m=1}^{n-2}m(n-1-m) = \frac{(n-1)(n-2)}3.
	\label{eq:endpoint-exponent-count}
\end{equation}

\paragraph{The finite Gamma product from Fermat norms.}
The connection coefficient can be identified sector by sector with the cohomology norm in Eq.~\eqref{eq:fermat-quillen-beta-norm}.
For $r>0$, $s>0$ and $r+s<1$, direct substitution gives
\begin{equation}
	\Xi(r,r+s) = \frac{J(1-r,1-s)}{\pi} = \frac{\Gamma(r)\Gamma(s)\Gamma(1-r-s)}{\Gamma(1-r)\Gamma(1-s)\Gamma(r+s)}.
	\label{eq:endpoint-fermat-norm-bridge}
\end{equation}
Here $J(a,b)$ is the plane integral defining the Serre-dual squared norm in Eq.~\eqref{eq:fermat-quillen-beta-norm}, different from the Euler period $J_{a,b}(x)$.
The stated domain makes the norm integral convergent at zero, one and infinity.
For integer labels the map
\begin{equation}
	(p,q)\longmapsto(p,q-p),\qquad 1\leq p<q\leq n-1,
	\label{eq:endpoint-triangle-bijection}
\end{equation}
is a bijection onto the positive adjacent root triangle: the inverse sends $(\ell_1,\ell_2)$ to $(p,q)=(\ell_1,\ell_1+\ell_2)$.
Since $\Xi(a,b)=\Xi(b,a)$, the two ordered halves contribute one inverse factor per unordered pair.
The bijection and the triangular identity~\eqref{eq:fermat-gamma-product} give
\begin{equation}
	\prod_{k_1\ne k_2}\Xi(k_1/n,k_2/n)^{-1/2} = \prod_{1\leq p<q\leq n-1}\Xi(p/n,q/n)^{-1} = \calP_n.
	\label{eq:endpoint-gamma-product}
\end{equation}

For one real boson, $\Delta_n^{(3)}=(n^2-1)/12$ and $g_d/4-(n-1)(n-2)/6=\Delta_n^{(3)}$.
The off-diagonal asymptotic~\eqref{eq:endpoint-offdiagonal}, the diagonal logarithms~\eqref{eq:endpoint-diagonal}, and Eqs.~\eqref{eq:endpoint-exponent-count} and~\eqref{eq:endpoint-gamma-product} give Lemma~\ref{lem:quantum-degeneration}, with the positive square roots fixed by the original Gaussian determinant.

\paragraph{The finite fusion coefficient.}
Insert Eqs.~\eqref{eq:endpoint-harmonic-limit} and \eqref{eq:endpoint-quantum-limit} into the normalized ratio \eqref{eq:assembly-hodge-form}.
Multiplication by $y^{\Delta_n^{(3)}}$ converts $\calR_{n}^{(3)}$ to $\calF_{n}^{(3)}$ and removes the remaining power.
Keeping the constants before simplifying,
\begin{align}
	\lim_{y\to0^+}\calF_{n}^{(3)}(1-y;\eta)
	&= \frac{(4\pi)^{g_d/2}}{n^{n-1}} \frac{\sqrt n}{(2\pi n)^{(n-1)/2}\,2^{(n-1)/2}} \calP_n\calH_{F_n}(\eta)^2\notag\\
	&= \frac{(4\pi)^{g_F}}{n^{(3n-4)/2}} \calP_n\calH_{F_n}(\eta)^2.
	\label{eq:endpoint-finite-assembly}
\end{align}
The $n-1$ logarithms from the quantum factor cancel the determinant growth of the $n-1$ neck directions.
The equality $g_d-(n-1)=2g_F$ then leaves precisely the two component factors.

Comparison with the direct coefficient~\eqref{eq:fermat-direct-normalized-coefficient} proves Theorem~\ref{thm:absolute-fusion}.
In particular,
\begin{equation}
	\calR_{n}^{(3)}(x;\eta) = \frac{|C_{n}^{(3)}(\eta)|^2}{(1-x)^{\Delta_n^{(3)}}}[1+o(1)].
	\label{eq:endpoint-ratio-limit}
\end{equation}
The convergence in Eq.~\eqref{eq:endpoint-ratio-limit} is uniform for $\eta$ in compact subsets of $(0,\infty)$ at fixed $n$.
The two components give the conjugate three-point vertices.

\subsection{Strict adjacency and continuum fusion}\label{sec:symmetries-limits}

The resolved fusion limit contains two conjugate three-point vertices, whereas strict adjacency contains one.
The corresponding normalized moments therefore contain different powers of the same unit-normalized coefficient.

\begin{corollary}[Genuine multi-entropy at the fusion endpoint]\label{cor:genuine-fusion-endpoint}
	After taking the continuum limit at positive interval separation, the fusion limit is
	\begin{equation}
		\lim_{x\to1^-}\mathrm{GM}_{n,\mathrm{disj}}^{(3)}(x;\eta)
		=\frac{\log|C_n^{(3)}(\eta)|^2}{n(1-n)}
		=2\kappa_n^{(3)}(\eta).
		\label{eq:limits-disjoint-genuine-endpoint}
	\end{equation}
	The strictly adjacent observable satisfies $\mathrm{GM}_{n,\mathrm{adj}}^{(3)}=\kappa_n^{(3)}(\eta)$ after its own continuum limit.
\end{corollary}

Apply the finite-function identity~\eqref{eq:disjoint-genuine-finite-function} to the unit ordinary residue~\eqref{eq:bipartite-crossed-residue} and the tripartite residue in Theorem~\ref{thm:absolute-fusion}.
Since the compact-boson coefficient is strictly positive, taking its logarithm gives Eq.~\eqref{eq:limits-disjoint-genuine-endpoint}.

\paragraph{The extra vertex in the fusion limit.}
The same conclusion follows directly with a common cutoff $\epsilon$.
Take
\begin{equation}
	A=[-\ell_A,0],\qquad B_d=[d,\ell_B],\qquad B_0=[0,\ell_B],
	\qquad O_d=(A\cup B_d)^c,
	\label{eq:limits-positive-gap-geometry}
\end{equation}
with fixed $\ell_A,\ell_B>0$ and $0<d<\ell_B$, and let $O_0$ be the complement at $d=0$.
Use the cycle-factorized endpoint normalization of Section~\ref{sec:twist-correlators} and take $\epsilon\to0$ at positive $d$ before $d\to0$.
The tripartite OPE multiplies the adjacent moment by $(C_n^{(3)})^*$ and the $n/2$ power of the ordinary gap moment.
The ordinary two-interval OPE contributes one power of the same gap moment, with unit leading coefficient.
These gap factors cancel in the genuine moment ratio~\eqref{eq:disjoint-genuine-moment-ratio}, while the remaining interval moments approach their adjacent values.
Consequently,
\begin{equation}
	\lim_{d\to0}\lim_{\epsilon\to0}
	\exp\!\left\{n(1-n)\left[
	\begin{aligned}
		&\mathrm{GM}_{n,\mathrm{disj}}^{(3)}(A:B_d:O_d)\\[-2pt]
		&{}-\mathrm{GM}_{n,\mathrm{adj}}^{(3)}(A:B_0:O_0)
	\end{aligned}
	\right]\right\}
	=(C_n^{(3)}(\eta))^*.
	\label{eq:limits-common-genuine-comparison}
\end{equation}
Setting $d=0$ before removing the cutoff instead defines strict adjacency.

\section{Comparison with known amplitudes}\label{sec:low-index-amplitudes}
The sphere at adjacent $n=2$ and the tori at disjoint $n=2$ and adjacent $n=3$ provide consistency checks of the higher-genus formulas.
These comparisons use a bosonic seed CFT with $c_L=c_R=c$, sphere vacuum state of unit norm and the canonical twist normalization of Section~\ref{sec:twist-correlators-normalization}.

\subsection{The adjacent coefficient at \texorpdfstring{$n=2$}{n=2}}\label{sec:adjacent-low-index}

At $n=2$ the cover for adjacent intervals is a sphere, so the normalized coefficient depends only on the seed CFT central charge~\cite{Harper:2024ker,Harper:2025uui}.
Its sign is fixed by the regular covering construction.
The known normalized sphere amplitude is~\footnote{The sphere correlator in Ref.~\cite{Harper:2024ker} is initially stated up to proportionality, and its finite adjacent constant is then fixed by a vertex-removal prescription.
Here cyclic two-point calibration fixes that constant directly.
The normalized geometric factor in Ref.~\cite{Ma:2026frw} gives the same result.}
\begin{equation}
	\left\langle\prod_{i=1}^{3} \calT_{\varsigma_i}(z_i)\right\rangle_{n=2} = \frac{2^{-c/2}}{|z_{12}z_{23}z_{31}|^{c/4}}, \qquad C_2^{(3)}=2^{-c/2}, \qquad \kappa_2^{(3)}=\frac c4\log2.
	\label{eq:adjacent-n2-coefficient}
\end{equation}

For one real compact boson at finite $\eta=R^2>0$, $F_2$ is a sphere and every circle-valued map has trivial winding class.
Hence
\begin{equation}
	g_F = 0, \qquad Z_{\rm cl}^{\bbR}(\Omega_{F,2};\eta) = \calH_{F_2}(\eta) = 1.
	\label{eq:fermat-boson-sphere}
\end{equation}

The Gamma product $\calP_2$ also equals one.
Thus Eq.~\eqref{eq:fermat-direct-normalized-coefficient} gives $C_2^{(3)}=2^{-1/2}$, reproducing the sphere value at $c=1$.
Appendix~\ref{sec:covering-regularization} gives covering-map and character-spectral derivations.

\subsection{Four-point amplitudes at \texorpdfstring{$n=2$}{n=2}}\label{sec:torus-comparison}

At $n=2$, the ordinary two-interval cover and the tripartite cover for disjoint intervals are tori whose moduli differ by a factor of two.
We first express both amplitudes through the full modular-invariant torus partition function $Z_{\calC}$ of the seed CFT $\calC$~\cite{Calabrese:2009ez,Harper:2024ker,Harper:2025uui}.
The compact-boson specialization then matches the general formulas.
For $\Im\tau>0$, define
\begin{equation}
	Z_{\calC}(\tau,\bar\tau) \coloneqq \Tr_{\calH_{\calC}}\!\left[ e^{2\pi i\tau(L_0-c/24)} e^{-2\pi i\bar\tau(\bar L_0-c/24)}\right].
	\label{eq:models-torus-partition}
\end{equation}
We abbreviate this trace to $Z_{\calC}(\tau)$, with $\bar\tau=\tau^*$.

\paragraph{The two genus-one periods.}
For $n=2$, only $\bmk=(1,1)$ occurs.
Equations~\eqref{eq:period-raw-euler-values} give
\begin{subequations}\label{eq:period-genus-one-raw}
	\begin{align}
		\calA_{1,1}&=-4K(x),\label{eq:period-genus-one-raw-A}\\
		\calB_{1,1}&=8iK(1-x).\label{eq:period-genus-one-raw-B}
	\end{align}
\end{subequations}
The canonical cycles are $A_{1,1}=-\alpha_{0,0}$ and $B_{1,1}=\beta_{0,0}$, so their A-period is $4K(x)$.
Writing $\tau(x)\coloneqq iK(1-x)/K(x)$ for the ordinary two-interval modulus, we obtain
\begin{subequations}\label{eq:models-torus-periods}
	\begin{align}
		\Omega_2^{(2)}(x)&=\tau(x),\\
		\tau_2^{(3)}(x)&\coloneqq\Omega_2^{(3)}(x)=2\tau(x).
		\label{eq:period-genus-one-matrix}
	\end{align}
\end{subequations}
The factor of two follows from the lifted commutator defining $\beta$.

The realignment construction of Ref.~\cite{Yin:2022toc} at moment order $k=2$ has endpoint permutations $(12)(34)$ and $(23)(41)$.
They generate the same regular $(\bbZ_2)^2$ action after replica relabeling, so its modulus $k\tau(x)$ gives an independent route to $2\tau(x)$.
For $k>2$ that realignment cover still has genus one, while the present $n^2$-copy cover has genus $(n-1)^2$.

Write $\eta(\tau)$ for the Dedekind function of Appendix~\ref{sec:math-gaussian-theta}.
In our conventions, the known covering-space result of Refs.~\cite{Harper:2024ker, Harper:2025uui} reads
\begin{equation}
	\calR_2^{(3)}(x) = \left|\frac{\eta(2\tau)}{\vartheta_4(0\mid2\tau)}\right|^{2c} Z_{\calC}(2\tau).
	\label{eq:models-torus-ratio}
\end{equation}
The fixed-frame function and the finite function are then
\begin{subequations}
	\label{eq:models-torus-GRF}
	\begin{align}
		\calG_2^{(3)}(x,x)
		&=2^{-4c/3}[x(1-x)]^{-c/6}Z_{\calC}(2\tau),
		\label{eq:models-torus-GRF-fixed-frame}\\
		\calF_2^{(3)}(x)
		&=2^{-4c/3}x^{c/3}(1-x)^{c/12}Z_{\calC}(2\tau).
		\label{eq:models-torus-GRF-finite}
	\end{align}
\end{subequations}
For comparison, the ordinary two-interval torus gives a finite function with a symmetric prefactor~\cite{Calabrese:2009ez}.
\begin{equation}
	\calF_2^{(2)}(x) = 2^{-2c/3}[x(1-x)]^{c/6}Z_{\calC}(\tau).
	\label{eq:discussion-torus-functions}
\end{equation}

\paragraph{Compact-boson evaluation.}
With the normalized circle measure $\dd{\phi_0}/(2\pi R)$, each integer momentum--winding pair contributes one state with the weights in Eq.~\eqref{eq:ward-radius-spectrum}.
Combining these states with the oscillator trace gives the unit-vacuum torus partition function for one real compact boson.
For $\tau=\tau_1+i\tau_2$ with $\tau_2>0$, the standard momentum--winding trace is~\cite{Headrick:2012fk}
\begin{align}
	Z(\tau;\eta) &= \frac1{|\eta(\tau)|^2} \sum_{w,m\in\bbZ} e^{-\pi\tau_2(\eta w^2+\eta^{-1}m^2)+2\pi i\tau_1mw}
	\label{eq:fermat-boson-torus-trace}\\
	&= \frac{\Theta(\bm0\mid T_\eta(\tau))}{|\eta(\tau)|^2}.
	\label{eq:models-torus-seeds}
\end{align}
The theta representation uses the genus-one matrix~\eqref{eq:classical-theta-matrix}.

For $\Omega=it$ with $t=2K(1-x)/K(x)$, the general compact-boson formula~\eqref{eq:assembly-hodge-form} specializes to
\begin{subequations}
	\label{eq:assembly-genus-one}
	\begin{align}
		\calR_2^{(3)}(x;\eta)
		&= \frac{\pi}{2K(x)}(1-x)^{-1/4} \vartheta_3(0\mid i\eta t)\vartheta_3(0\mid it/\eta),
		\label{eq:assembly-genus-one-ratio}\\
		\calF_2^{(3)}(x;\eta)
		&= \frac{\pi}{2K(x)} \vartheta_3(0\mid i\eta t)\vartheta_3(0\mid it/\eta).
		\label{eq:assembly-genus-one-finite}
	\end{align}
\end{subequations}
For this genus-one marking, the reference characteristic is $[0;1/2]$.
The identity
\begin{equation}
	\vartheta_4(0\mid it)^2=(1-x)^{1/4}\frac{2K(x)}{\pi}
	\label{eq:torus-elliptic-theta-identity}
\end{equation}
converts the general theta-ratio formula directly into Eq.~\eqref{eq:assembly-genus-one-ratio}.
Substituting the compact-boson trace into Eq.~\eqref{eq:models-torus-ratio} reproduces Eq.~\eqref{eq:assembly-genus-one-ratio}.

\paragraph{Vacuum limits of the amplitudes.}
For the vacuum limits, additionally assume a unique normalizable vacuum, a discrete spectrum of finite multiplicity with an isolated lowest positive full dimension $\Delta_{\min}$, a convergent thermal trace, and modular invariance of the full trace.
The modular asymptotics are derived in Appendix~\ref{sec:appendix-torus-check}.
Section~\ref{sec:noncompact-limit} instead treats correlation-ratio densities with a continuous momentum spectrum.

The identity limit is $\calR_2^{(3)}(x)\to1$ as $x\to0$.
In the fusion limit the ordinary and tripartite finite functions obey
\begin{subequations}
	\label{eq:discussion-n2-residues}
	\begin{align}
		\calF_2^{(2)}(1) &= 1,
		\label{eq:discussion-n2-residues-bipartite}\\
		\calF_2^{(3)}(1) &= 2^{-c}=|C_2^{(3)}|^2.
		\label{eq:discussion-n2-residues-tripartite}
	\end{align}
\end{subequations}
The values at $x=1$ are limits from disjoint configurations.
The positive sphere coefficient in Eq.~\eqref{eq:adjacent-n2-coefficient} fixes the sign, and the residues give the resolved endpoint $2\kappa_2^{(3)}$.

\paragraph{The period in the neck marking.}
The waist marking gives $\widetilde\Omega_2^{(3)}=-1/\Omega_2^{(3)}$.
Its elliptic asymptotic reproduces the logarithmic neck coefficient in Eq.~\eqref{eq:degeneration-period-blocks} with $\widehat G_2=(2)$, as shown in Appendix~\ref{sec:appendix-torus-check}.
The finite $\log16$ term remains even when the plumbing coefficient is one, because the regularized component integrals contribute to the finite neck block.

\paragraph{Bipartite and tripartite crossing.}
At $n=2$, the ordinary period is $\Omega_2^{(2)}(x)=\tau(x)$ and the tripartite period is $\Omega_2^{(3)}(x)=2\tau(x)$, as in Eq.~\eqref{eq:models-torus-periods}.
They satisfy
\begin{subequations}
	\label{eq:discussion-crossed-periods}
	\begin{align}
		\tau(1-x)&=-\frac1{\tau(x)},
		\label{eq:discussion-crossed-periods-bipartite}\\
		\Omega_2^{(3)}(1-x)&=-\frac4{\Omega_2^{(3)}(x)}.
		\label{eq:discussion-crossed-periods-tripartite}
	\end{align}
\end{subequations}
The first transformation is modular $S$.
Together with the symmetric prefactor in Eq.~\eqref{eq:discussion-torus-functions}, modular invariance of the full trace recovers the ordinary two-interval symmetry $\calF_2^{(2)}(1-x)=\calF_2^{(2)}(x)$ \cite{Calabrese:2009ez}.
The second follows from $\Omega_2^{(3)}=2\tau$ and is not the modular $S$ transformation of the tripartite torus.
This illustrates Eq.~\eqref{eq:discussion-crossed-tuple}, where tripartite crossing relates different labelled correlators.

\subsection{The adjacent coefficient at \texorpdfstring{$n=3$}{n=3}}\label{sec:low-genus-boson-checks}

At $n=3$, the degree-nine cover for adjacent intervals is a triangular torus with $\tau_\triangle=e^{i\pi/3}$.
With the full seed CFT trace~\eqref{eq:models-torus-partition}, the known coefficient and, for $C_3^{(3)}>0$, its entropy constant are~\cite{Harper:2024ker}
\begin{equation}
	C_3^{(3)}=3^{-3c/2}Z_{\calC}(\tau_\triangle), \qquad
	\kappa_3^{(3)}=\frac c4\log3 - \frac16\log Z_{\calC}(\tau_\triangle).
	\label{eq:adjacent-n3-coefficient}
\end{equation}
The finite external two-point normalizers give the term proportional to $\log3$.

\paragraph{Compact-boson comparison.}
At $n=3$, the canonical cycles are $A_1=e_{0,0}$ and $B_1=e_{1,0}$.
Their orbit periods give
\begin{equation}
	\Omega_{F,3}=\zeta_3=-\frac12+\frac{\sqrt3}{2}i.
	\label{eq:fermat-cubic-period}
\end{equation}
The integral shear $B_1\mapsto B_1+A_1$ sends this modulus to $\tau_\triangle=\Omega_{F,3}+1$ and preserves the winding sum.
For one real compact boson at finite $\eta=R^2>0$, Appendix~\ref{sec:appendix-triangle-check} converts the trace to the harmonic factor and gives
\begin{equation}
	C_{3}^{(3)}(\eta) = \frac{3^{-3/2}\calH_{F_3}(\eta)}{\sqrt{\Im\tau_\triangle} |\eta(\tau_\triangle)|^2}.
	\label{eq:adjacent-n3-boson-coefficient}
\end{equation}
The special value in Eq.~\eqref{eq:math-triangular-eta-value} gives
\begin{equation}
	\frac{3^{-3/2}}{\sqrt{\Im\tau_\triangle} |\eta(\tau_\triangle)|^2} = \frac{2\sqrt\pi}{3^{5/4}} \left[\frac{\Gamma(2/3)}{\Gamma(1/3)}\right]^{3/2},
	\label{eq:low-index-torus-gamma-match}
\end{equation}
Together with $\calP_3=[\Gamma(2/3)/\Gamma(1/3)]^3$, this equality reproduces the direct determinant formula~\eqref{eq:fermat-direct-normalized-coefficient} at $n=3$.

\section{Lattice tests of the continuum amplitudes}\label{sec:model-comparisons}
We test the CFT amplitudes in XXZ, XX and Ising spin chains, and their noncompact limit in a harmonic chain.

\subsection{Finite-size observables}
For a normalized pure spin state on a ring of $L$ sites, define the disjoint ratio and adjacent coefficient by
\begin{subequations}
	\label{eq:numerical-normalized-amplitudes}
	\begin{align}
		\calR_{n,L}^{(3)}(A:B:O)
		&\coloneqq \frac{\widehat Z_{n,\mathrm{disj}}^{(3)}(A:B:O)}{[\widehat Z_n^{(2)}(A:A^c)\widehat Z_n^{(2)}(B:B^c)]^n},
		\label{eq:numerical-disjoint-ratio}\\
		C_{n,L}^{(3)}(A:B:O)
		&\coloneqq \frac{\widehat Z_{n,\mathrm{adj}}^{(3)}(A:B:O)}{[\widehat Z_n^{(2)}(A:A^c)\widehat Z_n^{(2)}(B:B^c) \widehat Z_n^{(2)}(O:O^c)]^{n/2}}.
		\label{eq:numerical-adjacent-coefficient}
	\end{align}
\end{subequations}
All factors use the same finite state, geometry and spin sector.
The CFT predictions for the $L\to\infty$ limits at fixed arc fractions are $\calR_n^{(3)}(x)$ and $C_n^{(3)}$, respectively.
We display the adjacent quantity as $\kappa_{n,L}^{(3)}\coloneqq\log C_{n,L}^{(3)}/[n(1-n)]$.

For disjoint intervals of lengths $\ell_A,\ell_B$ and separating arc $d$, the circle cross ratio is
\begin{equation}
	x \coloneqq \frac{\sin(\pi\ell_A/L)\sin(\pi\ell_B/L)}{\sin[\pi(\ell_A+d)/L]\sin[\pi(\ell_B+d)/L]}.
	\label{eq:numerical-circle-cross-ratio}
\end{equation}
Changing $L$ at fixed arc fractions leaves $x$ fixed.

\subsection{Radius dependence in the XXZ chain}
For the zero-field spin-$\tfrac12$ XXZ chain with anisotropy $\Delta$, take
\begin{equation}
	H_{\rm XXZ} = J\sum_j \left(S_j^xS_{j+1}^x + S_j^yS_{j+1}^y + \Delta S_j^zS_{j+1}^z\right), \qquad J>0.
	\label{eq:ward-xxz-hamiltonian}
\end{equation}
In the gapless regime $-1<\Delta<1$, the Luttinger parameter $K$ fixes the real compactification parameter $\eta=R^2$ in the action of Eq.~\eqref{eq:ward-real-action} through~\footnote{This is the convention used for the XXZ comparison in Ref.~\cite{Calabrese:2009ez}.}
\begin{equation}
	K = \frac{\pi}{2[\pi-\arccos\Delta]}, \qquad \eta=\frac1{2K} = 1-\frac{\arccos\Delta}\pi.
	\label{eq:ward-luttinger-dictionary}
\end{equation}
At the XX point, $K=1$ and $\eta=1/2$.
The periodic data use $J=1$, zero field and total $S^z=0$.
The anisotropies $\Delta=-1/2,0,1/2$ correspond to $\eta=1/3,1/2,2/3$, with the XX spin chain at $\Delta=0$.
The adjacent and disjoint predictions are Eqs.~\eqref{eq:fermat-direct-normalized-coefficient} and~\eqref{eq:assembly-normalized-ratio}, with $\eta$ fixed by Eq.~\eqref{eq:ward-luttinger-dictionary}.

% \paragraph{TTN evaluation.}
We evaluate the adjacent coefficient and disjoint ratio in normalized, periodic, charge-conserving TTN approximations to the XXZ ground state in the sector $S^z=0$.
Appendix~\ref{sec:lattice-methods-ttn} specifies the state family and contraction.
For small systems, we checked the TTN contraction against contractions obtained from exact diagonalization.
At the XX point, we also compared the TTN observables with determinant calculations.

\paragraph{Adjacent intervals in the XXZ chain.}
\begin{figure}[t]
	\centering
	\includegraphics[width=\textwidth]{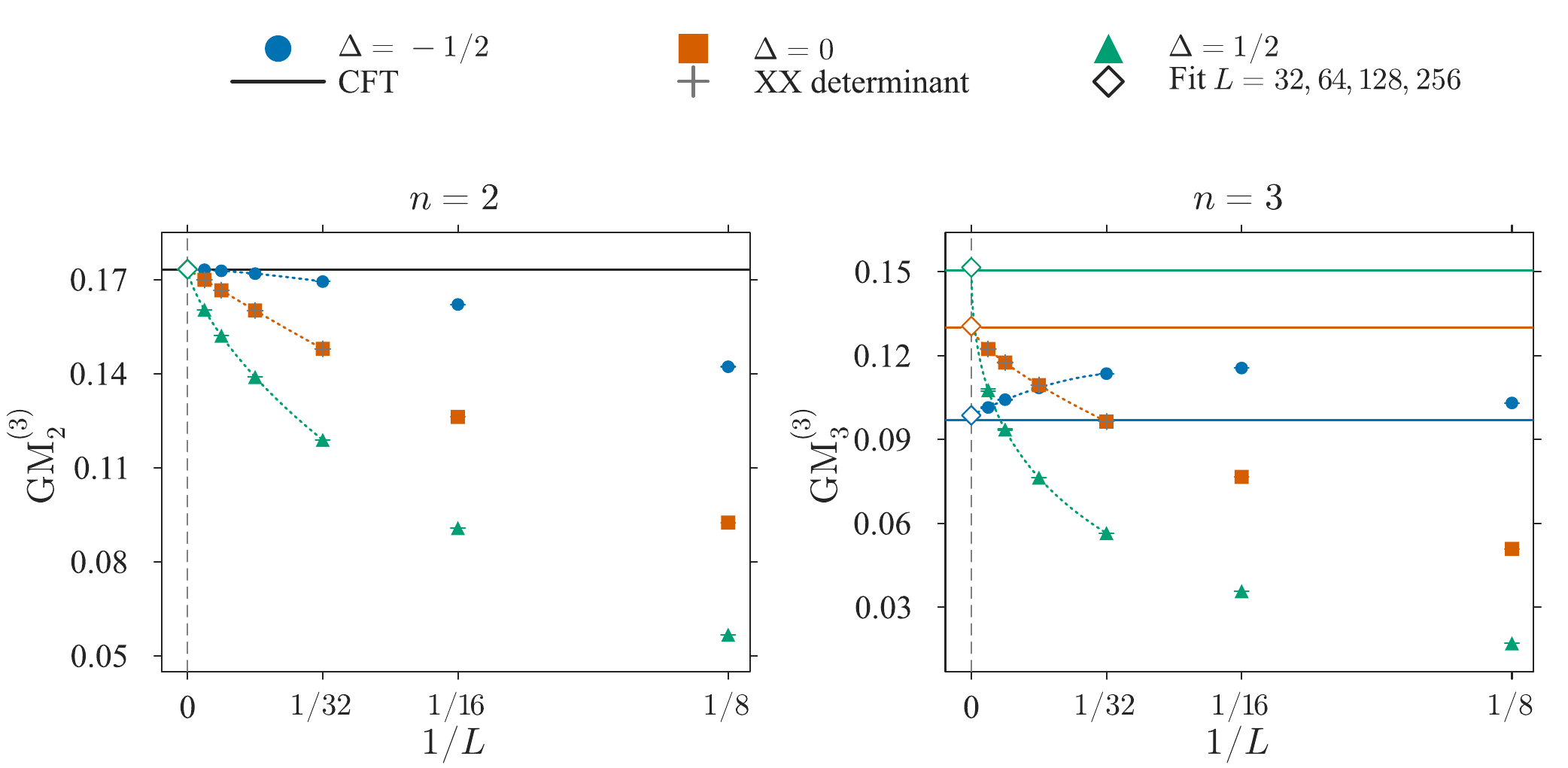}
	\caption{Adjacent genuine multi-entropy in periodic XXZ chains at arc proportions $1:1:2$.
	Colors and shapes distinguish $\Delta=-1/2,0,1/2$.
	Filled symbols show $L=8,16,32,64,128,256$.
	Solid horizontal lines are the compact-boson CFT constants at the fixed Hamiltonian radii.
	Open diamonds at $1/L=0$ are obtained by fitting the normalized amplitudes at $L=32,64,128,256$ to a quadratic in $L^{-2K/n}$ and then taking their logarithms.
	Dotted curves show these four-size fits. Gray pluses are matched finite XX determinant values.}
	\label{fig:numerical-adjacent-xxz}
\end{figure}

Figure~\ref{fig:numerical-adjacent-xxz} shows 36 finite XXZ values at $\Delta=-1/2,0,1/2$, $n=2,3$ and $L=8,16,32,64,128,256$, with $A:B:O=1:1:2$.
The ordinate is the genuine multi-entropy, $\GM_n^{(3)}=\log C_{n,L}^{(3)}/[n(1-n)]$, using the same-state ordinary-moment normalization of Eq.~\eqref{eq:numerical-adjacent-coefficient}.

At the free point, eight matched physical XX determinant values at $L=32,64,128,256$ check the finite TTN observables.
The relative difference between these paired TTN and determinant values is below $0.0059\%$.

At $L=32,64,128,256$, we fit the normalized amplitude $C_{n,L}^{(3)}$ to a quadratic in $u\coloneqq(16/L)^{2K/n}$ with equal weights.
The logarithm of its positive intercept gives the extrapolated genuine multi-entropy at $1/L=0$.
These fits have one residual degree of freedom.
We use the exponent $2K/n$ as an ansatz for the tripartite fits~\cite{Harper:2025uui}, motivated by ordinary R\'enyi corrections~\cite{Cardy:2010zs,Xavier:2011np,Calabrese:2010mtj}.

The horizontal lines use the continuum coefficients in Eq.~\eqref{eq:fermat-direct-normalized-coefficient}, with the Hamiltonian radii $\eta=1/3,1/2,2/3$ fixed by Eq.~\eqref{eq:ward-luttinger-dictionary}.
Thus the $n=2$ reference is $\kappa_2^{(3)}=\log2/4$ for all three anisotropies, whereas the $n=3$ references differ.
For the four-size fits, the largest absolute relative difference between the extrapolated amplitude and CFT is below $0.011\%$ at $n=2$ and $1.02\%$ at $n=3$.

\paragraph{Disjoint intervals in the XXZ chain.}
\begin{figure}[t]
	\centering
	\includegraphics[width=\textwidth]{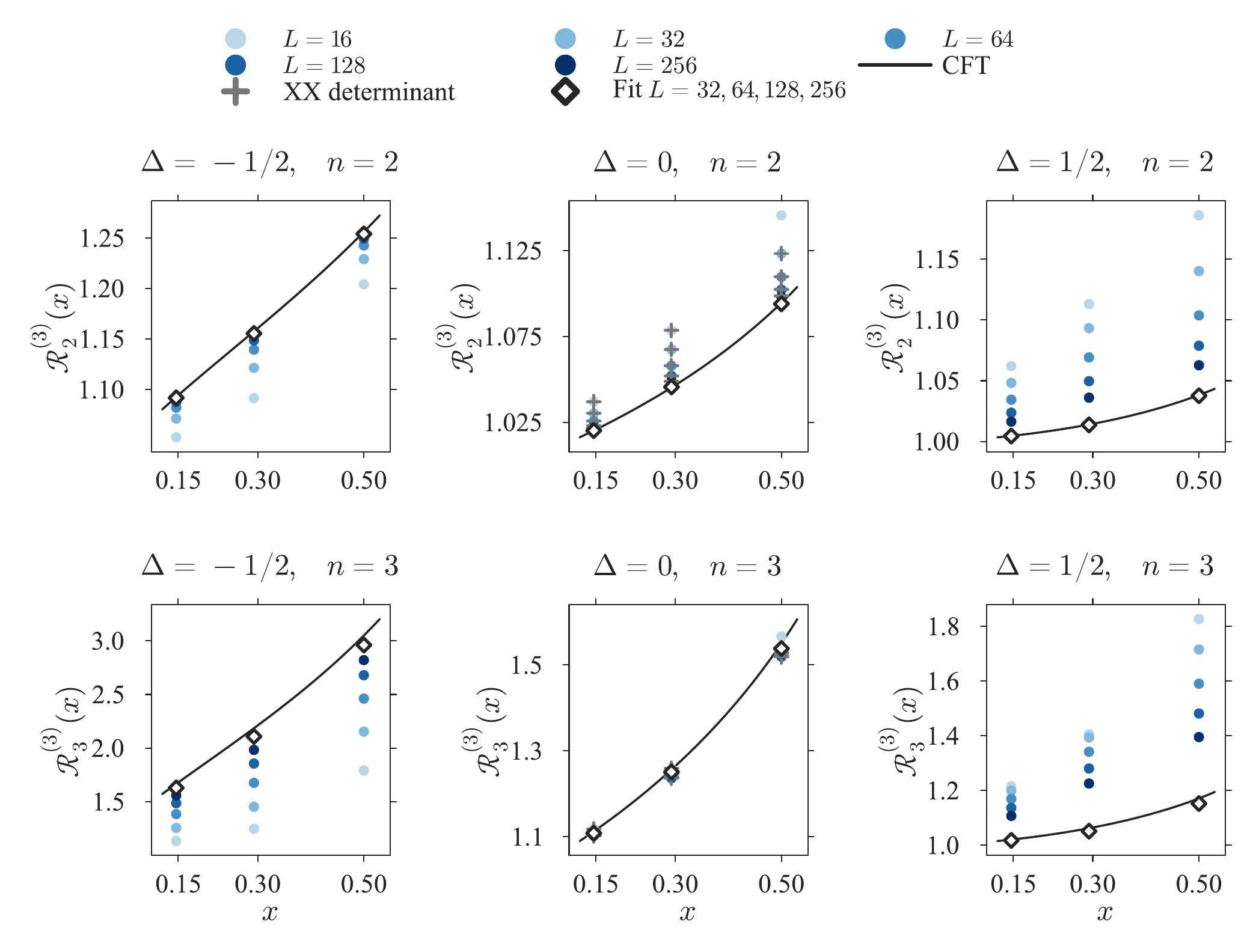}
	\caption{Absolute disjoint ratios in periodic XXZ chains, with rows $n=2,3$ and columns $\Delta=-1/2,0,1/2$.
	Black curves are the compact-boson CFT predictions at the fixed Hamiltonian radii.
	Light-to-dark blue circles distinguish increasing sizes $L=16,32,64,128,256$.
	Black open diamonds show the $L\to\infty$ extrapolations.
	Gray pluses in the free-point column are matched finite XX determinant values and do not enter the fits.}
	\label{fig:numerical-disjoint-ratios}
\end{figure}

The same TTN states give the absolute ratio in Eq.~\eqref{eq:numerical-disjoint-ratio} for disjoint intervals.
Figure~\ref{fig:numerical-disjoint-ratios} shows 90 finite values at $n=2,3$, $\Delta=-1/2,0,1/2$ and $L=16,32,64,128,256$.
The three arc proportions are $(1,3,1,3)/8$, $(1,1,1,5)/8$ and $(1,1,1,1)/4$, with cross ratios $x=(2-\sqrt2)/4$, $1-1/\sqrt2$ and $1/2$.
At the free point, the maximum absolute relative difference between TTN and determinant values is $0.193\%$.

We fit the normalized amplitude $\calR_{n,L}^{(3)}$ to a quadratic in $u$.
The CFT curves evaluate Eq.~\eqref{eq:assembly-normalized-ratio} at the Hamiltonian radii.
For the four-size fits shown in Figure~\ref{fig:numerical-disjoint-ratios}, the largest absolute relative differences from CFT are below $0.19\%$ at $n=2$ and $3.51\%$ at $n=3$ over the anisotropies and shapes.

\subsection{Higher-index spin-chain amplitudes}
\subsubsection{XX and Ising CFT amplitudes}\label{sec:compact-correlations}

% The XX and Ising comparisons use the ordinary and tripartite disjoint covers, whose genera are $g^{(2)}=n-1$ and $g_d=(n-1)^2$, respectively.
% The ordinary cyclic-cover period matrix is~\cite{Calabrese:2009ez}
% \begin{equation}
% 	(\Omega_n^{(2)}(x))_{rs} = \frac{2i}{n}\sum_{k=1}^{n-1}\sin\frac{\pi k}{n} \frac{{}_2F_1(k/n,1-k/n;1;1-x)}{{}_2F_1(k/n,1-k/n;1;x)} \cos\frac{2\pi k(r-s)}n, \quad 1\leq r,s\leq g^{(2)}.
% 	\label{eq:models-cct-period}
% \end{equation}
% The ordinary-cover period matrix is purely imaginary in this marking, so its momentum and winding sums factorize.
% In the real-field normalization of Eq.~\eqref{eq:common-real-action}, the ordinary two-interval function is
% \begin{equation}
% 	\calF_n^{(2)}(x;\eta)
% 	= \frac{\Theta(\bm0\mid\eta\Omega_n^{(2)}(x)) \Theta(\bm0\mid\eta^{-1}\Omega_n^{(2)}(x))}{\Theta(\bm0\mid\Omega_n^{(2)}(x))^2}.
% 	\label{eq:models-cct-finite}
% \end{equation}
% The corresponding theta quotient for ordinary multi-interval cyclic covers is given in Ref.~\cite{Coser:2013qda}.
% For the tripartite cover, Eq.~\eqref{eq:assembly-normalized-ratio} fixes the reference spin line and scalar factor in the physical replica marking.
% The conversion between $\calF$, $\calG$ and $\calR$ is given by Eqs.~\eqref{eq:real-cross-ratio-relations-2} and~\eqref{eq:real-cross-ratio-relations-3}.
% At $n=2$, the periods $\tau(x)$ and $2\tau(x)$ recover the independently normalized torus expression of Section~\ref{sec:torus-comparison}.
% Theta duplication also reduces the self-dual and XX-radius functions to elementary expressions.

The XX spin observable requires the modular Dirac spin sum, rather than a fixed fermion boundary condition~\cite{Headrick:2012fk,Mukhi:2017rex}.
Its tripartite ratio is
\begin{equation}
	\calR_{n,\rmD,\mathrm{sum}}^{(3)}(x) = \frac{2^{-g_d}\sum_\delta |\theta[\delta](\bm0\mid\Omega_n^{(3)}(x))|^2}{|\theta[\delta_*](\bm0\mid\Omega_n^{(3)}(x))|^2},
	\label{eq:models-dirac-sum}
\end{equation}
where the sum runs over the even half-characteristics of the tripartite disjoint cover.
The analogous spin sum on ordinary multi-interval cyclic covers appears in Ref.~\cite{Coser:2013qda}.
Odd theta constants vanish at the origin, so the sum can be extended to all half-characteristics $[\bm\epsilon;\bm\delta]$.
Let $\bm m,\bm m'\in\bbZ^g$ be the integer indices of the theta series and its complex conjugate.
Summing $\bm\delta$ imposes $\bm m-\bm m'\in2\bbZ^g$.
Set $\bmw=(\bm m-\bm m')/2$ and $\bm p=\bm m+\bm m'+2\bm\epsilon$.
The parity of $\bm p$ fixes $2\bm\epsilon$, so this is a bijection to $(\bmw,\bm p)\in\bbZ^{2g}$.
Its exponent is $-\pi[2\bmw^\top\Im\Omega\bmw+\bm p^\top\Im\Omega\bm p/2]+2\pi i\bmw^\top\Re\Omega\bm p$, giving
\begin{equation}
	2^{-g_n^{(3)}}\sum_\delta|\theta[\delta](\bm0\mid\Omega)|^2 = \Theta(\bm0\mid T_2(\Omega)), \qquad \calR_{n,\rmD,\mathrm{sum}}^{(3)}(x) = \calR_n^{(3)}(x;2)=\calR_n^{(3)}(x;1/2).
	\label{eq:models-dirac-bosonization}
\end{equation}
Here the last equality uses radius duality.
This identity fixes the compact-boson target for the XX spin moments at every integer $n\geq2$.

\paragraph{Ising spin sums.}\label{sec:ising-fermion-correlations}

The diagonal Ising CFT has $c=1/2$.
We use the sectorwise scalar--fermion determinant relation of Refs.~\cite{Headrick:2012fk,Mukhi:2017rex}.
In the convention of Appendix~\ref{sec:fixed-spin-reference}, the positive Majorana square root is proportional to $|\theta[\delta](\bm0\mid\Omega)|$, with a factor independent of $\delta$.

The auxiliary normalization of Appendix~\ref{sec:fixed-spin-reference} gives the denominator $|\theta[\delta_*](\bm0\mid\Omega)|$.

The diagonal Ising projection sums the even characteristics with a weight fixed by the identity-channel cusp $x\to0$.
The genus-$g$ spin sum is given in Ref.~\cite{Dijkgraaf:1987vp}.
Reference~\cite{Calabrese:2010he} applies it to the ordinary two-interval replica surface.
The spin sums in the present normalization are
\begin{subequations}
	\label{eq:models-ising}
	\begin{align}
		% \calF_{n,\rmI}^{(2)}(x)
		% &= \frac1{2^{g^{(2)}}} \sum_\delta \frac{|\theta[\delta](\bm0\mid\Omega_n^{(2)}(x))|}{\Theta(\bm0\mid\Omega_n^{(2)}(x))},
		% \label{eq:models-ising-bipartite-finite}\\
		\calR_{n,\rmI}^{(3)}(x)
		&= \frac1{2^{g_d}} \sum_\delta \frac{|\theta[\delta](\bm0\mid\Omega_n^{(3)}(x))|}{|\theta[\delta_*](\bm0\mid\Omega_n^{(3)}(x))|},
		\label{eq:models-ising-tripartite-ratio}\\
		\calF_{n,\rmI}^{(3)}(x)
		&= (1-x)^{(n^2-1)/24}\calR_{n,\rmI}^{(3)}(x).
		\label{eq:models-ising-tripartite-finite}
	\end{align}
\end{subequations}
% The first line is the two-interval Ising function of Ref.~\cite{Calabrese:2010he}.
% Equations~\eqref{eq:real-cross-ratio-relations-2} and~\eqref{eq:real-cross-ratio-relations-3}, with $c=1/2$, supply the remaining $\calR^{(2)}$ and both $\calG$ functions.
The sums run over even half-characteristics of the indicated genus.
The characteristic convention is given in Appendix~\ref{sec:math-thomae}.

% For either cover, let $g$ denote its genus.
The factor $2^{-g_d}$ is fixed by the identity-channel cusp $x\to0$.
For each fixed integer $n$, the smallest eigenvalue of $\Im\Omega$ diverges in the stated marking, by Eq.~\eqref{eq:assembly-hodge-cusp}.
Writing a characteristic as $[\bm\epsilon;\bm\delta]$, sectors with $\bm\epsilon\ne0$ therefore vanish, while the $2^g$ sectors with $\bm\epsilon=0$ tend to one.
The reference denominator has the same limit, giving $\calR\to1$.

\paragraph{The genus-one Ising trace.}
The diagonal Ising partition function is
\begin{equation}
	Z_{\rmI}(\tau) = \frac{|\vartheta_2(0\mid\tau)| + |\vartheta_3(0\mid\tau)| + |\vartheta_4(0\mid\tau)|}{2|\eta(\tau)|}.
	\label{eq:models-ising-diagonal-torus-trace}
\end{equation}
This is the Ising trace used in Ref.~\cite{Harper:2025uui}.
Its substitution in Eq.~\eqref{eq:models-torus-ratio} checks the tripartite spin sum at modulus $2\tau(x)$.
For $n=2$, all three even theta constants are positive on the imaginary axis.
Equation~\eqref{eq:models-ising} reduces to
\begin{subequations}
	\label{eq:models-ising-n2}
	\begin{align}
		% \calF_{2,\rmI}^{(2)}(x)
		% &= \frac{1+x^{1/4}+(1-x)^{1/4}}2,
		% \label{eq:models-ising-n2-bipartite-finite}\\
		\calR_{2,\rmI}^{(3)}(x)
		&= \frac{\vartheta_2(0\mid2\tau)+\vartheta_3(0\mid2\tau) + \vartheta_4(0\mid2\tau)}{2\vartheta_4(0\mid2\tau)},
		\label{eq:models-ising-n2-tripartite-ratio}\\
		\calF_{2,\rmI}^{(3)}(x)
		&= \frac{\vartheta_2(0\mid2\tau)+\vartheta_3(0\mid2\tau) + \vartheta_4(0\mid2\tau)}{2\vartheta_3(0\mid\tau)}.
		\label{eq:models-ising-n2-tripartite-finite}
	\end{align}
\end{subequations}
Here $\vartheta_4(0\mid2\tau)=(1-x)^{1/8}\vartheta_3(0\mid\tau)$.

\paragraph{The adjacent Ising coefficient.}
Applying the same sectorwise determinant relation~\cite{Headrick:2012fk,Mukhi:2017rex} to the Fermat cover gives the adjacent Ising coefficient.
Removing the momentum-winding theta sum from the real-boson coefficient \eqref{eq:fermat-direct-normalized-coefficient} leaves the scalar factor on the common smooth cap metric.
For one Majorana the sectorwise relation gives its positive square root, including the local anomaly and the external two-point norms.

The diagonal projection then gives a positive coefficient with the normalized sum of even theta constants.
With $g_F=(n-1)(n-2)/2$, the result is
\begin{equation}
	C_{n,\rmI}^{(3)} = \left[ \frac{(4\pi)^{g_F/2}}{n^{(3n-4)/4}} \sqrt{\calP_n\det\Im\Omega_{F,n}} \right]^{1/2} \frac1{2^{g_F}} \sum_\delta |\theta[\delta](\bm0\mid\Omega_{F,n})|.
	\label{eq:numerical-ising-adjacent-reference}
\end{equation}

The normalized four-point function gives a separate check of the spin-projection weight through its adjacent degeneration.
The vacuum term on each long neck leaves the two Fermat component partition functions.
The local connection factors are those of Section~\ref{sec:adjacent-factorization}, with the anomaly exponent halved for $c=1/2$.
In the adapted marking, $g_d=2g_F+n-1$.
Only zero upper characteristics on the $n-1$ neck cycles survive, and their $2^{n-1}$ lower characteristics give the same
vacuum term.
Hence the four-point projection factor becomes
\begin{equation}
	2^{-g_d}2^{n-1}=\left(2^{-g_F}\right)^2.
	\label{eq:ising-adjacent-spin-projection}
\end{equation}
Global even parity with zero upper neck characteristics allows even--even and odd--odd component characteristics.
The leading product in the odd--odd case vanishes because each odd theta constant vanishes at the origin; finite-neck corrections need not vanish identically.
Together with the scalar determinant factor, the surviving even--even sums give the leading fusion residue $|C_{n,\rmI}^{(3)}|^2$.

The coefficients at $n=2,3$ also agree with the independent sphere and torus calculations, giving $C_{2,\rmI}^{(3)}=2^{-1/4}$ and $C_{3,\rmI}^{(3)}=3^{-3/4}Z_{\rmI}(\tau_\triangle)$.

\subsubsection{Spin-state contractions}\label{sec:spin-state-contractions}
The critical Ising chain has
\begin{equation}
	H_{\rmI} = -\sum_{j=1}^{L}(\sigma_j^x\sigma_{j+1}^x+\sigma_j^z).
	\label{eq:numerical-ising-hamiltonian}
\end{equation}
We use the periodic even-parity ground state, with antiperiodic Jordan--Wigner fermions.
The XX chain uses the half-filled ground state with even fermion parity and antiperiodic Jordan--Wigner fermions for $L\equiv0\pmod4$.
A staggered on-site spin rotation relates the hopping convention used in the free calculation to the antiferromagnetic exchange in Eq.~\eqref{eq:ward-xxz-hamiltonian}.
The rotation commutes with every local replica permutation and therefore leaves the measured moments invariant.

Appendix~\ref{sec:lattice-methods-gaussian} gives the finite signed sum and its occupied-orbital determinant and ordered Pfaffian kernels.
We evaluate permutations of spin sites, retaining the complex phases and the Jordan--Wigner sign across the disjoint gap~\cite{2010EL.....8940001I,Fagotti:2010yr,Coser:2015dvp}.

In the constrained parity expansion used for the adjacent data, write the normalized signed classes as $C_{n,L}^{(3)}=P_+(L)-P_-(L)$.
The sums $P_+$ and $P_-$ collect terms with positive and negative expansion coefficients, corresponding to the even and odd Arf classes, respectively~\cite{Tasuki:2026fss}.

A sectorwise comparison between lattice pseudo-density matrices and continuum spin structures is available for ordinary two-interval R\'enyi entropies~\cite{Maric:2020ulk}.
For the regular tripartite family, we assume that each normalized kernel tends to the uninserted massless partition function in its marked spin structure.
Vanishing odd theta constants then set the odd-class intercept to zero.

\subsubsection{Adjacent coefficients}\label{sec:numerical-adjacent}
\begin{figure}[t]
	\centering
	\includegraphics[width=\textwidth]{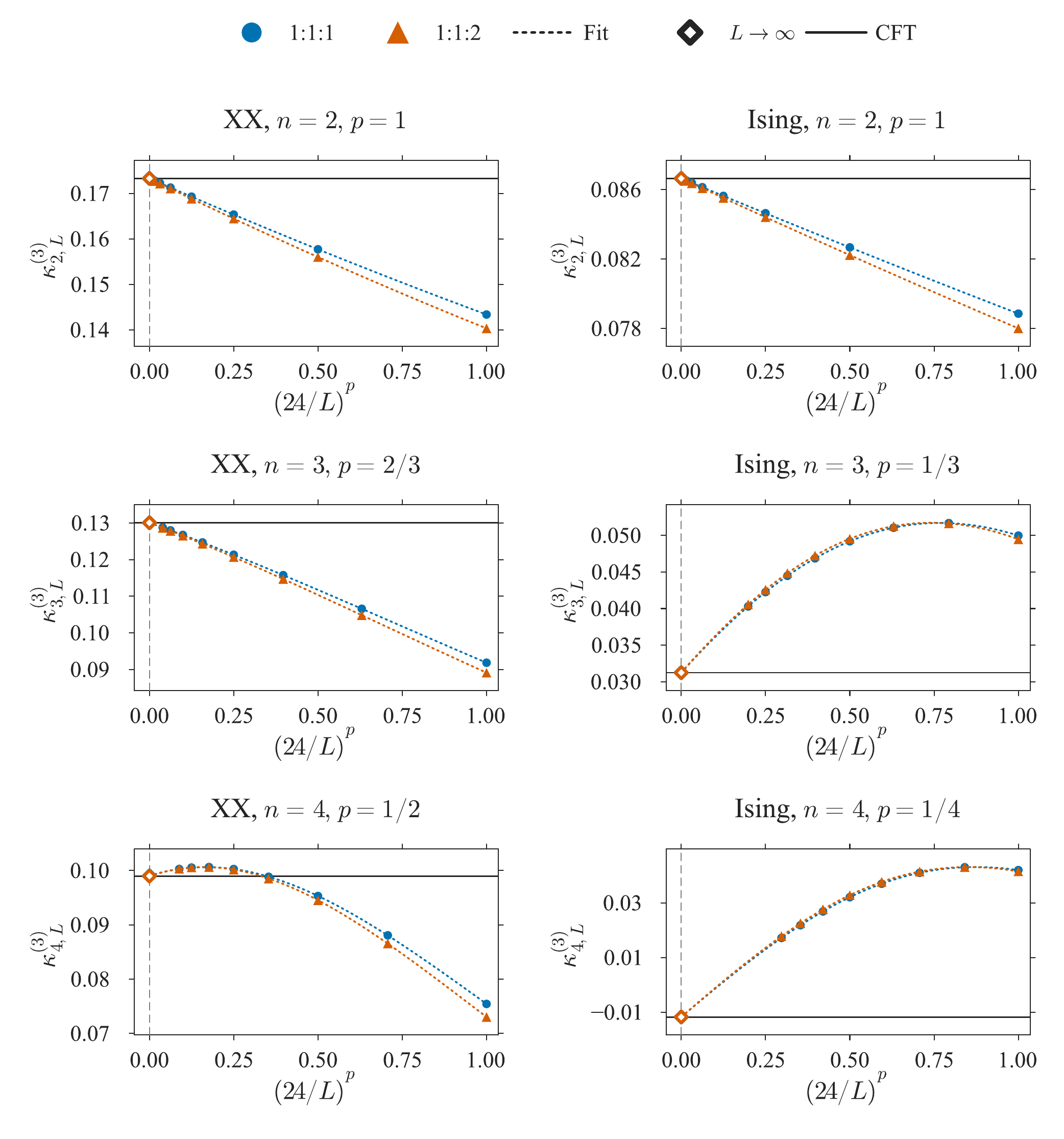}
	\caption{Adjacent genuine multi-entropy $\kappa_{n,L}^{(3)}$, with $n=2,3,4$ in the three rows and XX and Ising in the two columns.
	Markers distinguish $1:1:1$ and $1:1:2$ arc proportions.
	Each panel states the exponent $p$ of its horizontal axis $(24/L)^p$: $p=2/n$ for XX, $p=1$ for Ising at $n=2$, and $p=1/n$ for Ising at $n=3,4$.
	Dotted curves show the fit.
	Open diamonds show extrapolated values and horizontal lines show the CFT predictions.}
	\label{fig:numerical-adjacent-coefficients}
\end{figure}

Strict adjacency predicts a coefficient independent of the nonzero arc proportions.
We test this shape independence with the periodic XX and Ising states at $A:B:O=1:1:1$ and $1:1:2$.
Figure~\ref{fig:numerical-adjacent-coefficients} shows all 96 finite values at $n=2,3,4$ and $L=24 \times 2^j$, $j=0,\ldots,7$, with the same-state normalization of Eq.~\eqref{eq:numerical-adjacent-coefficient}.

We fit all eight displayed sizes with equal weights.
At $n>2$, we fit the even and odd components separately~\cite{Tasuki:2026fss}.
For XX, the even class is cubic in $u\coloneqq(24/L)^{2/n}$ and the odd class uses powers $u,u^2,u^3$ without an intercept.
For Ising at $n=3,4$, the even class uses $1,u^2,u^4,u^6$ and the odd class $u,u^3,u^5$, with $u\coloneqq(24/L)^{1/n}$.
At $n=2$ the odd class is absent, and the Ising fit is cubic in $24/L$.
For the Ising $n=4$ case, all measured $\kappa_{n,L}^{(3)}$ remain positive for both arc proportions through $L=3072$, while both fits give negative estimates of $\kappa_n^{(3)}$, matching the sign of the CFT reference.

\subsubsection{Disjoint correlation ratios}\label{sec:numerical-free-disjoint}
\begin{figure}[t]
	\centering
	\includegraphics[width=\textwidth]{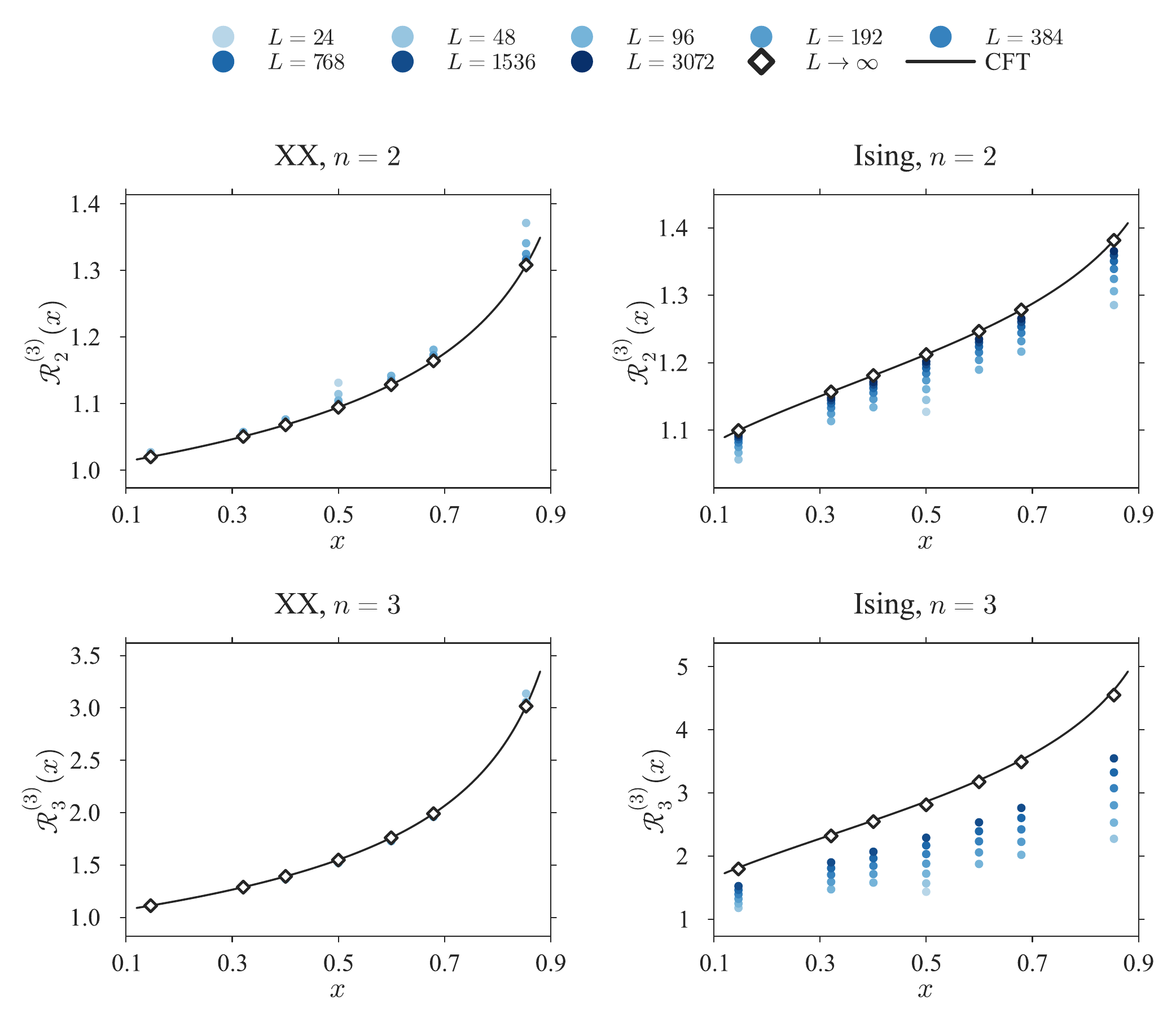}
	\caption{Absolute disjoint correlation ratios for physical XX and Ising spin chains at $n=2,3$.
	Light-to-dark blue circles distinguish increasing finite sizes, and open black diamonds show the quadratic extrapolations of the normalized amplitudes.
	Black curves are the CFT predictions.}
	\label{fig:numerical-free-disjoint-ratios}
\end{figure}

For disjoint intervals, changing the arc fractions probes the cross-ratio dependence of the absolute ratio \eqref{eq:numerical-disjoint-ratio}.
Figure~\ref{fig:numerical-free-disjoint-ratios} contains results for seven fixed shapes spanning $0.146447\lesssim x\lesssim0.853553$.
The geometry-dependent sequences extend to $L=3072$, or $1536$ for Ising at $n=3$.
In spatial order $(A,O_1,B,O_2)$, the shapes are $(1,1,1,1)/4$, $(1,3,1,3)/8$, $(3,1,3,1)/8$, and $(3,4,3,6)/16$, $(3,4,4,5)/16$, $(4,3,5,4)/16$, $(4,3,6,3)/16$.
The equal-quarter sequence starts at $L=24$, the two denominator-eight shapes at $L=48$, and the four denominator-sixteen shapes at $L=96$.

Each signed component is divided by the same-state denominator at each finite size before fitting.
Both components use equal-weight quadratics with free intercepts, in $u\coloneqq(24/L)^{2/n}$ for XX and $u\coloneqq(24/L)^{1/n}$ for Ising.

\subsection{The noncompact limit in the harmonic chain}
\subsubsection{Correlation-ratio densities from decompactification}\label{sec:noncompact-limit}

We remove the genus-dependent radius power $\eta^{g_q/2}$ and take $\eta\to\infty$ at fixed $0<x<1$, as in Eq.~\eqref{eq:models-density-definition} and Ref.~\cite{Calabrese:2009ez}.
Taking an endpoint limit of the resulting density probes a continuous low-momentum spectrum.

\paragraph{Poisson resummation.}
Poisson resummation of the Narain series gives the two-interval density of Ref.~\cite{Calabrese:2009ez} and its tripartite counterpart.

For either period matrix $\Omega=\calR+i\calI$, resumming the momentum vector gives
\begin{equation}
	\Theta(\bm0\mid T_\eta(\Omega)) = \frac{\eta^{g/2}}{\sqrt{\det\calI}} \sum_{\bmw_A,\bmw_B\in\bbZ^g} \exp\!\Bigl[-\pi\eta\Bigl\{ \bmw_A^{\top}\calI\bmw_A + (\bmw_B-\calR\bmw_A)^{\top}\calI^{-1}(\bmw_B-\calR\bmw_A)\Bigr\}\Bigr].
	\label{eq:models-decompactification-poisson}
\end{equation}
At fixed $x$, the quadratic form is positive definite and its zero vector is unique.
All other vectors are exponentially suppressed as $\eta\to\infty$, leaving the density factor $\eta^{g/2}/\sqrt{\det\calI}$.
The two covers give powers $g^{(2)}/2=(n-1)/2$ and $g_d/2=(n-1)^2/2$.

For $q=2,3$, define the correlation-ratio density by
\begin{equation}
	\widetilde{\calR}_{n,\mathrm{nc}}^{(q)}(x) \coloneqq \lim_{\eta\to\infty}\eta^{-g_q/2} \calR_{n}^{(q)}(x;\eta).
	\label{eq:models-density-definition}
\end{equation}
For related theta and period-determinant formulas on ordinary multi-interval cyclic covers, see Ref.~\cite{Coser:2013qda}.
Equation~\eqref{eq:assembly-normalized-ratio} then give
\begin{subequations}
	\label{eq:models-noncompact-ratios}
	\begin{align}
		\widetilde{\calR}_{n,\mathrm{nc}}^{(2)}(x)
		&=\frac{(1-x)^{-(n-1/n)/6}}{\sqrt{\det\Im\Omega_n^{(2)}(x)} \Theta(\bm0\mid\Omega_n^{(2)}(x))^2},
		\label{eq:models-noncompact-ratios-bipartite}\\
		\widetilde{\calR}_{n,\mathrm{nc}}^{(3)}(x)
		&=\frac1{\sqrt{\det\Im\Omega_n^{(3)}(x)} |\theta[\delta_*](\bm0\mid\Omega_n^{(3)}(x))|^2}.
		\label{eq:models-noncompact-ratios-tripartite}
	\end{align}
\end{subequations}
Define the finite and fixed-frame densities $\widetilde{\calF}_{n,\mathrm{nc}}^{(q)}$ and $\widetilde{\calG}_{n,\mathrm{nc}}^{(q)}$ by the same radius limit, retaining the usual coordinate prefactors.
The densities obey Eqs.~\eqref{eq:real-cross-ratio-relations-2} and~\eqref{eq:real-cross-ratio-relations-3} with $c=1$.
% In particular the two-interval determinant identity reduces the first finite function to
% \begin{equation}
% 	\widetilde{\calF}_{n,\mathrm{nc}}^{(2)}(x) = \prod_{k=1}^{n-1} \bigl[{}_2F_1(k/n,1-k/n;1;x) {}_2F_1(k/n,1-k/n;1;1-x)\bigr]^{-1/2}.
% 	\label{eq:models-cct-density-product}
% \end{equation}

\paragraph{The identity-channel limit after decompactification.}
At finite radius, $\calR\to1$ as $x\to0$.
The identity-channel limit $x\to0$ of the decompactified density follows from the period asymptotic $\Im\Omega_n^{(2)}=\log(1/x)C_{A_{n-1}}/(2\pi)+O(1)$, obtained from the Euler connection formula.
Using $\det C_{A_{n-1}}=n$ for $q=2$ and the character-period cusp in Eq.~\eqref{eq:assembly-hodge-cusp} for $q=3$ gives the identity-channel densities
\begin{subequations}
	\label{eq:models-density-cusp}
	\begin{align}
		\widetilde{\calR}_{n,\mathrm{nc}}^{(2)}(x)
		&\sim\frac{(2\pi)^{g^{(2)}/2}}{\sqrt n\,[\log(1/x)]^{g^{(2)}/2}},
		\label{eq:models-density-cusp-bipartite}\\
		\widetilde{\calR}_{n,\mathrm{nc}}^{(3)}(x)
		&\sim\frac{(2\pi)^{g_d/2}}{n^{n-1}[\log(1/x)]^{g_d/2}}.
		\label{eq:models-density-cusp-tripartite}
	\end{align}
\end{subequations}
Both densities tend to zero because the density limit replaces the discrete vacuum with a continuum of low-momentum states.
Related logarithmic factors and renormalization issues for noncompact-boson twist correlators are discussed in Ref.~\cite{Blondeau-Fournier:2016rtu}.

\paragraph{The noncompact torus density.}
At $n=2$ the determinants in Eq.~\eqref{eq:models-noncompact-ratios} reduce to the imaginary periods $\Im\tau$ and $2\Im\tau$.
Thus, at $n=2$,
\begin{subequations}
	\label{eq:models-density-n2}
	\begin{align}
		\widetilde{\calR}_{2,\mathrm{nc}}^{(2)}(x)
		&=\frac{\pi(1-x)^{-1/4}}{2\sqrt{K(x)K(1-x)}},
		\label{eq:models-density-n2-bipartite}\\
		\widetilde{\calR}_{2,\mathrm{nc}}^{(3)}(x)
		&=\frac1{\sqrt2}\widetilde{\calR}_{2,\mathrm{nc}}^{(2)}(x).
		\label{eq:models-density-n2-tripartite}
	\end{align}
\end{subequations}
The torus calculation in Ref.~\cite{Harper:2025uui} uses the volume-regulated noncompact partition function $Z_{\mathrm{nc}}^{(V)}(\tau)=V/[\sqrt{\Im\tau}|\eta(\tau)|^2]$.
Its volume-divided density is $z_{\mathrm{nc}}(\tau)\coloneqq Z_{\mathrm{nc}}^{(V)}(\tau)/V$.
Dividing by their target-space zero-mode volume $V$, evaluating the trace at $2\tau(x)$ and inserting it in the prefactor of Eq.~\eqref{eq:models-torus-ratio} reproduces the tripartite density in Eq.~\eqref{eq:models-density-n2-tripartite}.

\subsubsection{The harmonic-chain finite part}\label{sec:numerical-harmonic}
\begin{figure}[t]
	\centering
	\includegraphics[width=\textwidth]{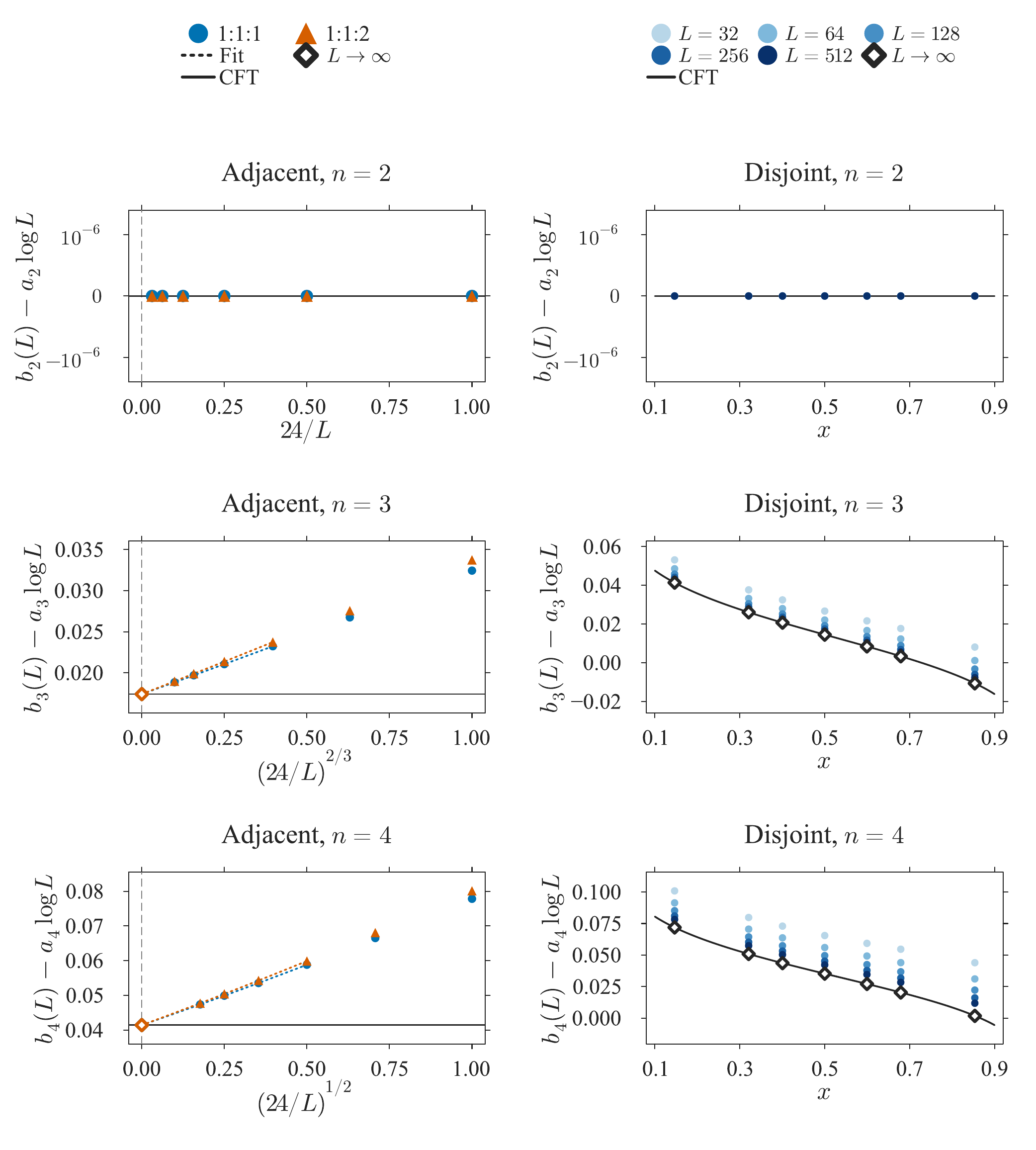}
	\caption{The finite parts $b_n(L)-a_n\log L$, with rows $n=2,3,4$ and adjacent/disjoint columns.
	The left column uses $(24/L)^{2/n}$, with markers distinguishing the two adjacent proportions.
	The right column uses the cross ratio $x$, with a light-to-dark sequence for $L=32,64,128,256,512$ at seven fixed geometries.
	Solid lines are the matched-Gaussian CFT values $B_n$ on the left and $B_n(x)$ on the right.
	At $n=3,4$, open diamonds show equal-weight quadratic extrapolations in $L^{-2/n}$ from the four largest sizes.}
	\label{fig:numerical-harmonic-finiteparts}
\end{figure}

We use the periodic oscillator Hamiltonian~\cite{Harper:2025uui},
\begin{equation}
	H=\frac12\sum_{j=1}^{L}
	[p_j^2+(q_{j+1}-q_j)^2+m^2q_j^2],\qquad
	[q_j,p_k]=i\delta_{jk},\qquad m>0.
	\label{eq:numerical-harmonic-chain}
\end{equation}
At positive mass, its normalized Gaussian ground state includes the zero coordinate.

The common zero mode fixes the logarithmic mass coefficient of the genuine subtraction.
The mass expansion and the finite part used in the size comparison are
\begin{align}
	\mathrm{GM}_{n,L}^{(3)}(m) &= a_n\log m+b_n(L)+O_L(m),
	\qquad a_n=\frac{n-2}{4n},\notag\\
	b_n(L)-a_n\log L
	&=\lim_{m\to0^+}\bigl[ \mathrm{GM}_{n,L}^{(3)}(m)-a_n\log(mL) \bigr].
	\label{eq:numerical-harmonic-finitepart}
\end{align}
The second line is the plotted ordinate.
Isolating the eigenvalue $m$ of the common uniform mode in the normalized massive overlap expresses the finite part in terms of nonsingular massless replica-Fourier blocks.
Sections~\ref{sec:lattice-harmonic-moments} and~\ref{sec:lattice-harmonic-mass-limit} derive the overlap and this reduction.

\paragraph{Matching the zero-mode profile.}
The action convention of Eq.~\eqref{eq:common-real-action} gives $q=\phi/\sqrt{2\pi}$ and $Q_0\coloneqq L^{-1/2}\sum_jq_j$.
A compact uniform zero coordinate has period $T=\sqrt{2\pi L}R$, whereas the oscillator ground-state density is $\sqrt{m/\pi}\exp(-mQ_0^2)$.
Matching their central densities fixes
\begin{subequations}
	\label{eq:numerical-harmonic-profile-match}
	\begin{align}
		T&=\sqrt{\pi/m},
		\label{eq:numerical-harmonic-profile-match-gaussian-period}\\
		\eta&=R^2=\frac1{2mL}.
		\label{eq:numerical-harmonic-profile-match-radius}
	\end{align}
\end{subequations}

The profile shape gives an additional shift in the genuine subtraction.
Writing the probability density as $p_T(Q_0)=T^{-1}f(Q_0/T)$, the connected $N$-copy coordinate integral is $T^{1-N}I_N$, where $I_N\coloneqq\int f(u)^N\dd{u}$.
The uniform profile has $f=1$ on an interval of unit length and zero elsewhere, hence $I_N=1$.
The Gaussian $f(u)=e^{-\pi u^2}$ gives $I_N=N^{-1/2}$.
The tripartite numerator has $n^2$ copies and each ordinary factor has $n$ copies.
Thus the profile contribution to the genuine subtraction is $[\log I_{n^2}-(3n/2)\log I_n]/[n(1-n)]$.
The resulting Gaussian-minus-uniform shift is
\begin{equation}
	\mathrm{GM}_{n,\rmG}^{(3)} - \mathrm{GM}_{n,\rmU}^{(3)} = -\frac{3n-4}{4n(n-1)}\log n.
	\label{eq:numerical-harmonic-profile-shift}
\end{equation}

\paragraph{Continuum finite parts.}
At fixed adjacent arc proportions, Eq.~\eqref{eq:models-decompactification-poisson} gives $\calH_{F_n}(\eta)\sim\eta^{g_F/2}$ as $\eta\to\infty$.
The plane-normalized coefficient in Eq.~\eqref{eq:fermat-direct-normalized-coefficient} then gives
\begin{equation}
	\log C_n^{(3)}(\eta) = \frac{g_F}{2}\log(4\pi\eta) - \frac{3n-4}{4}\log n + \frac12\log\calP_n+o(1).
	\label{eq:numerical-harmonic-adjacent-large-radius}
\end{equation}
Multiplying by $-1/[n(n-1)]$ and adding Eq.~\eqref{eq:numerical-harmonic-profile-shift} cancels the two $\log n$ terms.
With $\eta=1/(2mL)$ and $g_F/[2n(n-1)]=a_n$, the first term gives $a_n\log(mL)-a_n\log(2\pi)$.
Subtracting $a_n\log(mL)$ leaves the adjacent CFT finite part
\begin{equation}
	B_n=-a_n\log(2\pi)-\frac{\log\calP_n}{2n(n-1)}.
	\label{eq:numerical-harmonic-adjacent-reference}
\end{equation}
It is independent of the adjacent arc proportions.

For disjoint intervals, the density powers in Eq.~\eqref{eq:models-density-definition} give the coefficient of $\log\eta$ as $g^{(2)}/[4(n-1)]-g_d/[2n(n-1)]=-(n-2)/(4n)=-a_n$.
The substitution $\eta=1/(2mL)$ therefore contributes $a_n\log(mL)+a_n\log2$.
Subtracting the mass term and adding the profile shift~\eqref{eq:numerical-harmonic-profile-shift} gives
\begin{equation}
	B_n(x) = a_n\log2
	+\frac{\log\widetilde{\calR}_{n,\mathrm{nc}}^{(2)}(x)}{2(n-1)}
	-\frac{\log\widetilde{\calR}_{n,\mathrm{nc}}^{(3)}(x)}{n(n-1)}
	-\frac{3n-4}{4n(n-1)}\log n.
	\label{eq:numerical-harmonic-disjoint-reference}
\end{equation}
The hypergeometric form of $B_n(x)$ follows from Eq.~\eqref{eq:models-noncompact-ratios}.

\paragraph{Finite-size comparison.}
Figure~\ref{fig:numerical-harmonic-finiteparts} shows finite parts evaluated from the analytic fixed-size formula at $n=2,3,4$, for two adjacent proportions and seven disjoint geometries.

The adjacent series use $L=24,48,96,192,384,768$ and proportions $1:1:1$ and $1:1:2$.
The disjoint series use $L=32,64,128,256,512$ at the seven spatial fractions $(A,O_1,B,O_2)/L$ listed in Section~\ref{sec:numerical-free-disjoint}, spanning $0.146447\lesssim x\lesssim0.853553$.
At $n=3,4$, we fit the four largest sizes of each fixed-geometry series.

The exact pure-Gaussian factorization~\cite{Harper:2025uui,Camargo:2026xrd} gives $\mathrm{GM}_2^{(3)}=0$, so the displayed residuals at $n=2$ are roundoff.
At $n=2$, the maximum absolute residual is below $4.45\times10^{-16}$.

\section{Replica amplitudes beyond the compact boson}\label{sec:general-seed-amplitudes}
The compact-boson amplitudes are positive, but this property does not follow from unitarity alone for every replica geometry.
For disjoint intervals, two reflection cuts express the four-point function as a norm and an overlap, yielding a coarse-graining bound.
For adjacent intervals, the canonically sewn three-point coefficient is real but can be positive, zero or negative.

\subsection{Reflection bounds for the replica correlator}\label{sec:discussion-crossing}

Fix an integer $n\geq2$ and $0<x<1$, and write
\begin{equation}
	\Delta\coloneqq\Delta_n^{(3)}=n\Delta_n^{(2)}.
	\label{eq:discussion-twist-pair}
\end{equation}
Assume that the regular permutation cuts and caps realize positive Hilbert spaces of the unitary bosonic replica theory, with trivial discrete torsion. Adjunction reverses each cut, $\calT_g^\dagger=\calT_{g^{-1}}$.
This sewing uses the fixed replica labels of the bosonic permutation correlator.

For a scalar twist of full dimension $\Delta$, radial adjunction is
\begin{equation}
	[\calT_g(z,\bar z)]^\dagger = |z|^{-2\Delta}\calT_{g^{-1}}(1/\bar z,1/z),
	\label{eq:discussion-radial-adjoint}
\end{equation}
with the same local conformal factor as for a scalar primary~\cite{Simmons-Duffin:2016gjk}.

\paragraph{The positive mixed-sector norm.}
Put $s\coloneqq\sqrt{1-x}$ and define
\begin{equation}
	\ket{\chi(s)} \coloneqq \calT_{\sfg_1^{-1}}(-s)\calT_{\sfg_2}(0)\ket{0}.
	\label{eq:reflection-sewn-state}
\end{equation}
This state lies in the sector with monodromy $\sfg_1^{-1}\sfg_2$, whose inner product is also positive by the sewing assumption.
The coordinate $w=(z-1)/s$ sends the four insertions to $(-1/s,-s,0,\infty)$, and hence
\begin{equation}
	\begin{aligned}
		\calG_n^{(3)}(x,x) &= s^{-2\Delta} \langle \calT_{\sfg_1}(-1/s)\calT_{\sfg_1^{-1}}(-s)\calT_{\sfg_2}(0)\calT_{\sfg_2^{-1}}(\infty)\rangle\\
		&= \langle\chi(s)|\chi(s)\rangle\geq0.
	\end{aligned}
	\label{eq:discussion-mixed-norm}
\end{equation}
The reflection $z\mapsto1+(1-x)/(\bar z-1)$ exchanges $0$ with $x$ and $1$ with infinity, pairing each twist with its inverse.
The regular twists have zero conformal spin, so the conjugate chiral phases at the negative real insertion cancel and leave the factor $s^{-2\Delta}$.
The union of the lifted reflection cuts separates the sewn surface into two adjoint halves~\cite{Gadde:2025csh}.

\paragraph{The inverse-pair overlap.}
Put $r\coloneqq\sqrt{x}$ and form the inverse-pair states on the unit disk,
\begin{subequations}
	\label{eq:discussion-pair-states}
	\begin{align}
		\ket{\psi_1(r)} &\coloneqq \calT_{\sfg_1^{-1}}(r)\calT_{\sfg_1}(0)\ket{0},
		\label{eq:discussion-pair-states-a}\\
		\ket{\psi_2(r)} &\coloneqq \calT_{\sfg_2^{-1}}(r)\calT_{\sfg_2}(0)\ket{0}.
		\label{eq:discussion-pair-states-b}
	\end{align}
\end{subequations}
Both have trivial total monodromy and belong to the same untwisted $n^2$-copy Hilbert space.
Their overlap is
\begin{equation}
	\begin{aligned}
		\braket{\psi_2(r)}{\psi_1(r)} &= r^{-2\Delta} \langle \calT_{\sfg_1}(0)\calT_{\sfg_1^{-1}}(r)\calT_{\sfg_2}(1/r)\calT_{\sfg_2^{-1}}(\infty)\rangle\\
		&= \calG_n^{(3)}(x,x).
	\end{aligned}
	\label{eq:discussion-overlap}
\end{equation}
The rescaling $z=rw$ contributes $r^{-2\Delta}$, including the inverse factor at infinity, and equals the radial-adjunction factor.
In the original plane the reflection is $z\mapsto x/\bar z$.

Each diagonal norm uses only one generator.
Its permutation decomposes into $n$ disjoint length-$n$ cycles, so the cover splits into $n$ ordinary two-interval covers.
Cycle-factorizing unit two-point normalization gives
\begin{subequations}
	\label{eq:discussion-cauchy-schwarz}
	\begin{align}
		\|\psi_1(r)\|^2 &= \|\psi_2(r)\|^2 =[\calG_n^{(2)}(x,x)]^n,
		\label{eq:discussion-cauchy-schwarz-norms}\\
		|\calG_n^{(3)}(x,x)| &\leq [\calG_n^{(2)}(x,x)]^n,
		\label{eq:discussion-cauchy-schwarz-bound}
	\end{align}
\end{subequations}
where the last step uses the Cauchy--Schwarz inequality.

Multiplying by $x^{2\Delta}$ and using $\calR_n^{(q)}=x^{2\Delta_n^{(q)}}\calG_n^{(q)}$ gives
\begin{equation}
	0 \leq \calR_n^{(3)}(x) \leq [\calR_n^{(2)}(x)]^n.
	\label{eq:discussion-ratio-bound}
\end{equation}
Where $\calR_n^{(2)}(x)>0$ and $\calR_n^{(3)}(x)>0$, take their real logarithms.
The factor $n(1-n)<0$ reverses the inequality, giving
\begin{equation}
	\begin{aligned}
		\mathrm{GM}_{n,\mathrm{disj}}^{(3)}(x) &= \frac{\log\calR_n^{(3)}(x) - \frac n2\log\calR_n^{(2)}(x)}{n(1-n)}\\
		&\geq -\frac{\log\calR_n^{(2)}(x)}{2(n-1)} = -\frac12 I_n(A:B).
	\end{aligned}
	\label{eq:discussion-genuine-bound}
\end{equation}
Nonnegative genuine entropy requires the stronger inequality $\calR_n^{(3)}(x)\leq[\calR_n^{(2)}(x)]^{n/2}$.

At $n=2$ the bound also follows for an arbitrary finite-dimensional pure tripartite state from the reflected-purity identity~\eqref{eq:discussion-reflected-genuine}.
Section~\ref{sec:second-index-reflected-entropy} derives that state-level statement and its relation to coarse-graining.

\paragraph{Squared OPE coefficients and canonical phases.}
Expanding the state $\ket{\chi(s)}$ in an orthonormal basis of its twisted sector expresses its squared norm as a sum of squared moduli.
Under the isolated-leading-state and dominated-OPE assumptions of Eq.~\eqref{eq:crossed-residue}, a nonzero leading adjacent exchange consequently contributes
\begin{equation}
	\calG_n^{(3)}(x,x) \sim |C_n^{(3)}|^2(1-x)^{-\Delta_n^{(3)}}, \qquad x\longrightarrow1^-.
	\label{eq:discussion-ope-square}
\end{equation}
Reflection conjugates the coefficient of one three-point piece.
It is therefore $C_n^{(3)}C_n^{(3)*}$, not the sign of either factor, that positivity controls.

\paragraph{The two endpoint limits of the bound.}
The bound is saturated as $x\to0$, where both $I_n$ and $\mathrm{GM}_{n,\mathrm{disj}}^{(3)}$ vanish.
For $c>0$ and an isolated ordinary vacuum channel, $I_n$ diverges as $x\to1$ and the lower bound tends to $-\infty$.

\subsection{Signs of adjacent replica amplitudes}\label{sec:cft-amplitude-signs}
At $n=3$, Eq.~\eqref{eq:adjacent-n3-coefficient} relates the sign of the adjacent coefficient to the seed CFT partition function at the triangular modulus.

\paragraph{The triangular-torus trace.}
At $n=3$, the sign can be read directly from the full seed CFT spectrum.
Write $\Delta\coloneqq L_0+\bar L_0$ and $s\coloneqq L_0-\bar L_0$ for the dimension and spin operators.
The spin eigenvalues are integers in the bosonic theory.
Substituting $\tau_\triangle=1/2+i\sqrt3/2$ into Eqs.~\eqref{eq:models-torus-partition} and~\eqref{eq:adjacent-n3-coefficient} gives
\begin{equation}
	C_3^{(3)} = 3^{-3c/2}\Tr_{\calH_{\calC}} \left[e^{-\pi\sqrt3(\Delta-c/12)}(-1)^s\right].
	\label{eq:discussion-triangular-spin-parity}
\end{equation}
The trace includes descendants and subtracts the Boltzmann-weighted odd-spin contribution from the even-spin contribution~\cite{Gliozzi:2019ewk}.
On a spatial circle, the spin-parity insertion translates states by half the circumference.
Its expectation value can therefore vanish or be negative without a negative-norm state.

\paragraph{Positive and vanishing coefficients.}
The real compact boson has a positive winding sum~\eqref{eq:classical-winding-sum}, including its unit zero-winding term.
Its Gaussian measure and the reduced scalar determinants in Eq.~\eqref{eq:direct-fermat-spectral-coefficient} are positive.
Thus $C_n^{(3)}>0$ at every fixed integer $n\geq2$ and finite radius.

At $n=3$, a diagonal rational seed CFT has $Z(\tau,\bar\tau)=\sum_i|\chi_i(\tau)|^2$, where $\chi_i$ are its characters, and hence $C_3^{(3)}\geq0$.

The diagonal level-one $E_8$ theory has $c=8$ and
\begin{equation}
	Z_{E_8}(\tau,\bar\tau) = \left|\frac{E_4(\tau)}{\eta(\tau)^8}\right|^2,
	\label{eq:discussion-e8-character}
\end{equation}
where $E_4$ is the normalized weight-four Eisenstein series~\cite{LeonAndonayre:2024bot}.
The modular transformation $\tau\mapsto(\tau-1)/\tau$ fixes the triangular modulus. Since its weight-four multiplier is not one, modularity gives $E_4(\tau_\triangle)=\tau_\triangle^4E_4(\tau_\triangle)=0$.
The Dedekind function has no upper-half-plane zeros, so this unitary compact seed CFT has $C_3^{(3)}=0$.
A four-point degeneration must then identify the first nonzero twisted coupling.

For unitary diagonal Virasoro minimal models, the vacuum character instead guarantees strict positivity at $n=3$.
Set $q_\tau\coloneqq e^{2\pi i\tau}$ and write the reduced vacuum character as $1+\sum_{m\geq1}a_mq_\tau^m$. Its coefficients obey $0\leq a_m\leq p(m)$, where $p(m)$ counts integer partitions, because the vacuum module is a quotient of a Verma module. At $a=|q_\tau|=e^{-\pi\sqrt3}$,
\begin{equation}
	\left|\sum_{m\geq1}a_mq_\tau^m\right|
	\leq\prod_{m\geq1}(1-a^m)^{-1}-1
	\leq\exp\!\left[\frac{a}{(1-a)^2}\right]-1<0.005.
	\label{eq:discussion-minimal-positive}
\end{equation}
The vacuum character is therefore nonzero, so its absolute square already makes the diagonal trace and $C_3^{(3)}$ positive.

\paragraph{A negative coefficient in a Narain CFT.}
For a rank-24 even unimodular lattice $\Lambda$, let $N_\Lambda$ count its weight-one currents, consisting of 24 oscillator currents and one for each norm-two vector.
Its character is a modular function with a single simple pole at the cusp, so comparison of the first two Fourier coefficients gives
\begin{equation}
	\chi_\Lambda(\tau) = j(\tau) + N_\Lambda - 744.
	\label{eq:discussion-root-count-character}
\end{equation}
This is the character formula of Ref.~\cite{Gaberdiel:2009rd}, with its explicit factor of $q_\tau$ removed to restore the vacuum Casimir energy.~\footnote{The convention $J=j-744$ in Ref.~\cite{Kachru:2016nty} instead writes the Leech character as $J+24$.}
Here $j(\tau)=E_4(\tau)^3/\eta(\tau)^{24}$ vanishes at the triangular modulus.
At that modulus a left and right pair gives
\begin{equation}
	Z_{\calC}(\tau_\triangle) = (N_{\Lambda_L}-744)(N_{\Lambda_R}-744).
	\label{eq:discussion-lattice-seed-triangular-trace}
\end{equation}
The sign is negative when the two current counts lie on opposite sides of 744.

Take the Leech lattice on the left, with $N_{\Lambda_L}=24$, and $D_{24}^{+}$ on the right, with $N_{\Lambda_R}=24+2\times24\times23=1128$.
Their orthogonal sum with the right quadratic form reversed is even unimodular of signature $(24,24)$ and defines a unitary compact bosonic torus CFT with $c_L=c_R=24$~\cite{Yin:2017yyn}.
Substituting these current counts in Eq.~\eqref{eq:discussion-root-count-character} gives
\begin{subequations}
	\label{eq:discussion-lattice-characters}
	\begin{align}
		\chi_{\mathrm{Leech}}(\tau) &= j(\tau)-720,
		\label{eq:discussion-lattice-characters-leech}\\
		\chi_{D_{24}^{+}}(\tau) &= j(\tau)+384.
		\label{eq:discussion-lattice-characters-d24}
	\end{align}
\end{subequations}
The first is the Leech character of Ref.~\cite{Yin:2017yyn}. The second also follows from the lattice theta series of Ref.~\cite{Xu:2000}.
Their product with right complex conjugation is
\begin{equation}
	Z_{\calC}(\tau,\bar\tau) = [j(\tau)-720][\overline{j(\tau)}+384].
	\label{eq:discussion-asymmetric-partition}
\end{equation}
The two characters begin with $q_\tau^{-1}+24+\cdots$ and $q_\tau^{-1}+1128+\cdots$, respectively.
Their state multiplicities are nonnegative.

Equations~\eqref{eq:discussion-asymmetric-partition} and~\eqref{eq:adjacent-n3-coefficient}, evaluated at $\tau_\triangle$, give
\begin{equation}
	Z_{\calC}(\tau_\triangle) = -720 \times 384=-276480, \qquad C_3^{(3)}=-276480 \times 3^{-36}<0.
	\label{eq:discussion-negative-seed}
\end{equation}

\section{Discussion}\label{sec:discussion}
We have computed the vacuum tripartite R\'enyi multi-entropy of the compact boson for adjacent and disjoint intervals at every integer $n\geq2$.
In the fusion limit, the residue $|C_n^{(3)}|^2$ of the twist four-point function $\calG_n^{(3)}$ explains the factor of two between the limit and strict adjacency~\cite{Harper:2025uui}.

\paragraph{Lattice tests.}
The XXZ extrapolations support the compact-boson amplitudes at radii fixed by the Hamiltonian.
They distinguish the radius-independent adjacent constant at $n=2$ from the radius-dependent adjacent constant at $n=3$ and disjoint ratios.
The higher-index XX and Ising moments test the compact-boson and Ising CFT amplitudes, respectively.
For the harmonic chain, the lattice finite parts approach the noncompact CFT values after matching the zero-mode scale and including the Gaussian profile correction.

A general treatment of the determinant and Pfaffian methods in Section~\ref{sec:lattice-numerics}, including the relation between parity sectors and continuum spin structures, will be presented separately~\cite{Tasuki:2026fss}.

\paragraph{Seed CFT data.}
The replica index determines which seed CFT data can enter the amplitudes.
For adjacent intervals at $n=2$, the sphere cover leaves only the central-charge dependence.
At $n=3$, the triangular torus probes the spectrum and its degeneracies.
At $n\geq4$, the higher-genus partition function also contains contractions of OPE coefficients~\cite{Cardy:2017qhl}.
For disjoint intervals, spectral dependence already appears at $n=2$, where the cover is a torus.

The meromorphic lattice theories based on $E_8\times E_8$ and $\Spin(32)/\bbZ_2$ have identical vacuum amplitudes on Riemann surfaces through genus four and first differ at genus five~\cite{Gaberdiel:2009rd}.
The adjacent and disjoint replica genera, $(n-1)(n-2)/2$ and $(n-1)^2$, therefore exclude a distinction for $n\leq4$ and $n\leq3$, respectively.
The first cases to test are the adjacent $n=5$ Fermat curve and the disjoint $n=4$ replica family.

For a contiguous tripartition of a strip with conformal boundary
conditions, the $n=2$ replica cover is an annulus, so the genuine multi-entropy probes the boundary excitation spectrum.
This boundary-CFT extension, including discussions on higher R\'enyi cases and dynamics under quenches, will be reported separately~\cite{Tasuki:2026bcf}.

For two-interval bipartite R\'enyi entropies, the fusion limit of a crossing asymmetry measures the failure of Haag duality through the Jones index~\cite{Benedetti:2024dku,Benedetti:2026nlt}.
For tripartite multi-entropy, crossing relates correlators with different replica permutation assignments, so an asymmetry at fixed permutation labels can occur even when Haag duality holds.
Establishing an analogous relation to the Jones index for tripartite multi-entropy is therefore nontrivial.

\paragraph{Positivity and signs of replica amplitudes.}
For the disjoint setups, the vacuum multi-entropy is real-valued, and reflection positivity gives the coarse-graining bound.
The adjacent coefficient enters the fusion channel through $|C_n^{(3)}|^2$, so the reflection positivity does not fix the raw sign of $C_n^{(3)}$.

At $n=3$, the adjacent coefficient measures the imbalance between Boltzmann-weighted even-spin and odd-spin states, as expressed by Eq.~\eqref{eq:discussion-triangular-spin-parity}.
The Leech--$D_{24}^{+}$ example has a negative coefficient, while the diagonal $E_{8,1}$ example has an exact cancellation.
A negative replica moment makes the logarithmic multi-entropy complex, so the unmodified definition of the logarithmic quantity does not seem well-suited for discussing inequalities or monotonicity in the context of resource theory.

\paragraph{Continuation in the R\'enyi index.}
The compact-boson formulas involve theta functions whose dimensions depend on the integer replica genera, so they do not directly define the von Neumann limit $n\to1$.
The negative finite-state replica moment rules out a universal representation in terms of traces of powers of positive operators.
For the compact boson vacuum, a possible route is to construct a spectral or generating-function representation and establish the analyticity and growth conditions needed to determine the continuation~\cite{DHoker:2020bcv,Shyani:2020oja}.

% --------------------------------------------------
\acknowledgments
The author is grateful to Kosei Fujiki, Shunsuke Furukawa, Masahiro Hoshino, Norihiro Iizuka, Kohei Kawabata, Chen-Te Ma, Haruki Shimizu, Tadashi Takayanagi and Kotaro Tamaoka for valuable discussions.
This work benefited from the use of ChatGPT and Codex (GPT-6 Astra, OpenAI) to search the literature, cross-check calculations, develop and verify code, and improve the clarity of the manuscript.
The author takes full responsibility for the content of this paper.
This work is supported by MEXT KAKENHI Grant-in-Aid for Transformative Research Areas (A) through the ``Extreme Universe'' collaboration: Grant Number~21H05182.
KT is supported by Grant-in-Aid for JSPS Fellows No.~25KJ1455.

% --------------------------------------------------
\appendix

\section{Geometry of Riemann surfaces}\label{sec:math-appendix}
This appendix fixes the conventions used in Sections~\ref{sec:replica-geometry}--\ref{sec:low-index-amplitudes} and derives the homology and period identities entering the amplitudes.

\subsection{Riemann surfaces and their periods}\label{sec:math-riemann}

For standard background on compact Riemann surfaces we follow Ref.~\cite{Eynard:2018tcr}.
We use the complex orientation and fix the integral intersection pairing and Hodge normalization below.
Let $\Sigma$ be a connected compact Riemann surface of genus $g_\Sigma$.
Its first integral homology is free abelian of rank $2g_\Sigma$,
\begin{subequations}
    \label{eq:math-dimensions}
    \begin{align}
        H_1(\Sigma,\bbZ)&\simeq\bbZ^{2g_\Sigma},
        \label{eq:math-dimensions-homology}\\
        \dim_{\bbC}H^{1,0}(\Sigma)&=g_\Sigma.
        \label{eq:math-dimensions-holomorphic}
    \end{align}
\end{subequations}
Here $H^{1,0}(\Sigma)$ is the space of holomorphic one-forms; its dimension fixes the count of regular character differentials.
For a degree-$d$ branched cover $f:\Sigma\to\Sigma_0$, the genus is related to the base genus $g_0$ and the ramification indices $e_P$ by the Riemann--Hurwitz formula
\begin{equation}
    2g_\Sigma-2=d(2g_0-2)+\sum_{P\in\Sigma}(e_P-1).
    \label{eq:math-riemann-hurwitz}
\end{equation}

The unimodular intersection pairing identifies when independent cycles generate the full integral homology.
Choose a reference integral symplectic basis $\{A_i,B_i\}_{i=1}^{g_\Sigma}$ with intersection form
\begin{subequations}
    \label{eq:math-canonical-intersection}
    \begin{align}
        A_i\circ A_j&=B_i\circ B_j=0,
        \label{eq:math-canonical-intersection-isotropic}\\
        A_i\circ B_j&=\delta_{ij}.
        \label{eq:math-canonical-intersection-mixed}
    \end{align}
\end{subequations}
The intersection pairing on $H_1(\Sigma,\bbZ)$ is alternating and unimodular.
For $2g_\Sigma$ candidate independent integral cycles, let $T$ contain their coordinate columns in this reference basis and let $J$ be their intersection matrix.
Then $\det J=(\det T)^2$, so $\sqrt{\det J}$ is the index of their integer span in integral homology.
In particular, $\det J=1$ is equivalent to their generating the full integral homology.

\paragraph{Integral symplectic reduction.}
For a primitive vector $a$ in a unimodular alternating lattice, B\'ezout's identity supplies an integer partner $b$ with $a\circ b=1$.
The projection
\begin{equation}
    c^\perp=c-(c\circ b)a+(c\circ a)b, \qquad a\circ c^\perp=b\circ c^\perp=0
    \label{eq:math-integral-symplectic-projection}
\end{equation}
splits off this canonical pair.
Its orthogonal complement is again unimodular, so iteration gives
an integral symplectic basis.

The cover-specific ribbon orders and local intersection tests are given in Appendix~\ref{sec:replica-intersections}.

\paragraph{Period normalization.}
The A-periods fix the normalized holomorphic one-forms $\{\nu_i\}$, and their B-periods define the period matrix.
\begin{subequations}
    \label{eq:classical-normalized-periods}
    \begin{align}
        \oint_{A_j}\nu_i&=\delta_{ij},
        \label{eq:classical-normalized-periods-a}\\
        \oint_{B_j}\nu_i&=\Omega_{ij},
        \label{eq:classical-normalized-periods-b}
    \end{align}
\end{subequations}
The Riemann bilinear relations imply
\begin{subequations}
    \label{eq:math-period-properties}
    \begin{align}
        \Omega&=\Omega^{\top},
        \label{eq:math-period-properties-symmetry}\\
        \Im\Omega>0.
        \label{eq:math-period-properties-positivity}
    \end{align}
\end{subequations}
Thus $\Omega$ belongs to the Siegel upper half-space of degree $g_\Sigma$.

For smooth closed one-forms $\lambda$ and $\mu$, the first Riemann bilinear relation in this marking is
\begin{equation}
    \int_\Sigma\lambda\wedge\mu = \sum_{k=1}^{g_\Sigma} \left[
        \left(\oint_{A_k}\lambda\right)\left(\oint_{B_k}\mu\right)
        - \left(\oint_{B_k}\lambda\right)\left(\oint_{A_k}\mu\right)
    \right].
    \label{eq:math-bilinear}
\end{equation}
Applying this identity to $\lambda=\nu_i$ and $\mu=\overline{\nu_j}$ gives
\begin{equation}
    \int_\Sigma\nu_i\wedge\overline{\nu_j}
    =\overline{\Omega}_{ji}-\Omega_{ij}
    =-2i\,\left(\Im\Omega\right)_{ij}.
    \label{eq:classical-hodge-pairing}
\end{equation}
With the complex orientation and local coordinate $z$, use
\begin{subequations}
    \label{eq:math-hodge-signs}
    \begin{align}
        \dd[2]{z}&\coloneqq\frac{i}{2}\dd{z}\wedge\dd{\bar z},
        \label{eq:math-hodge-signs-area}\\
        *\dd{z}&=-i\dd{z},
        \label{eq:math-hodge-signs-holomorphic}\\
        *\dd{\bar z}&=+i\dd{\bar z}.
        \label{eq:math-hodge-signs-antiholomorphic}
    \end{align}
\end{subequations}
Consequently $*\nu_i=-i\nu_i$ and $*\overline{\nu_i}=+i\overline{\nu_i}$.
These conventions give $\int_\Sigma\lambda\wedge *\lambda>0$ for every nonzero real harmonic one-form $\lambda$.

\subsection{Special functions for period integrals}\label{sec:math-special}

The period calculations use Euler integrals on fixed real branches, and the endpoint limits use their connection formulas.
For $a,b\in\bbC$ with positive real parts, the beta function is defined by the convergent Euler integral
\begin{equation}
	B(a,b) \coloneqq\int_0^1 t^{a-1}(1-t)^{b-1}\dd{t} = \frac{\Gamma(a)\Gamma(b)}{\Gamma(a+b)}.
	\label{eq:math-beta}
\end{equation}
The gamma function obeys the reflection identity~\cite[Eq.~5.5.3]{DLMF}
\begin{equation}
	\Gamma(a)\Gamma(1-a)=\frac{\pi}{\sin(\pi a)}, \qquad a\notin\bbZ.
	\label{eq:math-gamma-reflection}
\end{equation}

For $\Re c>\Re b>0$ and $x\in\bbC\setminus[1,\infty)$, the Gauss hypergeometric function has
the Euler representation
\begin{equation}
	{}_2F_1(a,b;c;x) = \frac{\Gamma(c)}{\Gamma(b)\Gamma(c-b)} \int_0^1 t^{b-1}(1-t)^{c-b-1}(1-xt)^{-a}\dd{t},
	\label{eq:math-euler-hypergeometric}
\end{equation}
where the powers use the branch continuous from $x=0$~\cite[Eq.~15.6.1]{DLMF}.

For $0<y<1$ and $c-a-b\notin\bbZ$, the connection formula is
\begin{align}
	{}_2F_1(a,b;c;1-y)
	={}&\frac{\Gamma(c)\Gamma(c-a-b)}{\Gamma(c-a)\Gamma(c-b)} {}_2F_1(a,b;a+b-c+1;y)\notag\\
	&+y^{c-a-b}\frac{\Gamma(c)\Gamma(a+b-c)}{\Gamma(a)\Gamma(b)} {}_2F_1(c-a,c-b;c-a-b+1;y).
	\label{eq:math-hypergeometric-connection}
\end{align}
Parameters are taken away from poles, with other regular cases obtained by continuation~\cite[Eq.~15.10.21]{DLMF}.
In the zero-balanced case $c=a+b$, for fixed $a,b>0$, the limiting connection formula gives
\begin{equation}
	{}_2F_1(a,b;a+b;1-y) = \frac{\Gamma(a+b)}{\Gamma(a)\Gamma(b)} \bigl[\log(1/y)+O(1)\bigr],\qquad y\to0^+.
	\label{eq:math-zero-balanced}
\end{equation}
It follows from the $m=0$ logarithmic series in Ref.~\cite[Eq.~15.8.10]{DLMF}, multiplying its normalized hypergeometric function by $\Gamma(a+b)$.

For the complete elliptic integral, the argument $m$ is the parameter, equal to the square of the modulus.
We use
\begin{equation}
	K(m) \coloneqq \int_0^{\pi/2}\frac{\dd{\theta}}{\sqrt{1-m\sin^2\theta}}
	=\frac{\pi}{2}\,{}_2F_1\!\left(\frac12,\frac12;1;m\right).
	\label{eq:period-elliptic-K}
\end{equation}
For $0<m<1$, $K(m)$ is real and positive.

\subsection{Poisson summation and theta identities}\label{sec:math-gaussian-theta}

Poisson summation converts winding integers into momentum integers in the compact-boson Gaussian.
For a Schwartz function $f$ on $\bbR^g$, use
\begin{subequations}
	\label{eq:classical-poisson-convention}
	\begin{align}
		\widehat f(\bmm)
		&\coloneqq \int_{\bbR^g}\dd[g]{y} f(\bmy)e^{2\pi i\bmm^{\top}\bmy},
		\label{eq:classical-poisson-convention-transform}\\
		\sum_{\bmj\in\bbZ^g}f(\bmj)
		&= \sum_{\bmm\in\bbZ^g}\widehat f(\bmm).
		\label{eq:classical-poisson-convention-summation}
	\end{align}
\end{subequations}
Let $Q=Q^{\top}>0$ be a real $g\times g$ matrix and let $\bma\in\bbR^g$.
For $f(\bmy)=\exp[-\pi(\bmy-\bma)^{\top}Q(\bmy-\bma)]$, completing the square gives, for $\bmm\in\bbR^g$,
\begin{equation}
	\widehat f(\bmm) = \frac{1}{\sqrt{\det Q}} \exp\!\left[-\pi\bmm^{\top}Q^{-1}\bmm + 2\pi i\bmm^{\top}\bma\right].
	\label{eq:math-gaussian-transform}
\end{equation}

Substitution into the Poisson summation formula gives
\begin{equation}
	\sum_{\bmj\in\bbZ^g} e^{-\pi(\bmj-\bma)^{\top}Q(\bmj-\bma)} = \frac{1}{\sqrt{\det Q}} \sum_{\bmm\in\bbZ^g} e^{-\pi\bmm^{\top}Q^{-1}\bmm+2\pi i\bmm^{\top}\bma}.
	\label{eq:math-shifted-poisson}
\end{equation}

For $\bmz\in\bbC^d$ and a symmetric complex $d\times d$ matrix $\Pi$ with positive-definite imaginary part, the Riemann theta function is~\cite{Calabrese:2009ez}
\begin{equation}
	\Theta(\bmz\mid\Pi) \coloneqq \sum_{\bmN\in\bbZ^d} \exp\!\left(\pi i\bmN^{\top}\Pi\bmN + 2\pi i\bmN^{\top}\bmz\right).
	\label{eq:classical-theta-definition}
\end{equation}
The positivity condition makes the series absolutely convergent.

For a genus-$g$ marking and real characteristic vectors, define
\begin{equation}
	\theta\!\begin{bmatrix}\bm\epsilon\\\bm\delta\end{bmatrix} (\bmz\mid\Omega)
	\coloneqq
	\sum_{\bmm\in\bbZ^g}
	\exp\!\Bigl\{
		\pi i(\bmm+\bm\epsilon)^{\top}\Omega(\bmm+\bm\epsilon) + 2\pi i(\bmm+\bm\epsilon)^{\top}(\bmz+\bm\delta)
	\Bigr\}.
	\label{eq:math-thomae-characteristic}
\end{equation}
The zero characteristic gives $\Theta$ in Eq.~\eqref{eq:classical-theta-definition}.
For half-integer characteristics,
\begin{equation}
	\theta\!\begin{bmatrix}\bm\epsilon\\\bm\delta\end{bmatrix} (-\bmz\mid\Omega)
	= (-1)^{4\bm\epsilon\cdot\bm\delta} \theta\!\begin{bmatrix}\bm\epsilon\\\bm\delta\end{bmatrix} (\bmz\mid\Omega).
	\label{eq:math-theta-characteristic-parity}
\end{equation}
An odd theta constant vanishes at the origin, whereas even parity alone does not ensure nonvanishing.

\paragraph{Genus-one identities.}
In genus one, we use the Jacobi notation
\begin{subequations}
	\label{eq:math-jacobi-characteristic-dictionary}
	\begin{align}
		\vartheta_3(z\mid\tau)&\coloneqq \theta[\genfrac{}{}{0pt}{}{0}{0}](z\mid\tau),
		\label{eq:math-jacobi-characteristic-dictionary-theta-three}\\
		\vartheta_2(z\mid\tau)&\coloneqq \theta[\genfrac{}{}{0pt}{}{1/2}{0}](z\mid\tau),
		\label{eq:math-jacobi-characteristic-dictionary-theta-two}\\
		\vartheta_4(z\mid\tau)&\coloneqq \theta[\genfrac{}{}{0pt}{}{0}{1/2}](z\mid\tau).
		\label{eq:math-jacobi-characteristic-dictionary-theta-four}
	\end{align}
\end{subequations}
The genus-one series is
\begin{equation}
	\vartheta_3(z\mid\tau) = \sum_{j\in\bbZ} \exp\!\left(\pi i\tau j^2+2\pi i jz\right), \qquad z\in\bbC, \quad \Im\tau>0,
	\label{eq:math-jacobi-theta}
\end{equation}
so that $\Theta(0\mid\diag(\tau_1,\tau_2)) = \vartheta_3(0\mid\tau_1)\vartheta_3(0\mid\tau_2)$.

For $\tau=iK(1-x)/K(x)$ on $0<x<1$, with $x=k^2$ the elliptic parameter, the elliptic identities are
\begin{subequations}
	\label{eq:math-elliptic-duplication}
	\begin{align}
		\vartheta_3(0\mid\tau)^2&=\frac{2K(x)}\pi,
		\label{eq:math-elliptic-duplication-period}\\
		\frac{\vartheta_2(0\mid\tau)^4}{\vartheta_3(0\mid\tau)^4}&=x,
		\label{eq:math-elliptic-duplication-lambda}\\
		\frac{\vartheta_4(0\mid\tau)^4}{\vartheta_3(0\mid\tau)^4}&=1-x.
		\label{eq:math-elliptic-duplication-complement}
	\end{align}
\end{subequations}
The sum, difference and product identities at doubled modulus are
\begin{subequations}
	\label{eq:math-theta-doubling}
	\begin{align}
		\vartheta_3(0\mid2\tau)^2
		&=\tfrac12[\vartheta_3(0\mid\tau)^2+\vartheta_4(0\mid\tau)^2],
		\label{eq:math-elliptic-duplication-doubled-three}\\
		\vartheta_2(0\mid2\tau)^2
		&=\tfrac12[\vartheta_3(0\mid\tau)^2-\vartheta_4(0\mid\tau)^2],
		\label{eq:math-elliptic-duplication-doubled-two}\\
		\vartheta_4(0\mid2\tau)^2
		&=\vartheta_3(0\mid\tau)\vartheta_4(0\mid\tau).
		\label{eq:math-elliptic-duplication-doubled-four}
	\end{align}
\end{subequations}

\paragraph{Dedekind eta at the triangular torus.}
For $\Im\tau>0$, the Dedekind function and its theta-constant identity are
\begin{subequations}
	\label{eq:math-dedekind-identity}
	\begin{align}
		\eta(\tau)
		&\coloneqq e^{\pi i\tau/12}\prod_{m=1}^{\infty}(1-e^{2\pi i m\tau}),
		\label{eq:math-dedekind-identity-product}\\
		2\eta(\tau)^3
		&=\vartheta_2(0\mid\tau)\vartheta_3(0\mid\tau) \vartheta_4(0\mid\tau).
		\label{eq:math-dedekind-identity-theta}
	\end{align}
\end{subequations}

At $\tau_\triangle=e^{i\pi/3}$ the discriminant $-3$ case of the Chowla--Selberg formula gives~\cite[Eq.~(96)]{Zagier:2008modular}
\begin{equation}
	\left[4\pi\sqrt3\,\Im\tau_\triangle |\eta(\tau_\triangle)|^4\right]^{1/3} = \frac{\Gamma(1/3)}{\Gamma(2/3)}.
	\label{eq:math-triangular-chowla-selberg}
\end{equation}
Using $\Im\tau_\triangle=\sqrt3/2$ and Eq.~\eqref{eq:math-gamma-reflection} then yields
\begin{equation}
	|\eta(\tau_\triangle)|^2 = \frac1{\sqrt{6\pi}} \left[\frac{\Gamma(1/3)}{\Gamma(2/3)}\right]^{3/2} = \frac{3^{1/4}\Gamma(1/3)^3}{4\pi^2}.
	\label{eq:math-triangular-eta-value}
\end{equation}

\subsection{Thomae formulas for theta constants}\label{sec:math-thomae}
The abelian-cover Thomae theorem separates a marked theta constant into an algebraic branch factor and an A-period determinant.
We use the theta conventions of Appendix~\ref{sec:math-gaussian-theta} to evaluate the replica branch exponents and prove the reference spin data stated in Section~\ref{sec:theta-ratio}.

A half-characteristic represents a spin line $L=\calO(D)$ with $2D\sim K_\Sigma$ and $\deg D=g-1$.
Non-speciality here means $h^0(\Sigma,L)=0$, equivalently a nonzero theta constant by Riemann's vanishing theorem.

For a meromorphic differential $\varpi$ with even-order divisor, let $\ind_\varpi(C)$ be the winding of $\varpi(\dot C)$ along a simple closed curve that avoids its zeros and poles.
Johnson's quadratic refinement is~\cite{Johnson:1980spin}
\begin{subequations}
	\label{eq:math-thomae-quadratic}
	\begin{align}
		q_\varpi(C)&\coloneqq \ind_{\varpi}(C)+1
		\pmod 2,
		\label{eq:math-thomae-quadratic-definition}\\
		q_\varpi(C+C')&=q_\varpi(C)+q_\varpi(C')+C\circ C'
		\pmod2.
		\label{eq:math-thomae-quadratic-polarization}
	\end{align}
\end{subequations}
The marking fixes the characteristic convention by
\begin{subequations}
	\label{eq:math-thomae-characteristic-quadratic}
	\begin{align}
		q_\varpi(A_i)&=2\epsilon_i,
		\label{eq:math-thomae-characteristic-quadratic-a}\\
		q_\varpi(B_i)&=2\delta_i,
		\label{eq:math-thomae-characteristic-quadratic-b}\\
		q_\varpi(\bma\cdot\bmA+\bmb\cdot\bmB)
		&=\bma\cdot\bmb+2\bma\cdot\bm\epsilon + 2\bmb\cdot\bm\delta \pmod2,
		\label{eq:math-thomae-characteristic-quadratic-general}
	\end{align}
\end{subequations}
with $\bma,\bmb\in\bbZ_2^g$.

\paragraph{The abelian-cover formula.}
Let $f:X\to\bbP^1$ be a connected abelian Galois cover with group of order $N$ and exponent $m$, and let $g_X$ be its genus.
Choose a canonical marking and the theorem's monic algebraic character differentials $\omega_\ell$, and put
\begin{subequations}
	\label{eq:math-thomae-period-blocks}
	\begin{align}
		\sfA_{\ell i}&\coloneqq\oint_{A_i}\omega_\ell,
		\label{eq:math-thomae-period-blocks-a}\\
		\sfB_{\ell i}&\coloneqq\oint_{B_i}\omega_\ell,
		\label{eq:math-thomae-period-blocks-b}\\
		\Omega&\coloneqq\sfA^{-1}\sfB.
		\label{eq:math-thomae-period-blocks-normalized}
	\end{align}
\end{subequations}
For an invariant non-special divisor $D$ of degree $g_X-1$ and its Abel--Jacobi characteristic $e_D$, the theorem states
\begin{equation}
	\theta[e_D](0\mid\Omega)^{4m} = \alpha_{e_D}\,(\det\sfA)^{2m} \prod_{(\sigma,j)<(\rho,i)} \left(\lambda_{\sigma,j}-\lambda_{\rho,i}\right)^{\rho_{\sigma,j;\rho,i}(e_D)}.
	\label{eq:math-thomae-abelian-structure}
\end{equation}
Here $\lambda_{\sigma,j}$ are branch values of monodromy $\sigma$, the integer exponents are expressed in terms of generalized Dedekind sums determined by the cover and divisor, and $\alpha_{e_D}$ is constant on a connected marked moduli chart~\cite[Theorem~6.6]{KopeliovichZemel:2019thomae}.
The characteristic can be torsion without being half-integral.
The half-canonical case used here has the spin interpretation.
A branch value at infinity is treated in its reciprocal coordinate, absorbing its factors into the local-coordinate normalization~\cite[Proposition~7.5]{KopeliovichZemel:2019thomae}.

With fixed marking, branch order and a chosen local $2m$th root,
\begin{equation}
	\theta[e_D](0\mid\Omega(\lambda))^2 = c_{e_D,\frm}\,\det\sfA(\lambda) \prod_{a<b}(\lambda_a-\lambda_b)^{\,\varrho_{ab}(e_D)}.
	\label{eq:math-thomae-local}
\end{equation}
The constant records the chosen phases.
For $(\bbZ_n)^2$, reordering branch values can change the sign of the integer-power constant when $n\equiv2\pmod4$, precisely when the 2-Sylow subgroup is a Klein four group~\cite[Remark~6.7]{KopeliovichZemel:2019thomae}.

Under $\widetilde{\bm\omega}=M(\lambda)\bm\omega$,
\begin{equation}
	\widetilde{\sfA}=M(\lambda)\sfA, \qquad \det\widetilde{\sfA} = \det M(\lambda)\det\sfA.
	\label{eq:math-thomae-basis-change}
\end{equation}
A moduli-dependent $\det M$ therefore changes the branch factors in this representation and must be retained even though the normalized period matrix is unchanged.

\paragraph{The replica branch exponents.}
To apply the formula with initially unramified infinity, place the fourth branch value at $\Lambda>1$ and take
\begin{equation}
	U^n=\frac{z}{z-x},\qquad V^n=\frac{z-1}{z-\Lambda}.
	\label{eq:math-thomae-finite-cover}
\end{equation}
The four ordered monodromies are $(\sfg_1,\sfg_1^{-1},\sfg_2,\sfg_2^{-1})$ at $(0,x,1,\Lambda)$.
Here the group has order $N=n^2$, exponent $m=n$ and genus $g=(n-1)^2$.
Let $P_a$ be the reduced degree-$n$ branch fiber and let $F_\infty$ be the full degree-$N$ unramified fiber. The normalized divisor is
\begin{equation}
	D_\Lambda\coloneqq(n-1)(P_x+P_\Lambda)-F_\infty, \qquad \deg D_\Lambda=g-1.
	\label{eq:math-thomae-finite-divisor}
\end{equation}
Its four normalized coefficients are $(0,n-1,0,n-1)$, and the subtracted full fiber has coefficient one.
Write $[u]_n$ for the least nonnegative residue of $u$ modulo $n$.
For $0\leq r,s<n$, a character has residues $(r,[-r]_n,s,[-s]_n)$ and the fractional-order sum is $t_\chi=\mathbf1_{r\ne0}+\mathbf1_{s\ne0}$.
The non-speciality criterion counts a contribution only at $x$ when $r\ne0$ and at $\Lambda$ when $s\ne0$, so its count equals $t_\chi$ for every character, including the axis and trivial characters~\cite[Theorem~4.5]{KopeliovichZemel:2019divisors}.
Moreover $\div((UV)^{-(n-1)}\dd{z})=2D_\Lambda$.
After rescaling $V$ by $(-\Lambda)^{1/n}$ and taking $\Lambda\to\infty$, this gives the physical spin line $L_*$.

For branch monodromies $\sigma,\rho$ of orders $o(\sigma),o(\rho)$, define
\begin{equation}
	d\coloneqq|\langle\sigma\rangle\cap\langle\rho\rangle|, \qquad \rho^{o(\rho)/d}=(\sigma^{o(\sigma)/d})^h.
	\label{eq:math-thomae-inertia-data}
\end{equation}
Write $\beta_\sigma,\beta_\rho$ for the normalized divisor coefficients at the chosen branch points and set $\xi=\beta_\rho-h\beta_\sigma$ modulo $d$.
Dividing the integer exponent in Eq.~\eqref{eq:math-thomae-abelian-structure} by $2m$ gives the following exponent in $\theta^2/\det\sfA$~\cite{KopeliovichZemel:2019thomae}.
\begin{equation}
	\varrho_{\sigma,\rho} = \frac{N}{o(\sigma)o(\rho)} \left[2\phi_{h,d}(\xi)+\phi_{h,d}(0) + \frac{(o(\sigma)-1)(o(\rho)-1)}4\right],
	\label{eq:math-thomae-pair-exponent}
\end{equation}
where the generalized Dedekind sum is
\begin{equation}
	\phi_{h,d}(\xi) \coloneqq \sum_{j=1}^{d-1} \frac{e^{2\pi i j\xi/d}}{(1-e^{2\pi i jh/d})(1-e^{-2\pi i j/d})}.
	\label{eq:math-thomae-dedekind-sum}
\end{equation}

For independent directions $d=1$, the sum is empty and $\varrho=g/4$.
For the inverse pairs $(0,x)$ and $(1,\Lambda)$, $d=n$, $h=-1$ and $\xi=n-1$.
For $1\leq j<n$, differentiating the finite geometric series gives
\begin{equation}
 \sum_{u=0}^{n-1}\left(u-\frac{n-1}{2}\right)\zeta_n^{-ju}
 =-\frac{n}{1-\zeta_n^{-j}}.
 \label{eq:math-thomae-centered-fourier}
\end{equation}
The centered sequence has zero Fourier zero mode.
Inserting its inverse transform twice, with $h$ a unit modulo $n$ as required by Eq.~\eqref{eq:math-thomae-inertia-data}, gives the same sum in centered-residue form:
\begin{equation}
	\phi_{h,n}(\xi) = \frac1n\sum_{u=0}^{n-1} \left(u-\frac{n-1}2\right) \left([hu+\xi]_n-\frac{n-1}2\right).
	\label{eq:math-thomae-centered-sum}
\end{equation}
At $h=-1$ and $\xi=n-1$, the two factors are negatives of one another.
At $\xi=0$, separate the $u=0$ term and use $[-u]_n=n-u$ for $u>0$.
The two elementary sums are
\begin{equation}
	\phi_{-1,n}(n-1) = -\frac{n^2-1}{12}, \qquad \phi_{-1,n}(0)=-\frac{(n-1)(n-5)}{12}.
	\label{eq:math-thomae-inverse-sums}
\end{equation}
Substitution into Eq.~\eqref{eq:math-thomae-pair-exponent} yields
\begin{equation}
	\begin{aligned}
		\varrho_{0x}=\varrho_{1\Lambda}
		&=-\frac{n^2-1}{6}-\frac{(n-1)(n-5)}{12} + \frac{(n-1)^2}{4}=0,\\
		\varrho_{01}=\varrho_{0\Lambda}
		&=\varrho_{x1}=\varrho_{x\Lambda}=\frac g4.
	\end{aligned}
	\label{eq:math-thomae-replica-exponents}
\end{equation}

The monic denominators for the finite family are
\begin{equation}
	y_{\bmk,\Lambda}^{\,n} = z^{n-k_1}(z-x)^{k_1}(z-1)^{n-k_2}(z-\Lambda)^{k_2}, \qquad \omega_{\bmk,\Lambda}\coloneqq\frac{\dd{z}}{y_{\bmk,\Lambda}},
	\label{eq:math-thomae-finite-frame}
\end{equation}
with $1\leq k_1,k_2<n$.
The denominator has the inverse character to the differential, as in the source's character-basis construction.
Choose its roots continuously so that $(-\Lambda)^{k_2/n}\omega_{\bmk,\Lambda}$ tends to the physical representative $\varpi_{\bmk}$.
This gives the frame scaling in Eq.~\eqref{eq:assembly-thomae-frame-limit} and fixes the phases used in the branch product.
The characteristic vectors in Ref.~\cite{KopeliovichZemel:2019thomae} are twice those in Eq.~\eqref{eq:math-thomae-characteristic}.
Thus $[\bm0;\tfrac12\bm1_g]$ corresponds to the source's vectors $(\bm0,\bm1_g)$.

\paragraph{The replica reference spin line.}
For the physical cover and roots in Eq.~\eqref{eq:disjoint-physical-frame}, consider the differential and divisor defined in Eqs.~\eqref{eq:assembly-reference-differential} and~\eqref{eq:assembly-reference-divisor}.
The differential has orders $0,2(n-1),0,-2$ at the fibers over $0,x,1,\infty$, respectively.
For example, $u_0$ has a simple pole in the local coordinate at $x$, whereas $\dd{z}$ has a zero of order $n-1$.
These orders give $\div\varpi_*=2D_*$ and $\deg D_*=n(n-2)=g-1$.
For the spin line $L_*=\calO(D_*)$, write $s_*$ for its meromorphic section with $s_*^2=\varpi_*$.

This line has no holomorphic section.
A meromorphic section is $fs_*$, where $f$ is a meromorphic function, and it is holomorphic exactly when $\div f+D_*\geq0$.
Since $D_*$ is a sum of complete branch fibers, the deck group preserves this space of functions.
Its character eigenspaces have the form $R(z)u_0^rv_0^s$, with $R$ rational and $0\leq r,s<n$.
The orders of $u_0^rv_0^ss_*$ at the four fibers are
\begin{equation}
	r,\qquad n-1-r,\qquad s,\qquad -1-s.
	\label{eq:assembly-reference-section-orders}
\end{equation}
A pole of $R$ at a finite branch value lowers the corresponding order by a positive multiple of $n$, so none is allowed.
Nor can $R$ have a pole elsewhere at finite $z$.
Hence $R$ is a polynomial.
A polynomial of degree $d\geq0$ gives order $-1-s-nd<0$ at infinity, excluding every nonzero holomorphic section.
Therefore $H^0(\Sigma,L_*)=0$ and the associated theta constant is nonzero.
This valuation argument is a characterwise specialization of the invariant-divisor criterion in Ref.~\cite[Corollary~3.5 and Theorem~4.5]{KopeliovichZemel:2019divisors}.

In the algebraic coordinates $u,v$ of Eq.~\eqref{eq:disjoint-cover-equations}, $P_x$ is the full fiber of the cyclic projection over $u=0$, while $P_1$ and $P_\infty$ are the reduced zero and pole divisors of $v$.
The zero-characteristic line in Ref.~\cite[Eq.~(3.11) and Theorem~6.1]{Enolski:2006sdf} is represented by $(n-1)P_1-P_x$.
The principal-divisor identity
\begin{equation}
	D_*-\bigl[(n-1)P_1-P_x\bigr] = \div\!\left(\frac{u^n v}{1-u^n}\right)
	\label{eq:cyclic-reference-spin-comparison}
\end{equation}
identifies it with $L_*$.
The source's zero characteristic and our $\delta_*$ describe the same spin line in different canonical markings.

\paragraph{The characteristic in the canonical marking.}
Use the canonical cycles of Section~\ref{sec:disjoint-integral-homology}.
For a simple closed representative $C$ avoiding the divisor, we evaluate the quadratic refinement of Eq.~\eqref{eq:math-thomae-quadratic-definition} from the winding of $\varpi_*(\dot C)$.
Along a lifted primitive curve, its coefficient contributes
\begin{equation}
	\dd{\log(\varpi_*/\dd{z})} = \frac{n-1}{n} [\dd{\log(z-x)}-\dd{\log z}-\dd{\log(z-1)}].
	\label{eq:assembly-reference-winding}
\end{equation}
The coefficient windings about $0$ and $x$ cancel for alpha.
For the beta commutator, each signed winding cancels separately.
The coefficient therefore has zero total winding on both curves.

Smooth the connecting paths in the cyclic order of Eq.~\eqref{eq:homology-ribbon-order}, using the rounded representatives of Fig.~\ref{fig:disjoint-cycle-lifts}(b,c).
For alpha, follow the lower edge to the right: the four corners each rotate the tangent by $+\pi/2$, giving a total $2\pi$.
For beta, start on the lower outer arc and follow the four constant-sheet portions in panel (c).
The lower portion on $\bmr$ turns from downward to rightward and then upward, contributing $+\pi$.
On $\bmr+(0,1)$, the successive directions are upward, leftward, downward, leftward and upward; the signed turns are $+\pi/2,+\pi/2,-\pi/2,-\pi/2$, summing to zero.
On $\bmr+(1,1)$, upward, rightward and downward directions give $-\pi$.
On $\bmr+(1,0)$, downward, leftward, upward, leftward and downward directions give $-\pi/2,-\pi/2,+\pi/2,+\pi/2$, again summing to zero.
The cut crossings are unramified and keep the projected tangent continuous; intersections of the projection on distinct sheets introduce no turns.
Thus the tangent indices are one for alpha and zero for beta.

Together with the zero coefficient windings, $q_{\varpi_*}=\ind_{\varpi_*}+1$ modulo two gives, on every translated sheet,
\begin{equation}
	q_{\varpi_*}(\alpha_{\bmr})=0,\qquad
	q_{\varpi_*}(\beta_{\bmr})=1.
	\label{eq:assembly-primitive-quadratic-values}
\end{equation}

The quadratic addition rule~\eqref{eq:math-thomae-quadratic-polarization} transports these primitive values to the canonical basis.
All alpha--alpha intersections vanish, so their prefix sums have $q_{\varpi_*}(A_{I,J})=0$.
Every A-cycle in the correction to $B_{I,J}$ has a different index from $(I,J)$, and hence has zero intersection with $\beta_{I-1,J-1}$ by Eq.~\eqref{eq:homology-prefix-pairing}.
The corrections have zero mutual intersections as well.
Thus Eq.~\eqref{eq:assembly-reference-characteristic} follows from the characteristic convention~\eqref{eq:math-thomae-characteristic-quadratic}.

\subsection{Integral homology of the replica covers}\label{sec:replica-intersections}

\paragraph{The local ribbon pairing.}
For graph cycles, thicken the vertices to oriented discs and the edges to strips.
Alternating passage ends give one signed crossing, and nonalternating ends give zero.
For a chain $C$, let $\varepsilon_j(C;P)$ count departures minus arrivals through the $j$th end in counterclockwise order at a vertex $P$.
Closure gives $\sum_j\varepsilon_j(C;P)=0$.
Summing over the graph vertices gives
\begin{equation}
	C\circ C' = \frac12\sum_P\sum_{i<j} \bigl[\varepsilon_i(C;P)\varepsilon_j(C';P) - \varepsilon_i(C';P)\varepsilon_j(C;P)\bigr].
	\label{eq:homology-local-pairing}
\end{equation}
The half factor averages the two transverse separations along a shared edge.
Its two endpoints supply an integer total, and the zero-sum condition makes the result invariant under cyclic reindexing.

On the root graph of Eq.~\eqref{eq:fermat-pham-arc}, take closed chains $C=\sum c_{ij}\lambda_{i,j}$ and $D=\sum d_{ij}\lambda_{i,j}$.
Their outgoing coefficients are $c_{ij},d_{ij}$ at $b_j$ and $-c_{ij},-d_{ij}$ at $w_i$.
Applying Eq.~\eqref{eq:homology-local-pairing} in the two root-label orders gives
\begin{equation}
	C\circ D = \frac12 \sum_j\sum_{i<i'} (c_{ij}d_{i'j}-d_{ij}c_{i'j}) + \frac12\sum_i\sum_{j<j'} (c_{ij}d_{ij'}-d_{ij}c_{ij'}).
	\label{eq:fermat-ribbon-pairing}
\end{equation}
Both sums use the order $0,\ldots,n-1$.
The outgoing coefficients sum to zero at every vertex.

\subsubsection{Intersections on the cover for disjoint intervals}
\label{sec:appendix-disjoint-intersections}

\paragraph{The physical contour in root coordinates.}
The disjoint figures in this appendix use the physical $z$ plane and its replica labels $\bmr$, with the branch and access paths of Section~\ref{sec:disjoint-integral-homology}.
Let $\lambda^{(0)}$ be the lift of $[x,1]$ on the reference sheet, oriented from $x$ to $1$.
Under Eq.~\eqref{eq:common-base-coordinate-forward} it becomes $\lambda$.
Contracting the four small loops of beta to their access paths gives the chain
\begin{subequations}
	\label{eq:homology-square-chain-transport}
	\begin{align}
		\beta_{\bmr} &= \lambda^{(0)}_{\bmr}-\lambda^{(0)}_{\bmr+(1,0)}-\lambda^{(0)}_{\bmr+(0,1)}+\lambda^{(0)}_{\bmr+(1,1)},
		\label{eq:homology-square-chain-transport-square}\\
		e_{i,j} &= (1-\sfg_1^{-1})(1-\sfg_2)\lambda^{(0)}_{-i,j} = -\beta_{-i-1,j}.
		\label{eq:homology-square-chain-transport-pham}
	\end{align}
\end{subequations}
For example, the total access-path coefficients of beta at $1$ and $x$ are $(1-\sfg_1)(1-\sfg_2)$ and its negative.
Paths back to the base point cancel, leaving their difference $\lambda^{(0)}$.
In the physical root coordinates, the local coordinate at $x$ is $u_0^{-1}$, so increasing the first physical replica label turns clockwise.
In the common coordinates it is $u$, and increasing its root label turns counterclockwise.
The signed shift in the second line accounts for this change of labels without changing the oriented surface.

For the root Pochhammer word in Eq.~\eqref{eq:common-square-word}, let $a_0,a_1$ be the chosen access chains to the left and right endpoints, so $\lambda=a_1-a_0$.
Reading the word from the right and recording each sheet gives the contracted chain
\begin{align}
	C_E &= (1-\sfh_2)a_1 + \sfh_2(1-\sfh_1)a_0 + \sfh_1\sfh_2(1-\sfh_2^{-1})a_1 + \sfh_1(1-\sfh_1^{-1})a_0
	\notag\\
	&= (1-\sfh_1)(1-\sfh_2)(a_1-a_0).
	\label{eq:homology-root-access-chain}
\end{align}
Thus the root word gives the oriented square, while the retained physical beta word gives the shifted negative square above.

\paragraph{Disjoint ribbon incidences.}
The lifted generators and their sheet increments are given in Eq.~\eqref{eq:homology-loop-increments}.
Products follow the rightmost path first, as in Eq.~\eqref{eq:homology-lifted-words}.

On the positively oriented boundary of a small disc around each lift of the disjoint base point $z_*$, the outgoing and incoming ends of the positively oriented generators occur in the order
\begin{equation}
	(0^+,0^-,1^+,1^-,x^+,x^-).
	\label{eq:homology-ribbon-order}
\end{equation}
The superscripts identify generator-edge ends.
An inverse generator leaves through a minus end.
The two alpha incidences and four beta incidences follow by traversing the words in Eq.~\eqref{eq:homology-lifted-words} from right to left.
Table~\ref{eq:homology-half-edge-incidence} records the resulting departures and arrivals on each sheet.

\begin{table}[t]
	\centering
	\begin{tabular}{@{}lccc@{}}
		\toprule
		Curve & Sheet at the vertex & Departure $(+1)$ & Arrival $(-1)$\\
		\midrule
		$\alpha_{\bmr}$ & $(r_1,r_2)$ & $x^+$ & $0^-$\\
		& $(r_1-1,r_2)$ & $0^+$ & $x^-$\\
		\addlinespace
		$\beta_{\bmr}$ & $(r_1,r_2)$ & $1^+$ & $x^-$\\
		& $(r_1,r_2+1)$ & $x^-$ & $1^-$\\
		& $(r_1+1,r_2+1)$ & $1^-$ & $x^+$\\
		& $(r_1+1,r_2)$ & $x^+$ & $1^+$\\
		\bottomrule
	\end{tabular}
	\caption{Half-edge incidences of the two lifted words.
    A superscript identifies the half-edge's place in the ribbon order, and the column fixes the sign of its contribution to $\varepsilon_j$.}
	\label{eq:homology-half-edge-incidence}
\end{table}

For a mixed example, take $\alpha_{0,0}$ and $\beta_{0,0}$ at $n>2$.
Their only common vertex is $(0,0)$, where the ordered incidence vectors are
\begin{subequations}
	\label{eq:homology-ribbon-mixed-incidences}
	\begin{align}
		(\varepsilon_j(\alpha_{0,0};(0,0)))_{j=1}^6&=(0,-1,0,0,1,0),
		\label{eq:homology-ribbon-mixed-incidences-alpha}\\
		(\varepsilon_j(\beta_{0,0};(0,0)))_{j=1}^6&=(0,0,1,0,0,-1).
		\label{eq:homology-ribbon-mixed-incidences-beta}
	\end{align}
\end{subequations}
The nonzero summands of Eq.~\eqref{eq:homology-local-pairing} have index pairs $(2,3),(2,6),(3,5),(5,6)$, respectively, and give
\begin{equation}
	\alpha_{0,0}\circ\beta_{0,0} = \frac12(-1+1-1-1)=-1.
	\label{eq:homology-ribbon-mixed-evaluation}
\end{equation}
This is the zero-displacement coefficient of Eq.~\eqref{eq:homology-alpha-beta}.

For the beta pairing, $\beta_{0,0}$ and $\beta_{-1,0}$ share the two vertices $(0,0)$ and $(0,1)$ at $n>2$.
At $(0,0)$ the incidence vectors are $(0,0,1,0,0,-1)$ and $(0,0,-1,0,1,0)$, with nonzero index pairs $(3,5),(3,6),(5,6)$.
At $(0,1)$ they are $(0,0,0,-1,0,1)$ and $(0,0,0,1,-1,0)$, with nonzero index pairs $(4,5),(4,6),(5,6)$.
Each vertex contributes $\tfrac12(1-1+1)$, so
\begin{equation}
	\beta_{0,0}\circ\beta_{-1,0} = \frac12(1-1+1)+\frac12(1-1+1) = 1.
	\label{eq:homology-ribbon-beta-evaluation}
\end{equation}
This gives the positive $(-1,0)$ displacement in Eq.~\eqref{eq:homology-beta-beta}.
At $n=2$, adding coincident displacement classes cancels the opposite beta contributions while leaving the mixed zero-displacement pairing $-1$.

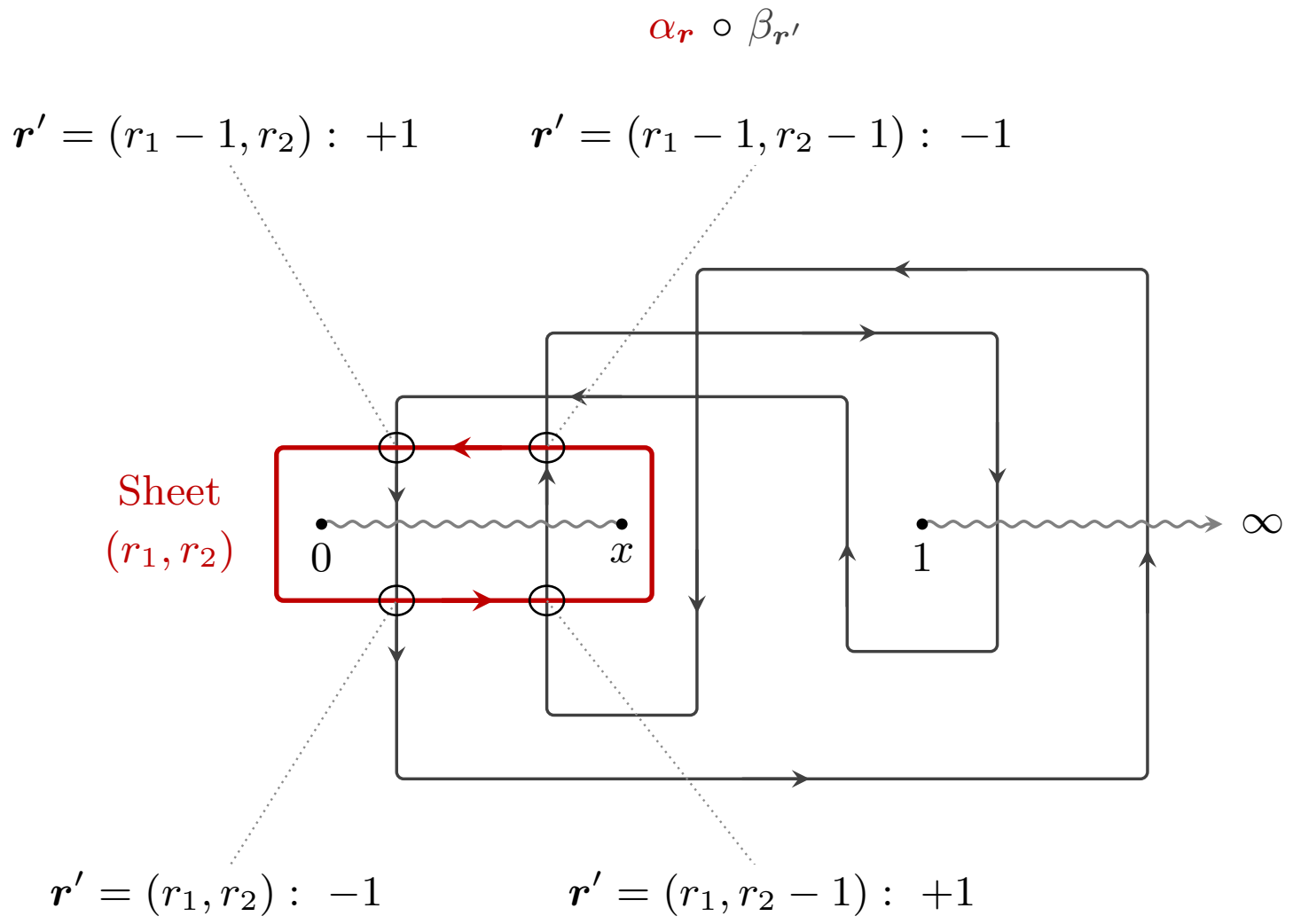
\begin{figure}[t]
	\centering
	\begin{tikzpicture}[x=1.12cm,y=.95cm,>=stealth,font=\footnotesize,
	reference/.style={draw=black!75,line width=.65pt},
	lifted/.style={draw=red!75!black,line width=1pt},
	leader/.style={draw=black!45,densely dotted,line width=.45pt},
	crossing/.style={draw=black,line width=.5pt},
	labelbox/.style={fill=white,inner sep=1.5pt}]
	\node at (2.7,3.9) {\textcolor{red!75!black}{$\alpha_{\bmr}$}
		$\circ$ \textcolor{black!75}{$\beta_{\bmr'}$}};
	\draw[reference,rounded corners=1.5pt]
	(.5,-2)--(5.5,-2)--(5.5,2)--(2.5,2)--(2.5,-1.5)
	--(1.5,-1.5)--(1.5,1.5)--(4.5,1.5)--(4.5,-1)
	--(3.5,-1)--(3.5,1)--(.5,1)--cycle;
	\foreach \xa/\ya/\xb/\yb in {
		2.8/-2/3.25/-2,5.5/-.7/5.5/-.2,4.3/2/3.8/2,
		2.5/-.2/2.5/-.7,1.5/.15/1.5/.45,3.2/1.5/3.7/1.5,
		4.5/.8/4.5/.3,3.5/-.65/3.5/-.15,2.0/1/1.65/1,
		.5/.45/.5/.15,.5/-.75/.5/-1.1}{
		\draw[reference,->] (\xa,\ya)--(\xb,\yb);
	}
	\draw[black!50,line width=.65pt,decorate,decoration={snake,amplitude=.6pt,segment length=5pt}] (0,0)--(2,0);
	\draw[black!50,line width=.65pt,->,decorate,decoration={snake,amplitude=.6pt,segment length=5pt,post length=3pt}] (4,0)--(6,0);
	\foreach \xx/\lab in {0/0,2/x,4/1}{
		\fill (\xx,0) circle (1.2pt);
		\node[below=3pt,fill=white,inner sep=1pt] at (\xx,0) {$\lab$};
	}
	\node[right] at (6,0) {$\infty$};
	\draw[lifted,rounded corners=1.5pt]
	(-.3,-.6)--(2.2,-.6)--(2.2,.6)--(-.3,.6)--cycle;
	\draw[lifted,->] (.8,-.6)--(1.15,-.6);
	\draw[lifted,->] (1.2,.6)--(.85,.6);
	\node[left,red!75!black,align=center] at (-.45,0) {Sheet\\$(r_1,r_2)$};
	\draw[leader] (-.7,3)--(.5,.6);
	\draw[leader] (3,3)--(1.5,.6);
	\draw[leader] (-.7,-2.85)--(.5,-.6);
	\draw[leader] (3,-2.85)--(1.5,-.6);
	\node[labelbox] at (-.7,3.05) {$\bmr'=(r_1-1,r_2):\ +1$};
	\node[labelbox] at (3,3.05) {$\bmr'=(r_1-1,r_2-1):\ -1$};
	\node[labelbox] at (-.7,-2.9) {$\bmr'=(r_1,r_2):\ -1$};
	\node[labelbox] at (3,-2.9) {$\bmr'=(r_1,r_2-1):\ +1$};
	\foreach \xx/\yy in {.5/.6,1.5/.6,.5/-.6,1.5/-.6}{
		\draw[crossing] (\xx,\yy) circle (.115);
	}
\end{tikzpicture}
	\caption{The four intersections of $\alpha_{\bmr}$ (red) with $\beta_{\bmr'}$ (black).
	The red contour encloses the first cut counterclockwise and lies entirely on sheet $\bmr$.
	Each circled crossing is labelled by the starting sheet $\bmr'$ of the black lift for which the two curves meet, followed by its contribution to $\alpha_{\bmr}\circ\beta_{\bmr'}$.
	The black lift has sheet $\bmr'$ on its lower outer arc.
	Signs use the red tangent followed by the black tangent, with the complex orientation.
	The grey wavy lines are the branch cuts.
	All replica labels are modulo $n$.}
	\label{fig:disjoint-alpha-beta-intersections}
\end{figure}

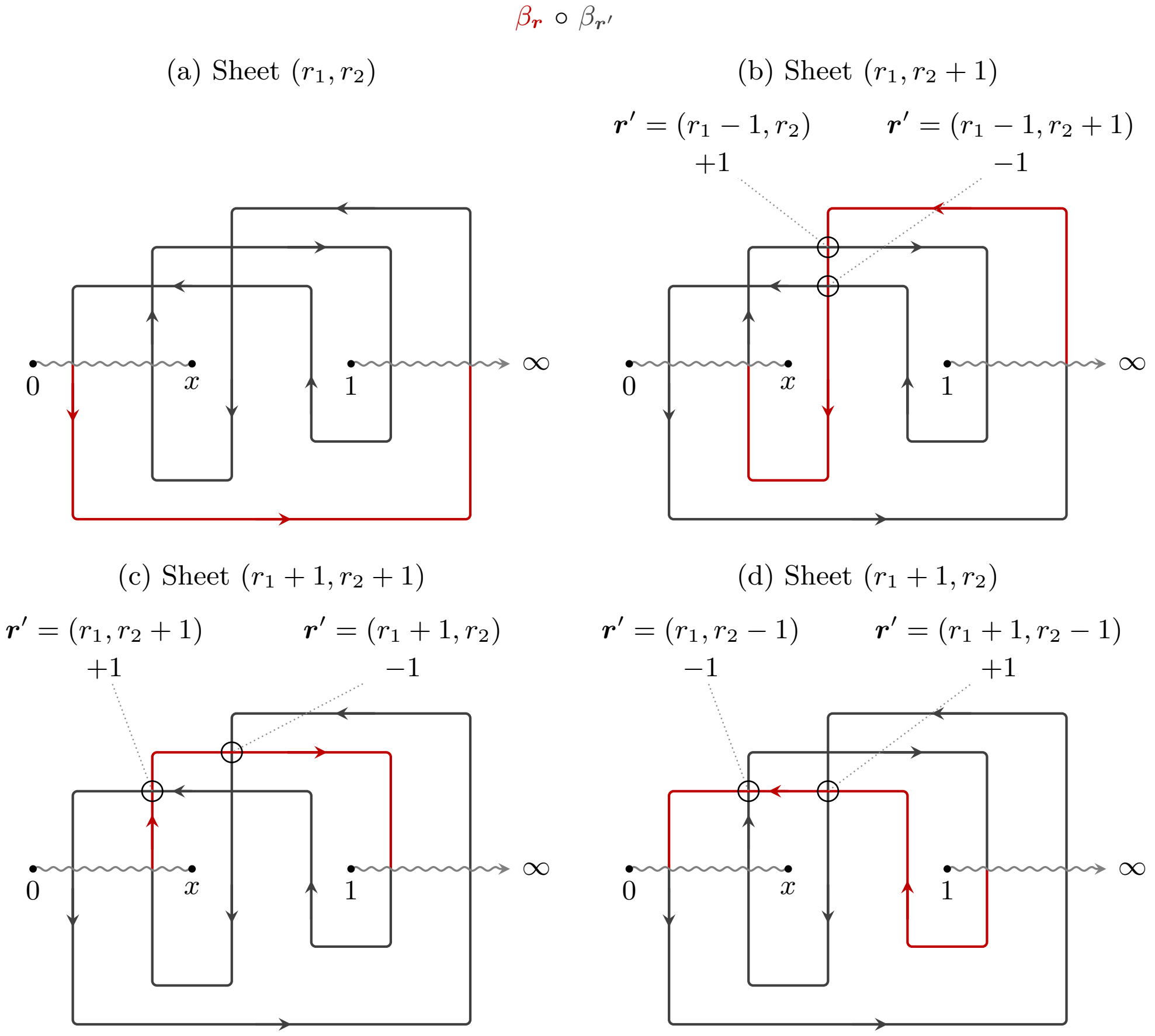
\begin{figure}[t]
	\centering
	\begin{tikzpicture}[x=.90cm,y=.88cm,>=stealth,font=\footnotesize,
	contour/.style={line width=.8pt,rounded corners=1.5pt},
	leader/.style={draw=black!45,densely dotted,line width=.45pt},
	crossing/.style={draw=black,line width=.5pt},
	labelbox/.style={fill=white,inner sep=1.5pt}]
	\node at (6.7,4.45) {\textcolor{red!75!black}{$\beta_{\bmr}$}
	$\circ$ \textcolor{black!75}{$\beta_{\bmr'}$}};
	\foreach \panel/\sx/\sy in {0/0/0,1/7.5/0,2/0/-6.5,3/7.5/-6.5}{
		\begin{scope}[shift={(\sx,\sy)}]
			\def\colorA{black!75}\def\colorB{black!75}
			\def\colorC{black!75}\def\colorD{black!75}
			\ifcase\panel\def\colorA{red!75!black}
			\or\def\colorB{red!75!black}
			\or\def\colorC{red!75!black}
			\or\def\colorD{red!75!black}\fi
			\draw[contour,draw=\colorA]
			(.5,0)--(.5,-2)--(5.5,-2)--(5.5,0);
			\draw[contour,draw=\colorB]
			(5.5,0)--(5.5,2)--(2.5,2)--(2.5,-1.5)--(1.5,-1.5)--(1.5,0);
			\draw[contour,draw=\colorC]
			(1.5,0)--(1.5,1.5)--(4.5,1.5)--(4.5,0);
			\draw[contour,draw=\colorD]
			(4.5,0)--(4.5,-1)--(3.5,-1)--(3.5,1)--(.5,1)--(.5,0);
			\draw[draw=\colorA,line width=.8pt,->] (2.8,-2)--(3.25,-2);
			\draw[draw=\colorA,line width=.8pt,->] (.5,-.25)--(.5,-.75);
			\draw[draw=\colorB,line width=.8pt,->] (4.3,2)--(3.8,2);
			\draw[draw=\colorB,line width=.8pt,->] (2.5,-.2)--(2.5,-.7);
			\draw[draw=\colorC,line width=.8pt,->] (1.5,.25)--(1.5,.7);
			\draw[draw=\colorC,line width=.8pt,->] (3.2,1.5)--(3.7,1.5);
			\draw[draw=\colorD,line width=.8pt,->] (3.5,-.65)--(3.5,-.15);
			\draw[draw=\colorD,line width=.8pt,->] (2.0,1)--(1.75,1);
			\draw[black!50,line width=.65pt,decorate,decoration={snake,amplitude=.6pt,segment length=5pt}] (0,0)--(2,0);
			\draw[black!50,line width=.65pt,->,decorate,decoration={snake,amplitude=.6pt,segment length=5pt,post length=3pt}] (4,0)--(6,0);
			\foreach \xx/\lab in {0/0,2/x,4/1}{
				\fill (\xx,0) circle (1.2pt);
				\node[below=3pt,fill=white,inner sep=1pt] at (\xx,0) {$\lab$};
			}
			\node[right] at (6,0) {$\infty$};
		\end{scope}
	}
	\begin{scope}
		\node at (3,3.75) {(a) Sheet $(r_1,r_2)$};
	\end{scope}
	\begin{scope}[xshift=6.75cm]
			\node at (3,3.75) {(b) Sheet $(r_1,r_2+1)$};
			\draw[leader] (1.05,2.65)--(2.5,1.5);
			\draw[leader] (4.8,2.65)--(2.5,1);
			\node[labelbox,align=center] at (1.05,2.85) {$\bmr'=(r_1-1,r_2)$\\$+1$};
			\node[labelbox,align=center] at (4.8,2.85) {$\bmr'=(r_1-1,r_2+1)$\\$-1$};
			\draw[crossing] (2.5,1.5) circle (.13);
			\draw[crossing] (2.5,1) circle (.13);
		\end{scope}
		\begin{scope}[yshift=-5.72cm]
			\node at (3,3.75) {(c) Sheet $(r_1+1,r_2+1)$};
			\draw[leader] (.9,2.65)--(1.5,1);
			\draw[leader] (4.65,2.65)--(2.5,1.5);
			\node[labelbox,align=center] at (.9,2.85) {$\bmr'=(r_1,r_2+1)$\\$+1$};
			\node[labelbox,align=center] at (4.65,2.85) {$\bmr'=(r_1+1,r_2)$\\$-1$};
			\draw[crossing] (1.5,1) circle (.13);
			\draw[crossing] (2.5,1.5) circle (.13);
		\end{scope}
		\begin{scope}[xshift=6.75cm,yshift=-5.72cm]
			\node at (3,3.75) {(d) Sheet $(r_1+1,r_2)$};
			\draw[leader] (.9,2.65)--(1.5,1);
			\draw[leader] (4.65,2.65)--(2.5,1);
			\node[labelbox,align=center] at (.9,2.85) {$\bmr'=(r_1,r_2-1)$\\$-1$};
			\node[labelbox,align=center] at (4.65,2.85) {$\bmr'=(r_1+1,r_2-1)$\\$+1$};
			\draw[crossing] (1.5,1) circle (.13);
			\draw[crossing] (2.5,1) circle (.13);
	\end{scope}
\end{tikzpicture}
	\caption{Intersections of lifted Pochhammer contours.
	Each panel draws the common projection once and highlights one constant-sheet portion of $\beta_{\bmr}$ in red.
	At each circled crossing, the black branch is assigned to $\beta_{\bmr'}$.
	The label gives the starting sheet $\bmr'$ for which the two branches meet, followed by its contribution to $\beta_{\bmr}\circ\beta_{\bmr'}$.
	The starting label of each lift is its sheet on the lower outer arc.
	The grey wavy lines are the branch cuts.
	The sign is positive when the red tangent followed by the black tangent agrees with the orientation of the complex plane.}
	\label{fig:disjoint-beta-intersections}
\end{figure}

\paragraph{Intersections in the cut plane.}
The intersection numbers follow by matching sheet labels at the projected crossings and reading the orientation of the two tangents.
Fix the starting label $\bmr$ of the first curve and denote the starting label of the second by $\bmr'$.
A point where the projections cross contributes only when the two lifted portions lie on the same sheet.
In Figs.~\ref{fig:disjoint-alpha-beta-intersections} and~\ref{fig:disjoint-beta-intersections}, the first tangent is red and the second is black.
Their ordered pair gives $+1$ when it agrees with the complex orientation and $-1$ otherwise.
Two alpha curves can be separated into parallel contours, so their intersection number vanishes.

The alpha--beta crossings in Fig.~\ref{fig:disjoint-alpha-beta-intersections} determine the mixed pairing.
Use the constant-sheet representative of $\alpha_{\bmr}$ in Fig.~\ref{fig:disjoint-cycle-lifts}(b).
Following the red contour counterclockwise, start at the lower-left crossing, where the black curve lies on sheet $\bmr'$, so the lifts meet when $\bmr'=\bmr$.
The rightward red tangent followed by the downward black tangent gives $-1$.
At the lower-right crossing, the black sheet is $(r_1',r_2'+1)$, giving $\bmr'=(r_1,r_2-1)$ with sign $+1$.
At the upper-right crossing, it is $(r_1'+1,r_2'+1)$, giving $\bmr'=(r_1-1,r_2-1)$ with sign $-1$.
Finally, at the upper-left crossing, it is $(r_1'+1,r_2')$, giving $\bmr'=(r_1-1,r_2)$ with sign $+1$.

The six beta--beta crossing contributions in Fig.~\ref{fig:disjoint-beta-intersections} are counted by the same rule.
Each panel highlights one constant-sheet portion of $\beta_{\bmr}$ in red, with its sheet label above the panel.
At a circled crossing, the black branch belongs to $\beta_{\bmr'}$, and equality of the two sheet labels fixes its starting label $\bmr'$.
For example, the upper crossing in panel (b) lies on sheets $(r_1,r_2+1)$ and $(r_1'+1,r_2'+1)$, so it contributes when $\bmr'=(r_1-1,r_2)$.
The downward red tangent followed by the rightward black tangent gives $+1$.
The other circled crossings give the remaining starting labels and signs displayed in the figure.

Summing the four mixed crossings and the six beta crossings gives Eqs.~\eqref{eq:homology-alpha-beta} and~\eqref{eq:homology-beta-beta}.
The ribbon incidence calculation above reproduces the same signs at the lifts of $z_*$.

\subsubsection{Integral bases on the Fermat curve}
\label{sec:appendix-fermat-bases}

\paragraph{Integral completeness of the rectangular Pham cycles.}
On the Fermat cover of Section~\ref{sec:fermat-integral-homology}, let $e_{i,j}$ be the closed Pham squares in Eq.~\eqref{eq:fermat-pham-square}.
For $n\geq3$, choose the rectangular family
\begin{equation}
	e_{i,j},\qquad 0\leq i\leq n-2,\quad 0\leq j\leq n-3, \qquad J_{(i,j),(k,l)}\coloneqq e_{i,j}\circ e_{k,l}.
	\label{eq:fermat-rectangular-basis}
\end{equation}
Order the pairs with the second index varying fastest.
There are $(n-1)(n-2)=2g_F$ cycles.
Their determinant gives the lattice index by Appendix~\ref{sec:math-riemann}.
This rectangle is equivalent to the integral basis of Ref.~\cite[Theorem~4.1]{Otsubo:2016fermat}.
The positive-real arc in that construction has the orientation of $\lambda$ here, and the deck generators act in the same root directions.
With the second excluded index chosen to be $n-1$, the source basis is obtained by successive partial sums in both indices of this rectangle.
In notation of Ref.~\cite{Otsubo:2016fermat} this is the explicit identity
\begin{equation}
	\kappa^{r,s}=\sum_{i=0}^{r-1}\sum_{j=0}^{s-1}e_{i,j},
	\qquad 1\leq r\leq n-1,\quad 1\leq s\leq n-2.
	\label{eq:fermat-otsubo-prefix-sums}
\end{equation}
It follows by factoring $1-\sfh_1^r=(1-\sfh_1)\sum_{i=0}^{r-1}\sfh_1^i$ and the analogous second-direction factor.
See also Ref.~\cite{Ejder:2018fermat}.
This change of generators is integral unitriangular.

To evaluate Eq.~\eqref{eq:fermat-intersection-kernel}, let $(U_m)_{ab}=\delta_{b,a+1}$ for $1\leq a,b\leq m$, and set $L_m=I_m-U_m$.
The rectangle contains no modular wraparound in the nearest-neighbour displacements of that kernel.
Its restricted pairing is
\begin{equation}
	J = L_{n-1}^{\top}\otimes L_{n-2}^{\top} - L_{n-1}\otimes L_{n-2}.
	\label{eq:fermat-rectangular-gram}
\end{equation}
Both $L_m$ and its inverse are integral with determinant one.
Multiplication on the left by $L_{n-1}^{-\top}\otimes I_{n-2}$ and on the right by $I_{n-1}\otimes L_{n-2}^{-1}$ therefore preserves the determinant and gives
\begin{equation}
	I_{n-1}\otimes(L_{n-2}^{\top}L_{n-2}^{-1}) - (L_{n-1}^{-\top}L_{n-1})\otimes I_{n-2}.
	\label{eq:fermat-determinant-reduction}
\end{equation}

The characteristic polynomial of $L_m^{-\top}L_m$ is $\det(tL_m^{\top}-L_m)$.
Expanding this tridiagonal determinant gives
\begin{equation}
	\begin{aligned}
		\chi_0(t) &= 1,\qquad \chi_1(t)=t-1,\\
		\chi_m(t) &= (t-1)\chi_{m-1}(t)+t\chi_{m-2}(t) = \frac{t^{m+1}+(-1)^m}{t+1}.
	\end{aligned}
	\label{eq:fermat-characteristic-polynomial}
\end{equation}
Cancellation at $t=-1$ extends the quotient to a polynomial.
Transposing $tL_m^{\top}-L_m$ gives the same polynomial for $L_m^{\top}L_m^{-1}$.
For every root $r$ of $\chi_{n-2}$, one has $r^{n-1}=(-1)^{n-1}$ and $r\ne-1$, hence $\chi_{n-1}(r)=(-1)^{n-1}$.
Taking the determinant of Eq.~\eqref{eq:fermat-determinant-reduction} therefore yields
\begin{equation}
	\det J = \prod_{\chi_{n-2}(r)=0}\chi_{n-1}(r) = (-1)^{(n-1)(n-2)}=1.
	\label{eq:fermat-unimodularity}
\end{equation}
Thus the rectangular cycles are an integral basis of $H_1(F_n,\bbZ)$.

\paragraph{The interval-cycle pairing.}
\label{par:fermat-interval-pairing}
We derive the mixed interval pairing from the cycle definitions in Eq.~\eqref{eq:fermat-explicit-cycles} and the common intersection kernel.
For $0\leq a\leq b\leq n-3$ and $0\leq c\leq d\leq n-3$, the interval sum for $A_{a,b}\circ\widetilde B_{c,d}$ is the first line of Eq.~\eqref{eq:fermat-explicit-cross-pairing}.
Let $H_q$ count pairs $(j,l)$ in the two intervals with $l-j=q$, and write $\sigma=c+d+1-a-b$.
The kernel selects $H_0-H_1$, $H_1-H_{-1}$ or $-H_0+H_{-1}$ when $\sigma$ is congruent to $1$, $0$ or $-1$, respectively, and vanishes otherwise.
If $H_q\ne0$, the interval bounds give
\begin{equation}
	q-(n-3)\leq\sigma-1\leq n-3+q.
	\label{eq:fermat-intersection-sigma-range}
\end{equation}
For $q=0,\pm1$, these bounds exclude every representative of the relevant residue class except $\sigma=1,0,-1$.
With $f(j)=\mathbf1_{\{c\leq j\leq d\}}$, the three remaining cases are
\begin{equation}
	\begin{aligned}
		H_0-H_1&=f(a)-f(b+1), &&\sigma=1,\\
		H_1-H_{-1}&=f(b)+f(b+1)-f(a-1)-f(a), &&\sigma=0,\\
		-H_0+H_{-1}&=f(a-1)-f(b), &&\sigma=-1.
	\end{aligned}
	\label{eq:fermat-intersection-boundary-differences}
\end{equation}
In these cases $c+d=a+b,a+b-1,a+b-2$, respectively.
Substituting the indicator values gives Eq.~\eqref{eq:fermat-explicit-cross-pairing}.

\paragraph{From rectangular to canonical cycles.}
The cycles $A_\mu,B_\mu$ of Eq.~\eqref{eq:fermat-explicit-cycles} have unique coordinate columns in the rectangular basis.
Compute the pairing of each canonical cycle with every rectangular cycle using Eq.~\eqref{eq:fermat-orbit-intersection}.
Multiplication by $J^{-1}$ then recovers its coordinate column.
Explicitly,
\begin{subequations}
	\label{eq:fermat-rectangular-coordinates}
	\begin{align}
		(T_A)_{\cdot,\mu} &= J^{-1} (e_{i,j}\circ A_\mu)_{(i,j)},
		\label{eq:fermat-rectangular-coordinates-a}\\
		(T_B)_{\cdot,\mu} &= J^{-1} (e_{i,j}\circ B_\mu)_{(i,j)}.
		\label{eq:fermat-rectangular-coordinates-b}
	\end{align}
\end{subequations}
The pairings on the right are finite interval sums of the explicit kernel, including when a translated square is outside the rectangle.
Since $\det J=1$, these coordinates are integers.
The canonical pairings then imply
\begin{equation}
	\begin{aligned}
		T &= (T_A\;T_B)\in M_{2g_F}(\bbZ),\\
		T^{\top}JT &= \begin{pmatrix}0&I_{g_F}\\-I_{g_F}&0\end{pmatrix},\\
		A_\mu &= \sum_{i,j}(T_A)_{(i,j),\mu}e_{i,j}, \qquad B_\mu=\sum_{i,j}(T_B)_{(i,j),\mu}e_{i,j}.
	\end{aligned}
	\label{eq:fermat-canonical-transform}
\end{equation}
All sums run over Eq.~\eqref{eq:fermat-rectangular-basis}.
Taking determinants and using $\det J=1$ gives $|\det T|=1$, hence $T\in GL(2g_F,\bbZ)$.

In the rescaled differential frame $\phi_{\bmell}=\omega_{\bmell}/b_{\bmell}$ of Eq.~\eqref{eq:fermat-period-blocks}, the rectangular character periods are $P_{\bmell,(i,j)}=\zeta_n^{i\ell_1+j\ell_2}$.
Consequently the reconstructed columns give
\begin{subequations}
	\label{eq:fermat-rectangular-period-blocks}
	\begin{align}
		\sfA&=PT_A,
		\label{eq:fermat-rectangular-period-blocks-a}\\
		\sfB&=PT_B,
		\label{eq:fermat-rectangular-period-blocks-b}\\
		\Omega_{F,n}&=(PT_A)^{-1}(PT_B).
		\label{eq:fermat-rectangular-period-blocks-normalized}
	\end{align}
\end{subequations}
In the original differential frame the nonzero beta factors $b_{\bmell}$ multiply both blocks on the left and cancel only in their quotient.

\subsection{Period identities for replica amplitudes}\label{sec:period-identities}

\paragraph{Complementary Euler periods.}
Fix an integer $n\geq2$, $0<x<1$, and a mixed physical character $\bmk=(k_1,k_2)$ with $1\leq k_1,k_2<n$.
Put $\bmk^*=(n-k_1,n-k_2)$ and retain the angles $a=k_1/n$, $b=k_2/n$ and phases $p_j=\zeta_n^{k_j}$ of Section~\ref{sec:disjoint-period-matrix}.
Apply Euler's transformation and the Gauss connection formula~\eqref{eq:math-hypergeometric-connection} to the Euler periods in Eqs.~\eqref{eq:period-I-euler} and~\eqref{eq:period-J-euler}.
Keeping their beta factors and using $\Gamma(a)\Gamma(1-a)=\pi/\sin\pi a$ gives~\cite[Eqs.~15.8.1 and~15.8.4]{DLMF}
\begin{subequations}\label{eq:assembly-complementary-period-connection}
	\begin{align}
		(1-x)^{b-a}I_{\bmk^*}&=I_{\bmk},
		\label{eq:assembly-complementary-period-connection-I}\\
		(1-x)^{b-a}J_{\bmk^*}&=J_{\bmk} + \sin(\pi a)[\cot(\pi a)-\cot(\pi b)]I_{\bmk}.
		\label{eq:assembly-complementary-period-connection-J}
	\end{align}
\end{subequations}
Dividing these relations and using $\sin(\pi a)[\cot(\pi a)-\cot(\pi b)]=-\sin\pi(a-b)/\sin\pi b$ gives
\begin{equation}
	\frac{J_{\bmk}(x)}{I_{\bmk}(x)} - \frac{J_{\bmk^*}(x)}{I_{\bmk^*}(x)} = \frac{\sin\pi(a-b)}{\sin\pi b}.
	\label{eq:period-euler-connection}
\end{equation}
The identities extend to $a=b$ by continuity.
The reference contour phases in Eq.~\eqref{eq:period-raw-euler-values} give
\begin{subequations}\label{eq:normalization-period-phase-identities}
	\begin{gather}
		\calA_{\bmk}=-2i e^{-i\pi b}\sin(\pi a)I_{\bmk}
		\label{eq:normalization-period-phase-identities-A}\\
		\tau_{\bmk}=-2i e^{i\pi(a+b)}\sin(\pi b)J_{\bmk}/I_{\bmk}.
		\label{eq:normalization-period-phase-identities-ratio}
	\end{gather}
\end{subequations}
Using these phases in the complementary Euler relation yields
\begin{equation}
	\tau_{\bmk}-p_1p_2\tau_{\bmk^*}=p_2-p_1.
	\label{eq:period-complementary-ratio}
\end{equation}

\paragraph{The two-angle quantum factor.}
For comparison with Ref.~\cite[Eq.~(4.1)]{Anastasopoulos:2011gn}, set its two angles to $a,b$ and write its beta and hypergeometric factors as follows:
\begin{subequations}\label{eq:period-two-angle-functions}
	\begin{align}
		B_1&=B(a,1-b),& B_2&=B(1-a,b),\\
		G_1(x)&={}_2F_1(a,1-b;1;x),&
		G_2(x)&={}_2F_1(1-a,b;1;x),\\
		H_1(x)&={}_2F_1(a,1-b;1+a-b;x),&
		H_2(x)&={}_2F_1(1-a,b;1-a+b;x).
	\end{align}
\end{subequations}
Our periods are
\begin{subequations}\label{eq:period-two-angle-euler-dictionary}
	\begin{align}
		I_{\bmk}(x)&=\frac{\pi}{\sin\pi a}G_1(x),&
		I_{\bmk^*}(x)&=\frac{\pi}{\sin\pi a}G_2(x),\\
		J_{\bmk}(x)&=(1-x)^{b-a}B_2H_2(1-x),&
		J_{\bmk^*}(x)&=(1-x)^{a-b}B_1H_1(1-x).
	\end{align}
\end{subequations}
Substituting these four expressions in Eq.~\eqref{eq:ward-bilinear} and using $G_2(x)=(1-x)^{a-b}G_1(x)$ gives
\begin{equation}
	\begin{aligned}
		&(1-x)^{a-b}(1-\bar x)^{b-a}\calI_{\bmk}(x,\bar x)\\
		&\qquad=B_1G_2(x)H_1(1-\bar x)+B_2G_1(\bar x)H_2(1-x) = 2\pi I_{\rm ABR}(x,\bar x).
	\end{aligned}
	\label{eq:period-two-angle-bilinear-dictionary}
\end{equation}
The equality holds on the ordered real interval and its local continuation with the same branches, with $x$ and $\bar x$ differentiated independently.

\paragraph{Disjoint complementary-character contractions.}
Use the row and column convention of Eq.~\eqref{eq:period-matrix-blocks}, and let $m_j,I_j\in\{1,\ldots,n-1\}$ for $j=1,2$.
The prefix sums $S_k(I)$ are defined in Eq.~\eqref{eq:period-prefix-sum}, and $\sfC_{\bmk,\bmI}$ is the correction in Eq.~\eqref{eq:period-B-correction}, fixed by the cycle intersections.
The finite root sum $\sum_{r=1}^{n-1}\zeta_n^{rh}=n\delta_n(h,0)-1$ gives the one-direction contraction
\begin{equation}
	\sum_{I=1}^{n-1}S_k(I)\zeta_n^{m(I-1)} = -\frac{n\zeta_n^k}{1-\zeta_n^k}\,\delta_{m,n-k}.
	\label{eq:period-complementary-sum}
\end{equation}
Taking the product in the two replica directions gives
\begin{equation}
	\sum_{I_1,I_2=1}^{n-1}S_{k_1}(I_1)S_{k_2}(I_2) \zeta_n^{m_1(I_1-1)+m_2(I_2-1)} = \frac{n^2p_1p_2\,\delta_{\bmm,\bmk^*}}{(1-p_1)(1-p_2)}.
	\label{eq:period-two-direction-contraction}
\end{equation}
This multiplies $\tau_{\bmm}$ in $\sfA\sfB^{\top}$.
Exchanging the characters gives the coefficient of $\tau_{\bmk}$ in $\sfB\sfA^\top$, whose numerator is $n^2$ on the delta's support.

The A-cycle correction supplies the remaining antisymmetric term.
Using $S_k(I-1)=S_k(I)-\zeta_n^{k(I-1)}$, rewrite Eq.~\eqref{eq:period-B-correction} as
\begin{equation}
	\sfC_{\bmk,\bmI} = -S_{k_1}(I_1)S_{k_2}(I_2) + (1-p_2)S_{k_1}(I_1)p_2^{I_2-1} + p_1^{I_1-1}p_2^{I_2}.
	\label{eq:period-correction-expanded}
\end{equation}
The first term cancels upon antisymmetrizing in $\bmk,\bmm$.
By Eq.~\eqref{eq:period-complementary-sum}, the second term contracts as
\begin{equation}
	\begin{aligned}
		&(1-p_2) \sum_{I_1=1}^{n-1}S_{k_1}(I_1)S_{m_1}(I_1)
				\sum_{I_2=1}^{n-1}p_2^{I_2-1}S_{m_2}(I_2)\\
		&\qquad =n\delta_{m_2,n-k_2}
				\sum_{I_1=1}^{n-1}S_{k_1}(I_1)S_{m_1}(I_1),
	\end{aligned}
	\label{eq:period-bilinear-character-sum}
\end{equation}
which is also symmetric and cancels.
The last term in Eq.~\eqref{eq:period-correction-expanded} gives
\begin{equation}
	\begin{aligned}
		&\sum_{I_1,I_2=1}^{n-1}\bigl[ \sfC_{\bmk,\bmI}S_{m_1}(I_1)S_{m_2}(I_2) - S_{k_1}(I_1)S_{k_2}(I_2)\sfC_{\bmm,\bmI} \bigr]\\
		&\quad=\left[\sum_{I_1=1}^{n-1}p_1^{I_1-1}S_{m_1}(I_1)\right]
		\left[\sum_{I_2=1}^{n-1}p_2^{I_2}S_{m_2}(I_2)\right]\\
		&\qquad-\left[\sum_{I_1=1}^{n-1}\zeta_n^{m_1(I_1-1)}S_{k_1}(I_1)\right]
		\left[\sum_{I_2=1}^{n-1}\zeta_n^{m_2 I_2}S_{k_2}(I_2)\right]\\
		&\quad=\frac{n^2(p_2-p_1)\,\delta_{\bmm,\bmk^*}}
		{(1-p_1)(1-p_2)}.
	\end{aligned}
	\label{eq:period-correction-contraction}
\end{equation}

The two-direction contraction and this correction give
\begin{equation}
	(\sfA\sfB^{\top}-\sfB\sfA^{\top})_{\bmk,\bmm}
	=\frac{n^2\calA_{\bmk}\calA_{\bmm} \delta_{\bmm,\bmk^*}}{(1-p_1)(1-p_2)} \times [p_1p_2\tau_{\bmk^*}-\tau_{\bmk}+p_2-p_1]=0.
	\label{eq:period-block-symmetry}
\end{equation}
The bracket vanishes by Eq.~\eqref{eq:period-complementary-ratio}.
Multiplying by $\sfA^{-1}$ and $\sfA^{-\top}$ proves the explicit symmetry stated in Eq.~\eqref{eq:period-symmetry}.

\paragraph{Diagonal Hodge contractions.}
The Hermitian bilinear relation in Eq.~\eqref{eq:period-positive-form} uses conjugate rather than complementary-character products.
Expanding the same geometric sums $S_k$ and the correction $\sfC$ gives
\begin{align}
	\sum_{I=1}^{n-1}S_k(I) \overline{p^{I-1}}
	&=\frac{n}{1-p^{-1}},\qquad p=e^{2\pi i k/n},
	\label{eq:period-diagonal-prefix-contraction}\\
	\Im\sum_{I_1,I_2=1}^{n-1} S_{k_1}(I_1)S_{k_2}(I_2) \overline{\sfC_{\bmk,\bmI}}
	&=\frac{n^2}{4}[\cot(\pi a)-\cot(\pi b)].
	\label{eq:period-diagonal-correction-contraction}
\end{align}
Substituting Eqs.~\eqref{eq:period-diagonal-prefix-contraction}--\eqref{eq:period-diagonal-correction-contraction} and the phases~\eqref{eq:normalization-period-phase-identities} into Eq.~\eqref{eq:period-positive-form} gives Eq.~\eqref{eq:assembly-hodge-period-contraction}.
The complementary Euler identities then convert this geometric contraction to the Ward bilinear.

\paragraph{The marked real part.}
For the disjoint cover, write $\Omega=\Omega_n^{(3)}(x)=\calR+i\calI$ with $g=g_d$.
In the specific marking of Section~\ref{sec:disjoint-period-matrix}, $2\calR$ is integral and independent of $x$.
The winding parity then resolves the coupled numerator in Eq.~\eqref{eq:assembly-normalized-ratio} into a finite sum of products of genus-$g$ theta constants.

To establish this property from the periods, use the physical phases of Eq.~\eqref{eq:period-character-parameters} and the complementary character $\bmk^*=(n-k_1,n-k_2)$.
The Euler integrals are real on $0<x<1$, so the phase in Eq.~\eqref{eq:period-character-ratio} gives $\overline{\tau_{\bmk^*}}=-p_1p_2\tau_{\bmk^*}$.
The connection relation~\eqref{eq:period-complementary-ratio} becomes
\begin{equation}
	\tau_{\bmk}(x)+\overline{\tau_{\bmk^*}(x)}=p_2-p_1.
	\label{eq:assembly-real-character-relation}
\end{equation}
Conjugation of the finite character sum sends $\bmk$ to $\bmk^*$.
Adding Eq.~\eqref{eq:period-canonical-explicit} to its conjugate therefore removes the Euler integrals.
For its A-cycle correction, Eq.~\eqref{eq:period-finite-inverse} turns each product of prefix sums in Eq.~\eqref{eq:period-B-correction} into two Kronecker deltas.
The result is
\begin{equation}
	\begin{aligned}
		2\calR_{\bmI\bmJ}
		&=\frac1{n^2}\sum_{k_1,k_2=1}^{n-1} D_{k_1}(I_1)D_{k_2}(I_2) p_1^{J_1-1}p_2^{J_2-1}(p_2-p_1)\\
		&\quad+2\bigl( \delta_{I_1,J_1}\delta_{I_2,J_2-1} - \delta_{I_1,J_1-1}\delta_{I_2,J_2} + \delta_{I_1,J_1-1}\delta_{I_2,J_2+1}\bigr).
	\end{aligned}
	\label{eq:assembly-real-period-entries}
\end{equation}
Here $1\leq I_i,J_i\leq n-1$, and the boundary prefix cycles vanish.
In each replica direction the remaining transform satisfies
\begin{equation}
	\frac1n\sum_{k=1}^{n-1}D_k(I)\zeta_n^{kr} = \delta_n(I-1,r)-\delta_n(I,r),\qquad r\in\bbZ,
	\label{eq:assembly-real-period-transform}
\end{equation}
where $\delta_n$ is the Kronecker delta modulo $n$ used in Eq.~\eqref{eq:period-finite-inverse}.
Indeed, $D_k(I)=\zeta_n^{-k(I-1)}-\zeta_n^{-kI}$, and the two Fourier sums have cancelling constant terms.
Expanding $p_2-p_1$ in Eq.~\eqref{eq:assembly-real-period-entries} leaves products of the integer differences in Eq.~\eqref{eq:assembly-real-period-transform}.
Thus
\begin{subequations}\label{eq:assembly-half-integral-real-part}
	\begin{gather}
	2\calR\in M_g(\bbZ)
	\label{eq:assembly-half-integral-real-part-integrality}\\
	\frac{\dd{\calR}}{\dd{x}}=0.
	\label{eq:assembly-half-integral-real-part-constancy}
	\end{gather}
\end{subequations}

The real part vanishes at $n=2$, whereas a half-integral off-diagonal entry survives for every $n\geq3$.
Taking $\bmI=(1,1)$ and $\bmJ=(1,2)$ in Eq.~\eqref{eq:assembly-real-period-entries} gives $-1$ from the character sum and $2$ from the correction, hence
\begin{equation}
	\calR_{(1,1),(1,2)}=\frac12.
	\label{eq:assembly-nontrivial-half-entry}
\end{equation}

\paragraph{The winding-parity decomposition.}
Half-integrality makes the mixed phase depend only on the lattice parities, since
\begin{equation}
	e^{2\pi i\bmm^{\top}\calR\bmw_A}
	=(-1)^{\bmm^{\top}(2\calR)\bmw_A}.
	\label{eq:assembly-parity-phase}
\end{equation}
To separate the two Gaussian sums, fix the winding parity by writing $\bmw_A=2\bmj+\bm\epsilon$, with $\bmj\in\bbZ^g$ and $\bm\epsilon\in\{0,1\}^g$.
The mixed phase is then independent of $\bmj$, since $2\calR$ is integral.
The absolutely convergent series of Eq.~\eqref{eq:classical-resummed-lattice} consequently gives
\begin{equation}
	\begin{aligned}
		\Theta(\bm0\mid T_\eta(\Omega))
		&=\sum_{\bm\epsilon\in\{0,1\}^g}
		\Bigl[\sum_{\bmj\in\bbZ^g}
			e^{-4\pi\eta(\bmj+\bm\epsilon/2)^{\top}
				\calI(\bmj+\bm\epsilon/2)}\Bigr]
		\Bigl[\sum_{\bmm\in\bbZ^g}
			e^{-\pi\eta^{-1}\bmm^{\top}\calI\bmm
				+2\pi i\bmm^{\top}\calR\bm\epsilon}\Bigr]\\
		&=\sum_{\bm\epsilon\in\{0,1\}^g}
		\theta\!\begin{bmatrix}\bm\epsilon/2\\\bm0\end{bmatrix}
			(\bm0\mid4i\eta\calI)\,
		\theta\!\begin{bmatrix}\bm0\\\calR\bm\epsilon\end{bmatrix}
			(\bm0\mid i\eta^{-1}\calI).
	\end{aligned}
	\label{eq:assembly-parity-theta-sum}
\end{equation}
Both factors have genus $g$, and $\calR\bm\epsilon$ is a half-characteristic modulo integers.
The factor $4$ in the first matrix comes from restricting the winding vector to one parity class.

The coupled sum reduces to the unrestricted product when $\calR$ is integral.
The mixed phase is then identically one and the unrestricted lattice sums give
\begin{equation}
	\Theta(\bm0\mid i\eta\calI)
	\Theta(\bm0\mid i\eta^{-1}\calI),
	\label{eq:theta-parity-two-classical-factors}
\end{equation}
as for the purely imaginary bipartite period matrix.
For $n\geq3$, the entry in Eq.~\eqref{eq:assembly-nontrivial-half-entry} instead gives a minus sign for the corresponding unit momentum and winding vectors.
The positive Gaussian weights then make the coupled sum strictly smaller than that unrestricted product.

\paragraph{The Fermat A-period inverse.}
For $\xi_r=\zeta_n^r$, $1\leq r\leq n-1$, finite orthogonality reads
\begin{equation}
	\sum_{r=1}^{n-1}\xi_r^h=n\delta_n(h,0)-1, \qquad h\in\bbZ.
	\label{eq:fermat-finite-root-sum}
\end{equation}
In Eq.~\eqref{eq:fermat-vandermonde-inverse-check}, the two exponents are $b-a$ and $b+1$, with $0\leq a,b\leq n-2$.
Only $a=b$ makes the first exponent divisible by $n$, while the second runs from $1$ to $n-1$.
The two constant terms therefore cancel, giving the stated inverse of $V$.

The pairs $(a,b+1)$ exhaust the two-element subsets of $\{0,\ldots,n-2\}$.
Thus, in the ordering of Eq.~\eqref{eq:fermat-schur-periods}, multiplying each A-period row by $\xi_s-\xi_r$ gives exactly the matrix of $2\times2$ minors of $V$.
The Cauchy--Binet identity makes the product of the minor matrices of $V$ and $V^{-1}$ the identity on these pairs.
Undoing the nonzero row rescaling gives Eq.~\eqref{eq:fermat-period-block-inverse}, including its factor $\xi_s-\xi_r$.
To restore the original differential frame from $\phi_{\bmell}=\omega_{\bmell}/b_{\bmell}$, multiply row $\bmell$ of both period blocks by $b_{\bmell}$ as in Eq.~\eqref{eq:fermat-orbit-periods}.
This common left multiplication cancels in $\sfA^{-1}\sfB$.

\section{Normalization of replica twist correlators}\label{sec:twist-normalization}
This appendix derives the scalar measure and determinant normalizations used in Sections~\ref{sec:disjoint-boson}--\ref{sec:low-index-amplitudes}.
It also contains the low-genus covering calculations and the auxiliary Dirac correlator entering the Ising formula.

\subsection{The real-boson measure on conical surfaces}\label{sec:math-conical-determinants}

For each connected cover, the scalar determinant uses the nonnegative Friedrichs Laplacian in the unit round pullback metric, with its single constant mode omitted:
\begin{equation}
	\zeta_\Delta(s)\coloneqq\sum_{\lambda>0}\lambda^{-s}, \qquad \det{}'\Delta=\exp[-\zeta_\Delta'(0)].
	\label{eq:math-spectral-zeta}
\end{equation}
The spectral sum converges for $\Re s>1$ and has a meromorphic continuation regular at $s=0$.
At a cone of angle $2\pi\alpha$, the Friedrichs domain selects the finite-energy radial solution and excludes the logarithm in the constant angular sector.
With smooth Gaussian curvature $K$ away from the cones,
\begin{equation}
	\zeta_\Delta(0) = \frac1{12\pi}\int K\dd{A} + \frac1{12}\sum_j(\alpha_j^{-1}-\alpha_j) - 1.
	\label{eq:math-conical-heat}
\end{equation}
The last term removes the constant kernel~\cite{Kalvin:2019fcs}.

After separating the harmonic winding representative, expand the single-valued part in a real $L^2$-orthonormal eigenbasis of the scalar Laplacian.
On a connected surface of area $A$, the constant eigenfunction is $A^{-1/2}$, so a shift of the field by $2\pi R$ gives its coefficient period $2\pi R\sqrt A$.
Use $\dd{a}/\sqrt{2\pi}$ for each real mode coefficient.
A nonzero mode with action $\lambda a^2/(4\pi)$ gives $(\lambda/(2\pi))^{-1/2}$.
Thus the connected real-boson measure is
\begin{equation}
	Z = \sqrt{2\pi}R\sqrt{A}\,(2\pi)^{\zeta_\Delta(0)/2} (\det{}'\Delta)^{-1/2}Z_{\rm cl}^{\bbR}.
	\label{eq:direct-fermat-mode-measure}
\end{equation}
Gluing uses the induced boundary $L^2$ measure.

\paragraph{The cyclic two-point reference.}
For $p_n(t)=t^n$ and unit round base metric $m_0$, separation of variables in $m_{p,n}=p_n^*m_0$ gives
\begin{equation}
	m_{p,n} = \dd{\theta}^2+n^2\sin^2\theta\dd{\varphi}^2, \qquad \lambda_{k,m} = \left(k+\frac{|m|}{n}\right) \left(k+\frac{|m|}{n}+1\right), \quad k\geq0,\quad m\in\bbZ.
	\label{eq:direct-fermat-football-spectrum}
\end{equation}
The zero pair $(k,m)=(0,0)$ is omitted.
In terms of the Barnes function
\begin{equation}
	\zeta_{\rm B}(s;a,b,c)\coloneqq \sum_{j,k\geq0}(aj+bk+c)^{-s}, \qquad a,b,c>0,
	\label{eq:math-barnes-double-zeta}
\end{equation}
continued from $\Re s>2$, the determinant is
\begin{subequations}
	\label{eq:direct-fermat-football-determinant}
	\begin{align}
		\log D_{p,n} &= \frac n2-\frac13\left(n+\frac1n\right)\log{n} - 4\zeta_{\rm B}'(0;n,1,1),
		\label{eq:direct-fermat-football-determinant-cover}\\
		\log D_0 &= \frac12-4\zeta_{\rm R}'(-1).
		\label{eq:direct-fermat-football-determinant-sphere}
	\end{align}
\end{subequations}
Primes on zeta functions denote differentiation in $s$.
Equation~\eqref{eq:direct-fermat-football-determinant-cover} follows from Kalvin's spindle formula~\cite[Lemma~2.1 and Eq.~(2.3)]{arXiv2011.05407} at $a=n$ and $K=1$, using $\log\det{}'\Delta=-\zeta_\Delta'(0)$.
See also Ref.~\cite{SpreaficoZerbini:2006sphere}.

\subsection{Quillen comparison with reference determinants}\label{sec:math-quillen}

Both scalar determinant calculations reduce to weighted line bundles on the unit round sphere.
The smooth anomaly fixes the metric dependence once the cohomology norms and the absolute spectra of the reference bundles are known.
For singular character weights, gluing introduces a local cap determinant and a global Dirichlet-to-Neumann determinant.
Appendix~\ref{sec:character-cap-comparison} controls the change of the global determinant as the caps shrink.

\paragraph{Smooth line-bundle determinant comparison.}
On $\bbP^1$, the line bundle $\calO(-m)$ has transition $e_\infty=z^{-m}e_0$ and degree $-m$.
For $m\geq1$ its zeroth cohomology vanishes and $\dim H^1=m-1$.
Serre duality identifies $H^1(\calO(-m))^*$ with $H^0(K_{\bbP^1}\otimes\calO(m))$.
We use the unit round metric and the smooth Hermitian reference $h_m=(1+|z|^2)^m$.

Let $L_h=2\bar\partial_h^*\bar\partial$ on sections and let $G_0(h),G_1(h)$ be the $L^2$ Gram matrices of fixed cohomology bases, using harmonic representatives in degree one.
Empty Gram determinants are one.
For $h=e^\varphi h_m$, the Quillen anomaly formula~\cite{Bismut:1988quillen} specializes to
\begin{equation}
	\log\frac{\det{}'L_h}{\det{}'L_{h_m}} - \log\frac{\det G_0(h)}{\det G_0(h_m)} + \log\frac{\det G_1(h)}{\det G_1(h_m)} = -\frac{1-m}{4\pi}\int\varphi\dd{A} - \frac1{8\pi}\int|\nabla\varphi|^2\dd{A}.
	\label{eq:math-quillen-anomaly}
\end{equation}
Here the factor two makes $L_{h=1}$ on the trivial line bundle the positive scalar Laplacian.
For a constant $\varphi$, the operator is unchanged and the Riemann--Roch theorem makes the left side $-(1-m)\varphi$, checking the Gram-factor signs.

\paragraph{Smooth reference spectra.}
For $\calO(-1)$, $L_{h_1}$ is one chiral block of the squared Dirac operator on the unit sphere.
Its eigenvalues are $j^2$ with multiplicity $2j$, $j=1,2,\ldots$ \cite{Camporesi:1995fb}.
For $\calO(-2)=K_{\bbP^1}$, the nonzero spectrum is the scalar spectrum $j(j+1)$ with multiplicity $2j+1$ by Hodge duality.
Thus
\begin{subequations}
	\label{eq:math-quillen-reference-determinants}
	\begin{align}
		\zeta_{L_{h_1}}(s)&=2\zeta_{\rm R}(2s-1),
		\label{eq:math-quillen-reference-determinants-zeta}\\
		D_1^\circ&=e^{-4\zeta_{\rm R}'(-1)},
		\label{eq:math-quillen-reference-determinants-degree-one}\\
		D_2^\circ&=D_0=e^{1/2-4\zeta_{\rm R}'(-1)}.
		\label{eq:math-quillen-reference-determinants-degree-two}
	\end{align}
\end{subequations}
The one-dimensional first cohomology of $\calO(-2)$ is Serre-dual to the space spanned by $\dd{z}\otimes e_0^{-1}$.
The squared norm of this dual section is proportional to $\int_{\bbC}(1+|z|^2)^{-2}\dd[2]{z}=\pi$.
The same proportionality factor multiplies $J(a,b)$ in Eq.~\eqref{eq:fermat-quillen-beta-norm} and cancels in the Gram-norm ratio.

\paragraph{Sphere Green logarithms.}
On the unit round sphere, define
\begin{equation}
	g_p(z)\coloneqq\log\sin^2\!\frac{d(z,p)}2,
	\qquad
	g_p(q)=\log\frac{|p-q|^2}{(1+|p|^2)(1+|q|^2)}
	\label{eq:math-sphere-green-logarithm}
\end{equation}
for finite $p,q$, with the corresponding limit at infinity.
Here $d$ is geodesic distance.
The positive scalar Laplacian gives $\Delta g_p=1-4\pi\delta_p$, where $\delta_p$ has unit mass with respect to $\dd{A}$.
Direct radial integration yields its mean, and integration by parts gives the pairing at distinct points:
\begin{subequations}
	\label{eq:fermat-sphere-green-pairing}
	\begin{align}
		\int g_p\dd{A} &= -4\pi,
		\label{eq:fermat-sphere-green-pairing-mean}\\
		\int\nabla g_p\mathbin{\cdot}\nabla g_q\dd{A} &= -4\pi[1+g_p(q)].
		\label{eq:fermat-sphere-green-pairing-pairing}
	\end{align}
\end{subequations}
The logarithmic self-pairing is regulated with a common round disk; its local finite and divergent terms cancel when the corresponding squared weights agree in a determinant quotient.

\paragraph{The complex beta integral.}
For $0<a,b<1$ and $a+b>1$, the Serre-dual norm is
\begin{equation}
	\int_{\bbC}\frac{\dd[2]{z}}{|z|^{2a}|1-z|^{2b}} = \pi\frac{\Gamma(a+b-1)\Gamma(1-a)\Gamma(1-b)}{\Gamma(a)\Gamma(b)\Gamma(2-a-b)}.
	\label{eq:math-complex-beta-integral}
\end{equation}
The inequalities ensure local integrability at $0,1,\infty$.
Schwinger parametrization yields $B(1-b,1-a)$ with prefactor $\pi\Gamma(a+b-1)/[\Gamma(a)\Gamma(b)]$.
Their product is the right-hand side of Eq.~\eqref{eq:math-complex-beta-integral}.

\paragraph{Gluing with interior conic points.}
The gluing formula factors a closed determinant into the Dirichlet piece determinants, a Dirichlet-to-Neumann determinant and a local collar constant~\cite{Burghelea:1992gluing}.
For scalar Friedrichs operators with conic points strictly inside the pieces, the same statement holds \cite[Proposition~2.8]{Kalvin:2019fcs}; the primed formula includes the constant-mode area divided by the total cutting length.

Invariant cuts, boundary restriction, Poisson extension and Friedrichs domains commute with the orthogonal deck-character projectors.
Let $L_{\chi,D}$ be the direct sum of the Dirichlet operators on the cut pieces, and $R_\chi(\lambda)$ the corresponding jump operator for $L_\chi+\lambda$.
The resolvent trace identity therefore restricts to every character,
\begin{equation}
	\dv{\lambda} \log\frac{\det(L_\chi+\lambda)}{\det(L_{\chi,D}+\lambda)\det R_\chi(\lambda)} = 0, \qquad\lambda>0.
	\label{eq:math-character-gluing-variation}
\end{equation}
The conic heat terms cancel between the full and cut operators.
The remaining large-parameter constant is local to the smooth collar, including its holonomy, so it cancels when the collar is unchanged under smoothing.
Nontrivial characters have no scalar kernel; the trivial sector and its omitted constant mode are kept unchanged.
The global boundary determinant still couples the caps through the exterior.
Its relative change, rather than its absolute value, is controlled in Appendix~\ref{sec:character-cap-comparison}.

\subsection{Determinant comparison under character-cap smoothing}\label{sec:character-cap-comparison}

Fix a nontrivial deck character, replica index, distinct branch values and disjoint round cutting discs of stereographic radius $\delta$.
A holomorphic frame change removes the regular factor from a local metric, leaving $h=r^{2\alpha}$ with $0<\alpha<1$. Smooth it by
\begin{equation}
	h_{\alpha,\varepsilon}(r) = \varepsilon^{2\alpha}H_\alpha(r/\varepsilon), \qquad h_{\alpha,\varepsilon}(r)=r^{2\alpha}\quad(r\geq\varepsilon).
	\label{eq:fermat-weight-smoothing}
\end{equation}
Here $H_\alpha$ is smooth, positive, constant near zero and agrees with $t^{2\alpha}$ near one.
For each nonzero weight, use matched normalized round-disc coordinates, holomorphic frames, cutting radius and smoothing profile.
The local frame change is unitary between the Hermitian $L^2$ spaces and conjugates the boundary maps by the same multiplier.

Let $L_\varepsilon=2\bar\partial_{h_\varepsilon}^*\bar\partial$ be the operator for the smoothed Hermitian metric, and let $L_{\rm F}$ be the singular Friedrichs realization on the same closed character line bundle.
At the fixed cuts $r=\delta$, let $R_\varepsilon=N_\varepsilon+N_{\rm ext}$ and $R_{\rm F}=N_{\rm F}+N_{\rm ext}$ be the corresponding jump operators, where $N_\varepsilon,N_{\rm F}$ are the cap Dirichlet-to-Neumann maps and $N_{\rm ext}$ is the common exterior map.
Apply the gluing formula of Appendix~\ref{sec:math-quillen}.
The exterior Dirichlet determinants and local collar constants cancel, leaving
\begin{equation}
	\frac{\det L_\varepsilon}{\det L_{\rm F}} = \prod_p C_{\alpha_p}(\varepsilon,\delta)\,\frac{\det_\zeta R_\varepsilon}{\det_\zeta R_{\rm F}}, \qquad R=N_{\rm cap}+N_{\rm ext}.
	\label{eq:fermat-cap-determinant-comparison}
\end{equation}
Here $C_\alpha$ is the Dirichlet cap-determinant ratio, determined by the weight and common profile.
The norm below is the trace norm on the boundary $L^2$ space.
\begin{lemma}[Relative boundary determinant under cap smoothing]\label{lem:relative-cap-boundary}
For this fixed geometry and smoothing data, as $0<\varepsilon<\delta$ tends to zero,
	\begin{equation}
		\left\|R_{\rm F}^{-1/2}(N_\varepsilon-N_{\rm F}) R_{\rm F}^{-1/2}\right\|_1 \leq C_{\bmk,\delta} \sum_{p:\,\alpha_p>0}\sum_{\ell\geq1} (\varepsilon/\delta)^{2(\ell-\alpha_p)} \longrightarrow0.
		\label{eq:fermat-cap-relative-limit}
	\end{equation}
	The constant is independent of $\varepsilon$.
	Consequently $\det_\zeta R_\varepsilon/\det_\zeta R_{\rm F}\to1$.
\end{lemma}
The regulator is removed at fixed distinct branch values and fixed cutting radius before any endpoint limit. Zero local weights remain unsmoothed; the trivial global sector retains its omitted constant mode and is excluded from this inverse-operator statement.

To prove the bound, write a cap mode as $u_j(r)e^{ij\theta}$.
For $j\geq0$, regularity selects $r^j$, which already equals the Friedrichs solution.
For $j=-\ell<0$, set $\nu=\ell-\alpha>0$.
Integration of the radial equation and multiplication by the flat-frame factor $r^\alpha$ give, outside the smoothed disc, $A_\ell r^\nu+B_\ell\varepsilon^{2\nu}r^{-\nu}$, with $B_\ell/A_\ell=2\nu\int_0^1t^{2\ell-1}H_\alpha(t)^{-1}\dd{t}-1$.
Positivity and smoothness of $H_\alpha$ bound this coefficient uniformly
in $\ell$, hence
\begin{equation}
	(N_\varepsilon-N_{\rm F})_j = O\!\left(\frac{\nu}{\delta}(\varepsilon/\delta)^{2\nu}\right).
	\label{eq:fermat-cap-dn-bound}
\end{equation}
Since $\nu\geq1-\alpha>0$, the mode estimates give a smoothing perturbation of the boundary map.
The positive jump operator $R_{\rm F}$ is invertible because a zero-energy global section would be parallel, contradicting the nontrivial holonomy.
Conjugating the perturbation by its order-$-1/2$ inverse half-powers gives a trace-class operator on boundary $L^2$ and the bound~\eqref{eq:fermat-cap-relative-limit}, even when the exterior mixes Fourier modes.

At the fixed cut the relative zeta determinant equals the Fredholm determinant of $I+R_{\rm F}^{-1/2}(N_\varepsilon-N_{\rm F})R_{\rm F}^{-1/2}$.
Their logarithmic variations agree and both are one at zero perturbation.
Trace-norm continuity proves the limit.

\subsection{Four-point normalization from character determinants}\label{sec:disjoint-quillen}

Put $g=(n-1)^2$.
Use the physical presentation $u_0^n=z/(z-x)$, $v_0^n=z-1$ of Eq.~\eqref{eq:disjoint-physical-frame}, with the integral marking used in Section~\ref{sec:four-point-normalization}.
For $\Omega=X+iY$, the winding sum in Eq.~\eqref{eq:classical-theta-result} is
\begin{equation}
	Z_{\rmc\rml}^{\bbR}(\Omega;\eta) = \eta^{-g/2}\sqrt{\det Y}\,\Theta(\bm0\mid T_\eta(\Omega)).
	\label{eq:quillen-winding-convention}
\end{equation}

\paragraph{The external pair-cover determinants.}
Equip the physical $z$ sphere with the unit round metric $m_0$ and every cover with its pullback.
Let $f:\Sigma_n^{(3)}(x)\to\bbP^1$ be the projection to $z$, and use the two cyclic quotient maps $f_u,f_v$.
Define
\begin{align}
	m_0 &= \frac{4|\dd{z}|^2}{(1+|z|^2)^2}, \qquad D_{d,n}(x) = \det{}'\Delta_{f^*m_0},\qquad D_0=\det{}'\Delta_{m_0},
	\nonumber\\
	D_{u,n}(x) &= \det{}'\Delta_{f_u^*m_0}, \qquad D_{v,n}=\det{}'\Delta_{f_v^*m_0}, \qquad f_u(u_0)=\frac{xu_0^n}{u_0^n-1}, \quad f_v(v_0)=1+v_0^n.
	\label{eq:quillen-four-cover-data}
\end{align}
The two degree-$n$ covers have branch values $(0,x)$ and $(1,\infty)$.

In terms of the antipodal reference $D_{p,n}$ of Eq.~\eqref{eq:direct-fermat-football-determinant-cover}, their displacement factors are
\begin{subequations}
	\label{eq:round-two-point-conversion}
	\begin{align}
		D_{u,n}(x) &= D_{p,n} \left(\frac{x^2}{1+x^2}\right)^{(n^2-1)/(6n)},
		\label{eq:round-two-point-conversion-u}\\
		D_{v,n} &= D_{p,n}\,2^{-(n^2-1)/(6n)}.
		\label{eq:round-two-point-conversion-v}
	\end{align}
\end{subequations}
After cancellation of the common cap, self-energy and smooth-reference terms, moving the branch pair from antipodal points to $p,q$ changes only the Green cross pairing in the degree-minus-one Quillen comparison of Appendix~\ref{sec:math-quillen}.
Its coefficient is $\sum_{k=1}^{n-1}(k/n)(1-k/n)=(n^2-1)/(6n)$, and the squared chordal separations of the two pairs are $x^2/(1+x^2)$ and $1/2$.

Let $Z_d,Z_u,Z_v,Z_0$ denote their connected real-boson path integrals, including the compact constant mode.
The four-point path integral is $Z_d/Z_0^{n^2}$.
The two external pair norms are $Z_u^n/Z_0^{n^2}$ and $Z_v^n/Z_0^{n^2}$, because each replica translation has $n$ length-$n$ cycles.
Therefore
\begin{equation}
	\calR_n^{(3)}(x;\eta) = \frac{Z_dZ_0^{n^2}}{Z_u^nZ_v^n}.
	\label{eq:quillen-four-path-integral-ratio}
\end{equation}
The Weyl factor at each endpoint occurs once in the numerator and once in its pair norm, so this ratio is also the plane correlation ratio.

For $0<a,b<1$, define the positive integral
\begin{equation}
	\scrJ_{a,b}(x)\coloneqq \int_{\bbC} \frac{\dd[2]{z}}{|z|^{2(1-a)}|z-x|^{2a}|z-1|^{2(1-b)}} > 0.
	\label{eq:quillen-four-hodge-norm}
\end{equation}
\begin{proposition}[Absolute disjoint scalar determinant quotient]\label{prop:disjoint-spectral-quotient}
	The scalar determinants defined in Eq.~\eqref{eq:quillen-four-cover-data} satisfy
	\begin{equation}
		\frac{D_{d,n}(x)D_0^{n^2}}{D_{u,n}(x)^nD_{v,n}^n} = (1-x)^{g/2}\prod_{k_1,k_2=1}^{n-1} \frac{\scrJ_{k_1/n,k_2/n}(x)}{\pi}.
		\label{eq:quillen-four-determinant-ratio}
	\end{equation}
\end{proposition}

\paragraph{Cohomology of the character bundles.}
Use the unitary deck-character decomposition, with $a=k_1/n$ and $b=k_2/n$.
For a mixed character $0<a,b<1$, the holomorphic local frame and its Hermitian norm are
\begin{subequations}
	\label{eq:quillen-four-character-metric}
	\begin{align}
		s_{a,b}&=(z-x)u_0^{n-k_1}v_0^{n-k_2},
		\label{eq:quillen-four-character-metric-section}\\
		h_{a,b}(z;x)&=|z|^{2(1-a)}|z-x|^{2a}|z-1|^{2(1-b)}.
		\label{eq:quillen-four-character-metric-metric}
	\end{align}
\end{subequations}
The four fractional weights are $(1-a,a,1-b,b)$, of total weight two; the frame $z^{-2}s_{a,b}$ at infinity consequently gives the extension $\calO(-2)$.
For an axis character, absorb the integral local factor into the holomorphic frame before extending across the branch points: remove $z-1$ at $b=0$ and $z$ at $a=0$.
The resulting axis metrics are
\begin{subequations}
	\label{eq:quillen-four-axis-metrics}
	\begin{align}
		h_{a,0}&=|z|^{2(1-a)}|z-x|^{2a},
		\label{eq:quillen-four-axis-metrics-u}\\
		h_{0,b}&=|z-1|^{2(1-b)},
		\label{eq:quillen-four-axis-metrics-v}
	\end{align}
\end{subequations}
and extend as $\calO(-1)$.
The trivial sector is $\calO$.
These sectors give
\begin{equation}
	f_*\calO_{\Sigma_n^{(3)}(x)} \simeq \calO\oplus\calO(-1)^{\oplus2(n-1)}\oplus\calO(-2)^{\oplus g}.
	\label{eq:quillen-four-pushforward}
\end{equation}
Away from the branch values the weighted operator $L_h=2\bar\partial_h^*\bar\partial$ is unitarily identified with the corresponding flat-character scalar Laplacian. Closing its Dirichlet energy selects the Friedrichs extension.

Write $D_{a,b}(x)$ for a character-sector determinant; only $D_{0,0}=D_0$ is primed.
Additivity of the spectral zeta function gives
\begin{equation}
	\frac{D_{d,n}D_0^{n^2}}{D_{u,n}^nD_{v,n}^n} = \prod_{k_1,k_2=1}^{n-1} \frac{D_{a,b}D_0}{D_{a,0}D_{0,b}}.
	\label{eq:quillen-four-character-quotient}
\end{equation}
Complexifying a real scalar to perform this character decomposition does not double its eigenvalue multiplicities.
The physical real-boson Gaussian factor is the inverse square root of this determinant quotient.

The axis bundles have no cohomology.
Each mixed bundle has a one-dimensional $H^1$, whose Serre dual is generated by $\dd{z}\otimes s_{a,b}^{-1}$.
Its squared norm is proportional to the integral $\scrJ_{a,b}(x)$ in Eq.~\eqref{eq:quillen-four-hodge-norm}.
The local weights $(1-a,a,1-b,b)$ are all below one, so the norm is integrable at all four branch values, including infinity.
For the reference metric $h_2=(1+|z|^2)^2$ the corresponding integral is $\pi$.
Fixing the Serre pairing in both metrics makes the ratio of $H^1$ Gram norms $G_1(h_{a,b})/G_1(h_2)=\pi/\scrJ_{a,b}$.
Thus Eq.~\eqref{eq:math-quillen-anomaly} contributes $+\log(\scrJ_{a,b}/\pi)$ to the scalar log determinant.

\paragraph{Cancellation of the character caps.}
By Lemma~\ref{lem:relative-cap-boundary}, the boundary-determinant ratio tends to one as the caps shrink.
In the quotient $D_{a,b}D_0/(D_{a,0}D_{0,b})$, the two weights at $0,x$ match those of $D_{a,0}$, and the two at $1,\infty$ match those of $D_{0,b}$.
Their Dirichlet cap factors therefore cancel at each point, including their finite parts.
The cohomology norms converge as well.
For the common smoothing profiles, a disk of radius $\varepsilon$ at weight $\alpha<1$ contributes $O(\varepsilon^{2-2\alpha})$ to the dual norm.
Thus the smoothed norms converge to $\scrJ_{a,b}(x)$ as the caps shrink, by the same local estimate used in the Fermat comparison after Eq.~\eqref{eq:fermat-quillen-beta-norm}.

\paragraph{The absolute sector determinant.}
Use the sphere Green logarithms $g_p$ and their pairings in Appendix~\ref{sec:math-quillen}.
Relative to $h_m=(1+|z|^2)^m$, the three potentials entering a mixed-sector quotient are
\begin{subequations}
\label{eq:quillen-four-potentials}
	\begin{align}
		\varphi_u&=(1-a)g_0+a g_x+a\log(1+x^2),
		\label{eq:quillen-four-potentials-u}\\
		\varphi_v&=(1-b)g_1+b g_\infty+(1-b)\log2,
		\label{eq:quillen-four-potentials-v}\\
		\varphi_{a,b}&=\varphi_u+\varphi_v.
		\label{eq:quillen-four-potentials-sum}
	\end{align}
\end{subequations}
The first two use $m=1$ and the last uses $m=2$.
Write $\calA_m$ for the right side of Eq.~\eqref{eq:math-quillen-anomaly}.
In the mixed-sector difference
\begin{equation}
	\calA_2(\varphi_u+\varphi_v) - \calA_1(\varphi_u) - \calA_1(\varphi_v),
	\label{eq:disjoint-quillen-mixed-anomaly-combination}
\end{equation}
the linear contribution is $-2+a\log(1+x^2)+(1-b)\log2$.
The gradient cross term is
\begin{align}
	-\frac1{4\pi}\int\nabla\varphi_u\mathbin{\cdot}\nabla\varphi_v\dd{A}
	&= 1+(1-a)(1-b)g_0(1)+(1-a)b\,g_0(\infty)
	\nonumber\\
	&\qquad+a(1-b)g_x(1)+ab\,g_x(\infty)
	\nonumber\\
	&= 1+2a(1-b)\log(1-x)-a\log(1+x^2)-(1-b)\log2.
	\label{eq:quillen-four-cross-pairing}
\end{align}
The round-coordinate terms cancel between these two contributions:
\begin{equation}
	\calA_2(\varphi_u+\varphi_v) - \calA_1(\varphi_u)-\calA_1(\varphi_v) = -1+2a(1-b)\log(1-x).
	\label{eq:quillen-four-anomaly-sum}
\end{equation}

The smooth reference determinants fix the finite constant in each character sector.
Their ratio is $D_0D_2^\circ/(D_1^\circ)^2=e$ by Eq.~\eqref{eq:math-quillen-reference-determinants}, cancelling the constant $-1$ in Eq.~\eqref{eq:quillen-four-anomaly-sum}.
Including the cohomology norm therefore proves the absolute sector identity
\begin{equation}
	\frac{D_{a,b}(x)D_0}{D_{a,0}(x)D_{0,b}} = (1-x)^{2a(1-b)}\frac{\scrJ_{a,b}(x)}\pi.
	\label{eq:quillen-four-sector-ratio}
\end{equation}

Multiplying these sector identities gives the full scalar quotient.
Since $\sum_{k_1,k_2=1}^{n-1}2a(1-b)=(n-1)^2/2=g/2$, Eq.~\eqref{eq:quillen-four-sector-ratio} proves Eq.~\eqref{eq:quillen-four-determinant-ratio}.

\paragraph{The normalized boson quotient.}
Apply the real-boson measure~\eqref{eq:direct-fermat-mode-measure} to the connected surfaces.
Only the disjoint cover has winding sectors.
Its area is $4\pi n^2$; the two pair-cover areas are $4\pi n$, and the base area is $4\pi$.
Equation~\eqref{eq:math-conical-heat}, with $4n$ cones of angle $2\pi n$ on the disjoint cover and two on each pair cover, gives
\begin{subequations}
	\label{eq:quillen-four-heat-counts}
	\begin{align}
		\zeta_d(0)&=\frac{n^2}{3} + \frac{4n}{12}(n^{-1}-n)-1=-\frac23,
		\label{eq:quillen-four-heat-counts-cover}\\
		\zeta_u(0)&=\zeta_v(0)=\frac{n+n^{-1}}6-1, \qquad \zeta_0(0)=-\frac23,
		\label{eq:quillen-four-heat-counts-references}\\
		1+n^2-2n&=g, \qquad \zeta_d(0)+n^2\zeta_0(0)-n\zeta_u(0)-n\zeta_v(0)=-g.
		\label{eq:quillen-four-heat-counts-combined}
	\end{align}
\end{subequations}
The signed count of connected constant-mode factors in the normalized quotient is $g$.
Their factors $(\sqrt{2\pi}R)^g$ combine with the spectral factor $(2\pi)^{-g/2}$ to leave $R^g=\eta^{g/2}$.
The area factor is $[(4\pi n^2)(4\pi)^{n^2}/(4\pi n)^{2n}]^{1/2}$, which reduces to $(4\pi)^{g/2}n^{1-n}$.
Thus Eq.~\eqref{eq:quillen-four-path-integral-ratio} becomes
\begin{equation}
	\calR_n^{(3)}(x;\eta) = \eta^{g/2}\frac{(4\pi)^{g/2}}{n^{n-1}} \left[\frac{D_{d,n}(x)D_0^{n^2}}{D_{u,n}(x)^nD_{v,n}^n}\right]^{-1/2} Z_{\rmc\rml}^{\bbR}(\Omega;\eta).
	\label{eq:quillen-four-measure}
\end{equation}

\paragraph{Agreement with the Ward--Takahashi result.}
The Hodge norm $h_{\bmk}=(i/2)\int_\Sigma\varpi_{\bmk}\wedge\overline{\varpi_{\bmk}}$, with $\varpi_{\bmk}=w_{\bmk}\dd{z}$, also has a base-integral representation.
Since $|w_{\bmk}|^2=h_{a,b}^{-1}$ and the cover has $n^2$ sheets, its base integral and the period evaluation~\eqref{eq:assembly-character-hodge-norm} give
\begin{subequations}
	\label{eq:quillen-four-period-norm}
	\begin{align}
		h_{\bmk} &= n^2\scrJ_{a,b}(x) = \pi n^2(1-x)^{b-a}\calI_{\bmk}(x,x),
		\label{eq:quillen-four-period-norm-character}\\
		\prod_{k_1,k_2=1}^{n-1}\frac{\scrJ_{a,b}(x)}{\pi} &= \prod_{k_1,k_2=1}^{n-1}\calI_{\bmk}(x,x).
		\label{eq:quillen-four-period-norm-product}
	\end{align}
\end{subequations}
The product identity uses $\sum_{k_1,k_2=1}^{n-1}(b-a)=0$.
Substituting it, the determinant ratio, and the Poisson-resummed winding sum into Eq.~\eqref{eq:quillen-four-measure} gives
\begin{equation}
	\calR_n^{(3)}(x;\eta) = \frac{(4\pi)^{g/2}}{n^{n-1}}(1-x)^{-g/4} \frac{\sqrt{\det Y}}{\displaystyle\prod_{k_1,k_2=1}^{n-1}\calI_{\bmk}(x,x)^{1/2}} \Theta(\bm0\mid T_\eta(\Omega)).
	\label{eq:quillen-four-ward-match}
\end{equation}

\subsection{Low-genus normalization calculations}\label{sec:covering-low-index}
We evaluate the $n=2$ adjacent sphere, the $n=2$ disjoint torus and the $n=3$ triangular torus, obtaining the amplitudes quoted in Section~\ref{sec:low-index-amplitudes}.
We take $c_L=c_R=c$.

\subsubsection{The sphere for adjacent intervals}\label{sec:covering-regularization}
The adjacent coefficient at $n=2$ follows from the Weyl anomaly on a degree-four sphere cover.
We first fix the finite external twist factors from a cyclic two-point cover, then apply them to the three-point cover.

In the flat-patch prescription of Ref.~\cite{Lunin:2000yv}, the base sphere is flat inside $|z|<\delta^{-1}$ and has metric $|\dd{z}|^2/(\delta^4|z|^4)$ outside.
Its partition function is
\begin{equation}
	Z_\delta=\calQ\,\delta^{-c/3}.
	\label{eq:cover-sphere-regulator}
\end{equation}
For $f^*g=e^\varphi\hat g$ on a degree-$d$ cover, the Liouville conformal anomaly relates the bare twists regulated by base discs of radius $\varepsilon$ to the partition function on the cover,
\begin{equation}
	\left\langle\prod_i\sigma_{\pi_i}^{\varepsilon}(z_i)\right\rangle_\delta = e^{S_L}\frac{Z_{\calC}[\Sigma,\hat g]}{(Z_\delta)^d}, \qquad S_L=\frac{c}{96\pi}\int_\Sigma\sqrt{\hat g}\,\left[(\hat\nabla\varphi)^2+2\hat R\varphi\right].
	\label{eq:cover-weyl-functional}
\end{equation}
For a local expansion $f(s)=z_i+a s^r+\cdots$ and a finite simple pole $f(s)=b/s+O(1)$, integration by parts gives respectively
\begin{equation}
	S_L^{\rm ram}(r,a) = -\frac{c(r-1)}{12} \left[\frac1r\log|a| + \log r + \frac{r-1}{r}\log\varepsilon\right].
	\label{eq:cover-ramification-term}
\end{equation}
\begin{equation}
	S_L^{\rm pole}(b) = -\frac c6\log|b|-\frac c3\log\delta.
	\label{eq:cover-pole-term}
\end{equation}
Use the analogous sphere metric on a genus-zero cover, with cutoff $\delta'$ and finite normalization $\calQ$.
When cover infinity is unramified, the outer annulus and curvature circle contribute $c\log(\delta'/\delta)/3$; together with the finite poles they cancel the cutoffs in $Z_{\delta'}/Z_\delta^d$.

The cyclic two-point map $f(t)=a t^p/[t^p-(t-1)^p]$ has two ramification coefficients of modulus $|a|$ and finite-pole product $|a|^{p-1}/p^{p+1}$.
Substitution yields
\begin{equation}
	\left\langle\sigma_{(p)}^{\varepsilon}(a) \sigma_{(p)^{-1}}^{\varepsilon}(0)\right\rangle = p^{c/3}\calQ^{1-p} \varepsilon^{-c(p-1)^2/(6p)} |a|^{-c(p-1/p)/6}.
	\label{eq:cover-cyclic-two-point}
\end{equation}
The finite cyclic normalizers also appear in Ref.~\cite[Eqs.~(142)--(143)]{Ma:2026frw}, whose two-flat-disks sphere partition function corresponds to $\calQ$ here.
Thus the unit-normalized length-$p$ twist is
\begin{equation}
	\calT_{(p)} = \lim_{\varepsilon\to0} p^{-c/6}\varepsilon^{c(p-1)^2/(12p)} \calQ^{(p-1)/2}\sigma_{(p)}^{\varepsilon}.
	\label{eq:adjacent-cycle-unit-normalization}
\end{equation}
Each adjacent insertion contains $n$ length-$n$ cycles and uses the product of these factors.
The factor $p^{-c/6}$ remains after the ultraviolet and sphere-normalization factors cancel.

For $n=2$, consider the degree-four map
\begin{equation}
	z(t) = u^2 = \frac{(t^2+1)^2}{4t^2}, \qquad \frac{\dd{z}}{\dd{t}} = \frac{(t^2-1)(t^2+1)}{2t^3}.
	\label{eq:adjacent-n2-rational-cover}
\end{equation}
It has two index-two ramification points above each of $0,1,\infty$.
The point $t=2$ is unramified, with $z(2)=25/16$.
Sending these two points to infinity puts all branch values at finite points and keeps cover infinity unramified.
Use
\begin{equation}
	\xi=\frac1{z-25/16}, \qquad \omega=\frac1{t-2}, \qquad \xi(\omega) = \frac{16\omega^2(2\omega+1)^2}{(5\omega+2)(3\omega+2)(4\omega+1)}.
	\label{eq:adjacent-n2-finite-frame}
\end{equation}
Its branch values are $(-16/25,-16/9,0)$.
The ramification points $t=0,\infty,1,-1,i,-i$ map to $\omega=-1/2,0,-1,-1/3,(-2-i)/5,(-2+i)/5$; the finite poles are $-2/5,-2/3,-1/4$.
Define $A_j$ by $\xi(\omega)=\xi_j+A_j(\omega-\omega_j)^2+\cdots$ and $B_p$ by $\xi(\omega)=B_p/(\omega-\omega_p)+O(1)$.
With $\xi_{ij}=\xi_i-\xi_j$, differentiation gives
\begin{subequations}
	\label{eq:adjacent-n2-local-products}
	\begin{align}
		\prod_{j=1}^{6}|A_j|&=\frac{2^{40}}{15^4},
		\label{eq:adjacent-n2-local-products-ramification}\\
		\prod_{p=1}^{3}|B_p|&=\frac{2^8}{15^5},
		\label{eq:adjacent-n2-local-products-poles}\\
		|\xi_{12}\xi_{23}\xi_{31}|&=\frac{2^{16}}{15^4}.
		\label{eq:adjacent-n2-local-products-distances}
	\end{align}
\end{subequations}
The local anomaly terms, after removing the three-point coordinate factor, give
\begin{equation}
	2^{-c/2}
	\left(\frac{2^{40}}{15^4}\right)^{-c/24}
	\left(\frac{2^8}{15^5}\right)^{-c/6}
	\left(\frac{2^{16}}{15^4}\right)^{c/4}
	=2^{c/2}.
	\label{eq:adjacent-n2-anomaly-constant}
\end{equation}
The remaining bare prefactor is $\calQ^{-3}\varepsilon^{-c/4}$, while the three external insertions supply $2^{-c}\varepsilon^{c/4}\calQ^3$.
Hence the normalized coefficient is $C_2^{(3)}=2^{-c/2}$, as in Eq.~\eqref{eq:adjacent-n2-coefficient}.

For one real compact boson, the character decomposition gives an independent spectral check at $c=1$.
Each of the three nontrivial characters is trivial at one branch value, so the three degree-two quotient surfaces have branch pairs $(0,\infty)$, $(0,1)$ and $(1,\infty)$.
The first pair is antipodal in $m_0$, and the other two have spherical separation $\pi/2$.
The two-cone formula of Ref.~\cite{Kalvin:2019fcs}, Proposition~3.1, gives $D_{p,2}$ for the first quotient and $2^{-1/4}D_{p,2}$ for each of the other two.
Multiplying the quotient spectra counts the trivial sector three times, hence
\begin{subequations}
	\label{eq:direct-fermat-n2-spectral-check}
	\begin{align}
		D_{F,2}&=2^{-1/2}\frac{D_{p,2}^{3}}{D_0^2},
		\label{eq:direct-fermat-n2-spectral-check-determinant}\\
		C_{2}^{(3)}&=2^{-1/2}.
		\label{eq:direct-fermat-n2-spectral-check-coefficient}
	\end{align}
\end{subequations}

The sphere calculation for strictly adjacent intervals also applies to a massless free fermion when the spin structure extends over the ramification points and the cover carries no operator insertions.
The capped sphere has a unique spin structure, so with sphere vacuum state of unit norm and the same two-point normalization the covering anomaly gives $C_2^{(3)}=2^{-c/2}$.
For the Dirac fermion ($c=1$), Ref.~\cite[Section~4.2, Eq.~(186)]{Harper:2024ker} obtains $\kappa_2^{(3)}=(\log 2)/4$ using the spin sum in the torus calculation for disjoint intervals.
An independent fermionic covariance-matrix calculation for three adjacent intervals on a ring confirms this value \cite[Appendix~D]{Liu:2024ulq}.
The auxiliary four-point sector of Appendix~\ref{sec:fixed-spin-reference} has a different fusion channel.

\subsubsection{The torus for disjoint intervals}
\label{sec:appendix-torus-check}
The torus partition function determines the disjoint four-point amplitude at $n=2$.

For the $n=2$ four-point cover, take $0<x<1$ and use the full seed CFT trace~\eqref{eq:models-torus-partition}.
Reference~\cite{Harper:2024ker} denotes the tripartite modulus by $\widetilde\tau=2\tau$, while Ref.~\cite{Harper:2025uui} uses $\tau$ for the doubled modulus and expresses the cross ratio through theta constants at half that modulus.
Both references use $\eta$ for the cross ratio. Here $\eta(\tau)$ denotes the Dedekind function, whereas the positive real parameter $\eta=R^2$ is the compactification parameter.

The genus-one identities of Appendix~\ref{sec:math-gaussian-theta} give
\begin{equation}
	\left|\frac{\eta(2\tau)}{\vartheta_4(0\mid2\tau)}\right|^{2c} = \left[\frac{\vartheta_2(0\mid2\tau)\vartheta_3(0\mid2\tau)}{2\vartheta_4(0\mid2\tau)^2}\right]^{2c/3} = 2^{-4c/3}x^{c/3}(1-x)^{-c/6}.
	\label{eq:models-torus-prefactor}
\end{equation}
This converts Eq.~\eqref{eq:models-torus-ratio} to the fixed-frame and finite functions~\eqref{eq:models-torus-GRF}.

For $n=2$, Eq.~\eqref{eq:period-genus-one-matrix} gives $\Omega=it$, with $t=2K(1-x)/K(x)>0$.
The two integers $(w,\ell)$ then have action
\begin{equation}
	S_{\rm cl}(w,\ell)=\pi\eta\left(tw^2+\frac{\ell^2}{t}\right).
	\label{eq:classical-torus-action}
\end{equation}
Using the Jacobi theta function in Eq.~\eqref{eq:math-jacobi-theta}, Eq.~\eqref{eq:classical-theta-result} reduces to
\begin{equation}
	Z_{\rm cl}^{\bbR}(it;\eta) = \sqrt{\frac{t}{\eta}}\sum_{w,m\in\bbZ} e^{-\pi t(\eta w^2+\eta^{-1}m^2)} = \sqrt{\frac{t}{\eta}}\,\vartheta_3(0\mid i\eta t)\vartheta_3(0\mid it/\eta).
	\label{eq:classical-torus-theta}
\end{equation}
The exponent is $-2\pi t(h_{m,w}+\bar h_{m,w})$, with the weights in Eq.~\eqref{eq:ward-radius-spectrum}.
This identifies the Poisson-dual integer with momentum in the stated compactification convention.

At genus one, the disjoint fluctuation bilinear reduces to
\begin{equation}
	\calI_{1,1}(x,x)=\frac8\pi K(x)K(1-x)=\frac4\pi K(x)^2\Im\tau_2^{(3)}(x).
	\label{eq:ward-genus-one-check}
\end{equation}
Substitution with $g=1$ and $\Delta_2^{(3)}=1/4$ gives Eq.~\eqref{eq:assembly-genus-one}.
In this case the reference characteristic is $[0;1/2]$, and Eq.~\eqref{eq:assembly-thomae-identity} becomes
\begin{equation}
	\vartheta_4(0\mid it)^2=(1-x)^{1/4}\frac{2K(x)}{\pi}.
	\label{eq:assembly-genus-one-thomae}
\end{equation}
As $x\to0$, $K(x)\to\pi/2$ and $t\to\infty$, so both theta factors tend to one and the normalization is explicit.

For the endpoint limits, impose the vacuum and spectrum conditions stated in Section~\ref{sec:torus-comparison}.
The vacuum and the modular $S$ transformation give
\begin{subequations}
	\label{eq:discussion-vacuum-dominance}
	\begin{align}
		Z_{\calC}(iY) &= e^{\pi cY/6}\bigl[1+O(e^{-2\pi\Delta_{\min}Y})\bigr],
		&&Y\longrightarrow\infty,
		\label{eq:discussion-vacuum-dominance-low-temperature}\\
		Z_{\calC}(it)=Z_{\calC}(i/t) &= e^{\pi c/(6t)}\bigl[1+O(e^{-2\pi\Delta_{\min}/t})\bigr],
		&&t\longrightarrow0^+.
		\label{eq:discussion-vacuum-dominance-high-temperature}
	\end{align}
\end{subequations}
The unique normalized vacuum fixes the unit coefficient in both lines.
As $x\to0$, $\tau(x)\sim(i/\pi)\log(16/x)$ gives $Z_{\calC}(2\tau)\sim(16/x)^{c/3}$.
The prefactor in Eq.~\eqref{eq:models-torus-prefactor} therefore gives $\calR_{2,\calC}^{(3)}\to1$.

The crossed torus limit determines the squared adjacent coefficient.
At $x\to1$, write $\tau(x)=it/2$ and combine the elliptic asymptotic with Eq.~\eqref{eq:discussion-vacuum-dominance-high-temperature} to obtain
\begin{subequations}
	\label{eq:discussion-cardy-endpoint}
	\begin{align}
		\frac{2\pi}{t} &= \log\frac{16}{1-x} + O\!\left((1-x)\log\frac{16}{1-x}\right),
		\label{eq:discussion-cardy-endpoint-period}\\
		Z_{\calC}(it) &= \left(\frac{16}{1-x}\right)^{c/12}[1+o(1)].
		\label{eq:discussion-cardy-endpoint-trace}
	\end{align}
\end{subequations}
For the ordinary cover, the same modular estimate applies at $t/2$.
The prefactors in Eqs.~\eqref{eq:discussion-torus-functions} and~\eqref{eq:models-torus-GRF} give the limits~\eqref{eq:discussion-n2-residues}.

For the pinching calculation, the component blocks are empty and their determinants are one.
Choose the waist marking $(\widehat A,\widehat B)=(B,-A)$ relative to Eq.~\eqref{eq:period-genus-one-matrix}.
It gives $\widetilde\Omega_2^{(3)}=-1/\tau_2^{(3)}$.
With $y=1-x\to0^+$ and the parameter convention~\eqref{eq:period-elliptic-K}, the zero-balanced hypergeometric continuation at $a=b=1/2$ gives~\cite[Eq.~15.8.10]{DLMF}
\begin{equation}
	K(y)=\frac\pi2\bigl[1+O(y)\bigr],\qquad
	K(1-y)=\frac12\log\frac{16}{y}+O(y\log(1/y)).
\end{equation}
Their ratio gives the full neck period, including its finite $\log16$.
\begin{equation}
	\widetilde\Omega_2^{(3)}(x) = -\frac1{\tau_2^{(3)}(x)} = \frac{iK(x)}{2K(1-x)} = \frac{i}{2\pi}\log\frac{16}{1-x}+o(1).
	\label{eq:degeneration-n2-check}
\end{equation}

\subsubsection{The triangular torus for adjacent intervals}
\label{sec:appendix-triangle-check}
For $n=3$, the degree-nine Fermat cover has three index-three points above each endpoint and triangular modulus $\tau_\triangle=e^{i\pi/3}$.
Its normalized amplitude is the known result~\cite{Harper:2024ker}
\begin{equation}
	\left\langle\prod_{i=1}^{3} \calT_{\varsigma_i}(z_i) \right\rangle_{n=3} = \frac{3^{-3c/2}Z_{\calC}(\tau_\triangle)}{|z_{12}z_{23}z_{31}|^{2c/3}}.
	\label{eq:adjacent-n3-normalized-correlator}
\end{equation}
Here $Z_{\calC}$ is the modular-invariant unit-vacuum torus trace.
Each length-three cycle supplies $3^{-c/6}$ by Eq.~\eqref{eq:adjacent-cycle-unit-normalization}.
There are three cycles per endpoint and three endpoints, giving $(3^{-c/6})^9=3^{-3c/2}$; the covering anomaly cancels their cutoff factors.
For the compact boson, the conversion to the harmonic factor can be performed term by term.

Equation~\eqref{eq:fermat-cubic-period} gives $\Omega_{F,3}=\zeta_3=-1/2+i\sqrt3/2$.
The integral shear $B_1\mapsto B_1+A_1$ changes the modulus to $\tau_\triangle=e^{i\pi/3}$ and the winding integer to $\ell\mapsto\ell+w$.
Thus $\ell-\calR{}w$ is unchanged in the raw action.
In the resummed expression, the shear multiplies each term by $e^{2\pi i m w}=1$.
Both markings consequently give
\begin{equation}
	\calH_{F_3}(\eta) = \sqrt{\frac{\sqrt3}{2}} \sum_{w,m\in\bbZ}(-1)^{mw} e^{-\frac{\pi\sqrt3}{2}(\eta w^2+\eta^{-1}m^2)}.
	\label{eq:fermat-boson-triangle}
\end{equation}
The real part of the triangular modulus gives the spin phase $(-1)^{mw}$; it does not change the unit state multiplicity of each integer momentum--winding pair.
Positivity follows from the unresummed winding sum in Eq.~\eqref{eq:classical-winding-sum}, rather than termwise positivity of this trace.

At the triangular modulus, the trace and harmonic factor differ by the oscillator and Gaussian period measure:
\begin{equation}
	Z(\tau_\triangle;\eta) = \frac{\calH_{F_3}(\eta)}{\sqrt{\Im\tau_\triangle} |\eta(\tau_\triangle)|^2} = \frac{\sqrt\eta\,Z_{\rm cl}^{\bbR}(\Omega_{F,3};\eta)}{\sqrt{\Im\tau_\triangle} |\eta(\tau_\triangle)|^2}.
	\label{eq:fermat-boson-torus-normalization}
\end{equation}

\subsection{The auxiliary Dirac correlator}\label{sec:fixed-spin-reference}
We use an auxiliary fixed-spin Dirac correlator to derive the Thomae identity and determine the common scalar factor in the Ising sum.

At fixed integer $n\geq2$ and $0<x<1$, the regular plane-vacuum lift uses commuting translations
\begin{equation}
	(\sfU_a\bm\psi)_{\bmr} = e^{-\pi i(n-1)/n}\psi_{\bmr+\bme_a}, \qquad \sfU_a^n=(-1)^{n-1},\qquad a=1,2,
	\label{eq:models-fixed-dirac-lifts}
\end{equation}
Inverse endpoints use the exact inverses; for even $n$ these differ from $\sfU_a^{n-1}$ by a minus sign.
Fourier transformation of all $n^2$ replicas gives
\begin{equation}
	a_{\bmk}=\frac{k_1-(n-1)/2}{n},\qquad
	b_{\bmk}=\frac{k_2-(n-1)/2}{n},\qquad
	(\bmq_1,\bmq_2,\bmq_3,\bmq_4) = (\bma,-\bma,\bmb,-\bmb).
	\label{eq:models-fixed-dirac-charges}
\end{equation}
Here $\bma=(a_{\bmk})$ and $\bmb=(b_{\bmk})$ are vectors over all $n^2$ Fourier modes $0\leq k_1,k_2<n$, and $\bmq_j$ is the corresponding charge vector at endpoint $z_j$.
The centered sums give $\bma\cdot\bmb=0$ and $\bma^2=\bmb^2=(n^2-1)/12=\Delta_n^{(3)}$ at $c=1$.
In the bosonization convention $\langle\varphi_{\bmk}(z)\varphi_{\bml}(w)\rangle = -\delta_{\bmk\bml}\log|z-w|^2$, the unit-normalized charge vertices therefore obey
\begin{equation}
	\left\langle\prod_{j=1}^{4}\calT_{\sigma_j}(z_j)\right\rangle = \prod_{i<j}|z_i-z_j|^{2\bmq_i\cdot\bmq_j} = |z_{12}z_{34}|^{-2\Delta_n^{(3)}}.
	\label{eq:models-fixed-dirac-factorization}
\end{equation}
At $n=2$, this lift reproduces the charge construction in Ref.~\cite{Harper:2024ker}, up to Fourier-mode relabelling.
The centered charges above give $\calR_{n,\rmD,\delta_*}^{(3)}=1$ for every integer $n\geq2$ in the specified lift.

\paragraph{The global spin line.}
Use the non-special line $L_*=\calO(D_*)$ of Eq.~\eqref{eq:assembly-reference-divisor}, written in the physical roots $u_0,v_0$ of Eq.~\eqref{eq:disjoint-physical-frame}, with
\begin{equation}
	s_* = (u_0v_0)^{-(n-1)/2}\sqrt{\dd{z}},\qquad
	u_0^{k_1}v_0^{k_2}s_* = u_0^{k_1-(n-1)/2}v_0^{k_2-(n-1)/2}\sqrt{\dd{z}}.
	\label{eq:models-lift-spin-line}
\end{equation}
Its character components have the phases of the chosen lifts.
Recombining their plane propagators gives the global kernel
\begin{equation}
	S_*(P,Q) = \frac{s_*(P)s_*(Q)}{n^2(z_P-z_Q)}
	\left(\sum_{k_1=0}^{n-1}u_{0,P}^{k_1}u_{0,Q}^{n-1-k_1}\right)
	\left(\sum_{k_2=0}^{n-1}v_{0,P}^{k_2}v_{0,Q}^{n-1-k_2}\right).
	\label{eq:models-fixed-dirac-kernel}
\end{equation}
The diagonal residue is one.
At distinct points with equal base coordinate, at least one root ratio is $\zeta_n^a\ne1$, and its sum $\sum_{k=0}^{n-1}\zeta_n^{ak}=0$ cancels the apparent pole.
The local half-differential powers in the first argument at $0,x,1$ are $k_1,n-1-k_1,k_2$, respectively; those in the second argument are the complementary powers $n-1-k_1,k_1,n-1-k_2$.
At infinity the denominator gives $n-1-k_2$ in the first argument and $k_2$ in the second.
All are nonnegative, so the kernel extends across the regular caps.
Non-speciality gives uniqueness and hence identifies the global line, whose characteristic is the $\delta_*$ determined in Eq.~\eqref{eq:assembly-reference-characteristic}.

\paragraph{The common scalar normalization.}
Use identical smooth deck-invariant caps in the coordinates $w_i^n=z-z_i$ and $w_\infty^n=z^{-1}$, with the pullback flat metric outside the caps.
For a metric $\dd{s}^2=\rho^2|\dd{w}|^2$, the half-differential norm is $\|\psi\|^2=\int|\psi(w)|^2\rho(w)\dd[2]{w}$.
Thus the cover density $\rho_{\Sigma,\epsilon}$ gives sheet density $\rho_{B,\epsilon}=\rho_{\Sigma,\epsilon}/|\dd{z}/\dd{w}|$.
Define the half-differential pushforward by
\begin{equation}
	(U_\epsilon\psi)_{\bmr}(z) = \left(\frac{\dd{w_{\bmr}}}{\dd{z}}\right)^{1/2} \psi(w_{\bmr}(z)).
	\label{eq:models-regulated-spin-map}
\end{equation}
This map is unitary for the spinor norms with one power of the respective conformal density.
Coordinate covariance intertwines the cover Dirac operator $D_{\Sigma,\epsilon}$ with its direct image $D_{B,\epsilon}$.
Taking the base domain to be the image of smooth cover spinors completed in graph norm therefore identifies the regular cap extensions as well as the operators.
The unitary replica Fourier transform preserves this domain.
Writing $D_{B,\epsilon}^{\bmk}$ for its character restrictions gives, at finite cap size
\begin{equation}
	\det_\zeta|D_{\Sigma,\epsilon}| = \prod_{k_1,k_2=0}^{n-1}\det_\zeta|D_{B,\epsilon}^{\bmk}|.
	\label{eq:models-regulated-spin-determinant}
\end{equation}
These are nonchiral absolute determinants; $H^0(\Sigma,L_*)=0$ excludes zero modes.
Dividing by $n$ copies of each ordinary cyclic two-point amplitude cancels the common local cap factors.
The cap limit is taken at fixed $x$, before the identity cusp.

The spin-structure dependence of the chiral Dirac determinant is expressed through theta functions~\cite{Alvarez-Gaume:1986rcs}; chiral bosonization gives the corresponding determinant formulas~\cite{Verlinde:1986kw}.
The spin-$1/2$ relation to the scalar determinant and theta norm is also given in Ref.~\cite[Eq.~(4.17)]{Alvarez-Gaume:1987wwg}.
These formulas determine the spin dependence on a common smooth metric.

For a nonspecial line bundle of degree $g_d-1$, the bosonization formula expresses the spin-$1/2$ determinant through a theta norm; see Eq.~(1.1) and Corollary~1.1 of Ref.~\cite{Wentworth:2010ic}.
The bosonization hypotheses hold because $\deg L_*=g_d-1$ and $H^0(\Sigma,L_*)=0$, while the Riemann--Roch theorem gives $H^1(\Sigma,L_*)=0$.
That formula uses the Arakelov metric and admissible bundle metrics.
Smooth changes to the common cap metric are governed by the Polyakov--Alvarez formula~\cite[Theorem~2.2 and Eq.~(2.18)]{Wentworth:2010ic}, with its fixed-basis cohomology norms retained.
For the induced spin metrics, flat changes of spin structure leave the local curvature data unchanged, so this metric anomaly is common to the sectors.
The sectorwise scalar--fermion determinant relation gives $|\theta[\delta_*](0\mid\Omega)|^2$ times the same $c=1$ scalar factor as the boson~\cite{Headrick:2012fk,Mukhi:2017rex}.
The scalar amplitude in Eq.~\eqref{eq:assembly-hypergeometric-ratio} supplies the full $x$ dependence.
The unit two-point limit fixes only the remaining $x$-independent normalization, yielding
\begin{equation}
	1 = \frac{(1-x)^{-g_d/4}}{\scrA_n(x)} |\theta[\delta_*](0\mid\Omega_n^{(3)}(x))|^2.
	\label{eq:models-fixed-dirac-physical-thomae}
\end{equation}

Taking the positive Majorana square root then fixes the common denominator $|\theta[\delta_*]|$ in the Ising sum.

For comparison, the ordinary two-interval Dirac sewing selects zero characteristic~\cite{Headrick:2012fk}; in the present conventions
\begin{subequations}
	\label{eq:models-fixed-dirac}
	\begin{align}
		\calF_{n,\rmD,0}^{(2)}(x) &= 1,&
		\calR_{n,\rmD,0}^{(2)}(x) &= (1-x)^{-(n-1/n)/6},
		\label{eq:models-fixed-dirac-bipartite}\\
		\calR_{n,\rmD,\delta_*}^{(3)}(x) &= 1,&
		\calF_{n,\rmD,\delta_*}^{(3)}(x) &= (1-x)^{(n^2-1)/12}.
		\label{eq:models-fixed-dirac-tripartite}
	\end{align}
\end{subequations}

The tripartite sector is auxiliary. Its fused charge is $\bmb-\bma$, of dimension $2\Delta_n^{(3)}$, so charge conservation excludes the regular adjacent ground twist of dimension $\Delta_n^{(3)}$.
It therefore has no nonzero leading adjacent residue.
At genus one it gives the parity-inserted trace $|\vartheta_4/\eta|^2$, while the modular-invariant observable sums over spin structures.

\section{Evaluation of lattice replica moments}\label{sec:lattice-numerics}
The lattice comparisons use normalized finite-state replica moments, but their evaluation depends on the state representation.
For the XXZ chain, TTN isometries reduce the spin-site permutation to a boundary contraction.
For the XX and Ising chains, signed Gaussian overlaps reproduce that spin permutation in the fermionic representation.
For the harmonic chain, Gaussian coordinate integrals give determinants whose common zero mode fixes the mass dependence.

\subsection{TTN contractions}\label{sec:lattice-methods-ttn}
The periodic XXZ states have $J=1$, zero field and $S^z=0$ in Eq.~\eqref{eq:ward-xxz-hamiltonian}.
Their charge-conserving binary trees pair consecutive sites, then neighboring subtrees, in the physical order $1,\ldots,L$~\cite{Shi:2005ctp,Singh:2011uone}.
Optimization uses TenNetLib.jl~\cite{Chanda:TenNetLib}, ITensors.jl and ITensorMPS.jl~\cite{Fishman:2020gel}, with adaptive-gauge TTN methods~\cite{Gerster:2014spa,Silvi:2017srb}.
The bond caps are $\chi=16$ at $L=8$, $\chi=256$ at $L=16,32,64,128$, and $\chi=320$ at $L=256$.

For the measured intervals, each party $A,B$ is an aligned subtree.
Let $V_A$ and $V_B$ be the isometries from the two virtual boundary spaces to their physical spin spaces, and compress $O$ to an orthonormal support with isometry $V_O$.
A normalized TTN state then gives
\begin{equation}
	\begin{aligned}
		\ket{\psi_\chi}&=(V_A\otimes V_B\otimes V_O) \sum_{a,b,o}T_{abo}\ket{a,b,o},\\
		\rho_{ab,a'b'}&=\sum_oT_{abo}\overline{T_{a'b'o}},\qquad \Tr\rho=1.
	\end{aligned}
	\label{eq:lattice-methods-boundary-density}
\end{equation}
Thus $\rho_{AB}^{\rm spin}=(V_A\otimes V_B)\rho(V_A\otimes V_B)^\dagger$.
A disconnected $O$ may retain several environment legs, jointly indexed by $o$.
The same isometry acts on every replica and intertwines its permutations.
Cancelling it with its adjoint leaves the exact boundary contraction for that checkpoint.
Writing $(B_{aa'})_{bb'}=\rho_{ab,a'b'}$, summing one replica direction gives
\begin{equation}
	(\sfT_n)_{\bma,\bma'}=\Tr_B(B_{a_1a'_1}\cdots B_{a_na'_n}),\qquad
	\widehat Z_n^{(3)}=\Tr\sfT_n^n,\qquad \bma=(a_1,\ldots,a_n).
	\label{eq:ttn-cyclic-transfer}
\end{equation}
Both replica axes use forward cyclic shifts, so reversing both leaves the contraction unchanged.

\subsection{Signed Gaussian kernels}\label{sec:lattice-methods-gaussian}
For the normalized pure states of the XX and Ising chains in the even fermion-parity sector, fix the Jordan--Wigner order $1,\ldots,L$, with $c_j=(\prod_{k<j}\sigma_k^z)\sigma_j^+$, so that the ordered occupation basis agrees with the spin basis.
For integer $n\geq2$ and $N=n^2$ copies, order $\bmr=(r_1,r_2)\in\bbZ_n^2$ by $1+r_1+nr_2$, then order physical sites within each copy.
There are $M=NL$ modes.
The source-to-target permutation $\pi$ shifts $A$ by $(1,0)$ and $B$ by $(0,1)$, and fixes $O$.
Exterior second quantization $\Gamma(W)$ fixes the vacuum and sends $c_j^\dagger$ to $\sum_i c_i^\dagger W_{ij}$.
For a source occupation $\bm\nu\in\bbF_2^M$, its reordering sign is $(-1)^{Q_\pi(\bm\nu)}$, where
\begin{equation}
	Q_\pi(\bm\nu)=\sum_{\substack{i<j\\\pi(i)>\pi(j)}}\nu_i\nu_j\pmod2,\qquad
	(W_S)_{\pi(j),j}=(-1)^{\mathbf1_{j\in S}},\qquad S\subseteq\{1,\ldots,M\}.
	\label{eq:lattice-methods-signed-map}
\end{equation}
All other entries of $W_S$ vanish.
Resolving the reordering sign on the full occupation cube gives the finite spin moment
\begin{align}
	w_S&=2^{-M}\sum_{\bm\nu\in\bbF_2^M} (-1)^{Q_\pi(\bm\nu)+\sum_{j\in S}\nu_j},\notag\\
	\widehat Z_n^{(3)}&=\sum_{S\subseteq\{1,\ldots,M\}}w_SK_S,\qquad
	K_S=\bra{\psi}^{\otimes N}\Gamma(W_S)\ket{\psi}^{\otimes N}.
	\label{eq:lattice-methods-full-mode-sum}
\end{align}
Indeed, Fourier inversion reconstructs $(-1)^{Q_\pi}$, which cancels the Fock reordering sign and leaves the ordinary spin permutation.
For disjoint intervals, $\pi$ uses the actual order $A|O_1|B|O_2$: inversions include the gap's Jordan--Wigner sign~\cite{Fagotti:2010yr,Coser:2015dvp}.
Replacing the spin reduced state by a restricted fermion covariance would omit this sign.

For interval permutations, $Q_\pi$ depends only on the occupation parities of the connected interval blocks in each copy.
Imposing the even-parity constraints of the replicated bra and ket and regrouping Eq.~\eqref{eq:lattice-methods-full-mode-sum} gives the constrained parity expansion used in Section~\ref{sec:spin-state-contractions}.
Thus $C_{n,L}^{(3)}=P_+(L)-P_-(L)$ where $P_+(L)$ and $P_-(L)$ collect its positive- and negative-coefficient terms, respectively.

\paragraph{The occupied-state determinant.}
The pure XX state fills the antiperiodic momenta $k=2\pi(m+1/2)/L$, $m=0,\ldots,L-1$, with $\cos k>0$, at $L\equiv0\pmod4$.
Its orthonormal occupied frame has entries $V_{jk}=L^{-1/2}e^{-ikj}$ and $L/2$ columns.
Set $V_{\rm rep}=I_N\otimes V$ in the same copied-mode order.
The overlap of its occupied wedges is
\begin{equation}
	K_S=\det(V_{\rm rep}^\dagger W_S V_{\rm rep}).
	\label{eq:lattice-methods-occupied-kernel}
\end{equation}
The determinant has size $NL/2$; it retains the full complex phase.

\paragraph{The paired-state Pfaffian.}
The even critical Ising state on the finite antiperiodic grid has a nonzero vacuum amplitude and admits the unblocked Thouless chart
\begin{equation}
	\ket{\psi} = \frac{\ket{F}}{\sqrt{\braket{F}}},\qquad
	\ket{F} = \exp\!\left(\frac12\sum_{jl}c_j^\dagger F_{jl}c_l^\dagger\right)\ket{0},
	\qquad F^{\top}=-F.
	\label{eq:lattice-methods-thouless-ket}
\end{equation}
In the physical-site order it is fixed by
\begin{equation}
	F_{jl}=\frac1L\sum_{k\in\calK_+}e^{-ik(j-l)} \frac{-i\sin k}{E_k+\xi_k},\qquad
	\xi_k=1-\cos k,\quad E_k=\sqrt{\xi_k^2+\sin^2 k},
	\label{eq:lattice-methods-ising-pairing}
\end{equation}
where $\calK_+$ is the full antiperiodic grid above; $E_k+\xi_k>0$ at every such momentum.
With $F_{\rm rep}=I_N\otimes F$, the normalized signed overlap is~\cite{Robledo:2009yd}
\begin{equation}
	K_S=
	\frac{\operatorname{pf}\begin{pmatrix}
		F_{\rm rep}^\dagger&-I_M\\ I_M&W_SF_{\rm rep}W_S^{\top}
	\end{pmatrix}}
	{\operatorname{pf}\begin{pmatrix}
		F_{\rm rep}^\dagger&-I_M\\ I_M&F_{\rm rep}
	\end{pmatrix}}.
	\label{eq:lattice-methods-normalized-pfaffian}
\end{equation}
Both skew matrices have size $2M$, with bra modes followed by ket modes in the fixed order.
The common ordering sign $(-1)^{M(M+1)/2}$ cancels between overlap and normalization.
The augmented formula allows singular $F$ within this unblocked chart; a blocked state needs a different overlap.
An unsigned determinant square root cannot fix the phase~\cite{Klich:2014paired}.
The real logarithm is taken after their signed sum has been
assembled.

\subsection{Harmonic-chain replica integrals}\label{sec:lattice-harmonic-moments}
For the harmonic chain, ordinary and tripartite replica moments reduce to Gaussian determinants, which we evaluate in replica-character blocks.

Take the periodic chain in Eq.~\eqref{eq:numerical-harmonic-chain}, with $L\geq3$, $m>0$ and three nonempty parties $A,B,O$.
The complement $O$ may be disconnected.
The normalized ground-state wavefunction and its real symmetric width are
\begin{subequations}
	\label{eq:lattice-harmonic-ground-state}
	\begin{align}
		\Psi_m(\bmq) &= \left(\frac{\det W_m}{\pi^L}\right)^{1/4} e^{-\bmq^\top W_m\bmq/2},
		\label{eq:lattice-harmonic-wavefunction}\\
		(W_m)_{jl} &= \frac1L\sum_{k=0}^{L-1} \sqrt{m^2+4\sin^2(\pi k/L)}\cos\frac{2\pi k(j-l)}L.
		\label{eq:lattice-harmonic-circulant}
	\end{align}
\end{subequations}
Its constant spatial mode has eigenvalue $m$, so the normalized state retains the zero coordinate.

Order coordinates first by replica and then by site, and let $U$ be the orthogonal pullback implementing the regional permutations.
For ordinary sewing, $N=n$ and $U$ shifts the replica label on $R$ while fixing $R^c$.
For tripartite sewing, $N=n^2$ and $(U\bmq)_{\bmr,j}=q_{\bmr-\bme_1,j}$ on $A$, $q_{\bmr-\bme_2,j}$ on $B$ and $q_{\bmr,j}$ on $O$.
Combining the widths of the two wavefunctions in their real Gaussian overlap gives~\cite{Harper:2025uui}
\begin{subequations}
	\label{eq:lattice-harmonic-full-integral}
	\begin{align}
		\widehat Z_N[U] &= \frac{(\det W_m)^{N/2}}{\sqrt{\det M_N}},
		\label{eq:lattice-harmonic-full-integral-moment}\\
		M_N &\coloneqq \frac12\bigl(W_N+U^\top W_NU\bigr),
		\label{eq:lattice-harmonic-full-integral-sewn-width}\\
		W_N &\coloneqq I_N\otimes W_m.
		\label{eq:lattice-harmonic-full-integral-replica-width}
	\end{align}
\end{subequations}
Both widths are positive definite and fix the positive square root.

Use the Fourier convention $\widetilde Q_{\bmk,j}=N^{-1/2}\sum_{\bmr}e^{2\pi i\bmk\cdot\bmr/n}q_{\bmr,j}$.
It diagonalizes these pullbacks with positive character phases.
Its tripartite blocks are
\begin{subequations}
	\label{eq:lattice-harmonic-character-blocks}
	\begin{align}
		D_{k,\ell}&\coloneqq\diag(e^{2\pi i k/n}|_A,e^{2\pi i\ell/n}|_B,1|_O),
		\label{eq:lattice-harmonic-character-blocks-phase}\\
		M_{k,\ell}(m)&\coloneqq\frac12(W_m+D_{k,\ell}^\dagger W_mD_{k,\ell}), \qquad k,\ell\in\bbZ_n.
		\label{eq:lattice-harmonic-character-blocks-width}
	\end{align}
\end{subequations}

The conjugate character to $\chi=(k,\ell)$ is $\bar\chi=(-k,-\ell)$ modulo $n$, with $M_{\bar\chi}=\overline{M_\chi}$.
They describe one pair of real Fourier coordinates.
Write a positive Hermitian block as $M_\chi=A_\chi+iB_\chi$, with $A_\chi$ real symmetric and $B_\chi$ real antisymmetric, and set $\widetilde{\bmQ}_\chi=(\bmx+i\bmy)/\sqrt2$.
The contribution of the pair is
\begin{subequations}
	\label{eq:lattice-harmonic-real-character-pair}
	\begin{align}
		2\widetilde{\bmQ}_\chi^\dagger M_\chi\widetilde{\bmQ}_\chi
		&=\begin{pmatrix}\bmx^\top&\bmy^\top\end{pmatrix}
		\begin{pmatrix}A_\chi&-B_\chi\\ B_\chi&A_\chi\end{pmatrix}
		\begin{pmatrix}\bmx\\\bmy\end{pmatrix},
		\label{eq:lattice-harmonic-real-character-pair-quadratic-form}\\
		\det\begin{pmatrix}A_\chi&-B_\chi\\ B_\chi&A_\chi\end{pmatrix}
		&=(\det M_\chi)^2.
		\label{eq:lattice-harmonic-real-character-pair-determinant}
	\end{align}
\end{subequations}
The real integral therefore gives $(\det M_\chi)^{-1}$, the product of the two conjugate factors.
A self-conjugate character satisfies $2k=2\ell=0$ modulo $n$ and has multiplicity one.
There is one such character for odd $n$ and four for even $n$, including the identity in each case.
The identity block is $W_m$, giving
\begin{equation}
	\log\widehat Z_n^{(3)} = \frac12\sum_{(k,\ell)\ne(0,0)} [\log\det W_m-\log\det M_{k,\ell}(m)].
	\label{eq:lattice-harmonic-q3-characters}
\end{equation}

For an ordinary moment, $D_{k,R}$ is $e^{2\pi ik/n}$ on $R$ and one on its complement.
Then
\begin{subequations}
	\label{eq:lattice-harmonic-q2-characters}
	\begin{align}
		M_{k,R}(m) &\coloneqq \frac12(W_m+D_{k,R}^\dagger W_mD_{k,R}),
		\label{eq:lattice-harmonic-q2-characters-width}\\
		\log\widehat Z_n^{(2)}(R) &= \frac12\sum_{k=1}^{n-1} [\log\det W_m-\log\det M_{k,R}(m)].
		\label{eq:lattice-harmonic-q2-characters-moment}
	\end{align}
\end{subequations}

The ordinary moments also follow from the Gaussian covariance construction~\cite{Peschel:2002yqj,Eisler:2009vye}, with $X_R=(W_m^{-1})_{RR}/2$ and $P_R=(W_m)_{RR}/2$, where the full width is inverted before restriction.
Since $X_R$ is positive definite, $X_RP_R$ is similar to the positive Hermitian matrix $X_R^{1/2}P_RX_R^{1/2}$.
These restricted covariances have symplectic eigenvalues $\nu_a=\sqrt{\lambda_a(X_RP_R)}\geq1/2$.
The product of the corresponding thermal oscillator moments gives
\begin{equation}
	\widehat Z_n^{(2)}(R) = \prod_{a=1}^{|R|} \left[(\nu_a+1/2)^n-(\nu_a-1/2)^n\right]^{-1}.
	\label{eq:lattice-harmonic-q2-covariance}
\end{equation}

The finite-state genuine subtraction is
\begin{equation}
	\GM_{n,L}^{(3)}(m) = -\frac{\log\widehat Z_n^{(3)}}{n(n-1)} + \frac1{2(n-1)}\sum_{R=A,B,O}\log\widehat Z_n^{(2)}(R).
	\label{eq:lattice-harmonic-GM}
\end{equation}
At $n=2$, the nonidentity characters flip $A$, $B$ and $A\cup B$.
The last flip differs from that on $O$ by an overall sign, which cancels in $D^\dagger W_mD$.
The three tripartite blocks are therefore the three ordinary purity blocks, and
\begin{equation}
	\widehat Z_2^{(3)} = \widehat Z_2^{(2)}(A)\widehat Z_2^{(2)}(B)\widehat Z_2^{(2)}(O), \qquad \GM_{2,L}^{(3)}(m)=0.
	\label{eq:lattice-harmonic-n2-identity}
\end{equation}
The vanishing excess in the harmonic chain was already found numerically in Ref.~\cite{Harper:2025uui}.
The argument above proves the identity for every real positive pure Gaussian width and every three-party spatial partition.
It is the specialization of the general pure bosonic Gaussian factorization in Ref.~\cite{Camargo:2026xrd}.

\subsection{The common zero mode in the mass limit}\label{sec:lattice-harmonic-mass-limit}

The normalized massive state becomes singular as its uniform oscillator softens.
Isolating the zero eigenvalues of the original and sewn widths determines the logarithmic mass term and leaves a finite expression in nonsingular replica blocks.

Take $m\to0^+$ at fixed $L$ with every party nonempty.
Let $u_0\coloneqq L^{-1/2}(1,\ldots,1)^\top$ and $P_0\coloneqq u_0u_0^\top$.
The massless width has kernel spanned by $u_0$, and its nonzero frequencies give
\begin{equation}
	W_m=W_0+mP_0+\frac{m^2}{2}W_0^++O_L(m^4),
	\label{eq:lattice-harmonic-width-expansion}
\end{equation}
where $W_0^+$ is the Moore--Penrose inverse.
Only the zero frequency contributes linearly.
The expansion is at fixed size because the first nonzero frequency vanishes as $L$ grows.

Positivity makes the kernel of the sewn width the intersection
\begin{equation}
	\ker M_{N,0} = \ker(I_N\otimes W_0) \cap U^\top\ker(I_N\otimes W_0).
	\label{eq:lattice-harmonic-common-kernel}
\end{equation}
A vector in the first kernel is spatially constant on each replica.
In the tripartite case its constants must satisfy $\alpha_{\bmr-\bme_1}=\alpha_{\bmr-\bme_2}=\alpha_{\bmr}$ because all three parties are nonempty.
The two translations generate the replica group, so every constant agrees.
Ordinary sewing similarly imposes $\alpha_{r-1}=\alpha_r$.
Thus both sewings have exactly one common zero mode,
\begin{equation}
	\ker M_{N,0} = \Span\{z_0\}, \qquad z_0 = N^{-1/2}(1,\ldots,1)^\top\otimes u_0, \qquad \rank M_{N,0}=NL-1.
	\label{eq:lattice-harmonic-rank}
\end{equation}

The vector $z_0$ remains an exact eigenvector of $M_N(m)$ with eigenvalue $m$.
The linear perturbation $[I_N\otimes P_0+U^\top(I_N\otimes P_0)U]/2$ can also act on the nonzero sewn subspace, producing a relative $O_L(m)$ determinant correction even though the original nonzero frequencies change only at order $m^2$.
Writing $\det{}'$ for the product of the positive massless eigenvalues gives
\begin{subequations}
	\label{eq:lattice-harmonic-pseudodet-asymptotics}
	\begin{align}
		\det W_m&=m\det{}'W_0[1+O_L(m^2)],
		\label{eq:lattice-harmonic-pseudodet-width}\\
		\det M_N(m)&=m\det{}'M_{N,0}[1+O_L(m)],
		\label{eq:lattice-harmonic-pseudodet-sewn-width}
	\end{align}
\end{subequations}
\begin{equation}
	\log\widehat Z_N[U]=\frac{N-1}{2}\log m + \frac N2\log\det{}'W_0-\frac12\log\det{}'M_{N,0} + O_L(m).
	\label{eq:lattice-harmonic-log-mass}
\end{equation}
The copy normalizations contribute $m^{N/2}$ and the common coordinate integral contributes $m^{-1/2}$.
For ordinary bosonic entropies, taking infinite volume before $m\to0$ can instead produce logarithm-of-logarithm behavior~\cite{Yazdi:2016cxn}.
Section~\ref{sec:numerical-harmonic} matches the Gaussian zero-mode profile to the noncompact CFT density.

For any nonidentity character, a null vector would have to lie in both $\ker W_0$ and $D^\dagger\ker W_0$.
Since every party is nonempty, the regional phases are not all equal and the corresponding block is positive definite.
The only singular block is the identity block, whose pseudodeterminant is
\begin{equation}
	\det{}'W_0=\prod_{k=1}^{L-1}2\sin(\pi k/L)=L.
	\label{eq:lattice-harmonic-pseudodet-L}
\end{equation}
Factoring $(z^L-1)/(z-1)$ at $z=1$ proves the last equality.

The mass-independent finite parts of the tripartite and ordinary logarithmic moments are
\begin{subequations}
	\label{eq:lattice-harmonic-log-constants}
	\begin{align}
		c_{n^2}&=\frac12\left[(n^2-1)\log L-
		\sum_{(k,\ell)\ne(0,0)}\log\det M_{k,\ell}(0)\right],
		\label{eq:lattice-harmonic-log-constants-tripartite}\\
		c_{n,R}&=\frac12\left[(n-1)\log L-
		\sum_{k=1}^{n-1}\log\det M_{k,R}(0)\right].
		\label{eq:lattice-harmonic-log-constants-ordinary}
	\end{align}
\end{subequations}

Substitution into Eq.~\eqref{eq:lattice-harmonic-GM} fixes
\begin{subequations}
	\label{eq:lattice-harmonic-mass-coefficients}
	\begin{align}
		\GM_{n,L}^{(3)}(m)&=a_n\log m+b_n(L)+O_L(m),
		\label{eq:lattice-harmonic-mass-coefficients-expansion}\\
		a_n&=-\frac{n^2-1}{2n(n-1)}+\frac34=\frac{n-2}{4n},
		\label{eq:lattice-harmonic-mass-coefficients-logarithmic}\\
		b_n(L)&=-\frac{c_{n^2}}{n(n-1)} + \frac1{2(n-1)}\sum_{R=A,B,O}c_{n,R}.
		\label{eq:lattice-harmonic-mass-coefficients-finite}
	\end{align}
\end{subequations}
The explicit $\log L$ terms combine to $a_n\log L$.
The finite part plotted in the main text is consequently
\begin{align}
	B_{n,L} &= \lim_{m\to0^+}[\GM_{n,L}^{(3)}(m)-a_n\log(mL)] = b_n(L)-a_n\log L\notag\\
	&= \frac1{2n(n-1)}\sum_{(k,\ell)\ne(0,0)}\log\det M_{k,\ell}(0) - \frac1{4(n-1)}\sum_{R=A,B,O}\sum_{k=1}^{n-1}\log\det M_{k,R}(0).
	\label{eq:lattice-harmonic-direct-finitepart}
\end{align}
Equation~\eqref{eq:lattice-harmonic-n2-identity} gives $B_{2,L}=0$.

% --------------------------------------------------
\bibliographystyle{JHEP}
% \nocite{*}
\bibliography{ref}

\end{document}